\documentclass[structabstract]{aa}  
\usepackage{amsmath, amssymb, graphicx, units, latexsym, tabularx, url,
  epsfig, longtable,subfigure}
\usepackage{threeparttable, array}
\usepackage[T1]{fontenc}
\usepackage[latin1]{inputenc}
\fontfamily{pplx}
\usepackage{txfonts}
\usepackage{microtype}
\DisableLigatures{encoding = *, family = * }
\newcolumntype{Y}{>{\centering\arraybackslash}X}
\begin{document}
\title{Testing the Low-Mass End of X-Ray Scaling Relations\\ with a Sample of \emph{Chandra} Galaxy Groups}
\author{H.~J.~Eckmiller
\and D.~S.~Hudson
\and T.~H.~Reiprich}
\institute{Argelander-Institut f\"ur Astronomie der Universit\"at Bonn, Auf dem H\"ugel 71, D-53121 Bonn, Germany\\
\email{heckmill@astro.uni-bonn.de}
   }
\date{Submitted 16 February 2011; Accepted 25 September 2011}
\abstract
{Well-determined scaling relations between X-ray observables and cluster mass
  are essential for using large cluster samples to constrain fundamental
  cosmological parameters. Scaling relations between cluster masses and
  observables, such as the luminosity-temperature, mass-temperature,
  luminosity-mass relations, have been investigated extensively, however the
  question of whether these relations hold true also for poor clusters and
  groups remains unsettled. Some evidence supports a ``break'' at the low end
  of the group/cluster mass range, possibly caused by the stronger influence
  of non-gravitational physics on low-mass systems.}
{The main goal of this work is to test local scaling relations for the low-mass range in
  order to check whether or not there is a systematic difference
  between clusters and groups, and to thereby extend this method of reliable and
  convenient cluster mass determination for future large samples down to the group regime.}
{We compiled a statistically complete sample of 112 X-ray galaxy groups, 26 of
  which have usable \emph{Chandra} data. Temperature, metallicity, and surface
  brightness profiles were created for these 26 groups, and used to determine
  the main physical quantities and scaling relations. We then compared the
  group properties to those of the \emph{HIFLUGCS} clusters, as well as several other
  group and cluster samples.}
{We present radial profiles for the individual objects and scaling
  relations of the whole sample ($L_{\text{x}}$-$T$, $M$-$T$,
  $L_{\text{x}}$-$M$, $M_{\text{g}}$-$M$, $M$-$Y_{\text{x}}$,
  $L_{\text{x}}$-$Y_{\text{x}}$, $f_{\text{g}}$-$T$). Temperature
  and metallicity profiles behave universally, except for the
    core regions. The $L_{\text{x}}$-$T$, $M$-$T$,
  $L_{\text{x}}$-$M$, $M_{\text{g}}$-$M$, $M$-$Y_{\text{x}}$, and
  $L_{\text{x}}$-$Y_{\text{x}}$ relations of the group sample are
  generally in good agreement with clusters. The $L_{\text{x}}$-$T$
  relation steepens for $T<\unit[3]{keV}$, which could point to a
  larger impact of heating mechanisms on cooler systems. We found a
  significant drop in the gas mass fraction below $\unit[\lesssim
  1]{keV}$, as well as a correlation with radius, which indicates the
  ICM is less dominant in groups compared to clusters and the
    galaxies have a stronger influence on the global properties of the
    system. In all relations the intrinsic scatter for
    groups is larger than for clusters, which appears not to be
    correlated with merger activity but could be due to scatter caused
    by baryonic physics in the group cores. We also demonstrate
    the importance of selection effects.}
{We have found some evidence for a similarity break between groups and
  clusters. However this does not have a strong effect on the scaling relations.}
\titlerunning{Testing X-Ray Scaling Relations with \emph{Chandra} Groups}
\authorrunning{Eckmiller et al.}
\maketitle
\renewcommand{\floatpagefraction}{0.6}
\section{Introduction}\label{sec:intro}
Clusters of galaxies are excellent tools for measuring fundamental
cosmological parameters such as the mean matter density $\Omega_{\text{m}}$ and the
amplitude of primordial density fluctuations, $\sigma_{8}$. These can readily
be constrained by comparing theoretical predictions to the observed cluster
mass function, but only if the masses are well-determined and not affected by
systematic bias.

X-ray emission from the intracluster medium (ICM\footnote{We use ICM
  to denote both intra\emph{cluster} and intra\emph{group} medium, to
  avoid confusion with the intergalactic medium, often abbreviated
  IGM.}) provides one of the main instruments to measure the mass, as
well as a wide range of additional physical gas properties, e.\ g.\
temperature, metallicity, gas density, entropy, luminosity, and also
to investigate the dynamical state. However, for samples of hundreds
of objects and future surveys encompassing up to $100,000$ clusters,
like \emph{eROSITA} (e.\ g.\ Predehl et al.\ 2010), it is convenient
and essential to use reliable mass-observable relations instead of
determining masses individually for each cluster.

The total gravitational mass of a cluster, consisting of both baryonic matter
and dark matter, is tightly correlated with observable quantities such as
temperature and luminosity, and relations between these have been studied
extensively with both observational and computational techniques
(e.\ g.\ Allen et al.\ 2001, Finoguenov et al.\ 2001, Reiprich \& B\"ohringer
2002, Borgani et al.\ 2004, Stanek et al.\ 2006, Nagai et al.\ 2007, Rykoff et
al.\ 2008, Hartley et al.\ 2008, Zhang et al.\ 2008, Lopes et al.\ 2009,
Vikhlinin et al.\ 2009, Ettori et al.\ 2010, Leauthaud et al.\ 2010, Mantz et
al.\ 2010, Plagge et al.\ 2010).

Groups and poor clusters of galaxies are more common than rich
clusters due to the steepness of the cluster mass function, and they contain a
total amount of hot gas comparable to or even larger than that of all rich
clusters combined. However, at the same time they are fainter and cooler, and
thus generally more difficult to detect and distinguish from the background,
especially at higher redshifts. So only recently, with the advent of highly
sensitive, high-resolution X-ray observatories like \emph{Chandra} and
\emph{XMM-Newton}, have groups been studied to a greater extent. Over the last
decade, many independent investigations have found indications that groups
cannot simply be treated as less massive, ``scaled-down'' cluster specimen,
but must be considered as a unique class of objects, for a number of reasons.

First of all, groups are different from clusters simply in that they are
cooler, less massive systems with shallower potential wells. But because
of this they are also expected to be more strongly affected by
non-gravitational mechanisms such as galactic winds or feedback from
supernovae, cosmic rays, and active galactic nuclei (AGN). These
influences are complex and difficult to reproduce in simulations, and
thus require careful cross-checking with observations. These processes
are suspected of systematically increasing intrinsic scatter and
changing global properties of groups.

Also, the matter composition in groups is different from that in
clusters. While in clusters the ICM strongly dominates over the
galactic component, in groups the situation is turned around, and the
combined mass of the member galaxies may even exceed that of the gas
(e.\ g.\ Giodini et al.\ 2009). Accordingly, e.\ g.\ Dell'Antonio et
al.\ (1994) and Mulchaey \& Zabludoff (1998) suggested that two
distinct X-ray emitting components coexist in groups, which have to be
considered separately, one of which is extended and consists of hot
diffuse gas bound to the group potential, and a second component,
which is associated with the halos of member galaxies. Jeltema et al.\
(2008a) also showed that the member galaxies have a higher chance of
retaining their gas halos in group environments because these have
lower densities and gas stripping is not nearly as efficient as it is
in clusters.

In terms of mass and richness, groups naturally stand as an intermediate class between
isolated field galaxies on one hand and rich clusters, typically containing
thousands of galaxies, on the other. In groups the average galaxy velocity
dispersions are much lower than in clusters, close to the relative velocities
of single galaxies, increasing the probability of galaxy-galaxy interactions
and mergers, making groups the most attractive environments to study these
activities.

As the physical emission processes in the ICM are strongly dependent on the
gas temperature, the X-ray emission spectra from low-temperature groups are
typically dominated by line emission, not bremsstrahlung continuum
emission which is dominant in hotter objects with $T\gtrsim\unit[2]{keV}$.

Many investigations have found or predicted a systematic difference
between the physical properties of groups compared to clusters, e.\
g.\ a flatter $L_{x}$-$\sigma$ relation (Mahdavi et al.\ 2000, Xue \&
Wu 2000b), or a steepening of the $L_{x}$-$T$ relation (e.\ g.\
Ponman et al.\ 1996, Xue \& Wu 2000b, Helsdon \& Ponman 2000, Dav\'e
et al.\ 2008). Pope (2009) predicted analytically that AGN feedback
should not affect the $L_{x}$-$\sigma$ relation above
$\sigma\sim\unit[500]{km~s^{-1}}$, corresponding to temperatures
around $\unit[1-2]{keV}$, which could explain a similarity break in
the $L_{x}$-$T$ relation. Sanderson et al.\ (2003) found a break and
gradual steepening in the $M$-$T$ relation measured over a wide range
of objects, spanning from elliptical galaxies to clusters. Finoguenov
et al. (2001) also reported a steeper $M$-$T$ slope than the cluster
relation when groups are included in the fit. Maughan et al.\
  (2011) found an observed steepening of the $L_{x}$-$T$ relation of
  relaxed clusters below $\unit[3.5]{keV}$ and argue it is caused by
  central heating that affects the ICM out to a larger radius in lower
  mass systems. This would also be in agreement with the findings of
  Mittal et al.\ (2011), who suggested that for systems below
  $\unit[2.5]{keV}$ AGN heating becomes a more important influence on
  the $L_{x}$-$T$ relation than ICM cooling, which is dominant in
  hotter clusters.

On the other hand, a number of studies do not support the discrepancy
between groups and clusters, but instead get consistent results for
the whole observed mass range, albeit often with larger scatter for
groups, as for the $L_{\text{x}}$-$\sigma$, $L_{\text{x}}$-$T$,
$M$-$T$, $\sigma$-$T$, $M$-$T$, and $M$-$Y_{\text{x}}$ relations (e.\ g.\
Mulchaey \& Zabludoff 1998, Osmond \& Ponman 2004, Sun et al.\ 2009).

Khosroshahi et al.\ (2007) investigated fossil groups in particular
and found that fossils are hotter and brighter for a given mass than
non-fossil groups, suggesting they follow more closely an extension of
cluster properties.  Similarly, Balogh et al.\ (2010) found a
dichotomy in the X-ray gas properties of low-mass clusters and groups
selected by $\sigma$. This might explain why observations of groups
seem to yield larger scatter and inconsistent results, namely there
could be different types of groups involved. Intracluster magnetic
fields may also affect the scaling relations more strongly for the
low-temperature range, resulting in an effective steepening, as
proposed by Colafrancesco \& Giordano (2007).

Another hint for a possible change in the properties of groups is the mass
fraction $f_{\text{g}}$ of the hot gas component compared to the total
mass. This parameter remains quite constant over the cluster regime, yet
e.\ g.\ Reiprich (2001), Gastaldello et al.\ (2007), and Dai et al.\ (2010)
detected a decrease in $f_{\text{g}}$ towards the low-mass, low-temperature
end. Croston et al.\ (2008) and Pratt et al.\ (2009) suggest that a dependence
of $f_{\text{g}}$ on temperature and mass may be responsible for the
steepening in scaling relations. Sun et al.\ (2009) find a lower gas fraction
as well, but only for the region within a small fiducial radius, so they argue the global
$f_{\text{g}}$ in groups may be decreased by lower gas fractions within the
central region alone. This would be in agreement with the findings of Zhang et
al.\ (2010), who detect a correlation of gas fraction and cluster
radius. However Sanderson et al.\ (2003) report a systematic trend in
$f_{\text{g}}$ with temperature in all \emph{but} the central regions.

The main goal of this work is to investigate whether or not there is
  a ``break'' by testing the main cluster scaling relations for a
  sample of galaxy groups. We used essentially the same reduction and
analysis pipelines that were applied to the \emph{HIFLUGCS} clusters
(Hudson et al. 2010), so the results of both samples are directly
comparable. We investigated the $L_{\text{x}}$-$T$, $M$-$T$,
$L_{\text{x}}$-$M$, $M_{\text{g}}$-$M$, $M$-$Y_{\text{x}}$,
$L_{\text{x}}$-$Y_{\text{x}}$, and $f_{\text{g}}$-$T$ relations, for
both $r_{2500}$ and $r_{500}$, where $Y_{\text{x}}$ is defined as the
product of temperature and gas mass, which has been introduced in
order to reduce scatter and bias from merger activity (Kravtsov et
al.\ 2006), and where the $r_{\Delta}$ is defined as the radius within
which the matter density is $\Delta$ times the critical density of the
Universe and which encloses the total matter $M_{\Delta}$.

Merging events are problematic for cluster scaling relations because
they have a large impact on the global properties of the system,
strongly boosting luminosity and temperature via shock-heating for a
certain time but also often biasing low the measured temperature due
to unthermalized gas, cool clumps and substructure (e.\ g.\ Mathiesen
\& Evrard 2001, Ricker \& Sarazin 2001, Ritchie \& Thomas 2002,
Hartley et al.\ 2008, Yang et al.\ 2009, B\"ohringer et al.\ 2010,
Takizawa et al.\ 2010). Another issue is the fact that X-ray mass
estimates must rely on the fundamental assumptions of hydrostatic
equilibrium and spherical symmetry, which are expected to be broken in
merging systems and may take several Gyr to re-establish (Ritchie \&
Thomas 2002). Tovmassian \& Plionis (2009) even state that groups in
general are not in equilibrium but still in the process of
virialization.

Including only regular, relaxed systems could in principle reduce scatter and
bias in scaling relations, but identifying mergers is not trivial, for
instance in the most extreme case a merger along the line-of-sight will appear
perfectly symmetric in the X-ray emission, so it may probably not be possible
to reliably determine the true dynamical state going by the surface brightness
distribution alone (Nagai et al. 2007a, Jeltema et al.\ 2008b, Ventimiglia et
al.\ 2008, Riemer-S\o rensen et al.\ 2009, Zhang et al. 2009). We did not
exclude any irregular groups, and so did not bias our sample with respect to
morphology. We do however investigate whether excluding unrelaxed objects
changes the scaling relations significantly (see section \ref{sec:compmorph}).

However we took care in the global temperature fits to exclude the central
region of groups with cooling cores that are strongly influenced by
complex baryonic physics not yet fully understood, which, if not
excluded, will introduce a large degree of scatter on the
luminosity-temperature and luminosity-mass relations (Markevitch 1998,
Loken et al.\ 2002, Sanderson et al.\ 2006, Chen et al.\ 2007, Mittal
et al.\ 2009, Pratt et al.\ 2009, Hudson et al.\ 2010).

This paper is organized as follows. We begin with the selection of the group
sample in section \ref{sec:sample}. In section \ref{sec:analysis} we outline
the analysis methods. We present our results in section \ref{sec:results} and
discuss and compare them to other work in section \ref{sec:discussion}. A
summary and our conclusions are given in section \ref{sec:summary}. Radial
profiles, images and notes for individual objects are included in the
appendix. Throughout this paper, we assume a flat $\Lambda$CDM cosmology with
$\Omega_{\text{m}}=0.3$, $\Omega_{\Lambda}=0.7$, and $h=0.7$, where
$H_{0}=\unit[100\,h]{km\,s^{-1}\,Mpc^{-1}}$ is the Hubble constant. X-ray
luminosities $L_{\text{x}}$ and fluxes $f_{\text{x}}$ are given in the range
of $\unit[0.1-2.4]{keV}$ (``\emph{ROSAT} band''). Logarithms are decadic.
%
%
%
\begin{figure*}[!htbp]
	\centering
		\includegraphics[width=0.49\textwidth]{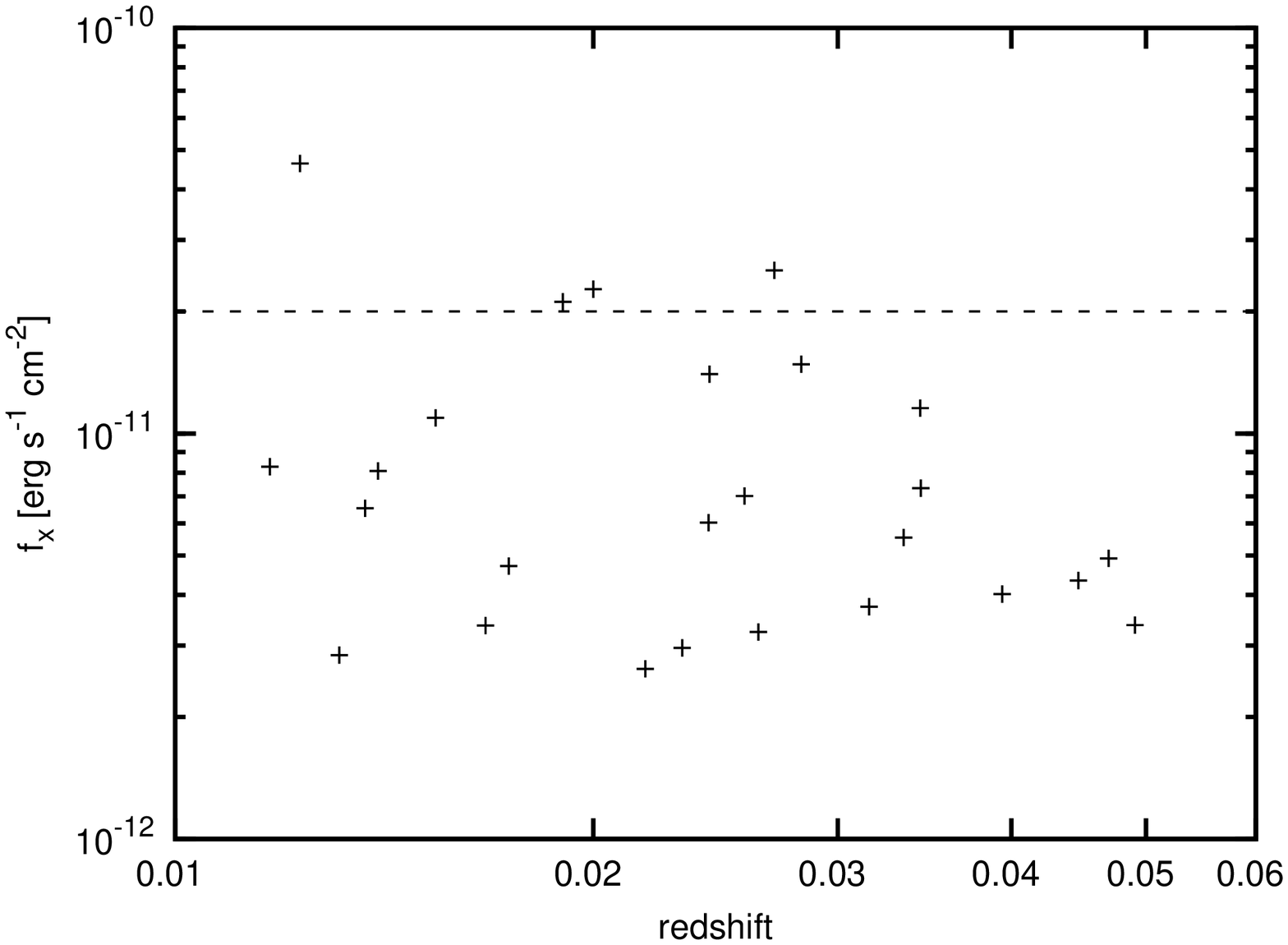}\quad
		\includegraphics[width=0.49\textwidth]{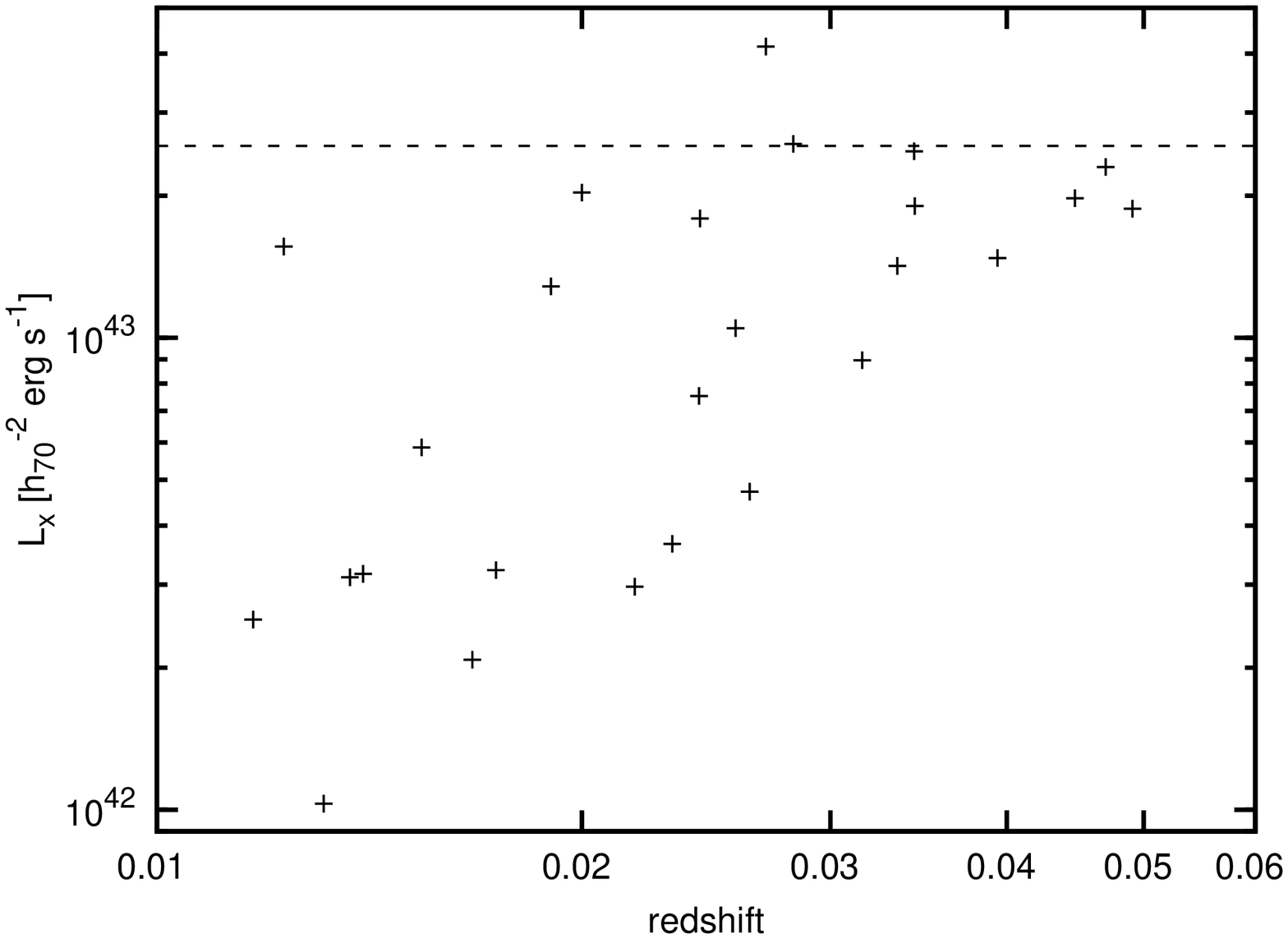}\quad
	\caption{\emph{Left:} X-ray fluxes of the 26 groups plotted against
          redshift. The dashed line is the \emph{HIFLUGCS}
          limit. \emph{Right:} X-ray luminosities plotted against redshift. The
          dashed line is the upper luminosity cut applied here. One object
          appears outside the limit due to conflicting luminosity measurements
          (section \ref{sec:sample}).}
	\label{fig:lz}
\end{figure*}
%
\section{Sample Selection}\label{sec:sample}
The group sample presented in this work was compiled from the X-ray selected,
highly complete \emph{HIFLUGCS}, \emph{NORAS}, and \emph{REFLEX} catalogues
(Reiprich \& B\"ohringer 2002, B\"ohringer et al.\ 2000, and B\"ohringer et
al.\ 2004, respectively). We selected groups by applying an upper limit to
the X-ray luminosity $L_{\text{x}}$, which was determined homogeneously for all three
parent catalogues, plus a lower redshift cut to exclude objects that are too
close to be observed out to sufficiently large radii:
\begin{equation*}
L_{\text{x}}<
5\cdot\unit[10^{43}\,h^{-2}_{50}]{erg\,s^{-1}}=2.55\cdot\unit[10^{43}\,h^{-2}_{70}]{erg\,s^{-1}},
\text{and } z >0.01.
\end{equation*}
This selection yields a statistically complete sample of 112 X-ray
selected galaxy groups. In this work only those groups were
investigated that have been observed with the \emph{Chandra} telescope
with sufficient exposure time ($\unit[\gtrsim 10]{ks}$). Of these 27
objects, listed in Table \ref{tab:sample}, one (IC4296) in the end had
to be excluded from the scaling relations for technical reasons, but
will still appear in some of the tables (for more details see section
\ref{sec:notes} in the appendix).  We do not consider any redshift
evolution for our scaling relations, as this is a local sample (median
redshift $0.025$). The fluxes and luminosities of the remaining 26
groups are plotted against redshift\footnote{From \emph{NED}
  (NASA/IPAC Extragalactic Database),
  \url{http://nedwww.ipac.caltech.edu}} in Fig.\ \ref{fig:lz}. The
effect of flux limits applied to the parent samples appears as a
deficiency of faint objects at high redshifts in the luminosity plot.

A few groups are included in more than one catalogue and thus have several
luminosity measurements. For those objects present in the \emph{HIFLUGCS}
sample, we adopted this luminosity measurement, for the others we preferred
the value obtained with \emph{REFLEX}. In one case (MKW8) the measured values
differ by $\unit[50]{\%}$ between \emph{HIFLUGCS} and \emph{NORAS}, where the
former measurement places the group outside the luminosity cut. We however
kept this object in the sample because the luminosity obtained with
\emph{NORAS} lies well within the cut.

As mentioned in the previous section, we avoided morphology bias by not
performing any morphological selection, however the results of this work may
be affected by ``archival bias'' as we so far have relied on what was
available in the \emph{Chandra} archives.

In future work we will continue to complete this sample, starting with
observational data taken with \emph{XMM-Newton}. For the full sample more
observations will be necessary, since some of
the groups on the selection list have not been observed with either
\emph{XMM-Newton} or \emph{Chandra}.
\begin{table*}
\begin{threeparttable}
\centering
\caption{Properties of the group sample}
\setlength\extrarowheight{2pt}
\begin{tabularx}{\textwidth}{Xcccccccl}\hline\hline
Group Name & RA & DEC & $z$ & $n_{\text{H}}$ & $f_{\text{x}}$ & $L_{\text{x}}$ & $L_{\text{x}}$ error \\ 
 & J2000 & J2000 & & $\unit[10^{20}]{cm^{-2}}$ & $\unit[10^{-12}]{erg\,s^{-1}\,cm^{-2}}$ &
$\unit[10^{44}\,h^{-2}_{70}]{erg\,s^{-1}}$ & \%\\ \hline
A0160  &  01 13 02.287  &  +15 29 46.37  & 0.0447 & 3.98 & 4.34 & 0.198 & 14.1 \\
A1177  &  11 09 44.387  &  +21 45 42.25  & 0.0316 & 5.51* & 3.74 & 0.090 & 14.3 \\ 
ESO552020  &  04 54 50.639  &  -18 06 46.99  & 0.033456 & 4.45 & 5.54 & 0.142 & 11.8 \\
HCG62  &  12 53 05.419  &  -09 12 17.54  & 0.0137 & 3.55 & 6.54 & 0.031 & 13.2 \\ 
HCG97  &  23 47 23.672  &  -02 18 13.70  & 0.0218 & 14.82* & 2.63 & 0.030 & 38.8 \\
IC1262  &  17 33 01.027  &  +43 45 32.85  & 0.034394 & 3.27* & 11.55 & 0.248 & 5.1 \\ 
IC1633  &  01 09 54.500  &  -45 54 22.67  & 0.02425 & 1.91 & 14.01 & 0.179 & 8.0 \\ 
\textit{IC4296} & \textit{13 36 37.460} & \textit{-33 58 37.99} &
\textit{0.012465} & \textit{4.14} & \textit{2.35} & \textit{0.008} &
\textit{16.4} \\
MKW4  &  12 04 26.365  &  +01 53 51.34  & 0.02 & 3.61* & 22.68 & 0.203 & 1.7 \\ 
MKW8  &  14 40 37.939  &  +03 27 50.15  & 0.027 & 2.34 & 25.25 & 0.414 & 8.4 \\ 
NGC326  &  00 58 21.121  &  +26 52 57.91  & 0.047 & 17.30* & 4.92 & 0.230 & 12.9 \\ 
NGC507  &  01 23 38.670  &  +33 15 24.93  & 0.01902 & 5.56 & 21.12 & 0.129 & 1.3 \\ 
NGC533  &  01 25 32.066  &  +01 45 22.28  & 0.017385 & 3.12 & 4.71 & 0.032 & 15.9 \\ 
NGC777  &  02 00 14.235  &  +31 25 24.63  & 0.016728 & 4.72 & 3.36 & 0.021 & 14.0 \\
NGC1132  &  02 52 50.838  &  -01 16 35.64  & 0.024213 & 5.56 & 6.03 & 0.075 & 16.2 \\ 
NGC1550  &  04 19 37.695  &  +02 24 30.46  & 0.0123 & 13.35* & 46.32 & 0.156 & 5.4 \\ 
NGC4325  &  12 23 06.592  &  +10 37 13.26  & 0.0257 & 5.95* & 7.01 & 0.105 & 7.8 \\ 
NGC4936  &  13 04 16.392  &  -30 32 08.23  & 0.0117 & 6.49 & 8.28 & 0.025 & 13.6 \\ 
NGC5129  &  13 24 09.943  &  +13 58 46.11  & 0.023176 & 11.04* & 2.96 & 0.037 & 17.1 \\ 
NGC5419  &  14 03 37.462  &  -33 59 09.24  & 0.014 & 5.16 & 8.08 & 0.032 & 14.4 \\ 
NGC6269  &  16 57 58.730  &  +27 51 22.43  & 0.034424 & 6.80* & 7.33 & 0.190 & 8.2 \\ 
NGC6338  &  17 15 21.347  &  +57 24 43.94  & 0.028236 & 2.29 & 14.81 & 0.257 & 15.4 \\ 
NGC6482  &  17 51 48.948  &  +23 04 21.76  & 0.013129 & 14.72* & 2.84 & 0.010 & 10.9 \\ 
RXCJ1022.0+3830  &  10 22 10.353  &  +38 31 25.57  & 0.0491 & 6.02* & 3.37 & 0.188 & 28.3 \\ 
RXCJ2214.8+1350  &  22 14 46.016  &  +13 50 50.80  & 0.0263 & 4.57 & 3.24 & 0.047 & 18.1 \\ 
S0463  &  04 28 38.623  &  -53 50 46.01  & 0.0394 & 2.40* & 4.02 & 0.148 & 16.5 \\ 
SS2B153  &  10 50 26.346  &  -12 50 44.69  & 0.0154 & 10.16* & 10.93 & 0.059 & 6.8 \\ \hline
\end{tabularx}
\label{tab:sample}
 \begin{tablenotes}
\item[] \emph{Notes:} Coordinates are given for the emission-weighted center. $L_{\text{x}}$ and
  $f_{\text{x}}$ are given in the $\unit[0.1-2.4]{keV}$ band. For
  \emph{REFLEX} and \emph{NORAS} objects we quote the corrected
  fluxes. Galactic $n_{\text{H}}$ values taken from the \emph{LAB} survey, fitted
  values are marked with an asterisk.
\end{tablenotes}
\end{threeparttable}
\end{table*}
%

%
%
%
\section{Data Analysis}\label{sec:analysis}
The observations used are listed in Table \ref{tab:obs} in the
appendix. In two cases, NGC507 and NGC1132, one of two available data
sets was badly flared, so we completely excluded it and relied solely
on the other one. For HCG62 there were four observations available,
however of the first two, one was from an early epoch and the other
had only $\unit[5]{ks}$, so we used the newer data sets, which
combined add up to $\unit[120]{ks}$.

All data sets were reduced using CIAO $3.4$ with CALDB $3.3$\footnote{ Snowden et
al.\ (2008) have shown that temperatures determined with
\emph{Chandra} are overestimated due to a calibration problem, which has been
only corrected in the \emph{Chandra CALDB} as of version 3.5.2. This issue is
however only relevant for temperatures $\gtrsim \unit[5]{keV}$, so we
do not expect it to affect our results, as the hottest object in our sample is
at $\sim \unit[3]{keV}$.}, following the
procedures given in the CIAO Science
Threads\footnote{\url{http://cxc.harvard.edu/ciao3.4/threads/index.html}} and
in M.\ Markevitch's
cookbook\footnote{\url{http://cxc.harvard.edu/contrib/maxim/acisbg/COOKBOOK}}.

In general the data analysis was carried out in the same way as described in
Hudson et al.\ (2006 and 2010), apart from minor modifications described below,
so only a brief outline of the methods will be given here.

Background subtraction was done using the ACIS blank-sky background files,
correcting for possible differences in the Galactic foreground emission
between science and background observations.

The emission peak (EP) and emission-weighted center (EWC) for each group were
determined from adaptively smoothed, background-subtracted, exposure-corrected
images. In most cases the EP and the EWC coincide, though in groups with
irregular morphology they can be separated by several arcminutes
(Table \ref{tab:morphology}). The temperature, metallicity, and
surface brightness profiles were all centered on the EWC, which we consider
more appropriate for studying the outer parts of the group and the overall
morphology, while the EP was used only to estimate the core region to be
excluded from the global temperature determination.
%
\subsection{Global Temperature} \label{globalt} The global
temperatures and metallicities were determined by fitting the whole
observed area to a single spectrum. The temperature may be strongly
biased by a cooler gas component at the center, detectable as a
central temperature drop. In order to minimize this bias, we
investigated temperature profiles centered on the EP to determine how
large a region should be excluded from the fit. The cutoff radii for
individual objects are given in Table \ref{tab:morphology}.

We slightly modified the core cutoff estimation method from the
\emph{HIFLUGCS} analysis (Hudson et al.\ 2010), as for the fainter
systems in our sample there are relatively few source counts and
radial bins. In cases where there were so few counts that the
temperature profile consisted of only 5 or fewer bins, no central
region was excluded at all. The other profiles were fit with a
two-component function, consisting of a powerlaw for the central
region and a constant beyond the cutoff radius, to account for the
flat plateau found in typical profiles at intermediate radii. In cases
where there were one or two bins at large radii significantly lower
than the plateau, they were ignored in this fit, as they are the least
important for the core region. The cutoff radius between central and
flat region was varied iteratively, starting from the center, until
the reduced $\chi^2$ of the combined fit reached a minimum. The outer
radius of the annulus with the best fit statistic was then set as the
cutoff radius. We did not simply use a fixed fraction of the virial
radius as core estimate because there is evidence the cutoff radius
does not scale uniformly with $r_{\text{vir}}$ (Hudson et al.\ 2010).
%
 \subsection{Temperature and Metallicity Prof\hspace{0mm}iles}\label{tprofiles}
For the radial temperature and metallicity profiles, successive annular
regions were created around the EWC. We required that each ring contain the
same number of source counts, $2500$, $5000$ or $10,000$, depending on the
brightness of the object and the depth of the available data. The outermost
annulli with fewer counts were still included in the
profiles if the spectral fit reasonably constrained the temperature.

Temperature $T$ and metallicity $Z$ (relative to the solar abundances
  measured by Anders \& Grevesse 1989) were fit simultaneously for each
annulus, with the hydrogen column density $n_{\text{H}}$ fixed to the
\emph{LAB}\footnote{The Leiden-Argentine-Bonn Galactic HI Survey (Kalberla et
  al.\ 2005),
  \url{http://www.astro.uni-bonn.de/~webaiub/english/} 
\url{tools_labsurvey.php}}
value. In some cases we found the $n_{\text{H}}$ value to be
significantly too low, so we left this parameter free in the fits (marked by
an asterisk in Table \ref{tab:sample}).

In our fit models we use one \emph{apec} component for the ICM, and
another to account for possible over- or undersubtraction of the CXB,
as estimated from the outer regions. In some cases we found it
necessary to include an unfolded powerlaw component as well, to
account for unresolved AGN emission or soft proton flares, especially
in the BI chips (S1 and S3). In cases where the reduced $\chi^2
\gtrsim 1.3$, slight model variations were used to improve the fit. In
some cases we tried an \emph{apec} model with separate abundances for
the most common elements. We found the reduced $\chi^2$ could be
improved slightly but the best fit parameters remained unchanged, so
we chose to keep the regular model. For several objects, the fit to
the innermost annuli was not very good, but could be significantly
improved by fitting a second \emph{apec} component, probably due to
the presence of multi-temperature gas at the center.

For the mass calculation we also determined the temperature gradient
by fitting a powerlaw function to the profile, excluding the same core
region as for the global temperature (see above). We used a simple
powerlaw function, which we found to describe the data reasonably well
(see the plots in section \ref{app:profiles} of the appendix):
\begin{equation}\label{eqn:tprofile}
\log \left( \frac{T}{T_{\text{mid}}} \right)=a \cdot \log \left( \frac{r}{r_{\text{mid}}} \right)+b,
\end{equation}
where $\log r_{\text{mid}}=(\log r_{\text{min}}+\log
r_{\text{max}})/2$ and $T_{\text{mid}}=(\log T_{\text{min}}+\log
T_{\text{max}})/2$.
%
\subsection{Surface Brightness Prof\hspace{0mm}iles} \label{sbps} Surface
brightness profiles (SBPs) were extracted from narrow annular regions
in the energy range of $\unit[0.1-2.0]{keV}$, with $\sim500$ source
counts per radial bin. Each profile was fit with a double
$\beta$-model (e.\ g.\ Cavaliere \& Fusco-Femiano 1976), which is the
superposition of two surface brightness components with respective
core radii $x_{\text{ci}}$, $\beta$ parameters $\beta_{\text{i}}$, and
central surface brightness $\Sigma_{0}= \Sigma_{01}+ \Sigma_{02}$:
\begin{equation}\label{eqndoublesigma}
\Sigma(x) = \Sigma_{01}\left(1+\frac{x^2}{x_{\text{c1}}^2}\right)^{-3\beta_1+\frac{1}{2}}+\Sigma_{02}\left(1+\frac{x^2}{x_{\text{c2}}^2}\right)^{-3\beta_2+\frac{1}{2}},
\end{equation}
and which yields the corresponding density profile
\begin{equation} \label{eqndoublebeta}
  n(r) = \sqrt{n^2_{01} \left( 1+\frac{r^2}{r_{\text{c1}}^2}\right)^{-3\beta_{1}} + n^2_{02} \left( 1+\frac{r^2}{r_{\text{c2}}^2}\right)^{-3\beta_{2}}},
\end{equation}
via these relations:
\begin{equation}
n^2_{01}=\frac{\Sigma_{12}LI_2}{\Sigma_{12}LI_2+LI_1}n^2_0,\quad\text{and}\quad n^2_{02}=\frac{LI_1}{\Sigma_{12}LI_2+LI_1}n^2_0.
\end{equation}
The central electron density, $n_0$, is given by
\begin{equation}
n_0=\left(\frac{10^{14}~4\pi~(\Sigma_{12}~LI_2 + LI_1)~D_{\text{A}}~D_{\text{L}}~\zeta~n}{\Sigma_{12}~LI_2~EI_1 + LI_1~EI_2} \right)^{1/2},
\end{equation}
where $D_{\text{A}}$ and $D_{\text{L}}$ are the angular diameter and luminosity distances, respectively, $\zeta$ is the ratio of electrons to protons, $n$ is the normalization of the \emph{apec} model, $EI_{\text{i}}$ is the emission integral for model i,
\begin{equation}
EI_{\text{i}}=2\pi \int \int^{R}_{0} x~\left(1+ \frac{x^2+l^2}{x^2_{\text{ci}}}  \right)^{-3\beta_{\text{i}}}~\text{d} x~\text{d} l,
\end{equation}
and
\begin{equation}
 LI_{\text{i}}=\int \left(1+\frac{z^2}{x_{\text{ci}}^2}\right)^{-3\beta_{\text{i}}}\text{d}z
\end{equation}
is the line integral along the line of sight (for $i=[1,2]$) and
$\Sigma_{12}$ is the ratio of $\Sigma_{01}$ and
$\Sigma_{02}$. $x_{\text{ci}}$ is the projected length of
$r_{\text{ci}}$.

In one case, NGC777, the data was only sufficient to fit a single
$\beta$-model. Since the slope of the outer $\beta$ component has the
greatest influence on the mass calculation, we put more weight on this
component by fitting the outer part separately and fixing the slope in
the overall fit. Some irregular or asymmetric groups have small humps
close to the center, which are likely caused by an offset between the
emission peak and the emission-weighted center, on which the SBPs are
centered, or by a secondary peak, e.\ g.\ in a merging system. In a
few cases we excluded these spikes when fitting the SBP, which
improved the fit statistic but did not significantly change the final
fit parameters of the $\beta$-model.
%
\subsection{Mass Determination}\label{masses}
The total mass within radius $r$ can be estimated assuming hydrostatic equilibrium and
spherical symmetry, using the density gradient as well as the
temperature gradient:

\begin{equation} \label{eqnHyEq}
M(<r)=-\frac{1}{G} \frac{k_{\text{B}} T r}{\mu m_{\text{p}}} \left(\frac{\text{d} \ln \rho_{\text{g}}}{\text{d} \ln r} + \frac{\text{d} \ln T}{\text{d}\ln r}\right),
\end{equation}
where $G$ is the gravitational constant, $\mu$ is the gas mean
molecular weight ($\mu \approx 0.6$), and $m_{\text{p}}$ is the proton
mass. Plugging in eqns.\ (\ref{eqn:tprofile}) and (\ref{eqndoublebeta}) for the temperature and density gradients, we arrive at:
\begin{equation} \label{eqndoubletotalmass}
 M(<r)=\frac{3r^3}{G} \frac{T(r)}{\mu m_{\text{p}}} \left(\frac{\Sigma_{12} LI_2 \zeta'_1 + LI_1 \zeta'_2}{\Sigma_{12}LI_2 \zeta_1 + LI_1 \zeta_2}-a\right) ,
\end{equation}
where $a$ is the slope of the temperature profile (eqn. \ref{eqn:tprofile}) and
\begin{equation}
 \zeta_{\text{i}}= \left( 1 +\frac{r^2}{r_{\text{ci}}^2}\right)^{-\frac{3\beta_{\text{i}}}{2}}\quad \text{and}\quad \zeta'_{\text{i}}=\frac{\beta_{\text{i}}}{r_{\text{ci}}^2}\left(1+\frac{r^2}{r_{\text{ci}}^2}\right)^{-\frac{3\beta_{\text{i}}}{2}-1}.
\end{equation}

In the double $\beta$-model fits $r_{\text{c1}}$, $r_{\text{c2}}$,
$\beta_1$ and $\beta_2$ are highly degenerate, resulting in
unphysically large errors (often $\sim \unit[100]{\%}$). For this
reason we used a Monte Carlo simulation to estimate the
  uncertainties. With this we created 1000 temperature and surface
  brightness profiles varied according to the statistical errors and
  used the standard deviation of the mass values derived from each
  variation as an estimate of the statistical uncertainty on total
  mass and gas mass.
 
The gas mass within $r$ can be calculated using eqn.\ (\ref{eqndoublebeta}) and
\begin{equation} \label{eqndoublegasmass}
 M_{\text{g}}(<r)= \mu m_{\text{p}} \left( 1+\frac{n_{\text{e}}}{n_{\text{p}}}\right) \int n(r)\, \text{d}V,
\end{equation}
where $n_{\text{e}}$ and $n_{\text{p}}$ are the electron and proton number densities.

The fiducial radii $r_{2500}$ and $r_{500}$ were determined
iteratively by integration of the total mass profiles.

%
\subsection{Scaling Relations}\label{sec:scaling}
Linear regressions for the scaling relations were calculated in
log-log space, using the bisector statistic of the \url{BCES_REGRESS}
code (Akritas \& Bershady 1996), which takes into account the
uncertainties on both the $X$ and $Y$ parameters. The errors were
estimated with $10\,000$ bootstrapping iterations. We used the
following functions for fitting:
\begin{eqnarray}
\log\left(\frac{L_{\text{x}}}{0.5\cdot\unit[10^{44}\,h^{-2}_{70}]{erg\,s^{-1}}}\right)&=&a \cdot \log\left(\frac{T}{\unit[3]{keV}}\right)+b,\\
\log\left(\frac{M_{\text{r}}}{\unit[10^{14}\,h^{-1}_{70}~M_{\odot}]{}}\right)&=&a \cdot \log\left(\frac{T}{\unit[3]{keV}}\right)+b,\\
\log\left(\frac{L_{\text{x}}}{0.5\cdot\unit[10^{44}\,h^{-2}_{70}]{erg\,s^{-1}}}\right)&=&a\cdot\log\left(\frac{M_{\text{r}}}{\unit[10^{14}\,h^{-1}_{70}~M_{\odot}]{}}\right)+b,\\
\log\left(\frac{M_{\text{r}}}{\unit[10^{14}\,h^{-1}_{70}~M_{\odot}]{}}\right)&=&a \cdot \log\left(\frac{Y_{\text{x}}}{\unit[10^{12}\,h^{-5/2}_{70}~M_{\odot}]{}}\right)+b,\\
\log\left(\frac{L_{\text{x}}}{0.5\cdot\unit[10^{44}\,h^{-2}_{70}]{erg\,s^{-1}}}\right)&=&a \cdot \log\left(\frac{Y_{\text{x}}}{\unit[10^{12}\,h^{-5/2}_{70}~M_{\odot}]{}}\right)+b,\\
\log\left(\frac{M_{\text{g,r}}}{\unit[10^{12}\,h^{-5/2}_{70}~M_{\odot}]{}}\right)&=&a\cdot\log\left(\frac{M_{\text{r}}}{\unit[10^{14}\,h^{-1}_{70}~M_{\odot}]{}}\right)+b,\\
\log\left(\frac{f_{\text{g,r}}}{0.1\cdot h^{-3/2_{70}}}\right)&=&a \cdot\log\left(\frac{T}{\unit[3]{keV}}\right)+b,
\end{eqnarray}
where $r$ is $r_{2500}$ or $r_{500}$, respectively.

The total logarithmic scatter is given by
$\sigma^{X}_{\text{tot}}=\langle ( \log X-(\log Y -b)/a)^2 \rangle$ and $\sigma^{Y}_{\text{tot}}=\langle ( \log Y-(a \cdot \log X +b))^2 \rangle$. The intrinsic logarithmic scatter is $\sigma^{X}_{\text{int}}=\left((\sigma^{X}_{\text{tot}})^2-(\sigma^{X}_{\text{stat}})^{2}-a^{2}\cdot(\sigma^{Y}_{\text{stat}})^{2}\right)^{1/2}$ and $\sigma^{Y}_{\text{int}}=\left((\sigma^{Y}_{\text{tot}})^2-(\sigma^{Y}_{\text{stat}})^{2}-a^{-2}\cdot(\sigma^{X}_{\text{stat}})^{2}\right)^{1/2}$, where the statistical errors are determined as $\sigma^{X}_{\text{stat}}=\langle\log_{10}(\text{e})\cdot\Delta X/X\rangle$ and $\sigma^{Y}_{\text{stat}}=\langle \log_{10}(\text{e})\cdot\Delta Y/Y\rangle$, respectively.
%
 \subsection{Morphology Selection} \label{sec:morphology}
\begin{table}[tbp]
  \caption{Morphological classification and core removal. An object is classified as morphology type 2
    (disturbed) if the separation between emission peak and
    emission-weighted center is $>\unit[30]{kpc}$ and/or if there are several
    distinct peaks visible in the X-ray emission. The last column is the low-temperature core excised when fitting the global temperature (see section \ref{globalt}).}
\begin{center}
\setlength\extrarowheight{2pt}
 \begin{tabularx}{\linewidth}{Xrrccc}\hline\hline
Name & \multicolumn{2}{c}{Separation} & Subpeak(s) & Type & Core\\
 & arcmin & kpc & & & arcmin \\  \hline
A0160  & 0.72 & 37.8 & Y & 2 & - \\ 
A1177  & 0.16 & 6.0 & Y & 2 & - \\ 
ESO552020  & 0.43 & 17.4 & N & 1 & - \\ 
HCG62  & 0.18 & 3.0 & N & 1 & 2.00 \\ 
HCG97  & 0.27 & 7.2 & N & 1 & - \\ 
IC1262  & 0.51 & 21.0 & N & 1 & 0.85 \\ 
IC1633  & 1.54 & 46.1 & N & 2 & - \\ 
MKW4  & 0.22 & 5.3 & N & 1 & 0.68 \\ 
MKW8  & 1.27 & 41.3 & N & 2 & - \\ 
NGC326  & 1.19 & 65.8 & N & 2 & - \\ 
NGC507  & 0.08 & 1.9 & N & 1 & 1.41 \\ 
NGC533  & 0.23 & 4.9 & N & 1 & 1.02 \\ 
NGC777  & 0.40 & 8.2 & N & 1 & - \\ 
NGC1132  & 0.27 & 7.8 & N & 1 & - \\ 
NGC1550  & 0.22 & 3.3 & N & 1 & 0.59 \\ 
NGC4325  & 0.04 & 1.3 & N & 1 & 0.67 \\ 
NGC4936  & 0.57 & 8.3 & N & 1 & - \\ 
NGC5129  & 0.16 & 4.4 & N & 1 & - \\ 
NGC5419  & 0.76 & 13.0 & N & 1 & - \\ 
NGC6269  & 0.22 & 9.0 & N & 1 & 1.62 \\ 
NGC6338  & 0.42 & 14.3 & Y & 2 & 1.92 \\ 
NGC6482  & 0.07 & 1.1 & N & 1 & - \\ 
RXCJ1022.0+3830  & 0.11 & 6.2 & N & 1 & - \\ 
RXCJ2214.8+1350  & 0.45 & 14.3 & N & 1 & - \\ 
S0463  & 3.24 & 151.7 & Y & 2 & - \\ 
SS2B153  & 0.08 & 1.5 & N & 1 & - \\  \hline
\end{tabularx}
\end{center}
\label{tab:morphology}
\end{table}
As discussed in the introduction, the dynamical state can have a strong
influence on the global properties of clusters and groups, and also on the
validity of the mass determination, which is based on the assumption of
hydrostatic equilibrium. We are aware that it is hard to reliably
determine the dynamical state from a projected image such as the surface
brightness distribution, but we nevertheless classify our groups
into different morphological categories, to at least get a qualitative
estimate of the influence of merging and substructure.

We sorted the groups into two morphological types (Table
\ref{tab:morphology}): Type 1 means regular or relaxed, a group is
considered irregular or unrelaxed (type 2) if the physical separation
between emission peak and emission-weighted center is
$>\unit[30]{kpc}$, and/or if there is more than one peak in the
central X-ray emission (unless it is clearly not interacting with the
ICM).
%
%
%
\section{Results}\label{sec:results}
%
Derived properties for individual groups are listed in Table
\ref{tab:derived}. The luminosity values used here are all taken from
the input catalogs and have been determined homogeneously from
\emph{ROSAT} observations (except \emph{LoCuSS}). The temperatures for
both the group sample and \emph{HIFLUGCS} were measured more recently
using \emph{Chandra} data. \emph{HIFLUGCS} masses were derived from
\emph{ROSAT} data.
\begin{table*}
\begin{threeparttable}
\centering
\caption{Derived properties of the group sample.}
\setlength\extrarowheight{2pt}
\begin{tabularx}{\textwidth}{lYYYYYY}\hline\hline
Group Name & $k_{\text{B}}~T$ & $Z$ & $r_{2500}$ & $M_{2500}$ & $r_{500}$ & $M_{500}$ \\
 & keV & $Z_{\odot}$ & $\unit[h^{-1}_{70}]{Mpc}$ & $10^{13}~h^{-1}_{70}~M_{\odot}$ & $\unit[h^{-1}_{70}]{Mpc}$ & $10^{13}~h^{-1}_{70}~ M_{\odot}$ \\ \hline
A0160 & $1.77\pm0.05$ & $0.17\pm0.02$ & $0.255\pm0.007$ & $2.35\pm0.19$ & $0.550\pm0.024$ & $4.79\pm0.61$ \\ 
A1177 & $1.61\pm0.04$ & $0.35\pm0.04$ & $0.280\pm0.015$ & $3.12\pm0.50$ & $0.625\pm0.047$ & $7.02\pm1.58$ \\ 
ESO552020 & $1.96\pm0.05$ & $0.39\pm0.04$ & $0.265\pm0.006$ & $2.66\pm0.18$ & $0.580\pm0.025$ & $5.58\pm0.75$ \\ 
HCG62 & $1.31\pm0.01$ & $0.22\pm0.01$ & $0.220\pm0.009$ & $1.54\pm0.17$ & $0.465\pm0.014$ & $2.88\pm0.28$ \\ 
HCG97 & $0.81\pm0.01$ & $0.15\pm0.01$ & $0.225\pm0.006$ & $1.66\pm0.12$ & $0.520\pm0.015$ & $4.03\pm0.30$ \\ 
IC1262 & $1.79\pm0.02$ & $0.24\pm0.01$ & $0.275\pm0.006$ & $2.96\pm0.20$ & $0.660\pm0.013$ & $8.28\pm0.48$ \\ 
IC1633 & $2.99\pm0.08$ & $0.35\pm0.04$ & $0.335\pm0.015$ & $5.35\pm0.74$ & $0.845\pm0.060$ & $17.29\pm3.91$ \\ 
MKW4 & $1.86\pm0.03$ & $0.49\pm0.03$ & $0.320\pm0.012$ & $4.67\pm0.52$ & $0.690\pm0.029$ & $9.48\pm1.18$ \\ 
MKW8 & $2.84\pm0.07$ & $0.41\pm0.04$ & $0.310\pm0.009$ & $4.25\pm0.39$ & $0.695\pm0.037$ & $9.65\pm1.61$ \\ 
NGC326 & $1.67\pm0.04$ & $0.21\pm0.02$ & $0.230\pm0.010$ & $1.72\pm0.23$ & $0.530\pm0.036$ & $4.29\pm0.89$ \\ 
NGC507 & $1.32\pm0.01$ & $0.28\pm0.01$ & $0.205\pm0.004$ & $1.23\pm0.07$ & $0.470\pm0.021$ & $2.98\pm0.41$ \\ 
NGC533 & $1.33\pm0.01$ & $0.28\pm0.02$ & $0.220\pm0.013$ & $1.52\pm0.24$ & $0.480\pm0.032$ & $3.20\pm0.58$ \\ 
NGC777 & $0.73\pm0.01$ & $0.32\pm0.06$ & $0.180\pm0.005$ & $0.83\pm0.05$ & $0.395\pm0.015$ & $1.76\pm0.13$ \\ 
NGC1132 & $1.08\pm0.01$ & $0.31\pm0.02$ & $0.190\pm0.007$ & $0.99\pm0.12$ & $0.440\pm0.023$ & $2.47\pm0.43$ \\ 
NGC1550 & $1.33\pm0.00$ & $0.28\pm0.00$ & $0.210\pm0.005$ & $1.32\pm0.09$ & $0.445\pm0.010$ & $2.52\pm0.19$ \\ 
NGC4325 & $0.98\pm0.01$ & $0.39\pm0.02$ & $0.195\pm0.003$ & $1.05\pm0.04$ & $0.430\pm0.006$ & $2.28\pm0.10$ \\ 
NGC4936 & $0.89\pm0.02$ & $0.17\pm0.02$ & $0.155\pm0.007$ & $0.54\pm0.08$ & $0.355\pm0.026$ & $1.30\pm0.31$ \\ 
NGC5129 & $0.81\pm0.01$ & $0.30\pm0.03$ & $0.210\pm0.009$ & $1.34\pm0.17$ & $0.490\pm0.027$ & $3.36\pm0.54$ \\ 
NGC5419 & $2.09\pm0.04$ & $0.32\pm0.03$ & $0.220\pm0.009$ & $1.52\pm0.18$ & $0.480\pm0.030$ & $3.20\pm0.61$ \\ 
NGC6269 & $1.87\pm0.06$ & $0.26\pm0.03$ & $0.225\pm0.009$ & $1.62\pm0.19$ & $0.570\pm0.025$ & $5.30\pm0.75$ \\ 
NGC6338 & $2.00\pm0.03$ & $0.17\pm0.01$ & $0.295\pm0.005$ & $3.62\pm0.19$ & $0.580\pm0.019$ & $5.61\pm0.54$ \\ 
NGC6482 & $0.62\pm0.01$ & $0.61\pm0.20$ & $0.125\pm0.002$ & $0.29\pm0.02$ & $0.265\pm0.005$ & $0.52\pm0.03$ \\ 
RXCJ1022.0+3830 & $1.74\pm0.04$ & $0.19\pm0.03$ & $0.260\pm0.012$ & $2.55\pm0.35$ & $0.590\pm0.034$ & $5.89\pm0.99$ \\ 
RXCJ2214.8+1350 & $1.34\pm0.01$ & $0.21\pm0.01$ & $0.230\pm0.006$ & $1.74\pm0.18$ & $0.605\pm0.026$ & $6.40\pm1.05$ \\ 
S0463 & $1.97\pm0.06$ & $0.15\pm0.02$ & $0.265\pm0.009$ & $2.71\pm0.27$ & $0.565\pm0.019$ & $5.22\pm0.54$ \\ 
SS2B153 & $0.81\pm0.00$ & $0.51\pm0.03$ & $0.180\pm0.005$ & $0.86\pm0.06$ & $0.400\pm0.014$ & $1.84\pm0.17$ \\ \hline \\ \hline
Group Name & $M_{\text{g,2500}}$ & $Y_{\text{x,2500}}$ & $f_{\text{g,2500}}$ & $M_{\text{g,500}}$ & $Y_{\text{x,500}}$ & $f_{\text{g,500}}$ \\ 
 & $10^{13}~h^{-5/2}_{70}~ M_{\odot}$ & $\unit[10^{13}~h^{-5/2}_{70}~ M_{\odot}]{keV}$ & $h^{-3/2}_{70}$ & $10^{13}~h^{-5/2}_{70}~ M_{\odot}$ & $\unit[10^{13}~h^{-5/2}_{70}~ M_{\odot}]{keV}$ & $h^{-3/2}_{70}$ \\ \hline
A0160 & $0.101\pm0.001$ & $0.179\pm0.004$ & $0.043\pm0.004$ & $0.419\pm0.009$ & $0.742\pm0.019$ & $0.087\pm0.011$ \\ 
A1177 & $0.095\pm0.005$ & $0.153\pm0.009$ & $0.030\pm0.005$ & $0.251\pm0.035$ & $0.405\pm0.069$ & $0.036\pm0.009$ \\ 
ESO552020 & $0.123\pm0.002$ & $0.240\pm0.005$ & $0.046\pm0.003$ & $0.456\pm0.011$ & $0.893\pm0.024$ & $0.082\pm0.011$ \\ 
HCG62 & $0.018\pm0.001$ & $0.023\pm0.002$ & $0.012\pm0.002$ & $0.049\pm0.009$ & $0.065\pm0.013$ & $0.017\pm0.003$ \\ 
HCG97 & $0.030\pm0.002$ & $0.025\pm0.001$ & $0.018\pm0.002$ & $0.069\pm0.007$ & $0.056\pm0.005$ & $0.017\pm0.002$ \\ 
IC1262 & $0.172\pm0.005$ & $0.307\pm0.009$ & $0.058\pm0.004$ & $0.545\pm0.030$ & $0.976\pm0.053$ & $0.066\pm0.005$ \\ 
IC1633 & $0.196\pm0.004$ & $0.586\pm0.014$ & $0.037\pm0.005$ & $0.829\pm0.045$ & $2.476\pm0.149$ & $0.048\pm0.011$ \\ 
MKW4 & $0.069\pm0.004$ & $0.128\pm0.007$ & $0.015\pm0.002$ & $0.128\pm0.003$ & $0.238\pm0.044$ & $0.013\pm0.002$ \\ 
MKW8 & $0.191\pm0.003$ & $0.541\pm0.012$ & $0.045\pm0.004$ & $0.749\pm0.022$ & $2.127\pm0.068$ & $0.078\pm0.013$ \\ 
NGC326 & $0.013\pm0.001$ & $0.021\pm0.001$ & $0.007\pm0.001$ & $0.067\pm0.003$ & $0.112\pm0.006$ & $0.016\pm0.003$ \\ 
NGC507 & $0.068\pm0.001$ & $0.090\pm0.002$ & $0.056\pm0.003$ & $0.291\pm0.007$ & $0.384\pm0.011$ & $0.098\pm0.014$ \\ 
NGC533 & $0.024\pm0.002$ & $0.032\pm0.003$ & $0.016\pm0.003$ & $0.080\pm0.013$ & $0.106\pm0.024$ & $0.025\pm0.006$ \\ 
NGC777 & $0.021\pm0.003$ & $0.015\pm0.002$ & $0.025\pm0.004$ & $0.053\pm0.008$ & $0.038\pm0.006$ & $0.030\pm0.005$ \\ 
NGC1132 & $0.039\pm0.001$ & $0.042\pm0.001$ & $0.040\pm0.005$ & $0.170\pm0.007$ & $0.183\pm0.008$ & $0.069\pm0.012$ \\ 
NGC1550 & $0.054\pm0.001$ & $0.071\pm0.001$ & $0.041\pm0.003$ & $0.172\pm0.007$ & $0.228\pm0.009$ & $0.068\pm0.006$ \\ 
NGC4325 & $0.043\pm0.001$ & $0.042\pm0.001$ & $0.041\pm0.002$ & $0.129\pm0.006$ & $0.126\pm0.006$ & $0.056\pm0.003$ \\ 
NGC4936 & $0.018\pm0.001$ & $0.016\pm0.001$ & $0.033\pm0.005$ & $0.087\pm0.004$ & $0.078\pm0.004$ & $0.067\pm0.016$ \\ 
NGC5129 & $0.020\pm0.001$ & $0.016\pm0.001$ & $0.015\pm0.002$ & $0.048\pm0.005$ & $0.039\pm0.004$ & $0.014\pm0.003$ \\ 
NGC5419 & $0.059\pm0.002$ & $0.123\pm0.004$ & $0.039\pm0.005$ & $0.314\pm0.012$ & $0.656\pm0.024$ & $0.098\pm0.019$ \\ 
NGC6269 & $0.082\pm0.002$ & $0.154\pm0.005$ & $0.051\pm0.006$ & $0.492\pm0.012$ & $0.919\pm0.029$ & $0.093\pm0.013$ \\ 
NGC6338 & $0.147\pm0.002$ & $0.295\pm0.005$ & $0.041\pm0.002$ & $0.410\pm0.011$ & $0.821\pm0.025$ & $0.073\pm0.007$ \\ 
NGC6482 & $0.007\pm0.001$ & $0.005\pm0.001$ & $0.026\pm0.002$ & $0.023\pm0.001$ & $0.014\pm0.001$ & $0.043\pm0.004$ \\ 
RXCJ1022.0+3830 & $0.112\pm0.003$ & $0.195\pm0.007$ & $0.044\pm0.006$ & $0.405\pm0.031$ & $0.706\pm0.071$ & $0.069\pm0.013$ \\ 
RXCJ2214.8+1350 & $0.053\pm0.001$ & $0.070\pm0.002$ & $0.030\pm0.003$ & $0.289\pm0.008$ & $0.386\pm0.011$ & $0.045\pm0.008$ \\ 
S0463 & $0.118\pm0.002$ & $0.232\pm0.006$ & $0.043\pm0.004$ & $0.503\pm0.018$ & $0.990\pm0.038$ & $0.096\pm0.011$ \\ 
SS2B153 & $0.029\pm0.001$ & $0.023\pm0.001$ & $0.033\pm0.002$ & $0.083\pm0.003$ & $0.068\pm0.002$ & $0.045\pm0.004$ \\ \hline
\end{tabularx}
\label{tab:derived}
 \begin{tablenotes}
\item[] \emph{Notes:} The statistical errors on the temperatures of NGC1550 and SS2B153 were smaller than $\unit[0.01]{keV}$.
\end{tablenotes}
\end{threeparttable}
\end{table*}
\begin{figure*}[!ht]
 \centering
  \includegraphics[width=0.31\textwidth]{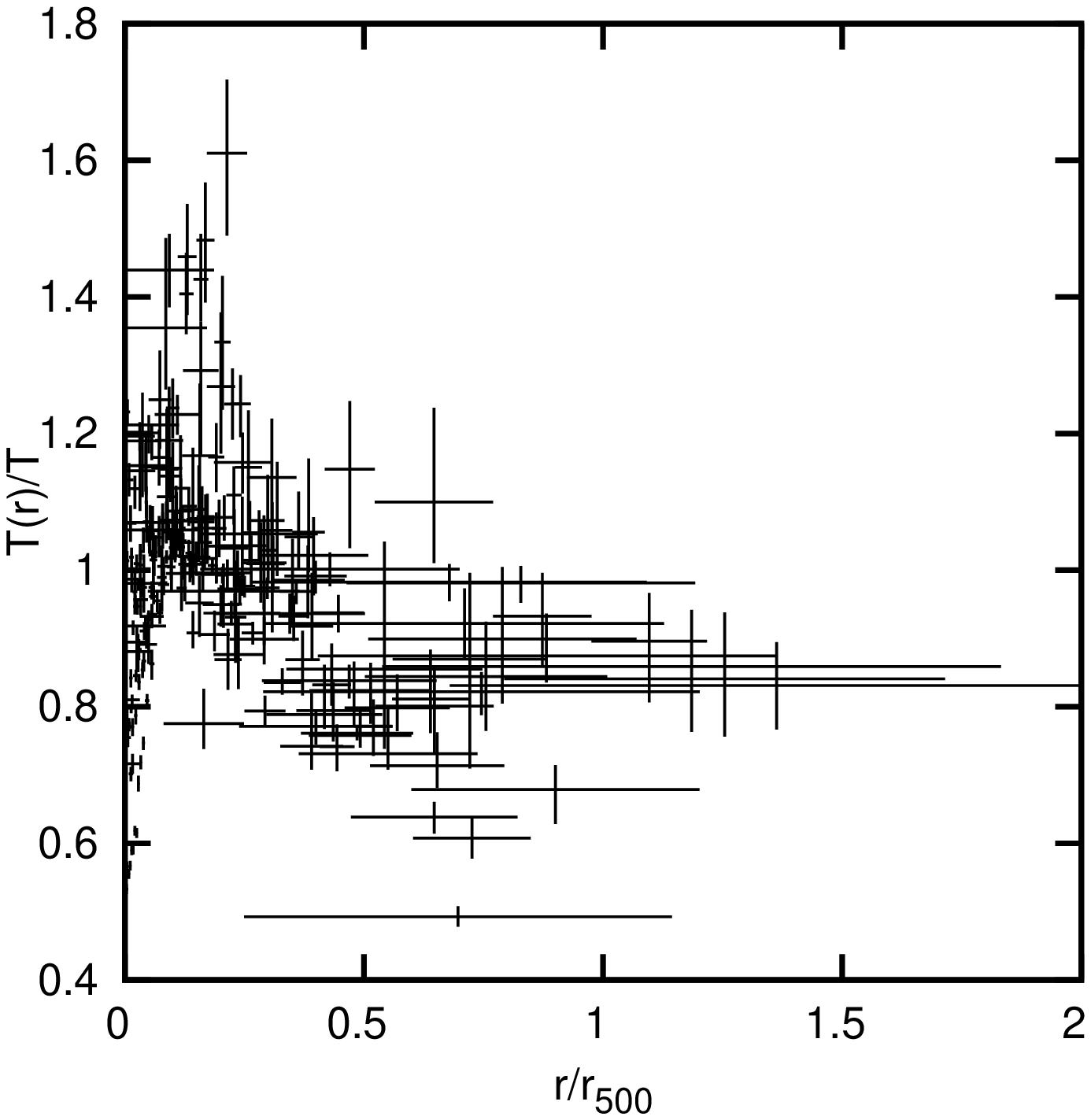}\quad
  \includegraphics[width=0.31\textwidth]{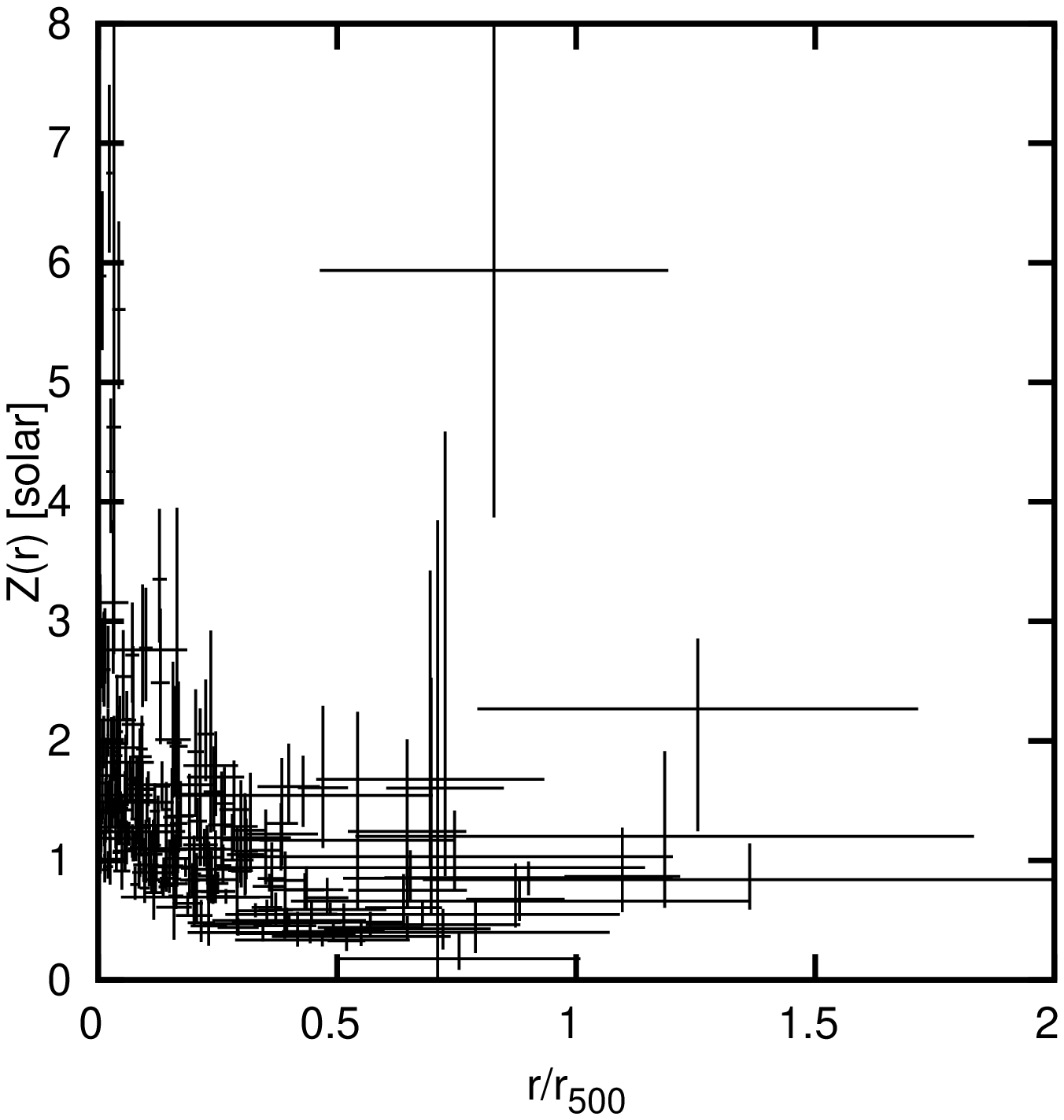}\qquad
  \includegraphics[width=0.31\textwidth]{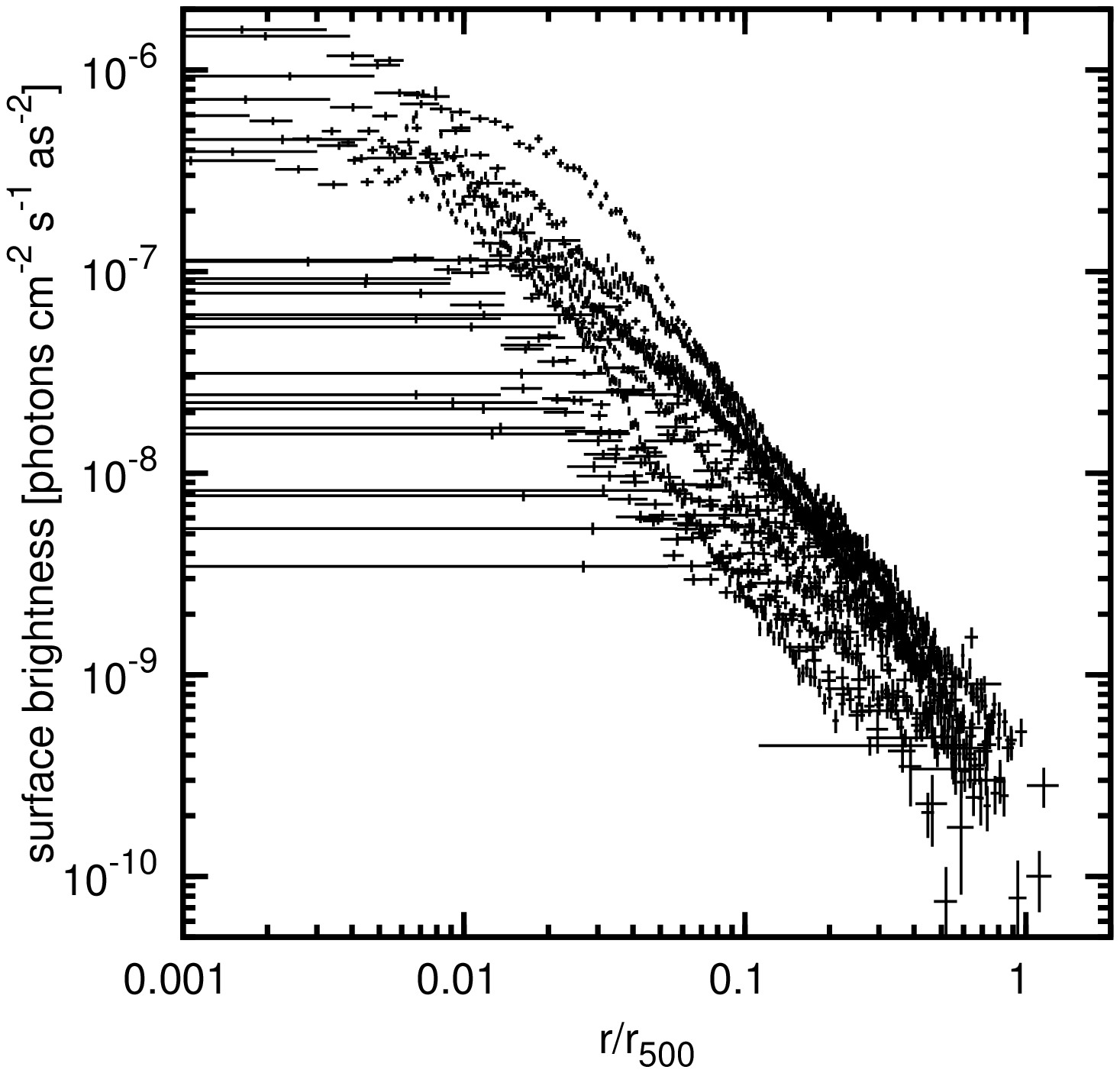}
  \includegraphics[width=0.31\textwidth]{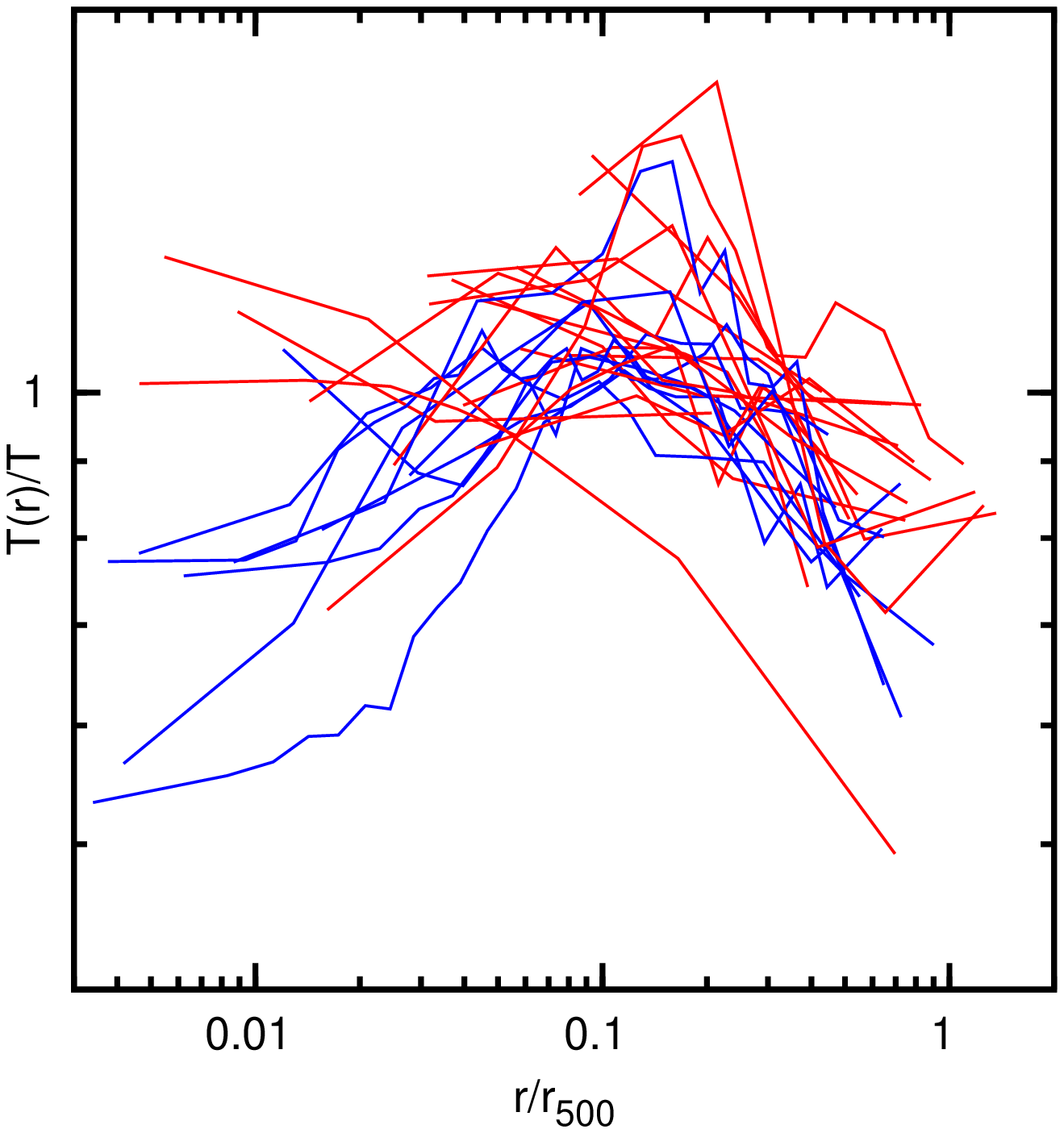}\quad
  \includegraphics[width=0.31\textwidth]{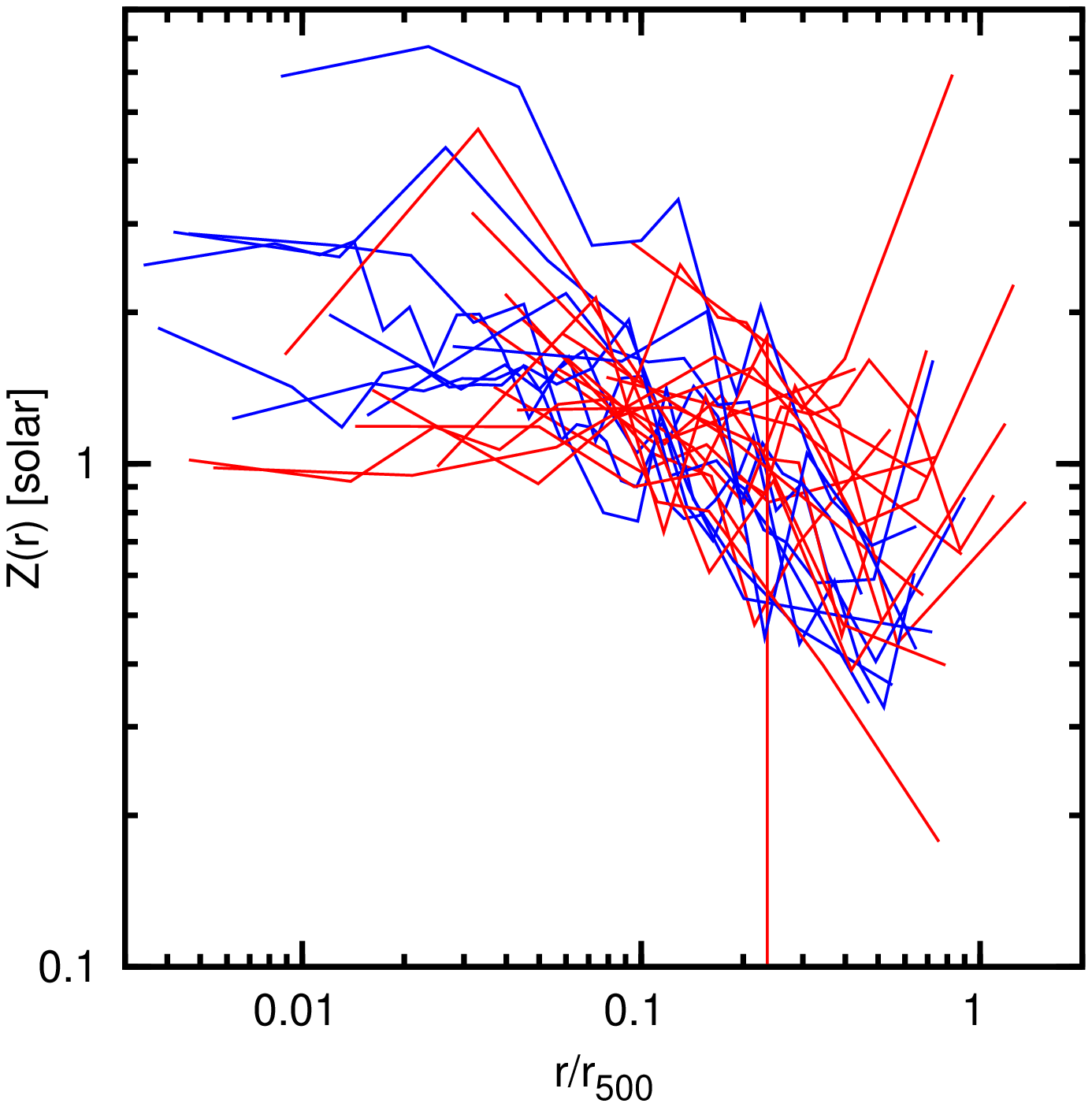}\qquad
  \includegraphics[width=0.31\textwidth]{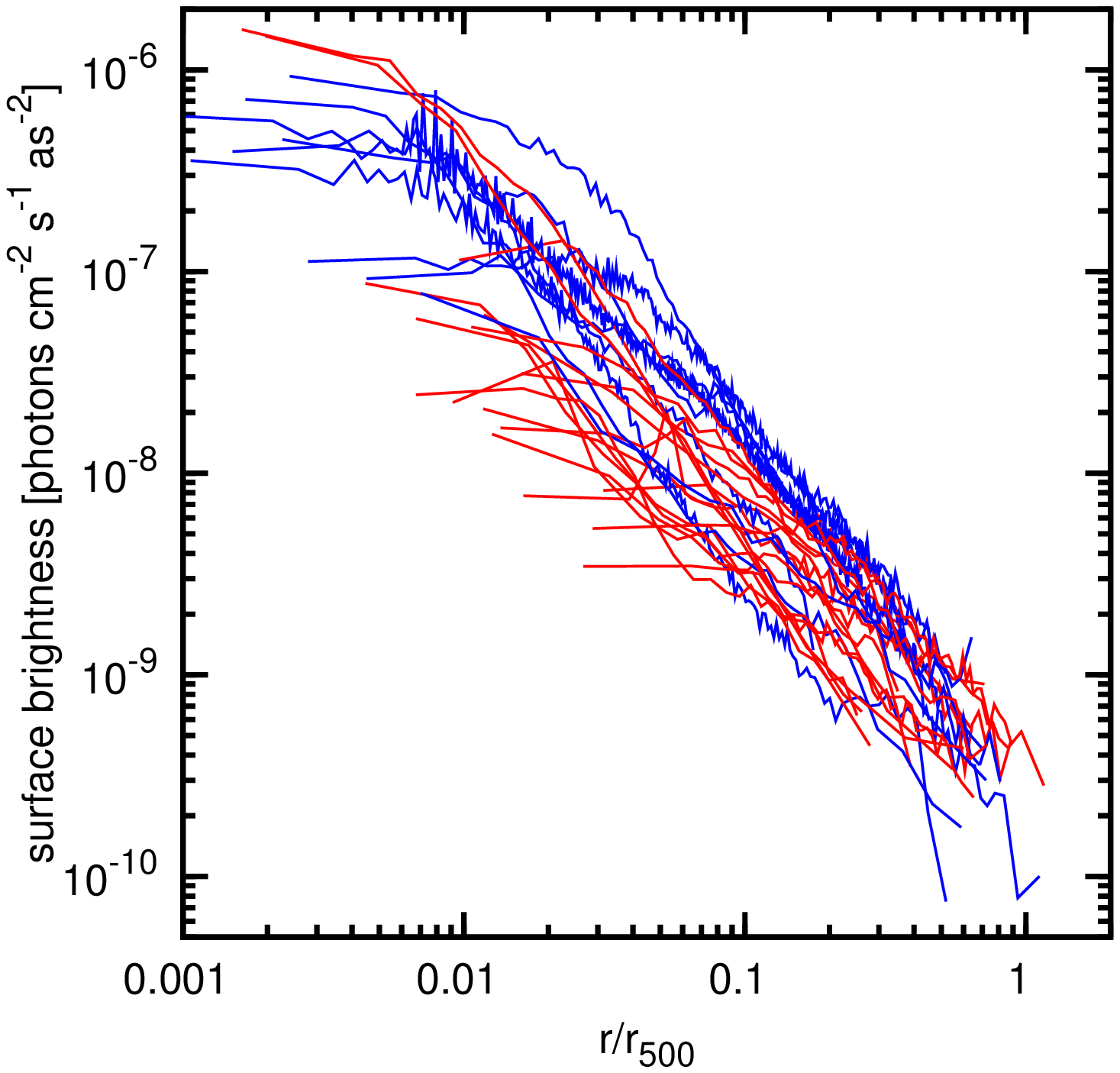}
  \caption{\emph{Top row, left to right:} Temperature profiles scaled
    by $r_{500}$ and global temperature (linear), metallicity profiles
    scaled by $r_{500}$ (linear), surface brightness profiles scaled
    by $r_{500}$ (logarithmic). \emph{Bottom row:} Same as above, but
    all plotted with lines, in logarithmic scale. Groups where a low-temperature core has been
    removed are shown in blue, others in red.}
 \label{fig:profiles}%
\end{figure*}
 \subsection{Radial Prof\hspace{0mm}iles}\label{sec:results:profiles}
 The projected temperature and metallicity profiles of the group
 sample are plotted in Fig.\ \ref{fig:profiles} (left and center),
 scaled by $r_{500}$, both in linear and logarithmic scales (top and
 bottom). In the lower part of the figure the objects where a
 low-temperature core was removed from the temperature analysis are
 shown as dotted lines. The temperature profiles behave quite
 universally for $r \gtrsim 0.05\,r_{500}$, with a flat plateau in the
 middle and decline beyond $r \sim 0.2 r_{500}$. Towards the inner
 regions however, there is an increase in scatter, and at small radii
 the profiles vary between being quite flat and showing a clear drop
 in temperature.

    The metallicity profiles have a larger scatter, especially in the
    outer bins, where $Z$ was not very well constrained by the data,
    but also show a universal decrease with radius. However
      objects with a central temperature drop tend to have higher
      metallicities in the center.

The observed surface brightness profiles are plotted in Fig.\
\ref{fig:profiles} (right), scaled by $r_{500}$. Individual temperature,
metallicity, and surface brightness profiles are shown in appendix
\ref{app:profiles}.
%
\subsection{Scaling Relations}\label{sec:results:scaling}
The best-fit results for all relations, including scatter, are listed
in Table \ref{tab:fits}.
   \begin{figure*}
   \centering
   \includegraphics[width=0.82\textwidth]{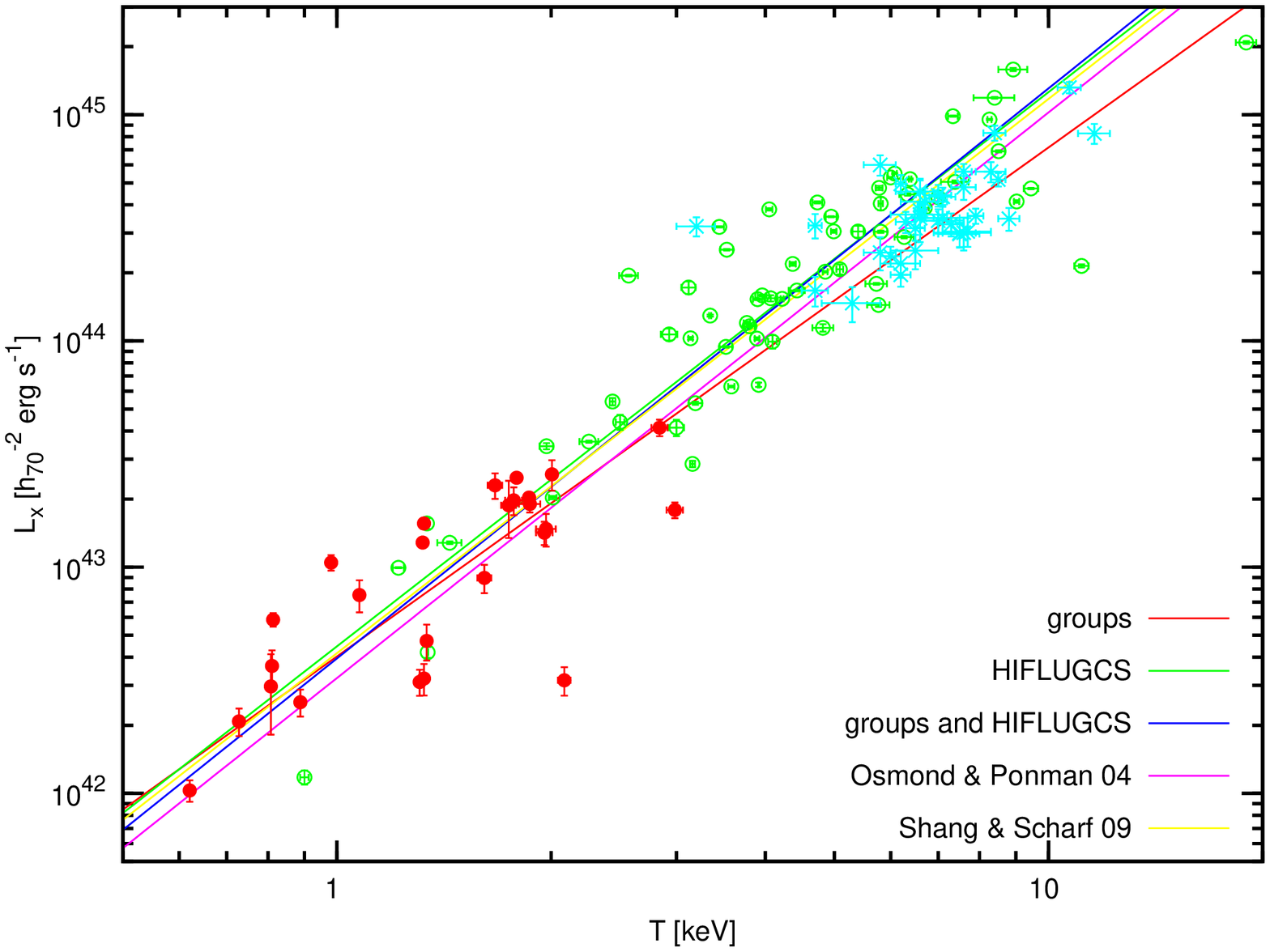}
   \caption{$L_{\text{x}}$-$T$ relation. Filled red circles are
     groups, open green circles are \emph{HIFLUGCS} clusters (Reiprich \&
     B\"ohringer 2002, Hudson et al.\ 2010), light blue asterisks are
     \emph{LoCuSS} clusters (Zhang et al.\ 2008).}
              \label{fig:ltrelation}%
    \end{figure*}
   \begin{figure*}
   \centering
   \includegraphics[width=0.82\textwidth]{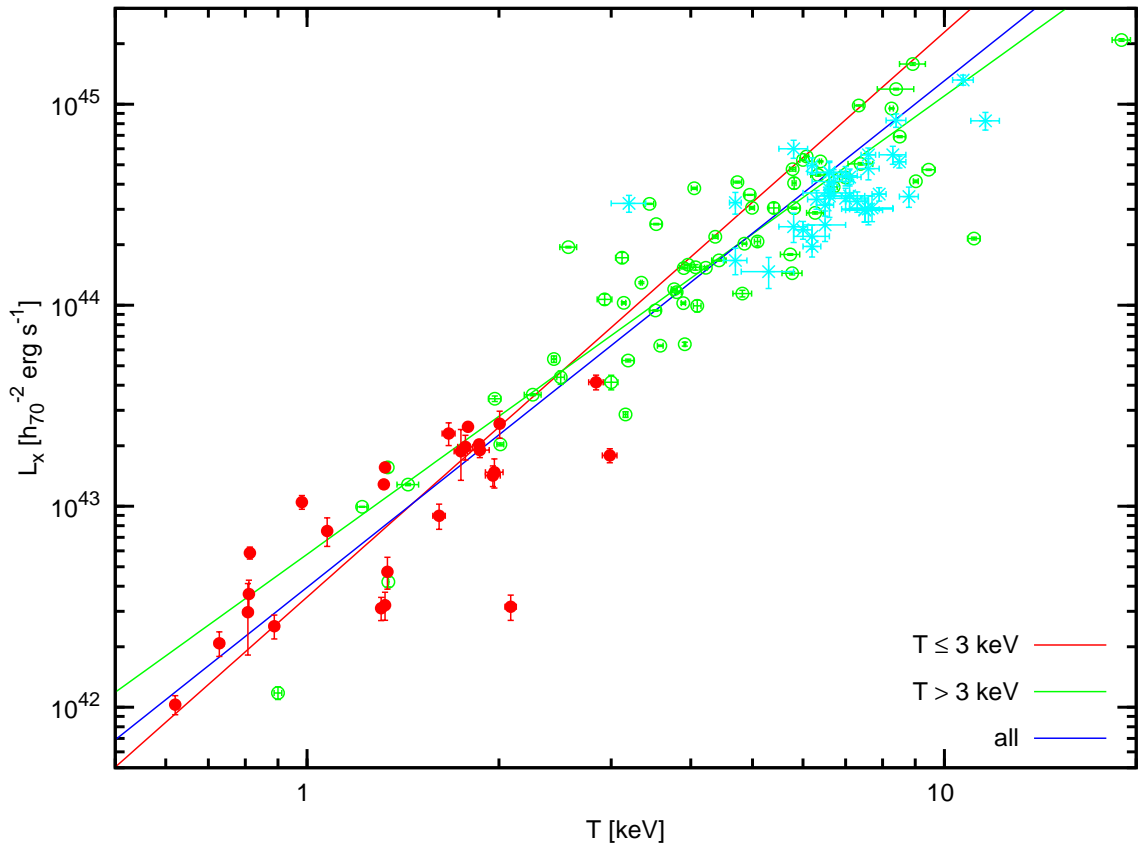}
   \caption{Same points as in Fig.\ \ref{fig:ltrelation},
     but the fits are cut at $\unit[3]{keV}$.}
              \label{fig:lt3krelation}%
    \end{figure*}
   \begin{figure*}
   \centering
   \includegraphics[width=0.82\textwidth]{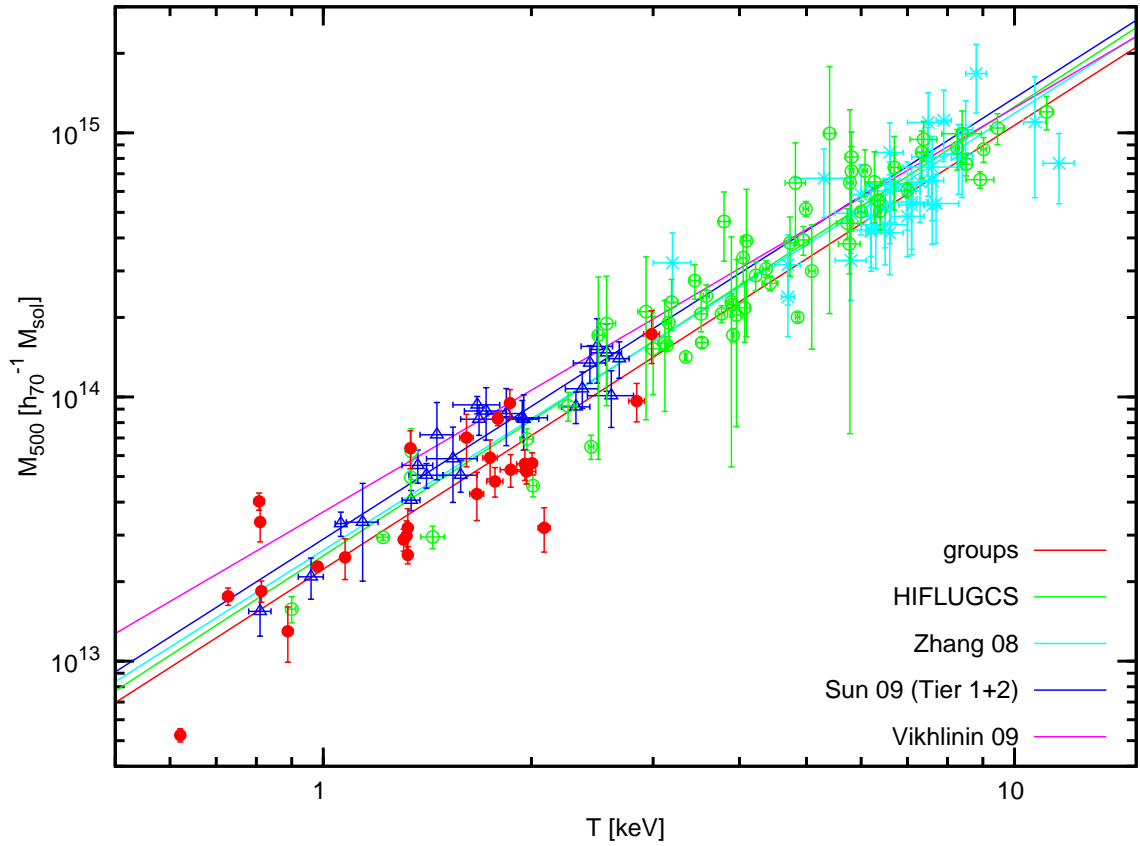}
   \caption{$M_{500}$-$T$ relation. Filled red circles are groups, open green circles
     are \emph{HIFLUGCS} clusters, blue open
     triangles are the groups from Sun et al.\ (2009), light blue asterisks are
     \emph{LoCuSS} clusters.}
              \label{fig:mtrelation}%
    \end{figure*}
   \begin{figure*}
   \centering
   \includegraphics[width=0.82\textwidth]{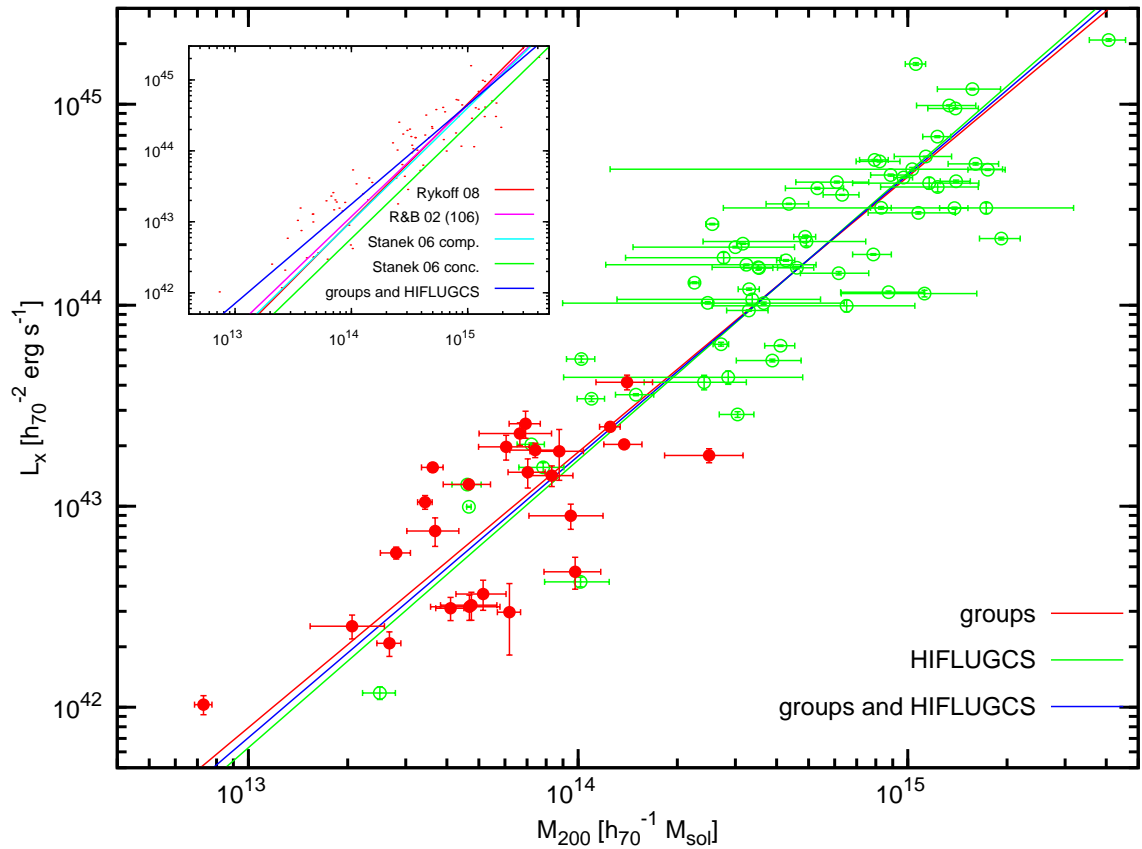}
   \caption{$L_{\text{x}}$-$M_{200}$ relation. Filled red circles are groups and
     open green circles are \emph{HIFLUGCS} clusters.}
              \label{fig:lmrelation}%
    \end{figure*}
   \begin{figure*}
   \centering
   \includegraphics[width=0.82\textwidth]{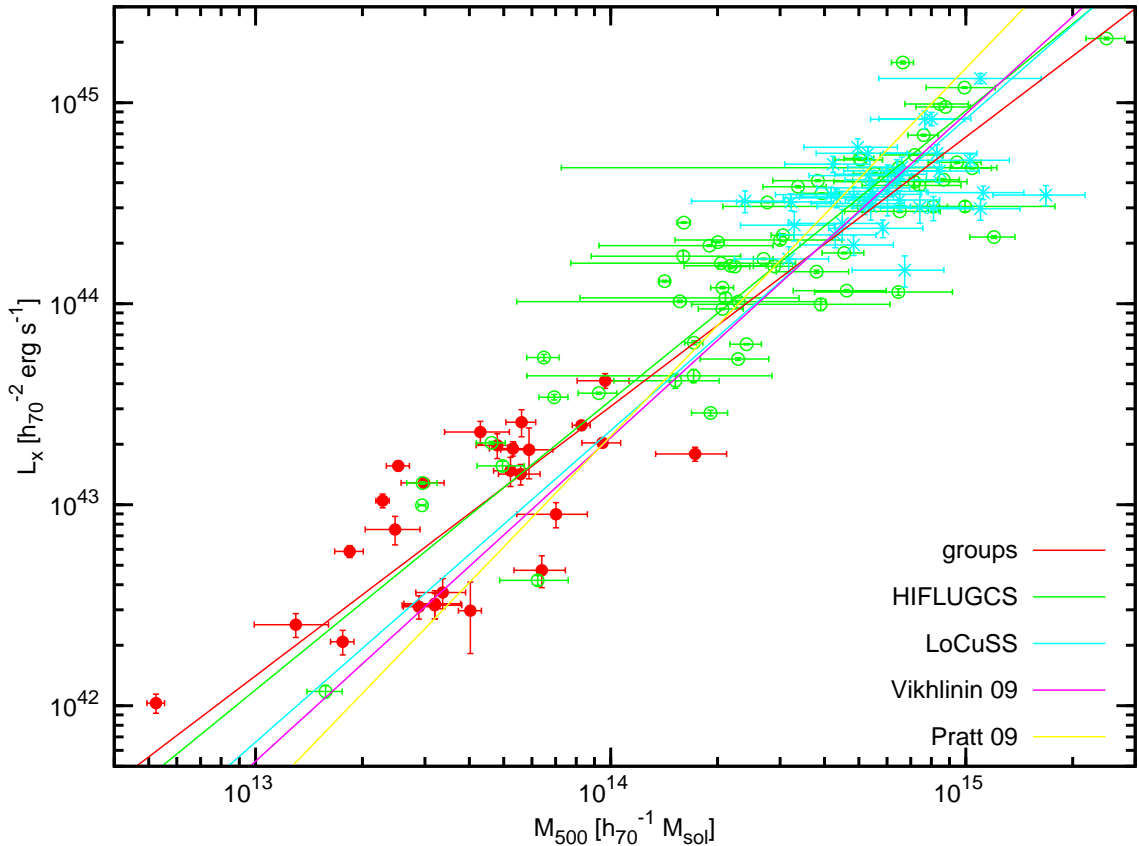}
   \caption{$L_{\text{x}}$-$M_{500}$ relation. Filled red circles are
     groups, open green circles are \emph{HIFLUGCS} clusters, and light blue asterisks are \emph{LoCuSS} clusters.}
              \label{fig:lm500relation}%
    \end{figure*}
\begin{table*}
\begin{threeparttable}
\caption{Fit results of all scaling relations}
\begin{center}
\setlength\extrarowheight{2pt}
\begin{tabularx}{\linewidth}{X X XXllll}
\hline\hline
Relation ($Y$-$X$) & Sample & a & b & $\sigma_{\text{tot}}$ (X) & $\sigma_{\text{tot}}$ (Y) & $\sigma_{\text{int}}$ (X) & $\sigma_{\text{int}}$ (Y) \\ \hline
$L_{\text{x}}$-$T$ & groups & $2.25 \pm 0.21$ & $-0.02 \pm 0.09$ & $0.122$ & $0.275$ & $0.119$ & $0.268$ \\ 
 & HIFLUGCS & $2.45 \pm 0.16$ & $0.12 \pm 0.04$ & $0.105$ & $0.258$ & $0.105$ & $0.257$ \\ 
 & groups \& HIFLUGCS & $2.52 \pm 0.10$ & $0.10 \pm 0.03$ & $0.109$ & $0.275$ & $0.108$ & $0.274$ \\ 
 & $T<\unit[3]{keV}$ & $2.81 \pm 0.27$ & $0.19 \pm 0.10$ & $0.110$ & $0.308$ & $0.108$ & $0.304$ \\ 
 & $T\geq\unit[3]{keV}$ & $2.28 \pm 0.25$ & $0.15 \pm 0.06$ & $0.107$ & $0.243$ & $0.106$ & $0.242$ \\ \hline
$M_{500}$-$T$ & groups & $1.68 \pm 0.20$ & $0.15 \pm 0.07$ & $0.104$ & $0.176$ & $0.098$ & $0.165$ \\ 
 & HIFLUGCS & $1.70 \pm 0.08$ & $0.21 \pm 0.02$ & $0.077$ & $0.131$ & $0.051$ & $0.087$ \\ 
 & groups \& HIFLUGCS & $1.75 \pm 0.06$ & $0.20 \pm 0.02$ & $0.084$ & $0.147$ & $0.067$ & $0.117$ \\ 
$M_{2500}$-$T$ & groups & $1.59 \pm 0.18$ & $-0.24 \pm 0.06$ & $0.097$ & $0.154$ & $0.093$ & $0.147$ \\ \hline
$L_{\text{x}}$-$M_{200}$ & groups & $1.37 \pm 0.20$ & $-0.43 \pm 0.08$ & $0.229$ & $0.313$ & $0.215$ & $0.293$ \\ 
 & HIFLUGCS & $1.43 \pm 0.10$ & $-0.47 \pm 0.08$ & $0.211$ & $0.301$ & $0.186$ & $0.265$ \\ 
 & groups \& HIFLUGCS & $1.40 \pm 0.06$ & $-0.45 \pm 0.04$ & $0.219$ & $0.307$ & $0.199$ & $0.278$ \\ 
$L_{\text{x}}$-$M_{500}$ & groups & $1.34 \pm 0.18$ & $-0.21 \pm 0.09$ & $0.226$ & $0.303$ & $0.214$ & $0.287$ \\ 
 & HIFLUGCS & $1.44 \pm 0.10$ & $-0.18 \pm 0.06$ & $0.202$ & $0.291$ & $0.176$ & $0.254$ \\ 
 & LoCuSS & $1.55 \pm 0.98$ & $-0.33 \pm 0.76$ & $0.167$ & $0.259$ & $0.099$ & $0.153$ \\ 
 & all & $1.44 \pm 0.06$ & $-0.18 \pm 0.04$ & $0.209$ & $0.301$ & $0.189$ & $0.272$ \\ 
$L_{\text{x}}$-$M_{2500}$ & groups & $1.42 \pm 0.16$ & $0.33 \pm 0.12$ & $0.203$ & $0.288$ & $0.194$ & $0.276$ \\ \hline
$M_{500}$-$Y_{\text{x},500}$ & groups & $0.53 \pm 0.06$ & $-0.62 \pm 0.05$ & $0.369$ & $0.194$ & $0.350$ & $0.184$ \\ 
 & HIFLUGCS & $0.55 \pm 0.04$ & $-0.63 \pm 0.09$ & $0.286$ & $0.157$ & $0.218$ & $0.120$ \\ 
 & groups \& HIFLUGCS & $0.55 \pm 0.02$ & $-0.63 \pm 0.04$ & $0.311$ & $0.170$ & $0.264$ & $0.144$ \\ 
$M_{2500}$-$Y_{\text{x},2500}$ & groups & $0.53 \pm 0.05$ & $-0.69 \pm 0.03$ & $0.283$ & $0.150$ & $0.270$ & $0.143$ \\ \hline
$L_{\text{x}}$-$Y_{\text{x},500}$ & groups & $0.71 \pm 0.05$ & $-1.05 \pm 0.06$ & $0.379$ & $0.269$ & $0.369$ & $0.262$ \\ 
 & HIFLUGCS & $0.79 \pm 0.03$ & $-1.10 \pm 0.05$ & $0.200$ & $0.158$ & $0.196$ & $0.155$ \\ 
 & groups \& HIFLUGCS & $0.78 \pm 0.02$ & $-1.08 \pm 0.05$ & $0.261$ & $0.203$ & $0.257$ & $0.200$ \\ 
$L_{\text{x}}$-$Y_{\text{x},2500}$ & groups & $0.76 \pm 0.05$ & $-0.66 \pm 0.05$ & $0.336$ & $0.256$ & $0.327$ & $0.249$ \\ \hline 
$M_{\text{g},500}$-$M_{500}$ & groups & $1.38 \pm 0.18$ & $0.83 \pm 0.09$ & $0.217$ & $0.300$ & $0.208$ & $0.288$ \\
 & HIFLUGCS & $1.26 \pm 0.12$ & $0.78 \pm 0.08$ & $0.192$ & $0.242$ & $0.161$ & $0.204$ \\ 
 & groups \& HIFLUGCS & $1.28 \pm 0.06$ & $0.78 \pm 0.04$ & $0.202$ & $0.259$ & $0.181$ & $0.231$ \\ 
$M_{\text{g},2500}$-$M_{2500}$ & groups & $1.34 \pm 0.16$ & $0.75 \pm 0.11$ & $0.175$ & $0.235$ & $0.169$ & $0.227$ \\ \hline
$f_{\text{g},500}$-$T$ & groups & $1.32 \pm 0.32$ & $0.11 \pm 0.12$ & $0.222$ & $0.294$ & $0.216$ & $0.286$ \\ 
 & HIFLUGCS & $0.83 \pm 0.42$ & $-0.22 \pm 0.08$ & $0.314$ & $0.262$ & $0.286$ & $0.239$ \\ 
 & groups \& HIFLUGCS & $0.79 \pm 0.09$ & $-0.17 \pm 0.03$ & $0.337$ & $0.268$ & $0.315$ & $0.250$ \\ 
$f_{\text{g},2500}$-$T$ & groups & $1.12 \pm 0.32$ & $-0.15 \pm 0.12$ & $0.226$ & $0.253$ & $0.222$ & $0.249$ \\ \hline 
\end{tabularx}
 \begin{tablenotes}
\item[] \emph{Notes:} See section \ref{sec:scaling} for fit functions.
\end{tablenotes}
\end{center}
\label{tab:fits}
\end{threeparttable}
\end{table*}
%
\subsubsection{$L_{\text{x}}$-$T$ Relation}
The luminosity-temperature relation for the group sample is plotted in
Fig.\ \ref{fig:ltrelation}, together with the $64$ \emph{HIFLUGCS}
clusters (Hudson et al.\ 2010) and the $37$ \emph{LoCuSS} clusters
from Zhang et al.\ (2008). Fitting groups and clusters separately
gives a slightly shallower slope for the groups ($2.25 \pm 0.21$
compared to $2.45 \pm 0.16$ for \emph{HIFLUGCS}), but they are
consistent within the errors. Fitting \emph{HIFLUGCS} together with
the group sample steepens the relation to $2.52 \pm 0.10$, but this
effect is again not significant considering the uncertainties. The
best fit group relation is also in good agreement with both the
relations found for the \emph{GEMS} group sample\footnote{with
  $L_{\text{x}}$ extrapolated to $r_{500}$} (Osmond \& Ponman 2004)
and a local \emph{Suzaku} cluster sample (Shang \& Scharf 2009).

We also investigated whether selecting groups and clusters by
temperature produces a different result from the luminosity cut
applied before. We applied cuts at different temperatures
  (stepsize $\unit[0.5]{keV}$), and the lowest temperature at which the
  difference in slope of cooler and hotter objects was significant was
  $\unit[3]{keV}$, so we chose this as a cut. The resulting fits are shown in Fig.\
\ref{fig:lt3krelation}. We found that systems below $\unit[3]{keV}$
follow a significantly steeper relation ($2.81 \pm 0.28$) than the
hotter systems ($2.28 \pm 0.24$).

It is also interesting to note that the groups have higher scatter
than the clusters both in $T$ and $L_{\text{x}}$. Similarly, when
selecting by temperature the intrinsic scatter in both
parameters is higher for the cooler systems ($0.268$ vs.\ $0.257$
  for $L_{\text{x}}$ and $0.119$ vs.\ $0.105$ for $T$). 

  We also tried fitting this relation with a broken powerlaw function,
  but this improves the fit only marginally and the data do not
  constrain the fit well due to the large scatter.
%
 \subsubsection{$M$-$T$ Relation}
 In Fig.\ \ref{fig:mtrelation} we show the fits to the $M_{500}$-$T$
 relation. We again compare the group sample to the \emph{HIFLUGCS}
 and the \emph{LoCuSS} clusters, as well as the group sample of Sun et
 al.\ (2009), and the cluster sample of Vikhlinin et al.\ (2009).

 The slopes of all these fits are very similar, and the best fit
 relations of groups and clusters are consistent with each other
 ($1.68\pm0.20$ vs.\ $1.70\pm0.08$). However the normalization of our
 group sample is $\unit[\sim10-30]{\%}$ lower than the others, in
 particular of the other group sample. Possible reasons for this will
 be discussed in section \ref{sec:discussion}.

 When fitting groups and clusters we once more find a slightly,
   but not significantly steeper best fit relation ($1.75\pm0.06$)
   compared to the pure cluster fit. Again we note that the scatter
 in groups is larger than in clusters ($0.165$ vs.\ $0.087$ for
   $M$ and $0.098$ vs.\ $0.051$ for $T$).

 For $r_{2500}$ we find a best fit $M$-$T$ relation of
   $a=1.59\pm0.18$ and $b=-0.24\pm0.06$.
%
 \subsubsection{$L_{\text{x}}$-$M$ Relation}
 The luminosity-mass relation is plotted in Figs.\
 \ref{fig:lmrelation} and \ref{fig:lm500relation} (for $M_{200}$ and
 $M_{500}$, respectively). Note that for our group sample the
 measurement of $M_{200}$ involved significant extrapolation beyond
 the actual data. We show the $L_{\text{x}}$-$M_{200}$ relation here
 primarily for comparison with other publications.
  
 All the three fits are very close together, and the slopes of the
   group sample and the \emph{HIFLUGCS} clusters are consistent with
   each other (in the $L_{\text{x}}$-$M_{200}$ relation $1.37 \pm
   0.20$ for groups, $1.43 \pm 0.10$ for \emph{HIFLUGCS}). The
 cluster fit is not changed significantly by including the groups
 ($1.40 \pm 0.06$). The same trend can also be seen for $M_{500}$.
 
 The $L_{\text{x}}$-$M_{200}$ relation of clusters and groups
 combined is shown together with several other cluster relations in
 the inset plot of Fig.\ \ref{fig:lmrelation}. The fit to the extended
 \emph{HIFLUGCS} sample (106 clusters, Reiprich \& B\"ohringer 2002),
 the stacked relation found by Rykoff et al.\ (2008), and the
 compromise model of Stanek et al.\ (2006) agree very well with each
 other, while Stanek et al.'s concordance model fit is significantly
 lower in normalization. See section \ref{sec:discussion} for a more
 detailed discussion.

In Fig.\ \ref{fig:lm500relation} we compare the group
$L_{\text{x}}$-$M_{500}$ relation to both the \emph{HIFLUGCS} and
\emph{LoCuSS} clusters, as well as Vikhlinin et al.\
(2009)\footnote{Luminosity in the $\unit[0.5-2.0]{keV}$ band was
  scaled by a factor of $1/0.62$ to match our energy band
  ($\unit[0.1-2.4]{keV}$).} and Pratt et al.\
(2009)\footnote{Malmquist bias corrected, $\unit[0.1-2.4]{keV}$ band,
  fit using the BCES Orthogonal method.}. The group relation is
shallower than all these cluster samples but consistent with
\emph{HIFLUGCS} (see discussion).

Once more we find the intrinsic scatter for groups to be larger
  than for the cluster samples, for instance the scatter in $M_{500}$
  is $0.287$ for the groups, but only $0.254$ for \emph{HIFLUGCS} and
  $0.153$ for \emph{LoCuSS}.

The best fit $L_{\text{x}}$-$M$ relation for $r_{2500}$ is
  $a=1.42\pm0.16$ and $b=0.33\pm0.12$.

%
\subsubsection{$Y_{\text{x}}$,  $M_{\text{g}}$ and $f_{\text{g}}$ Relations}
\begin{figure*}
   \centering
   \includegraphics[width=0.49\textwidth]{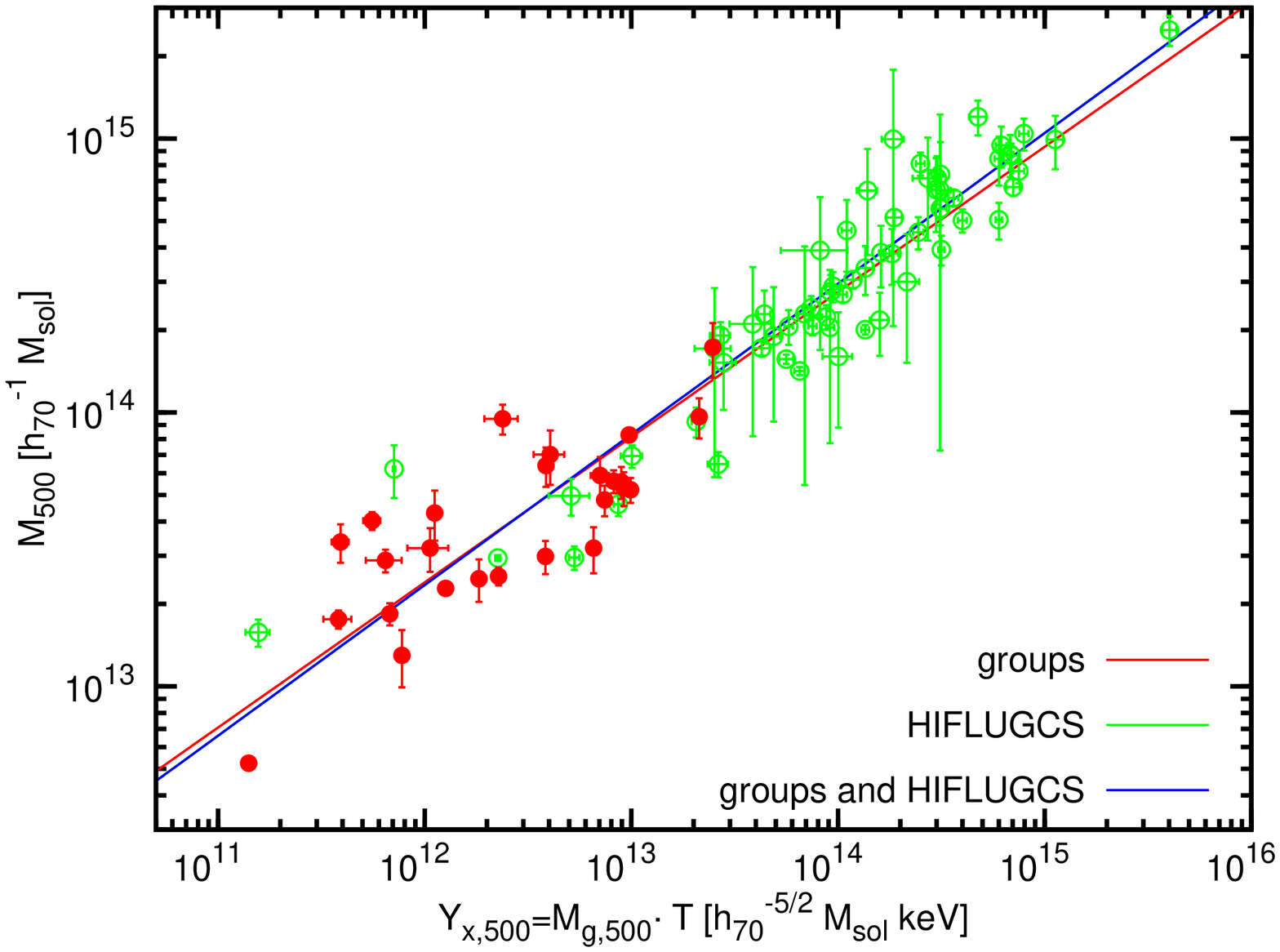}\quad
   \includegraphics[width=0.49\textwidth]{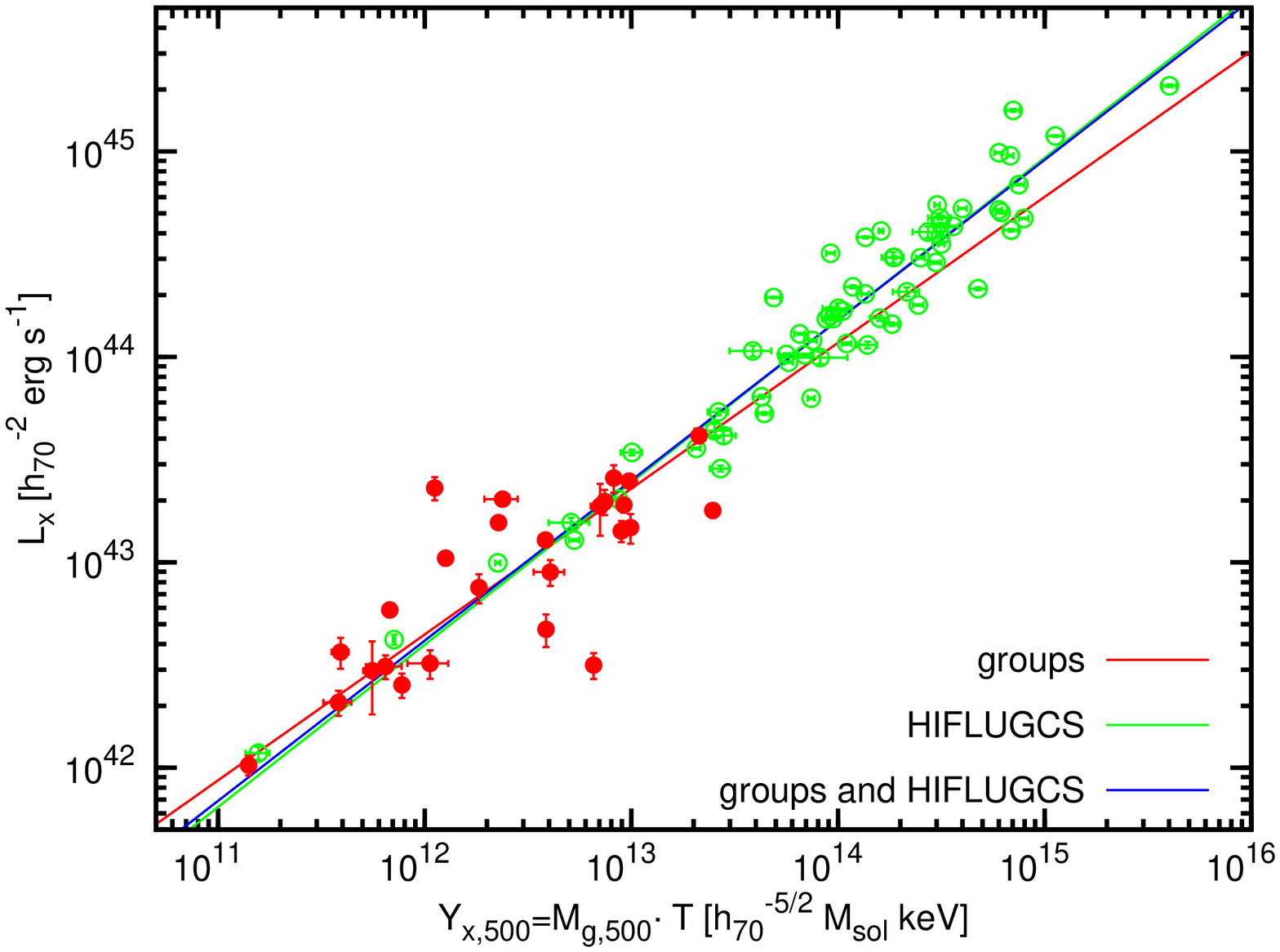}\\\medskip
   \includegraphics[width=0.49\textwidth]{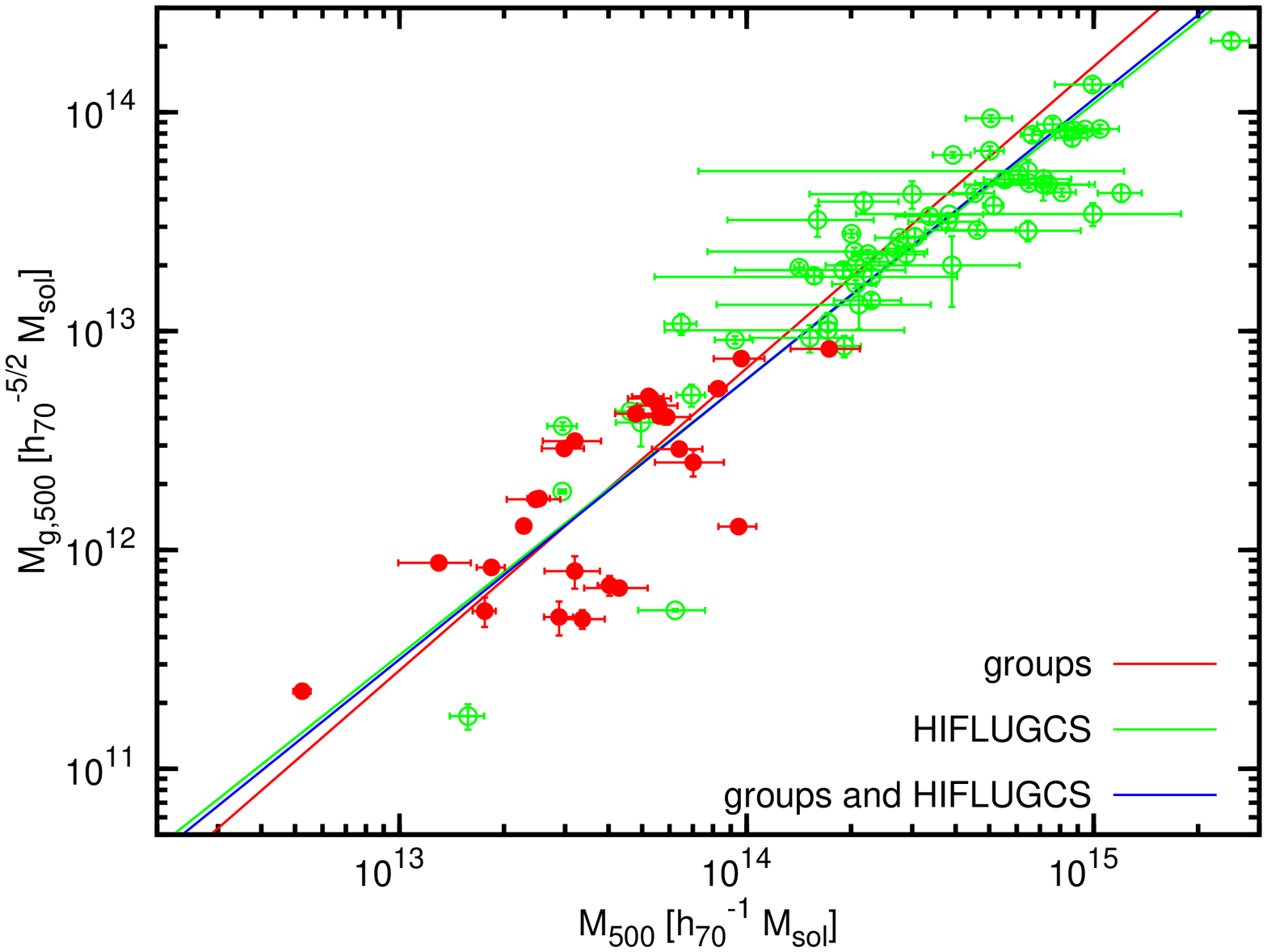}\quad
   \includegraphics[width=0.49\textwidth]{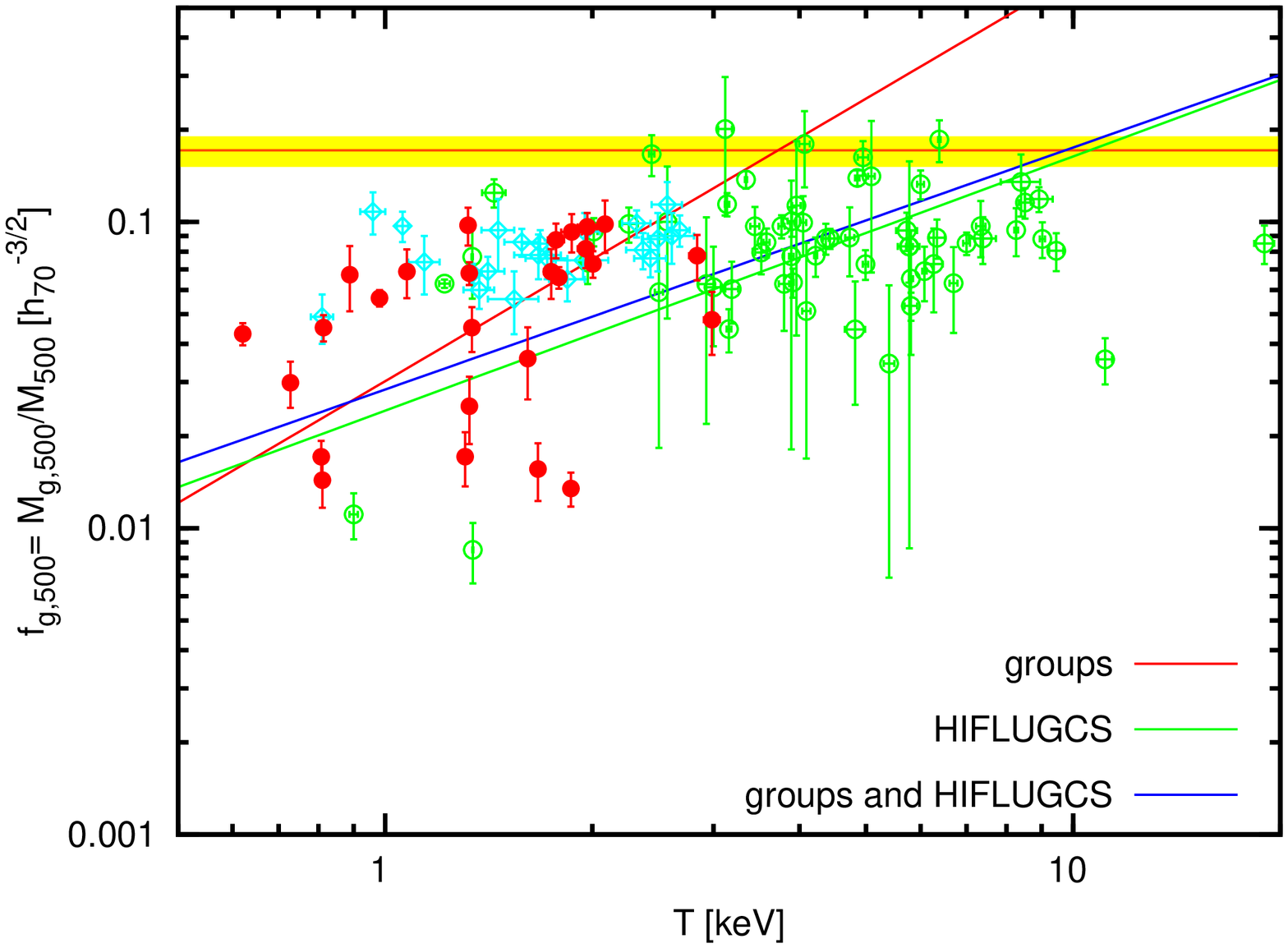}
   \caption{$Y_{\text{x}}$ and gas fraction relations at
     $r_{500}$. \emph{Top left:} $M$-$Y_{\text{x}}$
     relation. \emph{Top right:} $L_{\text{x}}$-$Y_{\text{x}}$
     relation. \emph{Lower left:} $M_{\text{g}}$-$M$
     relation. \emph{Lower right:} $f_{\text{g}}$-$T$
     relation. Filled red circles are groups, open green circles are
     \emph{HIFLUGCS} clusters (gas masses from Zhang et al.\ 2011), and light blue asterisks are the groups from Sun
     et al.\ (2009). The orange line with yellow error regions is
     the cosmic baryon fraction from \emph{WMAP}5 (Dunkley et al.\
     2009).}
              \label{fig:yxfgrelation}%
\end{figure*}
\begin{figure*}
   \centering
   \includegraphics[width=0.49\textwidth]{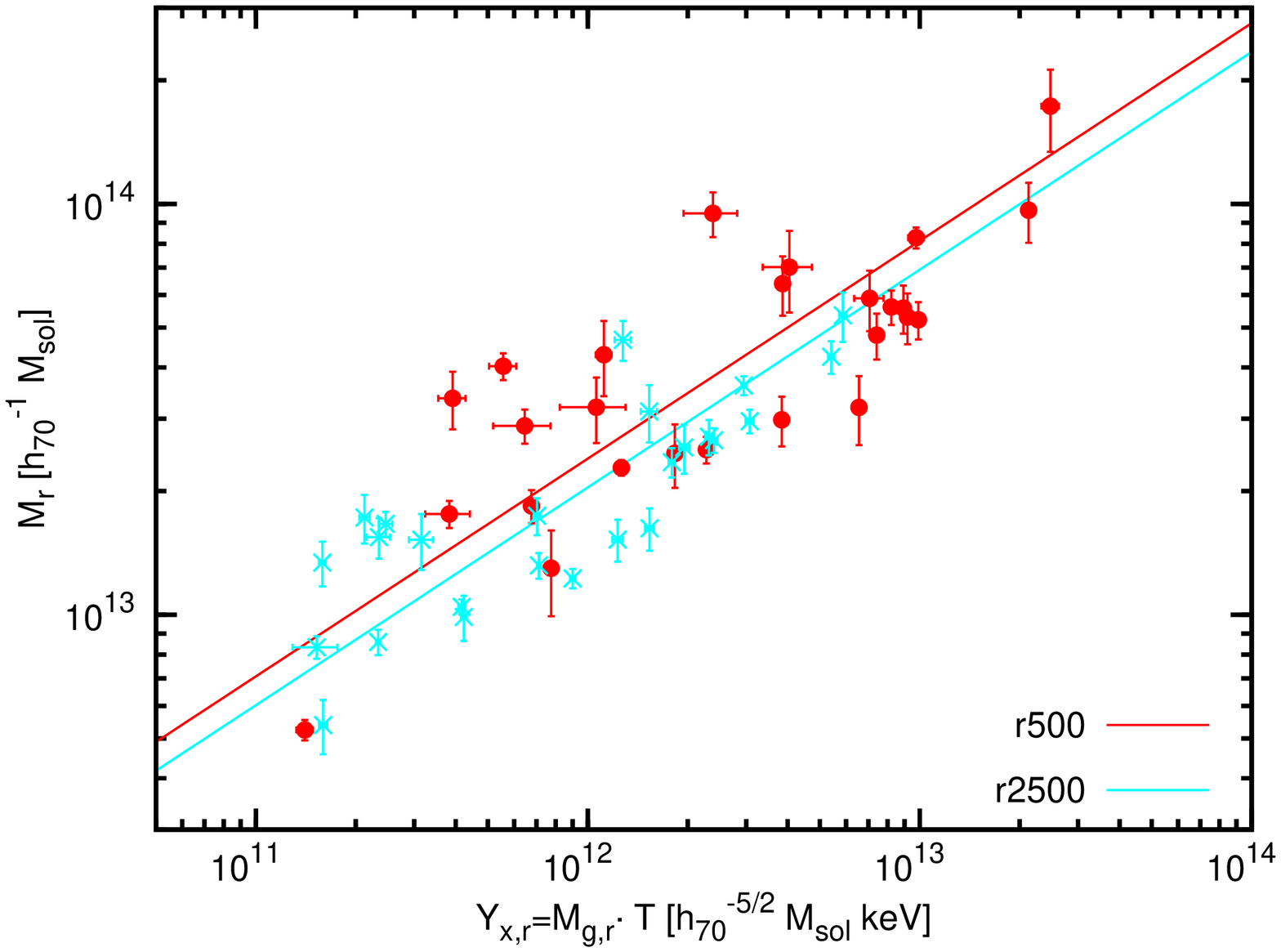}\quad
   \includegraphics[width=0.49\textwidth]{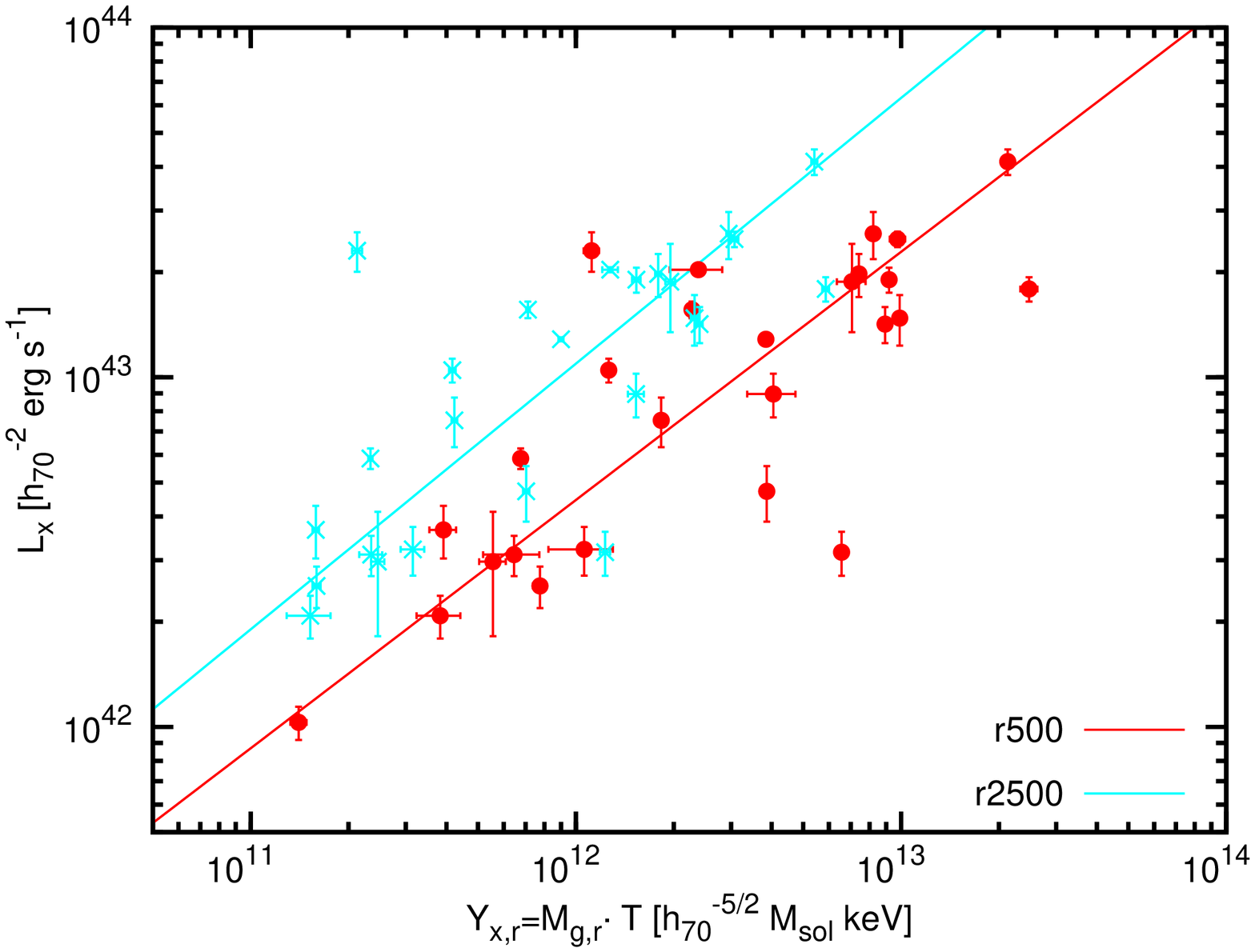}\\\medskip
   \includegraphics[width=0.49\textwidth]{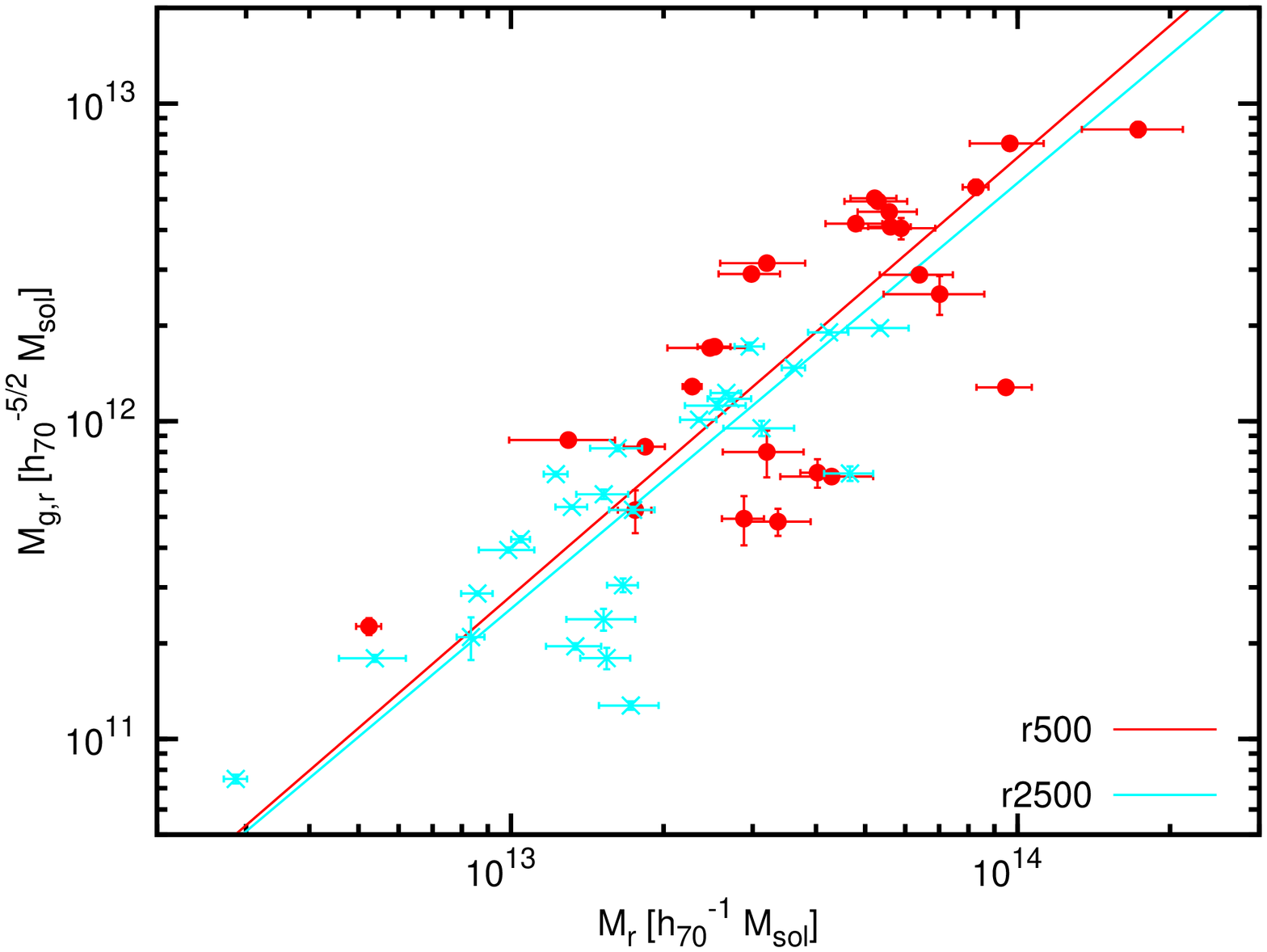}\quad
   \includegraphics[width=0.49\textwidth]{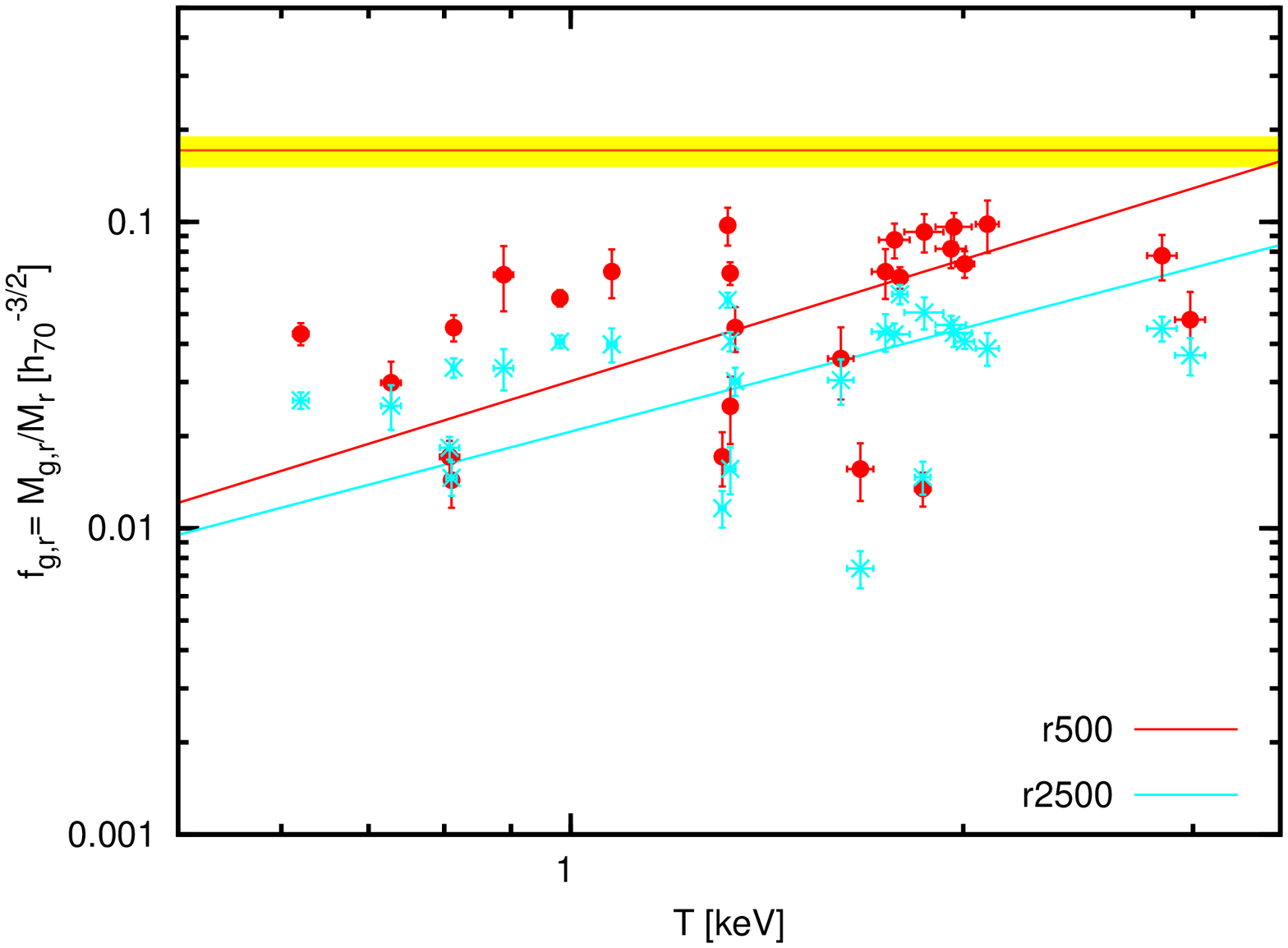}
   \caption{Same relations as in Fig. \ref{fig:yxfgrelation}, but at different
     radii. Filled red circles are for $r=r_{500}$
     and light blue asterisks are for $r=r_{2500}$.}
              \label{fig:yxfgrelation_radii}%
\end{figure*}
In Fig.\ \ref{fig:yxfgrelation} we show fit results to the
$M$-$Y_{\text{x}}$, $L_{\text{x}}$-$Y_{\text{x}}$,
$M_{\text{g}}$-$M$, and $f_{\text{g}}$-$T$ relations, all determined
for $r_{500}$. We compare the group results to the \emph{HIFLUGCS}
clusters (gas masses were measured by Zhang et al.\ 2011 for all except
2A0335, which is omitted here). The gas mass fractions of
Sun et al.'s group sample and the cosmic baryon fraction
$\Omega_{\text{b}}/\Omega_{\text{m}}$ measured from \emph{WMAP}
five-year data (Dunkley et al.\ 2009) are also included in the
$f_{\text{g}}$-$T$ plot for comparison.

For the $M$-$Y_{\text{x}}$ and $L_{\text{x}}$-$Y_{\text{x}}$
relations, the different samples are in good agreement, although the slope
  of the fit is slightly flatter for the groups than for the clusters
($0.53 \pm 0.06$ vs.\ $0.55 \pm 0.04$ and $0.71 \pm 0.05$ vs.\ $0.79
\pm 0.03$, respectively). The $M_{\text{g}}$-$M$ relation for groups
is a bit steeper than that of the clusters, but the slopes are
  consistent with each other within the uncertainties
  ($1.38\pm0.18$ vs.\ $1.26\pm0.12$).

The intrinsic scatter for these relations is much larger in
  groups, for example in the $L_{\text{x}}$-$Y_{\text{x}}$ relation,
  $Y_{\text{x}}$ has a scatter of $0.369$ for the groups, compared to
  $0.196$ for \emph{HIFLUGCS}, and similar trends can be seen for the
  other parameters and relations. Including groups in the fit does not
  significantly influence any of the cluster relations, for instance
  the best fit \emph{HIFLUGCS} $M$-$Y_{\text{x}}$ relation is
  $a=0.55\pm0.04$ and $b=-0.63\pm0.09$, compared to $a=0.55\pm0.02$ and
  $b=-0.63\pm0.04$ of the combined fit, and the difference is negligible for the
  $L_{\text{x}}$-$Y_{\text{x}}$ and $M_{\text{g}}$-$M$ relations as well.

There is a strong trend for lower $f_{\text{g}}$ at lower
temperatures. The slope of the group $f_{\text{g}}$-$T$ relation is
clearly different from zero ($1.32 \pm 0.32$), as are the shallower
fit to the \emph{HIFLUGCS} clusters ($0.83\pm0.42$), and the combined
fit of groups and clusters ($0.79 \pm 0.09$). The intrinsic
  scatter in $f_{\text{g}}$ is larger for the group sample than for
  \emph{HIFLUGCS} ($0.286$ vs.\ $0.239$, but it is smaller
  for $T$ ($0.216$ vs.\ $0.286$).

In Fig.\ \ref{fig:yxfgrelation_radii} we show the same relations as
before, but compare the radii $r_{500}$ and $r_{2500}$. The relations
measured out to $r_{2500}$ have smaller uncertainties and scatter,
compared to $r_{500}$.

The gas mass fractions within $r_{2500}$ are typically lower than
within $r_{500}$, and the slope of the $f_{\text{g}}$-$T$ relation
measured within $r_{2500}$ is flatter than for $r_{500}$, but also
significantly differs from zero ($1.12\pm0.32$).

We have inspected the other scaling relations in terms of
  $f_{\text{g}}$ but did not find any particular trends for low or
  high gas mass objects to behave differently. As an example, in
  Fig. \ref{fig:ccfgrelation} (left) we show the $L_{\text{x}}$-$T$
  relation with objects with gas mass fractions above and below the
  mean marked as different symbols.
%
\subsection{Morphology and Cool Cores}\label{sec:compmorph}
\begin{figure*}
  \centering
  \includegraphics[width=0.49\textwidth]{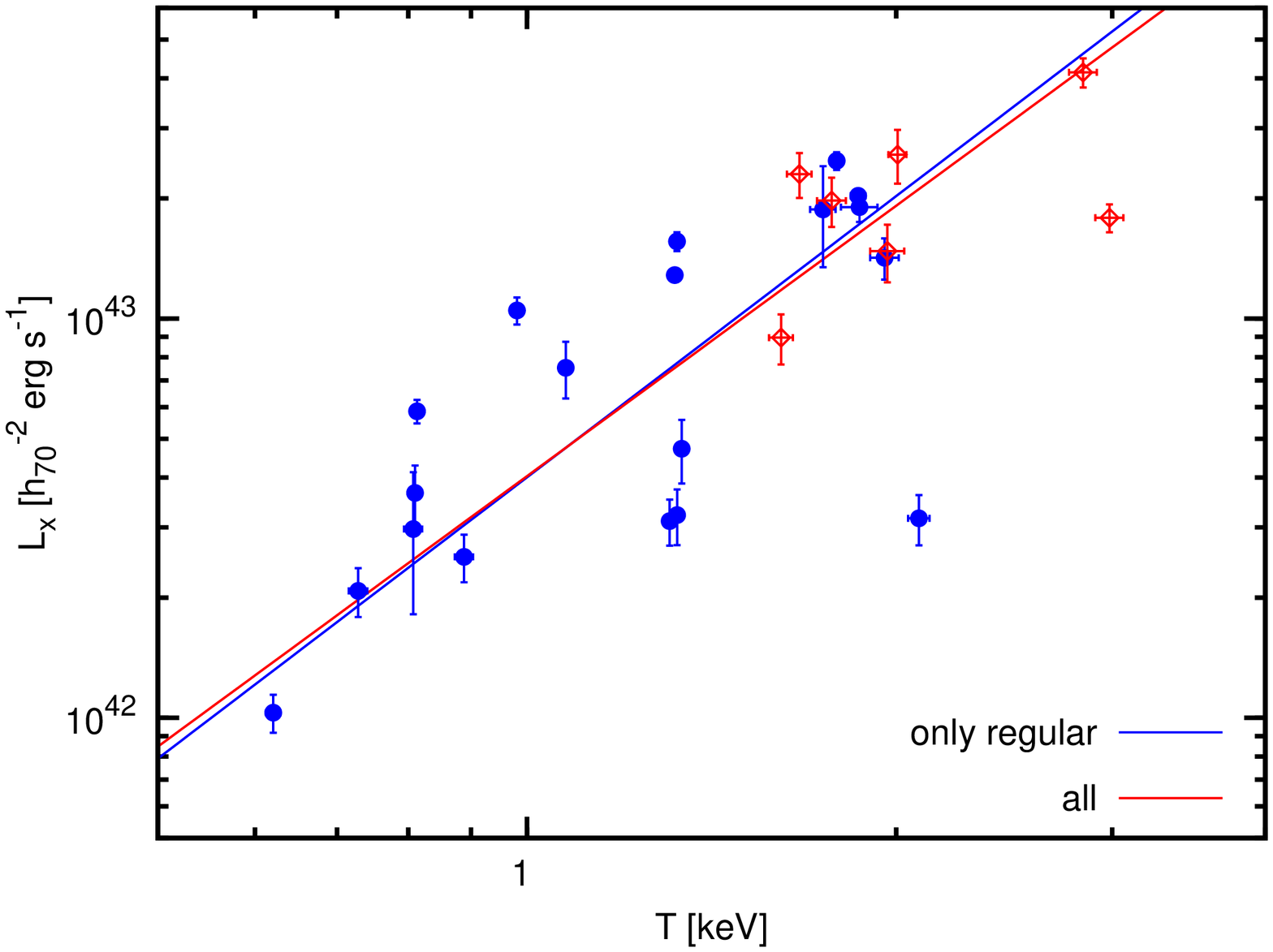}\quad
  \includegraphics[width=0.49\textwidth]{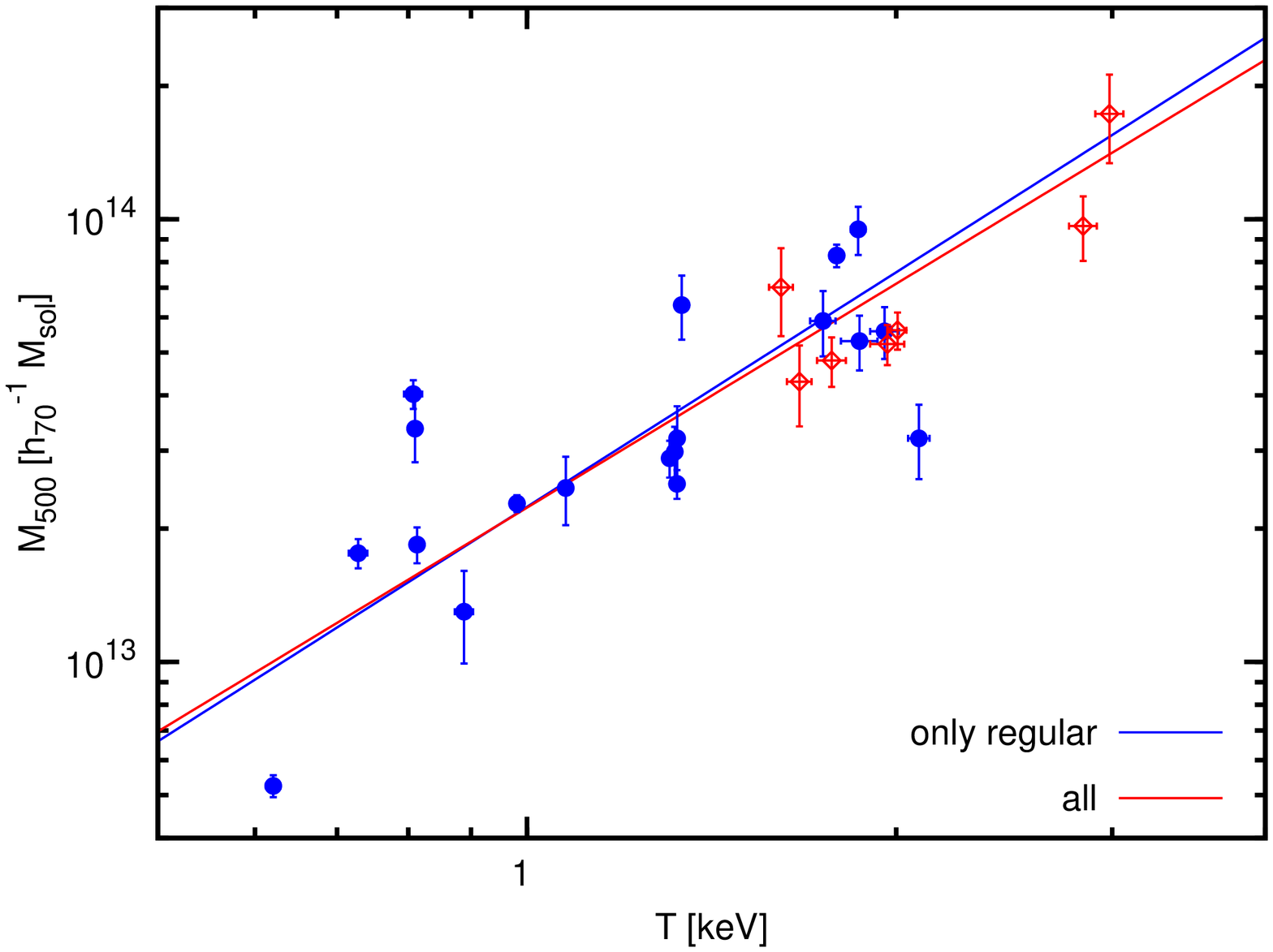}\\\medskip
  \includegraphics[width=0.49\textwidth]{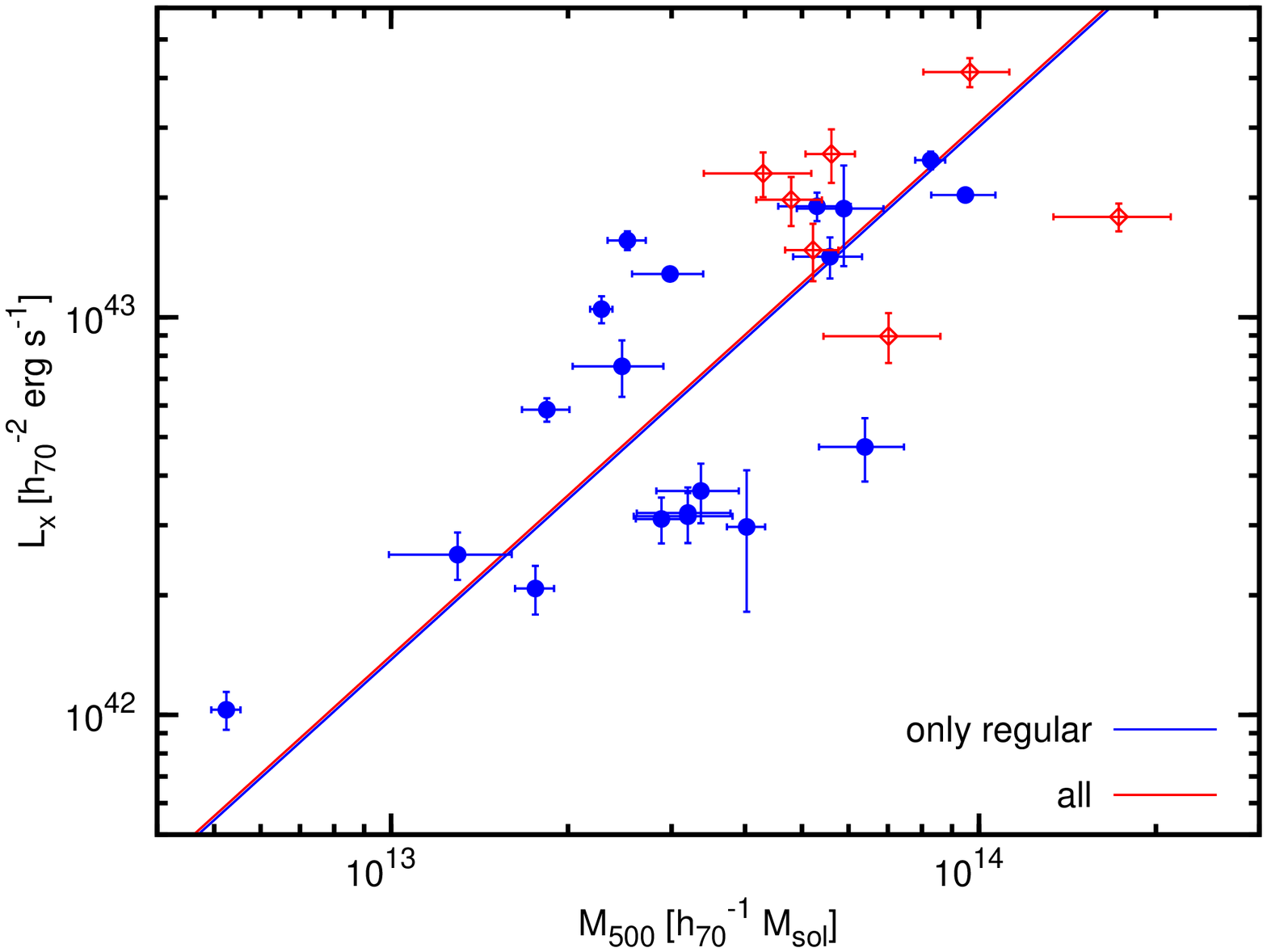}\quad
   \includegraphics[width=0.49\textwidth]{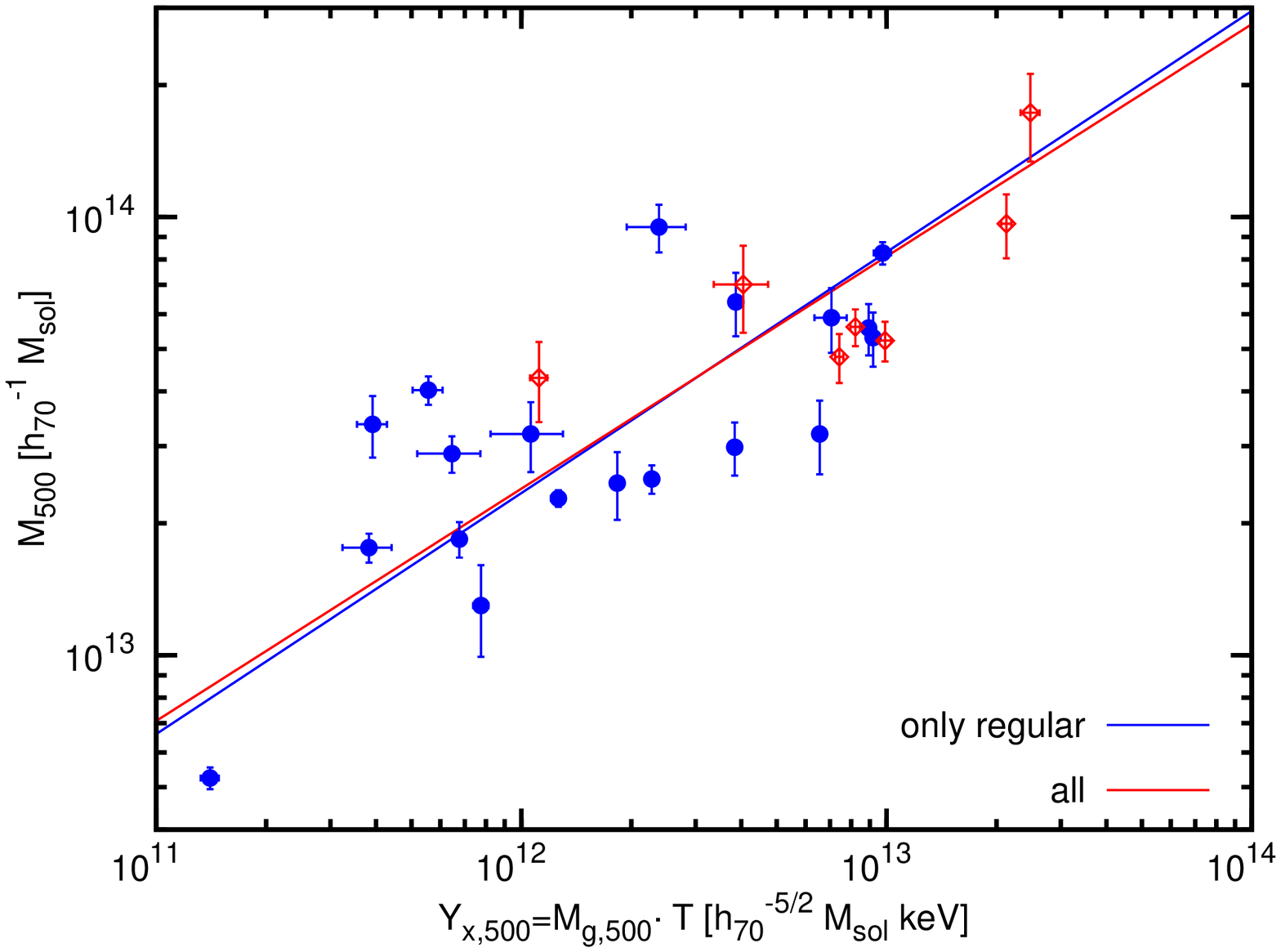}
   \caption{Scaling relations fit using different morphological
     selections. Filled blue circles are regular (relaxed) objects, open red diamonds are
     irregular (unrelaxed). \emph{Top left:} $L_{\text{x}}$-$T$. \emph{Top
       right:} $M$-$T$. \emph{Lower left:} $L_{\text{x}}$-$M$. \emph{Lower
       right:} $M$-$Y_{\text{x}}$. }
   \label{fig:morphrelation}%
\end{figure*}
\begin{figure*}
  \centering
  \includegraphics[width=0.49\textwidth]{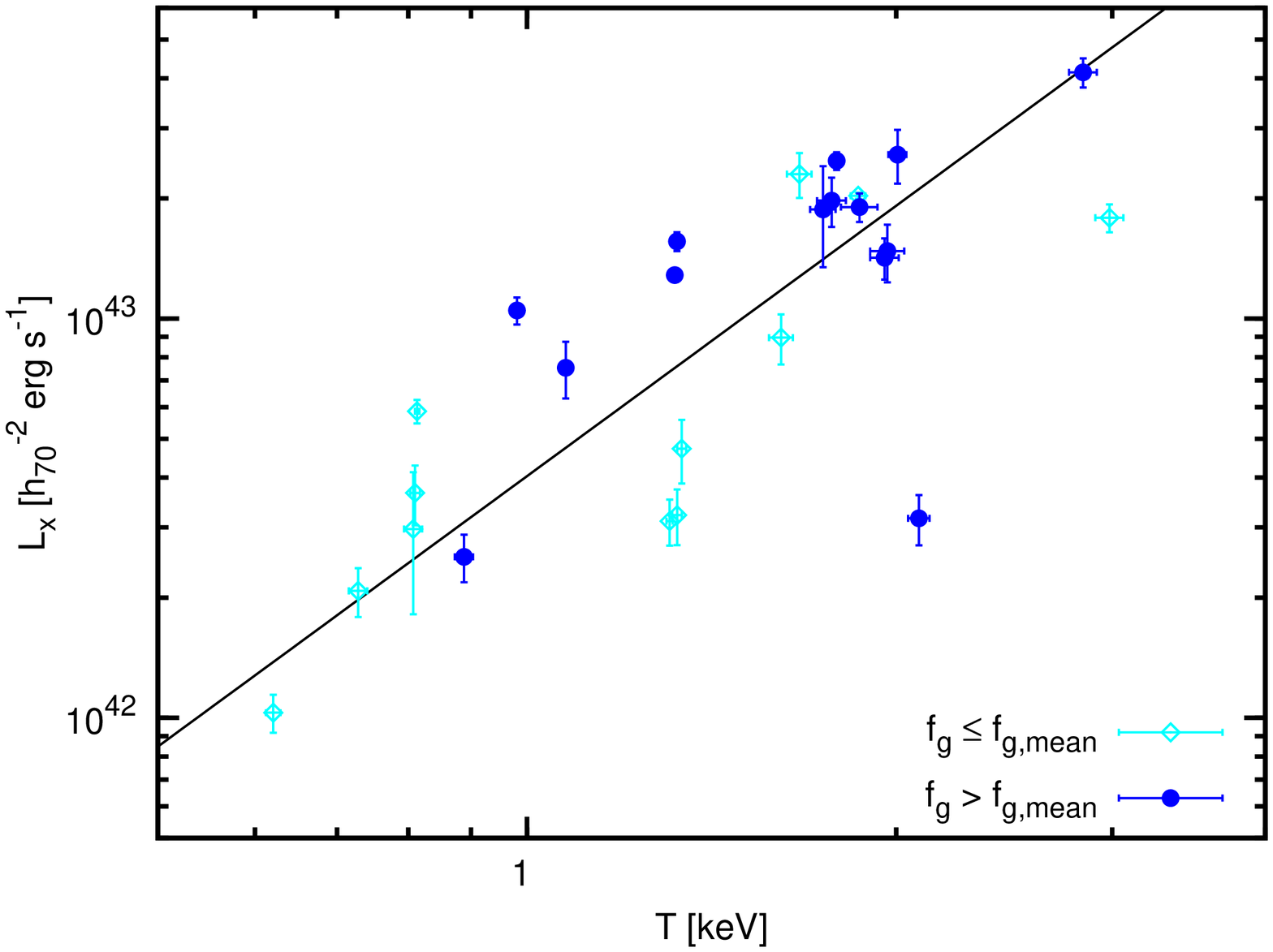}
  \includegraphics[width=0.49\textwidth]{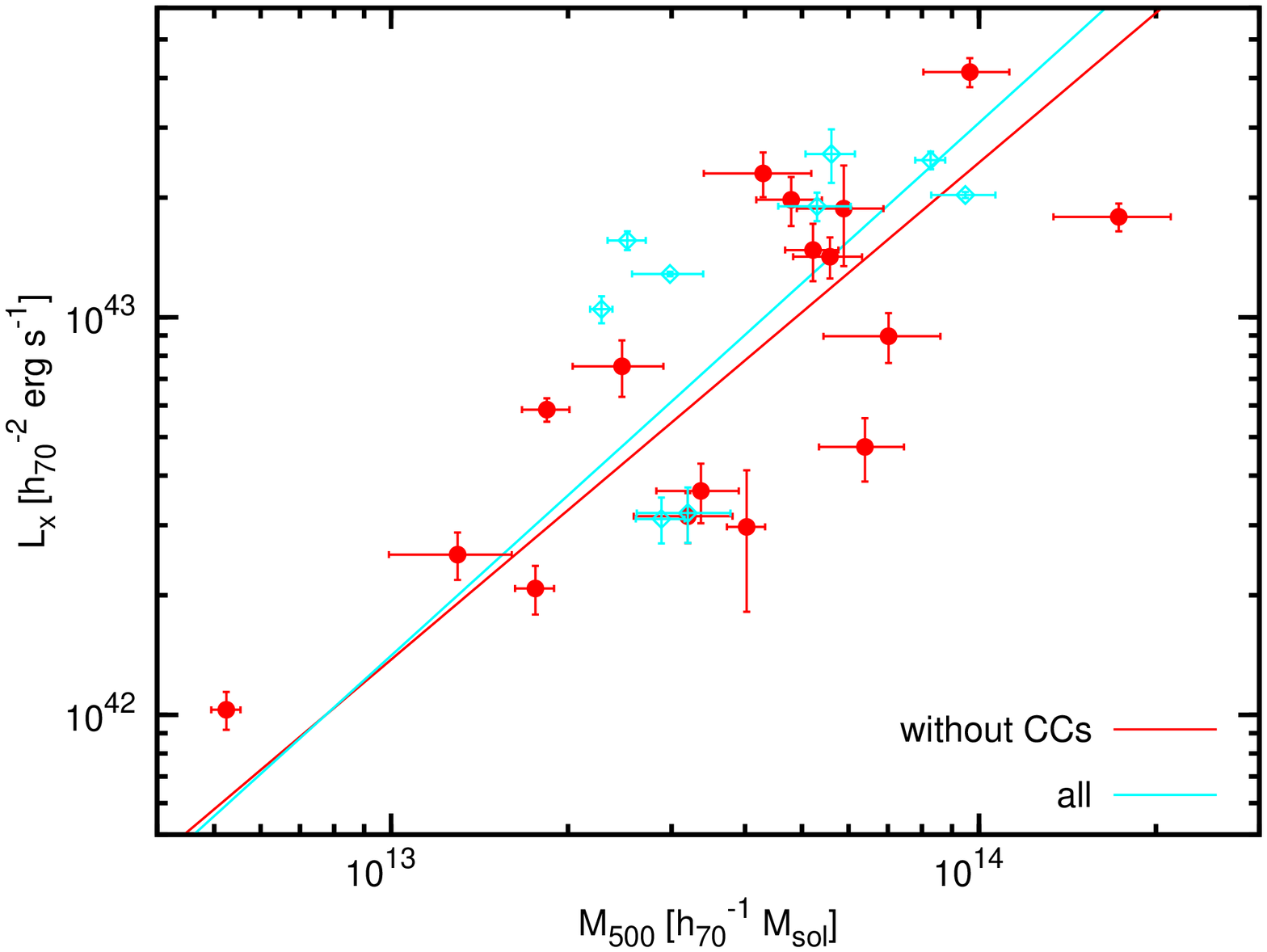}\quad
  \caption{\emph{Left:} $L_{\text{x}}$-$T$ relation, objects with gas
    mass fraction above and below the mean value are plotted as
    different symbols (lower $f_{\text{g}}$ are light blue open
    diamonds, others are filled blue circles). \emph{Right:}
    $L_{\text{x}}$-$M$ relation, comparison of fit with and without
    cool core objects (light blue open diamonds).}
  \label{fig:ccfgrelation}%
\end{figure*}
\begin{table*}
  \caption{Fit results selected by morphology and excluding cool core objects.}
  \begin{center}
    \setlength\extrarowheight{2pt}
    \begin{tabularx}{\linewidth}{X X XXllll}
      \hline\hline
      Relation ($Y$-$X$) & Sample & a & b & $\sigma_{\text{tot}}$ (X) & $\sigma_{\text{tot}}$ (Y) & $\sigma_{\text{int}}$ (X) & $\sigma_{\text{int}}$ (Y) \\ \hline
$L_{\text{x}}$-$T$ & regular & $2.34 \pm 0.32$ & $0.02 \pm 0.16$ & $0.128$ & $0.300$ & $0.126$ & $0.294$ \\ 
 & without CCs & $2.03 \pm 0.22$ & $-0.13 \pm 0.11$ & $0.125$ & $0.255$ & $0.121$ & $0.245$ \\ 
 & all & $2.25 \pm 0.21$ & $-0.02 \pm 0.09$ & $0.122$ & $0.275$ & $0.119$ & $0.268$ \\ \hline
$M_{500}$-$T$ & regular & $1.76 \pm 0.29$ & $0.19 \pm 0.12$ & $0.112$ & $0.197$ & $0.107$ & $0.189$ \\ 
 & without CCs & $1.63 \pm 0.24$ & $0.14 \pm 0.08$ & $0.122$ & $0.199$ & $0.115$ & $0.187$ \\ 
 & all & $1.68 \pm 0.20$ & $0.15 \pm 0.07$ & $0.104$ & $0.176$ & $0.098$ & $0.165$ \\ \hline
$L_{\text{x}}$-$M_{500}$ & regular & $1.34 \pm 0.19$ & $-0.22 \pm 0.11$ & $0.222$ & $0.298$ & $0.211$ & $0.283$ \\ 
 & without CCs & $1.25 \pm 0.22$ & $-0.31 \pm 0.12$ & $0.230$ & $0.287$ & $0.214$ & $0.267$ \\ 
 & all & $1.34 \pm 0.18$ & $-0.21 \pm 0.09$ & $0.226$ & $0.303$ & $0.214$ & $0.287$ \\ \hline
$M_{500}$-$Y_{\text{x},500}$ & regular & $0.55 \pm 0.08$ & $-0.63 \pm 0.05$ & $0.383$ & $0.210$ & $0.369$ & $0.202$ \\ 
 & without CCs & $0.52 \pm 0.07$ & $-0.62 \pm 0.06$ & $0.374$ & $0.196$ & $0.351$ & $0.184$ \\ 
 & all & $0.53 \pm 0.06$ & $-0.62 \pm 0.05$ & $0.369$ & $0.194$ & $0.350$ & $0.184$ \\ \hline
$f_{g,500}$-$T$ & regular & $1.45 \pm 0.62$ & $0.22 \pm 0.26$ & $0.203$ & $0.294$ & $0.198$ & $0.286$ \\ 
 & without CCs & $1.18 \pm 0.28$ & $0.08 \pm 0.11$ & $0.218$ & $0.258$ & $0.209$ & $0.247$ \\ 
 & all & $1.32 \pm 0.32$ & $0.11 \pm 0.12$ & $0.222$ & $0.294$ & $0.216$ & $0.286$ \\ \hline
\end{tabularx}
    \label{tab:fitsmorphcc}
  \end{center}
\end{table*}
We also tested several scaling relations for a possible influence of
morphology and dynamical state, using the selection criteria
introduced in section \ref{sec:morphology}. We compared the whole
sample to one where unrelaxed objects are excluded, and to one
  where the cool core objects are excluded. Table
\ref{tab:fitsmorphcc} contains the best-fit results for the
$L_{\text{x}}$-$T$, $M$-$T$, $L_{\text{x}}$-$M$, and
$M$-$Y_{\text{x}}$ relations, plotted in Fig.\
\ref{fig:morphrelation} for the morphology selection, as well as
  the $f_{\text{g}}$-$T$ relation.

We found that the morphology selection had no significant effect on
any of these relations. For all relations the fit for regular systems
has the larger uncertainty, mainly because there are fewer objects
constraining the fit. There is however a trend for these objects to be
hotter, e.\ g.\ none of the unrelaxed groups has a temperature below
$\unit[1.5]{keV}$. Merging systems are not scattered
  significantly more strongly than groups that appear to be regular,
  and excluding disturbed systems only slightly improves the scatter in
  the $L_{\text{x}}$-$M$ and the $f_{\text{g}}$-$T$ relations, while
  in the other relations the scatter actually increases, probably due
  to statistics. For example, the $M$-$T$ relation fit using only
  regular groups has an intrinsic scatter of $0.189$ in $M$ and
  $0.107$ in $T$, while for the complete relation the scatter is only
  $0.165$ and $0.098$, respectively.

We also compared the best-fit relations to the fits excluding
  objects with a cool core, in order to determine whether cool cores
  have an effect on global properties apart from the temperature
  (where cool cores have been excluded). The results are again shown
  in Table \ref{tab:fitsmorphcc}. We found that the cool cores do not
  significantly affect the relations, the largest impact is on the
  luminosity-temperature relation, which is flatter when the CC
  objects are excluded ($2.03\pm0.22$ compared to $2.25\pm0.21$), but
  the changes are not significant within the uncertainties. The
  $L_{\text{x}}$-$M$ relation is shown as an example in
  Fig. \ref{fig:ccfgrelation} (right), as a relation that does not
  explicitly contain $T$. Here the fits are also consistent within the
  errorbars ($a=1.25\pm0.22$ and $b=-0.31\pm0.12$ vs.\ $a=1.34\pm0.18$
  and $b=-0.21\pm0.09$ for the full sample). The main effect of
  excluding cool core objects is reducing the scatter in
  $L_{\text{x}}$ (from $0.287$ to $0.267$ in the $L_{\text{x}}$-$M$
  relation) and $f_{\text{g}}$ (from $0.286$ to $0.247$ in the
  $f_{\text{g}}$-$T$ relation). We observe that objects with a cool
  core tend to have a higher luminosity, for instance the mean $\Delta
  L_{\text{x}}=\log L_{\text{x}}-\left(a\cdot \log T +b\right)$ is
  $0.08$ for the cool core objects and $-0.04$ for the others, meaning
  the CC objects mostly lie above the $L_{\text{x}}$-$T$
  relation. This is a well-known phenomenon, but apart from this they
  do not particularly stand out in any of the scaling relations.
%
%
%
\section{Discussion}\label{sec:discussion}
In section \ref{sec:results:profiles} we have shown universal behavior
in the radial temperature and metallicity profiles scaled to a
characteristic radius. However, there is an increase in scatter and a
large variability in the inner parts of the temperature profiles
($\lesssim 0.05 r_{500}$), where some groups exhibit a drop in
temperature while others appear to remain flat down to the very
center. A variability of the inner temperature slopes has also
  been observed in the \emph{HIFLUGCS} clusters, by Hudson et al.\
  (2010). It is difficult to say whether this is due to the cool
  core/non-cool core bimodality often reported for clusters, since the
  profiles have only a small number of radial bins and in many cases
  do not extend down to the very center. Unfortunately, this is a
  typical problem when analyzing groups, as the profiles cannot be
  resolved as finely as in clusters, due to the much lower count rates
  even at the center, and the low signal to noise in the
  outskirts. Those profiles that do resolve the core however have
  clear trends for either constant temperature or a cool core. It is
  possible that the bimodality is either not as pronounced in groups
  as it is in clusters, or it is simply not clearly detectable in this
  sample because of the low count rates and would require more data
  with longer exposure times.

With regards to a systematic difference or ``break'' in the scaling
relations due to feedback or other non-gravitational processes, our
results are somewhat ambiguous (see section
\ref{sec:results:scaling}). When comparing fits of the group sample
with cluster relations we found that in virtually all parameters
  the intrinsic scatter is larger for groups, while the slopes of the
  scaling relations are still consistent with the cluster fits. The
  strong temperature-dependence of the gas mass fraction however
  indicates a systematic difference in the physical
  properties of clusters and groups. If groups generally have lower
  $f_{\text{g}}$ one would expect to see a stronger influence of
  the galactic component on the whole system and especially the
  central region via AGN feedback, compared to clusters, where the ICM
  component is dominant over the galaxies in terms of mass. 

  Our data do not show a clear break in a broken powerlaw sense, but
  the scaling relations are not completely consistent either. It is
  possible there is a gradual, continuous shift which may be harder to
  detect due to the increase in scatter, and which leads to different
  results depending on which objects are compared. For example, we
found that the $L_{\text{x}}$-$T$ slope of our group sample selected
by luminosity is consistent with that of the \emph{HIFLUGCS}
clusters, as well as the samples of Osmond \& Ponman (2004) and
  Shang \& Scharf (2009), but the relation for systems below
$\unit[3]{keV}$ is steeper than the slope for hotter objects.
  Using a broken powerlaw function however does not significantly
  improve the fit.  Fitting the groups together with the clusters
steepens the $L_{\text{x}}$-$T$ as well as the $M$-$T$ relation in
comparison to the pure cluster fits, which is in agreement with the
findings of \mbox{e.\ g.\ } Finoguenov et al.\ (2001). However the
best-fit normalization for groups is rather low, especially of the
$M$-$T$ relation, and this may be sufficient to cause the observed
steepening.

In section \ref{sec:results:scaling} we have noted that the
normalization of our $M_{500}$-$T$ relation is $\lesssim\unit[30]{\%}$
lower than, among others, the relation found by Vikhlinin et al.\
  (2009). However these authors used the $Y_{\text{x}}$-$M$ relation
  to estimate their masses, and their sampling only goes down to
  temperatures of $\sim\unit[2.5]{keV}$. But our $M$-$T$
  normalization is still lower than a comparable group relation
obtained by Sun et al.\ (2009), which is based solely on
  \emph{Chandra} data, like our sample. This discrepancy could be
caused by a multitude of effects, some of which we will briefly
discuss here.

Probably the most important known bias on the $M$-$T$ relation due to
selection is based on the difference between relaxed and unrelaxed
systems. As mentioned in the introduction, simulations indicate that
for a given mass merging clusters are observed to be cooler than
relaxed clusters, due to gas that has not yet been thermalized (e.\
g.\ Mathiesen \& Evrard 2001, Ventimiglia et al.\ 2008). Sun et al.'s
and Vikhlinin et al.'s samples are explicitly selected to be relaxed,
while our groups were not subjected to any selection by morphology or
dynamical state. However, following this chain of reasoning our groups
should be cooler than the others, not hotter, so this cannot be the
explanation. On the other hand, it is possible that this sample is
missing some of the fainter, cooler objects due to flux limits and
archive bias, and by chance appears to have an offset from the other
samples.

Another possible reason why observed group properties differ from
cluster scaling relations could be the limited radial extent out to
which group emission can be measured (e.\ g.\ Mulchaey 2000). While
group emission is in theory detectable out to large radii, the gas is
emitting at very low temperatures ($\lesssim \unit[0.5]{keV}$) and
hard to distinguish from the Galactic foreground, which is similar in
terms of both surface brightness and temperature. In addition, our
analysis is limited by \emph{Chandra}'s field-of-view. Due to
  these limitations, our profiles could only be traced out to
  $\sim 0.7 r_{500}$ on the average. There has been some
computational and observational evidence indicating that SBPs steepen
at large radii, even more strongly than a double $\beta$-model
accounts for (e.\ g.\ Vikhlinin et al.\ 2006), which could bias
low our mass measurements. This would however not explain an
  offset between our results and other work that relies on
  \emph{Chandra} data as well, like Sun et al.\ (2009). Furthermore,
  other investigators have found the cluster density profiles to be
  consistent with (Humphrey et al.\ 2011) or even flatter than
  predictions (Kawaharada et al.\ 2010, Simionescu et al.\ 2011, Urban
  et al.\ 2011).

Fortunately there is some overlap between our groups and those of Sun
et al.\ (2009) and also Gastaldello et al.\ (2007). Seven of our
groups are present in the former sample\footnote{A0160, A1177, MKW4,
  NGC1550, NGC6269, RXCJ1022, RXCJ2214}, and six in the
latter\footnote{ESO552020, MKW4, NGC533, NGC1550, NGC4325, NGC5129},
so we can perform a quantitative direct comparison between the
results of the different temperature and mass measurements. When
comparing $T$ values for the groups overlapping with Sun et al.'s work
(Gastaldello et al.\ have not published temperatures for their
sample), we found that on average our temperatures are in agreement
($10^{\langle\log T-\log
T_{\text{Sun}}\rangle}=1.09^{+0.13}_{-0.12}$). Differences in the
individual cluster temperatures may be caused by the different
temperature determinations applied. While we fitted the (projected)
global temperature using the whole observed area, Sun et al.\ used the
deprojection method developed by Vikhlinin et al.\ (2006a and 2006b)
to fit 3D temperature profiles. This method is likely to yield lower
temperatures because it puts more weight on cooler gas components at
large radii. Sun et al.\ excluded the inner core regions
$<0.15\,r_{500}$, which is comparable to our cuts, so we do not expect
cool gas at the group centers to make a large difference. It is
possible however that the different background treatments (blank-sky
vs.\ stowed background files) also have an effect on the temperature
measurements.

When comparing the masses, we find that our values are again
consistent with other work ($10^{\langle \log M-\log
  M_{\text{Sun}}\rangle}=1.10^{+0.87}_{-0.49}$ and $10^{\langle \log M
  -\log M_{\text{Gastaldello}}\rangle}=0.99^{+0.61}_{-0.38}$,
respectively), albeit with quite large statistical spread. So we
conclude that since the individual mass and temperature values
  are consistent the difference in normalization of the $M$-$T$
relation is most likely caused by incompleteness of the samples and/or
the high scatter in properties of galaxy groups. We also point
  out that the combined fit of the groups with the \emph{HIFLUGCS}
  clusters is in good agreement with the other relations.

Of the scaling relations investigated here, the relation between
luminosity and mass is the most useful for future cluster cataloguing
missions like \emph{eROSITA}. For tens of thousands of new
  detections $L_{\text{x}}$ is the easiest X-ray property to measure,
  and which can be determined from even a small number of
  photons. In this context it is convenient that our results
indicate that the cluster relation holds also for low-mass
objects. The best fit $L_{\text{x}}$-$M_{500}$ relation for groups in
the low-mass range agrees well with the \emph{HIFLUGCS} cluster
relation, while being shallower than the fit found for
the \emph{LoCuSS} clusters, as well as the relations found by
Vikhlinin et al.\ (2009) and Pratt et al.\ (2009). However we point
out that the \emph{LoCuSS} clusters all lie in a narrow range both in
luminosity and mass and therefore by themselves cannot constrain the
slope of the $L_{\text{x}}$-$M$ relation very well, and the authors
of the latter two publications did not measure the masses individually
but estimated $M_{500}$ from the $Y_{\text{x}}$-$M$ relation.

For comparison with previous publications we also show our results for
the $L_{\text{x}}$-$M_{200}$ relations, although the temperature and
surface brightness profiles had to be extrapolated considerably to
determine the mass within $r_{200}$ and we expect the
  uncertainties to be large. Again we found reasonable agreement
between the group relation and \emph{HIFLUGCS}, as well as the
relations found by Stanek et al.\ (2006, compromise model) and Rykoff
et al.\ (2008). Stanek et al.'s concordance model appears to
significantly underestimate the normalization, which can be explained
by the difference in the assumed cosmological models, as has been
argued by Reiprich (2006).

The $M$-$Y_{\text{x}}$, $L_{\text{x}}$-$Y_{\text{x}}$, and
$M_{\text{g}}$-$M$ relations also are consistent for groups and
  clusters, although the scatter is much larger for groups. Including
  groups into the relations does not change the fits of the
  \emph{HIFLUGCS} clusters.

We observed a strong correlation between temperature and gas mass
fraction, which is in agreement with e.\ g.\ Reiprich (2001),
Gastaldello et al.\ (2007), and Pratt et al.\ (2009), and could
explain naturally why the X-ray emitting gas in groups can only be
traced out to smaller physical radii than in clusters. The measured group
gas fractions are also lower than the typical cluster $f_{\text{g}}$
around $0.1$ (e.\ g.\ Vikhlinin et al.\ 2009a). Sun et al.\ (2009)
pointed out this may be caused by a central drop in $f_{\text{g}}$,
and might not be a global effect. This is in in principle in agreement
with our finding that the gas fraction increases with radius, which
should be considered when using $f_{\text{g}}$ to determine
cosmological parameters. We also find the $f_{\text{g}}$-$T$
relation within $r_{2500}$ to be significantly different from
constant, although the slope is flatter than for
$r_{500}$.

We have quantitatively confirmed the expectation that, compared to
clusters, groups have a larger intrinsic scatter in properties such as
luminosity or temperature, which clearly exceeds the statistical
uncertainties. This increases the scatter in derived parameters like
total mass and even the $Y_{\text{x}}$ parameter, which for clusters
is thought to be the most robust against scatter due to merger
activity. But for our sample the $Y_{\text{x}}$ scaling relations the
scatter is not reduced compared to other relations, but is even
actually larger. Interestingly, in the $f_{\text{g}}$-$T$ relation
the scatter in $T$ was actually quite a bit lower for the group sample
than for the clusters, but not for $f_{\text{g}}$. We assume this is
an effect of the much steeper slope.

We did not find dynamical state and morphology to have any significant
effect on the relations, and merging objects are apparently not
  responsible for the large scatter. Concerning the impact of cool
  cores, by excluding the core region in the temperature analysis we
  have in principle removed any possible bias on temperature. On the
  other hand it was not feasable to remove the cores from the
  luminosity measurements. The original \emph{ROSAT} observations have
  a low spatial resolution, comparable to the core regions removed for
  the temperature determination in our analysis, and \emph{Chandra}
  data do not extend out to $r_{500}$ due to the small field of
  view. In addition, for the purpose of applying our relation to, for
  example, \emph{eROSITA}, excluding the centers would not necessarily
  be useful since the clusters we compare our relation to will not
  have the cores removed, either.

  Overall, we found that objects with a central temperature drop tend
  to have higher luminosities and lie above for instance the
  $L_{\text{x}}$-$T$ relation. This is not surprising since cool,
  dense gas produces the most X-rays and consequently cool cores
  clusters are generally found to be brighter than non-cool cores, and are
  more likely to be detected. Excluding the cool core objects from the
  fits has however not significantly changed the best fit results for
  our scaling relations (see section \ref{sec:compmorph}).

  Therefore it seems more likely that scatter due to baryonic physics
  in the core regions, and not substructure and merger bias, is
  responsible for the large uncertainties in the fit results. For
  instance, the X-ray luminosities have been measured without applying
  any core exclusion and thus may be subject to large scatter due to
  galactic influences like AGN feedback. While this holds true for
  both the groups and the \emph{HIFLUGCS} clusters, it is possible
  either that core properties are generally more variable in groups,
  or that any variation in the strength of feedback processes such as
  heating has a stronger impact on cooler systems with lower gas
  masses (e.\ g.\ Mittal et al.\ 2011). The latter argument could
  explain both the increased scatter and the steepening of the
  $L$-$T$ relation for objects with $T<\unit[3]{keV}$.

We note that properties measured out to $r_{2500}$ have systematically
lower scatter than those determined out to $r_{500}$. This is not
surprising, since the necessary extrapolations add a large uncertainty
to the measurements, because it is not possible to reliably measure
whether group profiles at large radii are shaped like cluster profiles
or perhaps drop off more quickly. These limits may be improved by
including more data into the analysis, for instance observations taken
with \emph{XMM-Newton} or \emph{Suzaku}.

A comparison between groups and clusters in addition depends on the
criteria used to distinguish the two classes, as the transition is a
smooth one. We compiled our group sample using a luminosity limit, but
for instance for the $L_{\text{x}}$-$T$ relation we found more
conclusive results when dividing the objects by temperature, perhaps
partly due to the higher scatter in $L_{\text{x}}$ for the cooler
systems. This indicates differences in selection criteria may be a
reason why different authors find inconsistent and even contradictory
results. Therefore it could be useful, but is beyond the scope of this
paper, to test and compare a number of selection parameters, such as
temperature, luminosity, velocity dispersion, richness, virial radius,
gas mass, or total mass. In the next section, we test the
  influence of luminosity, flux, and redshift cuts on our sample.
%
\subsection{Selection Effects}\label{sec:malmquist}
\begin{figure}[ht]
\begin{center}
\resizebox{0.9\columnwidth}{!}{\includegraphics{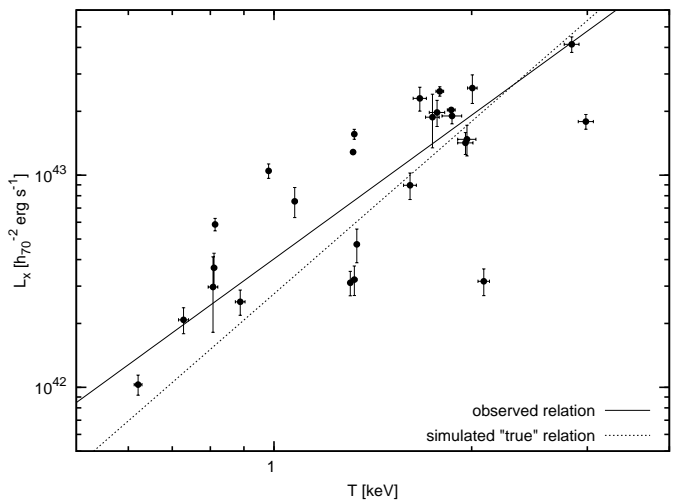}}
\end{center}
\caption{Observed $L_{\text{x}}$-$T$ relation and ``true''
  relation according to Monte Carlo simulations.}
\label{fig:lt_malmquist}
\end{figure}Since our sample is derived from flux-limited parent samples, we
expect our results to be susceptible to Malmquist bias, and perhaps
also ``archival bias.''  Another possible selection bias may result
from the luminosity cut applied in the construction of the group
sample, since systems scattered above the luminosity limit are
excluded, effectively placing more weight on fainter objects. However
these effects cannot be properly corrected for in a sample that is
incomplete, as is the case for our and most other published
samples. We hope to overcome these biases by using a statistically
complete sample, but even with this incomplete sample we can make
  an estimation of how large an effect the Malmquist bias has on the
  scaling relations, as well as the cuts in $L_{\text{x}}$ and $z$. We
  performed this test for the luminosity-temperature relation,
  following a similar procedure as described in Mittal et al.\ (2011).

  Malmquist bias is caused by the fact that using a flux limit to
  select a sample of objects favors bright objects because of
  intrinsic scatter. In order to measure this effect and the other
  aforementioned biases for our sample we created a simulated group
  sample and iteratively varied the input $L_{\text{x}}$-$T$ slope
  and normalization until the fake data reproduced the observed
  relation. For this we first created a mock cluster sample in the
  redshift range of 0.01 to 0.2, such that the number of objects grew
  as luminosity distance $D_{\text{L}}$ cubed, corresponding to a
  random homogeneous distribution in a spherical volume with radius
  $D_{\text{L}}$. We then used the powerlaw temperature function
  measured by Markevitch (1998) to assign temperatures to the objects,
  in the range $\unit[0.5-10]{keV}$. These temperatures were converted
  to luminosities using the respective $L_{\text{x}}$-$T$ relation to
  be tested, and applying the measured intrinsic and statistical
  scatter (we did not consider evolution effects, since our sample is
  nearby). We then applied a flux cut of $\unit[2.5 \cdot
  10^{-12}]{erg\,s^{-1}\,cm^{-2}}$ and a luminosity cut of
  $L_{\text{x}}<2.55\cdot\unit[10^{43}\,h^{-2}_{70}]{erg\,s^{-1}}$, to
  reproduce the selections made for the actual group sample. The total
  number of objects was chosen such that after the cuts the sample had
  on average the same number of objects as the real sample
  ($\sim20-30$). 

  The input $L_{\text{x}}$-$T$ parameters that produced the measured
  relation ($a=2.25$, $b=-0.02$) are $a=2.70$ and $b=0.03$. This means
  that the observed relation is higher in normalization in the range
  in which our group sample lies (see Fig.\ \ref{fig:lt_malmquist}),
  which is what is expected due to Malmquist bias. The observed
  relation is also flatter, which is probably an effect of the
  luminosity cut mentioned above. However at this point we cannot
  reliably correct for these biases since the sample is not
  statistically complete. For the purpose of comparing groups
  and clusters this implies that if the flattening due to the
  luminosity cut is a significant influence the actual group
  $L_{\text{x}}$-$T$ relation could actually be steeper than the cluster
  relation. This would also be supported by our observation that applying
  a temperature cut instead of the luminosity cut results in a
  significantly steeper group relation (see section \ref{sec:results:scaling}).
%
\subsection{Note on Model Choice}\label{sec:modelchoice}
We discuss briefly to what extent different model choices affect the
final mass result, and justify the direct comparison to the
\emph{HIFLUGCS} measurements. We found, in agreement with Xue \& Wu
(2000a), that fitting a double $\beta$-model instead of a single one
results in, on the average, higher values of $\beta$ and hence higher
masses. In most cases the double model gives a much better and more
accurate fit because it uses separate components to describe the
central region, which may have excess emission from a cooling core,
and the outer parts, which are more representative of the overall
density gradient.

We note that on the other hand including the temperature gradient
decreases the mass compared to the isothermal model. We found the mass values
determined using the single $\beta$ and isothermal model are on average lower
than the double $\beta$-model with the temperature gradient by a factor of
$10^{\langle \log M_{1\beta\text{, iso}} - \log M_{2\beta\text{, d}T} \rangle}
=0.88^{+0.88}_{-0.44}$, which is consistent with unity. In addition we have confirmed
that the best-fit slopes for the $L$-$M$ relation are consistent within the
errors.

In particular, we argue our mass measurements can be compared to the
\emph{HIFLUGCS} masses from Reiprich \& B\"ohringer (2002), even
though these were determined assuming an isothermal profile, fit with
single $\beta$-models only. On the other hand, our masses are mostly
calculated with double $\beta$-models, taking into account the
temperature gradients. This is the same procedure as applied for the
\emph{LoCuSS} clusters (Zhang et al.\ 08), which we also include in
the plots for comparison.
%
%
%
\section{Summary and Conclusions}\label{sec:summary}
Using a luminosity limit of
$L_{\text{x}}<2.55\cdot\unit[10^{43}\,h^{-2}_{70}]{erg\,s^{-1}}$ and a
redshift cut of $z >0.01$, we have compiled a statistically complete sample of
X-ray galaxy groups with the goal of investigating a possible
``break'' between scaling relations of groups and clusters, which are
important cosmological tools. For this work, we have reduced and analyzed
\emph{Chandra} observations for a subsample of 26 groups, extracted radial
temperature, metallicity, and surface brightness profiles, and determined the
global temperature, metallicity, total mass, gas mass, gas mass fraction, and
$Y_{\text{x}}$ parameter for each object. We have investigated the
$L_{\text{x}}$-$T$, $M$-$T$, $L_{\text{x}}$-$M$, $M_{\text{g}}$-$M$,
$M$-$Y_{\text{x}}$, $L_{\text{x}}$-$Y_{\text{x}}$, and $f_{\text{g}}$-$T$
scaling relations, and compared these to several cluster relations. We
summarize our results as follows:
\begin{enumerate}
\item The group temperature profiles scaled to $r_{500}$ decrease universally
  beyond a radius of $\gtrsim0.05\,r_{500}$, with larger scatter in the
  central regions.
\item The $L_{\text{x}}$-$T$ and $M$-$T$ relations are consistent
  for groups and clusters, with a slight steepening when groups and
  clusters are fitted together, in agreement with e.\ g.\ Finoguenov
  et al.\ (2001). Our $M$-$T$ normalization appears to be a bit lower
  than other comparable work, but on average our results are
  consistent and we attribute the offset to mainly scatter and selection effects.
\item The most useful relation to estimate masses for large cluster
  samples, $L_{\text{x}}$-$M$, is also found to be consistent within
  the errors for both groups and clusters.
\item The $L_{\text{x}}$-$T$ relation is steeper for groups selected
  via a temperature cut at $T<\unit[3]{keV}$, compared to hotter
  objects. This could indicate that cooler systems are more strongly
  affected by heating processes such as AGN feedback, supernovae, or
  cosmic rays.
\item We found a systematic drop of $f_{\text{g}}$ with temperature,
  significantly different from a constant relation, both within $r_{2500}$ and
  $r_{500}$. The gas fractions are lower than typical values for
  clusters ($\sim 0.1$, e.\ g.\ Vikhlinin et al.\ 2009a). The gas
  mass fractions are also lower at smaller radii, in agreement with
  the findings of Sun et al.\ (2009).
\item Groups generally have large scatter in all parameters and large
  uncertainties where radial profiles had to be extrapolated, which
  increases the uncertainties on the best-fit relations. This may be
  improved by including more objects and completing the group sample,
  and including more data, for instance taken with \emph{XMM-Newton}
  or \emph{Suzaku}. 
\item Dynamical state and cool cores have no significant effect
  on any of the scaling relations. This indicates that merging
  activity, dynamical state and cool cores do not have as strong
  an impact on groups as on high-mass systems, and the large scatter
  is probably due to a different effect, for instance the
    increasing influence of the galactic component on the ICM.
\item A quantitative test of the impact of selection effects on
    the $L_{\text{x}}$-$T$ relation showed that the observed group
    relation is higher in normalization and flatter than the actual
    relation. We argue that these effects are caused by Malmquist bias
    and the upper luminosity cut, respectively. However to reliably
    correct for these effects a statistically complete sample is
    necessary.
\end{enumerate}
In short, we have found some evidence for a systematic difference
between the group and cluster regimes, however the most commonly used
scaling relations do not seem to be strongly affected by this. The
strongest effects appear to be the lower gas fractions which points to
a less dominant role of the ICM in groups than in clusters and
  stronger influence of the galaxies, and the significantly larger
scatter in all relations, which is likely not caused by merging and
irregularity, but rather by non-gravitational galactic physics
in the core. This large scatter on group properties is highly
problematic, as it may generate spurious effects or mask out actual
trends. We did not find a hard powerlaw ``break'', but it is
  possible that there is a gradual change that is obscured by the
  scatter, and by bias due to selection effects. Therefore we will
work to continue completing our sample, to be able to eliminate
selection biases and to produce more conclusive results.
%
%
%
%
\begin{acknowledgements}
The authors would like to thank the anonymous referee for both
  general advice and detailed suggestions on how to improve this publication.

This research has made use of data obtained from the Chandra Data Archive and
the Chandra Source Catalog, and software provided by the Chandra X-ray Center
(CXC) in the application packages CIAO, ChIPS, and Sherpa.

This research has made use of the NASA/IPAC Extragalactic Database
(NED) which is operated by the Jet Propulsion Laboratory, California
Institute of Technology, under contract with the National Aeronautics
and Space Administration.

The authors acknowledge support from the Deutsche Forschungsgemeinschaft 
through Emmy Noether research grant RE 1462/2, priority program 1177 grant 
RE 1462/4, Heisenberg grant RE 1462/5, and grant 1462/6.
\end{acknowledgements}
\bibliographystyle{aa}
\nocite{Akritas-Bershady:96}
\nocite{Allen-Schmidt-Fabian:01}
\nocite{Anders-Grevesse:89}
\nocite{Baldi-Forman-Jones:09}
\nocite{Balogh-Mazzotta-Bower:10}
\nocite{Boehringer-Voges-Huchra:00}
\nocite{Boehringer-Schuecker-Guzzo:04}
\nocite{Boehringer-Pratt-Arnaud:10}
\nocite{Borgani-Murante-Springel:04}
\nocite{Cavaliere-Fusco-Femiano:76}
\nocite{Chen-Reiprich-Boehringer:07}
\nocite{Colafrancesco-Giordano:07}
\nocite{Croston-Pratt-Boehringer:08}
\nocite{Dai-Bregman-Kochanek:10}
\nocite{Dave-Oppenheimer-Sivandam:08}
\nocite{Dell'Antonio-Geller-Fabricant:94}
\nocite{Drake-Merrifield-Sakelliou:00}
\nocite{Dunkley-Komatsu-Nolta:09}
\nocite{Ettori-Gastaldello-Leccardi:09}
\nocite{Finoguenov-Reiprich-Boehringer:01}
\nocite{Forbes-Ponman-Pearce:06}
\nocite{Fukazawa-Nakazawa-Isobe:01}
\nocite{Fukazawa-Kawano-Kawashima:04}
\nocite{Gastaldello-Buote-Humphrey:07}
\nocite{Giodini-Pierini-Finoguenov:09}
\nocite{Gitti-OSullivan-Giacintucci:10}
\nocite{Gu-Xu-Gu:07}
\nocite{Hardcastle-Kraft-Worrall:07}
\nocite{Hartley-Gazzola-Pearce:08}
\nocite{Helsdon-Ponman:00}
\nocite{Hudson-Henriksen-Colafrancesco:03}
\nocite{Hudson-Henriksen:03}
\nocite{Hudson-Reiprich-Clarke:06}
\nocite{Hudson-Mittal-Reiprich:10}
\nocite{Humphrey-Buote-Brighenti:11}
\nocite{Hwang-Mushotzky-Burns:99}
\nocite{Jeltema-Binder-Mulchaey:08}
\nocite{Jeltema-Hallman-Burns:08}
\nocite{Jetha-Sakelliou-Hardcastle:05}
\nocite{Kalberla-Burton-Hartmann:05}
\nocite{Kawaharada-Makishima-Takahashi:03}
\nocite{Kawaharada-Makishima-Kitaguchi:09}
\nocite{Kawaharada-Okabe-Umetsu:10}
\nocite{Khosroshahi-Jones-Ponman:04}
\nocite{Khosroshahi-Ponman-Jones:07}
\nocite{Kraft-Forman-Churazov:04}
\nocite{Kravtsov-Vikhlinin-Nagai:06}
\nocite{Leauthaud-Finoguenov-Kneib:10}
\nocite{Loken-Norman-Nelson:02}
\nocite{Lopes-deCarvalho-Kohl:09}
\nocite{Mahdavi-Boehringer-Geller:97}
\nocite{Mahdavi-Boehringer-Geller:00}
\nocite{Markevitch:98}
\nocite{Mathiesen-Evrard:01}
\nocite{Maughan:07}
\nocite{Maughan-Giles-Randall:11}
\nocite{Mittal-Hudson-Reiprich:09}
\nocite{Mittal-Hicks-Reiprich:11}
\nocite{Morita-Ishisaki-Yamasaki:06}
\nocite{Mulchaey-Zabludoff:98}
\nocite{Mulchaey-Zabludoff:99}
\nocite{Mulchaey:00}
\nocite{Murgia-Parma-deRuiter:01}
\nocite{Nagai-Vikhlinin-Kravtsov:07}
\nocite{Nagai-Kravtsov-Vikhlinin:07}
\nocite{Nakazawa-Makishima-Fukazawa:07}
\nocite{Nevalainen-Markevitch-Forman:00}
\nocite{Okabe-Zhang-Finoguenov:10}
\nocite{Osmond-Ponman:04}
\nocite{O'Sullivan-Vrtilek-Read:03}
\nocite{O'Sullivan-Vrtilek-Harris:07}
\nocite{Paolillo-Fabbiano-Peres:03}
\nocite{Pellegrini-Venturi-Comastri:03}
\nocite{Plagge-Benson-Ade:10}
\nocite{Ponman-Bourner-Ebeling:96}
\nocite{Pope:09}
\nocite{Pratt-Croston-Arnaud:09}
\nocite{Predehl-Andritschke-Boehringer:10}
\nocite{Rasmussen-Ponman-Mulchaey:06}
\nocite{Rasmussen-Ponman:07}
\nocite{Reiprich:01}
\nocite{Reiprich-Boehringer:02}
\nocite{Reiprich:06}
\nocite{Ricker-Sarazin:01}
\nocite{Riemer-Sorensen-Paraficz-Ferreira:09}
\nocite{Ritchie-Thomas:02}
\nocite{Russell-Ponman-Sanderson:07}
\nocite{Rykoff-Evrard-McKay:08}
\nocite{Sanderson-Ponman-Finoguenov:03}
\nocite{Sanderson-Ponman-O'Sullivan:06}
\nocite{Sato-Matsushita-Ishisaki:09}
\nocite{Sato-Kawaharada-Nakazawa:10}
\nocite{Shang-Scharf:09}
\nocite{Simionescu-Allen-Mantz:11}
\nocite{Snowden-Mushotzky-Kuntz:08}
\nocite{Spavone-Iodice-Longo:06}
\nocite{Stanek-Evrard-Boehringer:06}
\nocite{Subrahmanyan-Beasley-Goss:03}
\nocite{Sun-Forman-Vikhlinin:03}
\nocite{Sun-Voit-Donahue:09}
\nocite{Takizawa-Nagino-Matsushita:10}
\nocite{Tokoi-Sato-Ishisaki:08}
\nocite{Tovmassian-Plionis:09}
\nocite{Trinchieri-Breitschwerdt-Pietsch:07}
\nocite{Urban-Werner-Simionescu:11}
\nocite{Ventimiglia-Voit-Donahue:08}
\nocite{Vikhlinin-Markevitch-Murray:05}
\nocite{Vikhlinin-Kravtsov-Forman:06}
\nocite{Vikhlinin:06}
\nocite{Vikhlinin-Burenin-Ebeling:09}
\nocite{Vikhlinin-Kravtsov-Burenin:09}
\nocite{Worrall-Birkinshaw-Kraft:07}
\nocite{Xue-Wu:00a}
\nocite{Xue-Wu:00b}
\nocite{Yang-Ricker-Sutter:09}
\nocite{Zhang-Finoguenov-Boehringer:08}
\nocite{Zhang-Reiprich-Finoguenov:09}
\nocite{Zhang-Okabe-Finoguenov:10}
\nocite{Zhang-Andernach-Caretta:11}
\bibliography{Eckmiller-Chandra-groups}

\begin{thebibliography}{119}
\expandafter\ifx\csname natexlab\endcsname\relax\def\natexlab#1{#1}\fi

\bibitem[{{Akritas} \& {Bershady}(1996)}]{Akritas-Bershady:96}
{Akritas}, M.~G. \& {Bershady}, M.~A. 1996, \apj, 470, 706

\bibitem[{{Allen} {et~al.}(2001){Allen}, {Schmidt}, \&
  {Fabian}}]{Allen-Schmidt-Fabian:01}
{Allen}, S.~W., {Schmidt}, R.~W., \& {Fabian}, A.~C. 2001, \mnras, 328, L37

\bibitem[{{Anders} \& {Grevesse}(1989)}]{Anders-Grevesse:89}
{Anders}, E. \& {Grevesse}, N. 1989, \gca, 53, 197

\bibitem[{{Baldi} {et~al.}(2009){Baldi}, {Forman}, {Jones}, {Nulsen}, {David},
  {Kraft}, \& {Simionescu}}]{Baldi-Forman-Jones:09}
{Baldi}, A., {Forman}, W., {Jones}, C., {et~al.} 2009, \apj, 694, 479

\bibitem[{{Balogh} {et~al.}(2010){Balogh}, {Mazzotta}, {Bower}, {Eke},
  {Bourdin}, {Lu}, \& {Theuns}}]{Balogh-Mazzotta-Bower:10}
{Balogh}, M.~L., {Mazzotta}, P., {Bower}, R.~G., {et~al.} 2010, \mnras, 1842

\bibitem[{{B{\"o}hringer} {et~al.}(2010){B{\"o}hringer}, {Pratt}, {Arnaud},
  {Borgani}, {Croston}, {Ponman}, {Ameglio}, {Temple}, \&
  {Dolag}}]{Boehringer-Pratt-Arnaud:10}
{B{\"o}hringer}, H., {Pratt}, G.~W., {Arnaud}, M., {et~al.} 2010, \aap, 514,
  A32+

\bibitem[{{B{\"o}hringer} {et~al.}(2004){B{\"o}hringer}, {Schuecker}, {Guzzo},
  {Collins}, {Voges}, {Cruddace}, {Ortiz-Gil}, {Chincarini}, {De Grandi},
  {Edge}, {MacGillivray}, {Neumann}, {Schindler}, \&
  {Shaver}}]{Boehringer-Schuecker-Guzzo:04}
{B{\"o}hringer}, H., {Schuecker}, P., {Guzzo}, L., {et~al.} 2004, \aap, 425,
  367

\bibitem[{{B{\"o}hringer} {et~al.}(2000){B{\"o}hringer}, {Voges}, {Huchra},
  {McLean}, {Giacconi}, {Rosati}, {Burg}, {Mader}, {Schuecker}, {Simi{\c c}},
  {Komossa}, {Reiprich}, {Retzlaff}, \&
  {Tr{\"u}mper}}]{Boehringer-Voges-Huchra:00}
{B{\"o}hringer}, H., {Voges}, W., {Huchra}, J.~P., {et~al.} 2000, \apjs, 129,
  435

\bibitem[{{Borgani} {et~al.}(2004){Borgani}, {Murante}, {Springel}, {Diaferio},
  {Dolag}, {Moscardini}, {Tormen}, {Tornatore}, \&
  {Tozzi}}]{Borgani-Murante-Springel:04}
{Borgani}, S., {Murante}, G., {Springel}, V., {et~al.} 2004, \mnras, 348, 1078

\bibitem[{{Cavaliere} \& {Fusco-Femiano}(1976)}]{Cavaliere-Fusco-Femiano:76}
{Cavaliere}, A. \& {Fusco-Femiano}, R. 1976, \aap, 49, 137

\bibitem[{{Chen} {et~al.}(2007){Chen}, {Reiprich}, {B{\"o}hringer}, {Ikebe}, \&
  {Zhang}}]{Chen-Reiprich-Boehringer:07}
{Chen}, Y., {Reiprich}, T.~H., {B{\"o}hringer}, H., {Ikebe}, Y., \& {Zhang},
  Y.-Y. 2007, \aap, 466, 805

\bibitem[{{Colafrancesco} \& {Giordano}(2007)}]{Colafrancesco-Giordano:07}
{Colafrancesco}, S. \& {Giordano}, F. 2007, \aap, 466, 421

\bibitem[{{Croston} {et~al.}(2008){Croston}, {Pratt}, {B{\"o}hringer},
  {Arnaud}, {Pointecouteau}, {Ponman}, {Sanderson}, {Temple}, {Bower}, \&
  {Donahue}}]{Croston-Pratt-Boehringer:08}
{Croston}, J.~H., {Pratt}, G.~W., {B{\"o}hringer}, H., {et~al.} 2008, \aap,
  487, 431

\bibitem[{{Dai} {et~al.}(2010){Dai}, {Bregman}, {Kochanek}, \&
  {Rasia}}]{Dai-Bregman-Kochanek:10}
{Dai}, X., {Bregman}, J.~N., {Kochanek}, C.~S., \& {Rasia}, E. 2010, \apj, 719,
  119

\bibitem[{{Dav{\'e}} {et~al.}(2008){Dav{\'e}}, {Oppenheimer}, \&
  {Sivanandam}}]{Dave-Oppenheimer-Sivandam:08}
{Dav{\'e}}, R., {Oppenheimer}, B.~D., \& {Sivanandam}, S. 2008, \mnras, 391,
  110

\bibitem[{{dell'Antonio} {et~al.}(1994){dell'Antonio}, {Geller}, \&
  {Fabricant}}]{Dell'Antonio-Geller-Fabricant:94}
{dell'Antonio}, I.~P., {Geller}, M.~J., \& {Fabricant}, D.~G. 1994, \aj, 107,
  427

\bibitem[{{Drake} {et~al.}(2000){Drake}, {Merrifield}, {Sakelliou}, \&
  {Pinkney}}]{Drake-Merrifield-Sakelliou:00}
{Drake}, N., {Merrifield}, M.~R., {Sakelliou}, I., \& {Pinkney}, J.~C. 2000,
  \mnras, 314, 768

\bibitem[{{Dunkley} {et~al.}(2009){Dunkley}, {Komatsu}, {Nolta}, {Spergel},
  {Larson}, {Hinshaw}, {Page}, {Bennett}, {Gold}, {Jarosik}, {Weiland},
  {Halpern}, {Hill}, {Kogut}, {Limon}, {Meyer}, {Tucker}, {Wollack}, \&
  {Wright}}]{Dunkley-Komatsu-Nolta:09}
{Dunkley}, J., {Komatsu}, E., {Nolta}, M.~R., {et~al.} 2009, \apjs, 180, 306

\bibitem[{{Ettori} {et~al.}(2010){Ettori}, {Gastaldello}, {Leccardi},
  {Molendi}, {Rossetti}, {Buote}, \&
  {Meneghetti}}]{Ettori-Gastaldello-Leccardi:09}
{Ettori}, S., {Gastaldello}, F., {Leccardi}, A., {et~al.} 2010, \aap, 524, A68+

\bibitem[{{Finoguenov} {et~al.}(2001){Finoguenov}, {Reiprich}, \&
  {B{\"o}hringer}}]{Finoguenov-Reiprich-Boehringer:01}
{Finoguenov}, A., {Reiprich}, T.~H., \& {B{\"o}hringer}, H. 2001, \aap, 368,
  749

\bibitem[{{Forbes} {et~al.}(2006){Forbes}, {Ponman}, {Pearce}, {Osmond},
  {Kilborn}, {Brough}, {Raychaudhury}, {Mundell}, {Miles}, \&
  {Kern}}]{Forbes-Ponman-Pearce:06}
{Forbes}, D.~A., {Ponman}, T., {Pearce}, F., {et~al.} 2006, Publications of the
  Astronomical Society of Australia, 23, 38

\bibitem[{{Fukazawa} {et~al.}(2004){Fukazawa}, {Kawano}, \&
  {Kawashima}}]{Fukazawa-Kawano-Kawashima:04}
{Fukazawa}, Y., {Kawano}, N., \& {Kawashima}, K. 2004, \apjl, 606, L109

\bibitem[{{Fukazawa} {et~al.}(2001){Fukazawa}, {Nakazawa}, {Isobe},
  {Makishima}, {Matsushita}, {Ohashi}, \& {Kamae}}]{Fukazawa-Nakazawa-Isobe:01}
{Fukazawa}, Y., {Nakazawa}, K., {Isobe}, N., {et~al.} 2001, \apjl, 546, L87

\bibitem[{{Gastaldello} {et~al.}(2007){Gastaldello}, {Buote}, {Humphrey},
  {Zappacosta}, {Bullock}, {Brighenti}, \&
  {Mathews}}]{Gastaldello-Buote-Humphrey:07}
{Gastaldello}, F., {Buote}, D.~A., {Humphrey}, P.~J., {et~al.} 2007, \apj, 669,
  158

\bibitem[{{Giodini} {et~al.}(2009){Giodini}, {Pierini}, {Finoguenov}, {Pratt},
  {Boehringer}, {Leauthaud}, {Guzzo}, {Aussel}, {Bolzonella}, {Capak}, {Elvis},
  {Hasinger}, {Ilbert}, {Kartaltepe}, {Koekemoer}, {Lilly}, {Massey},
  {McCracken}, {Rhodes}, {Salvato}, {Sanders}, {Scoville}, {Sasaki}, {Smolcic},
  {Taniguchi}, {Thompson}, \& {the COSMOS
  Collaboration}}]{Giodini-Pierini-Finoguenov:09}
{Giodini}, S., {Pierini}, D., {Finoguenov}, A., {et~al.} 2009, \apj, 703, 982

\bibitem[{{Gitti} {et~al.}(2010){Gitti}, {O'Sullivan}, {Giacintucci}, {David},
  {Vrtilek}, {Raychaudhury}, \& {Nulsen}}]{Gitti-OSullivan-Giacintucci:10}
{Gitti}, M., {O'Sullivan}, E., {Giacintucci}, S., {et~al.} 2010, \apj, 714, 758

\bibitem[{{Gu} {et~al.}(2007){Gu}, {Xu}, {Gu}, {An}, {Wang}, {Zhang}, \&
  {Wu}}]{Gu-Xu-Gu:07}
{Gu}, J., {Xu}, H., {Gu}, L., {et~al.} 2007, \apj, 659, 275

\bibitem[{{Hardcastle} {et~al.}(2007){Hardcastle}, {Kraft}, {Worrall},
  {Croston}, {Evans}, {Birkinshaw}, \& {Murray}}]{Hardcastle-Kraft-Worrall:07}
{Hardcastle}, M.~J., {Kraft}, R.~P., {Worrall}, D.~M., {et~al.} 2007, \apj,
  662, 166

\bibitem[{{Hartley} {et~al.}(2008){Hartley}, {Gazzola}, {Pearce}, {Kay}, \&
  {Thomas}}]{Hartley-Gazzola-Pearce:08}
{Hartley}, W.~G., {Gazzola}, L., {Pearce}, F.~R., {Kay}, S.~T., \& {Thomas},
  P.~A. 2008, \mnras, 386, 2015

\bibitem[{{Helsdon} \& {Ponman}(2000)}]{Helsdon-Ponman:00}
{Helsdon}, S.~F. \& {Ponman}, T.~J. 2000, \mnras, 315, 356

\bibitem[{{Hudson} \& {Henriksen}(2003)}]{Hudson-Henriksen:03}
{Hudson}, D.~S. \& {Henriksen}, M.~J. 2003, \apjl, 595, L1

\bibitem[{{Hudson} {et~al.}(2003){Hudson}, {Henriksen}, \&
  {Colafrancesco}}]{Hudson-Henriksen-Colafrancesco:03}
{Hudson}, D.~S., {Henriksen}, M.~J., \& {Colafrancesco}, S. 2003, \apj, 583,
  706

\bibitem[{{Hudson} {et~al.}(2010){Hudson}, {Mittal}, {Reiprich}, {Nulsen},
  {Andernach}, \& {Sarazin}}]{Hudson-Mittal-Reiprich:10}
{Hudson}, D.~S., {Mittal}, R., {Reiprich}, T.~H., {et~al.} 2010, \aap, 513,
  A37+

\bibitem[{{Hudson} {et~al.}(2006){Hudson}, {Reiprich}, {Clarke}, \&
  {Sarazin}}]{Hudson-Reiprich-Clarke:06}
{Hudson}, D.~S., {Reiprich}, T.~H., {Clarke}, T.~E., \& {Sarazin}, C.~L. 2006,
  \aap, 453, 433

\bibitem[{{Humphrey} {et~al.}(2011){Humphrey}, {Buote}, {Brighenti}, {Flohic},
  {Gastaldello}, \& {Mathews}}]{Humphrey-Buote-Brighenti:11}
{Humphrey}, P.~J., {Buote}, D.~A., {Brighenti}, F., {et~al.} 2011, ArXiv
  e-prints

\bibitem[{{Hwang} {et~al.}(1999){Hwang}, {Mushotzky}, {Burns}, {Fukazawa}, \&
  {White}}]{Hwang-Mushotzky-Burns:99}
{Hwang}, U., {Mushotzky}, R.~F., {Burns}, J.~O., {Fukazawa}, Y., \& {White},
  R.~A. 1999, \apj, 516, 604

\bibitem[{{Jeltema} {et~al.}(2008{\natexlab{a}}){Jeltema}, {Binder}, \&
  {Mulchaey}}]{Jeltema-Binder-Mulchaey:08}
{Jeltema}, T.~E., {Binder}, B., \& {Mulchaey}, J.~S. 2008{\natexlab{a}}, \apj,
  679, 1162

\bibitem[{{Jeltema} {et~al.}(2008{\natexlab{b}}){Jeltema}, {Hallman}, {Burns},
  \& {Motl}}]{Jeltema-Hallman-Burns:08}
{Jeltema}, T.~E., {Hallman}, E.~J., {Burns}, J.~O., \& {Motl}, P.~M.
  2008{\natexlab{b}}, \apj, 681, 167

\bibitem[{{Jetha} {et~al.}(2005){Jetha}, {Sakelliou}, {Hardcastle}, {Ponman},
  \& {Stevens}}]{Jetha-Sakelliou-Hardcastle:05}
{Jetha}, N.~N., {Sakelliou}, I., {Hardcastle}, M.~J., {Ponman}, T.~J., \&
  {Stevens}, I.~R. 2005, \mnras, 358, 1394

\bibitem[{{Kalberla} {et~al.}(2005){Kalberla}, {Burton}, {Hartmann}, {Arnal},
  {Bajaja}, {Morras}, \& {P{\"o}ppel}}]{Kalberla-Burton-Hartmann:05}
{Kalberla}, P.~M.~W., {Burton}, W.~B., {Hartmann}, D., {et~al.} 2005, \aap,
  440, 775

\bibitem[{{Kawaharada} {et~al.}(2009){Kawaharada}, {Makishima}, {Kitaguchi},
  {Okuyama}, {Nakazawa}, {Matsushita}, \&
  {Fukazawa}}]{Kawaharada-Makishima-Kitaguchi:09}
{Kawaharada}, M., {Makishima}, K., {Kitaguchi}, T., {et~al.} 2009, \apj, 691,
  971

\bibitem[{{Kawaharada} {et~al.}(2003){Kawaharada}, {Makishima}, {Takahashi},
  {Nakazawa}, {Matsushita}, {Shimasaku}, {Fukazawa}, \&
  {Xu}}]{Kawaharada-Makishima-Takahashi:03}
{Kawaharada}, M., {Makishima}, K., {Takahashi}, I., {et~al.} 2003, \pasj, 55,
  573

\bibitem[{{Kawaharada} {et~al.}(2010){Kawaharada}, {Okabe}, {Umetsu},
  {Takizawa}, {Matsushita}, {Fukazawa}, {Hamana}, {Miyazaki}, {Nakazawa}, \&
  {Ohashi}}]{Kawaharada-Okabe-Umetsu:10}
{Kawaharada}, M., {Okabe}, N., {Umetsu}, K., {et~al.} 2010, \apj, 714, 423

\bibitem[{{Khosroshahi} {et~al.}(2004){Khosroshahi}, {Jones}, \&
  {Ponman}}]{Khosroshahi-Jones-Ponman:04}
{Khosroshahi}, H.~G., {Jones}, L.~R., \& {Ponman}, T.~J. 2004, \mnras, 349,
  1240

\bibitem[{{Khosroshahi} {et~al.}(2007){Khosroshahi}, {Ponman}, \&
  {Jones}}]{Khosroshahi-Ponman-Jones:07}
{Khosroshahi}, H.~G., {Ponman}, T.~J., \& {Jones}, L.~R. 2007, \mnras, 377, 595

\bibitem[{{Kraft} {et~al.}(2004){Kraft}, {Forman}, {Churazov}, {Laslo},
  {Jones}, {Markevitch}, {Murray}, \& {Vikhlinin}}]{Kraft-Forman-Churazov:04}
{Kraft}, R.~P., {Forman}, W.~R., {Churazov}, E., {et~al.} 2004, \apj, 601, 221

\bibitem[{{Kravtsov} {et~al.}(2006){Kravtsov}, {Vikhlinin}, \&
  {Nagai}}]{Kravtsov-Vikhlinin-Nagai:06}
{Kravtsov}, A.~V., {Vikhlinin}, A., \& {Nagai}, D. 2006, \apj, 650, 128

\bibitem[{{Leauthaud} {et~al.}(2010){Leauthaud}, {Finoguenov}, {Kneib},
  {Taylor}, {Massey}, {Rhodes}, {Ilbert}, {Bundy}, {Tinker}, {George}, {Capak},
  {Koekemoer}, {Johnston}, {Zhang}, {Cappelluti}, {Ellis}, {Elvis}, {Giodini},
  {Heymans}, {Le F{\`e}vre}, {Lilly}, {McCracken}, {Mellier},
  {R{\'e}fr{\'e}gier}, {Salvato}, {Scoville}, {Smoot}, {Tanaka}, {Van
  Waerbeke}, \& {Wolk}}]{Leauthaud-Finoguenov-Kneib:10}
{Leauthaud}, A., {Finoguenov}, A., {Kneib}, J., {et~al.} 2010, \apj, 709, 97

\bibitem[{{Loken} {et~al.}(2002){Loken}, {Norman}, {Nelson}, {Burns}, {Bryan},
  \& {Motl}}]{Loken-Norman-Nelson:02}
{Loken}, C., {Norman}, M.~L., {Nelson}, E., {et~al.} 2002, \apj, 579, 571

\bibitem[{{Lopes} {et~al.}(2009){Lopes}, {de Carvalho}, {Kohl-Moreira}, \&
  {Jones}}]{Lopes-deCarvalho-Kohl:09}
{Lopes}, P.~A.~A., {de Carvalho}, R.~R., {Kohl-Moreira}, J.~L., \& {Jones}, C.
  2009, \mnras, 399, 2201

\bibitem[{{Mahdavi} {et~al.}(1997){Mahdavi}, {Boehringer}, {Geller}, \&
  {Ramella}}]{Mahdavi-Boehringer-Geller:97}
{Mahdavi}, A., {Boehringer}, H., {Geller}, M.~J., \& {Ramella}, M. 1997, \apj,
  483, 68

\bibitem[{{Mahdavi} {et~al.}(2000){Mahdavi}, {B{\"o}hringer}, {Geller}, \&
  {Ramella}}]{Mahdavi-Boehringer-Geller:00}
{Mahdavi}, A., {B{\"o}hringer}, H., {Geller}, M.~J., \& {Ramella}, M. 2000,
  \apj, 534, 114

\bibitem[{{Markevitch}(1998)}]{Markevitch:98}
{Markevitch}, M. 1998, \apj, 504, 27

\bibitem[{{Mathiesen} \& {Evrard}(2001)}]{Mathiesen-Evrard:01}
{Mathiesen}, B.~F. \& {Evrard}, A.~E. 2001, \apj, 546, 100

\bibitem[{{Maughan}(2007)}]{Maughan:07}
{Maughan}, B.~J. 2007, \apj, 668, 772

\bibitem[{{Maughan} {et~al.}(2011){Maughan}, {Giles}, {Randall}, {Jones}, \&
  {Forman}}]{Maughan-Giles-Randall:11}
{Maughan}, B.~J., {Giles}, P.~A., {Randall}, S.~W., {Jones}, C., \& {Forman},
  W.~R. 2011, ArXiv e-prints

\bibitem[{{Mittal} {et~al.}(2011){Mittal}, {Hicks}, {Reiprich}, \&
  {Jaritz}}]{Mittal-Hicks-Reiprich:11}
{Mittal}, R., {Hicks}, A., {Reiprich}, T.~H., \& {Jaritz}, V. 2011, \aap, 532,
  A133+

\bibitem[{{Mittal} {et~al.}(2009){Mittal}, {Hudson}, {Reiprich}, \&
  {Clarke}}]{Mittal-Hudson-Reiprich:09}
{Mittal}, R., {Hudson}, D.~S., {Reiprich}, T.~H., \& {Clarke}, T. 2009, \aap,
  501, 835

\bibitem[{{Morita} {et~al.}(2006){Morita}, {Ishisaki}, {Yamasaki}, {Ota},
  {Kawano}, {Fukazawa}, \& {Ohashi}}]{Morita-Ishisaki-Yamasaki:06}
{Morita}, U., {Ishisaki}, Y., {Yamasaki}, N.~Y., {et~al.} 2006, \pasj, 58, 719

\bibitem[{{Mulchaey}(2000)}]{Mulchaey:00}
{Mulchaey}, J.~S. 2000, \araa, 38, 289

\bibitem[{{Mulchaey} \& {Zabludoff}(1998)}]{Mulchaey-Zabludoff:98}
{Mulchaey}, J.~S. \& {Zabludoff}, A.~I. 1998, \apj, 496, 73

\bibitem[{{Mulchaey} \& {Zabludoff}(1999)}]{Mulchaey-Zabludoff:99}
{Mulchaey}, J.~S. \& {Zabludoff}, A.~I. 1999, \apj, 514, 133

\bibitem[{{Murgia} {et~al.}(2001){Murgia}, {Parma}, {de Ruiter}, {Bondi},
  {Ekers}, {Fanti}, \& {Fomalont}}]{Murgia-Parma-deRuiter:01}
{Murgia}, M., {Parma}, P., {de Ruiter}, H.~R., {et~al.} 2001, \aap, 380, 102

\bibitem[{{Nagai} {et~al.}(2007{\natexlab{a}}){Nagai}, {Kravtsov}, \&
  {Vikhlinin}}]{Nagai-Kravtsov-Vikhlinin:07}
{Nagai}, D., {Kravtsov}, A.~V., \& {Vikhlinin}, A. 2007{\natexlab{a}}, \apj,
  668, 1

\bibitem[{{Nagai} {et~al.}(2007{\natexlab{b}}){Nagai}, {Vikhlinin}, \&
  {Kravtsov}}]{Nagai-Vikhlinin-Kravtsov:07}
{Nagai}, D., {Vikhlinin}, A., \& {Kravtsov}, A.~V. 2007{\natexlab{b}}, \apj,
  655, 98

\bibitem[{{Nakazawa} {et~al.}(2007){Nakazawa}, {Makishima}, \&
  {Fukazawa}}]{Nakazawa-Makishima-Fukazawa:07}
{Nakazawa}, K., {Makishima}, K., \& {Fukazawa}, Y. 2007, \pasj, 59, 167

\bibitem[{{Nevalainen} {et~al.}(2000){Nevalainen}, {Markevitch}, \&
  {Forman}}]{Nevalainen-Markevitch-Forman:00}
{Nevalainen}, J., {Markevitch}, M., \& {Forman}, W. 2000, \apj, 532, 694

\bibitem[{{Okabe} {et~al.}(2010){Okabe}, {Zhang}, {Finoguenov}, {Takada},
  {Smith}, {Umetsu}, \& {Futamase}}]{Okabe-Zhang-Finoguenov:10}
{Okabe}, N., {Zhang}, Y., {Finoguenov}, A., {et~al.} 2010, \apj, 721, 875

\bibitem[{{Osmond} \& {Ponman}(2004)}]{Osmond-Ponman:04}
{Osmond}, J.~P.~F. \& {Ponman}, T.~J. 2004, \mnras, 350, 1511

\bibitem[{{O'Sullivan} {et~al.}(2007){O'Sullivan}, {Vrtilek}, {Harris}, \&
  {Ponman}}]{O'Sullivan-Vrtilek-Harris:07}
{O'Sullivan}, E., {Vrtilek}, J.~M., {Harris}, D.~E., \& {Ponman}, T.~J. 2007,
  \apj, 658, 299

\bibitem[{{O'Sullivan} {et~al.}(2003){O'Sullivan}, {Vrtilek}, {Read}, {David},
  \& {Ponman}}]{O'Sullivan-Vrtilek-Read:03}
{O'Sullivan}, E., {Vrtilek}, J.~M., {Read}, A.~M., {David}, L.~P., \& {Ponman},
  T.~J. 2003, \mnras, 346, 525

\bibitem[{{Paolillo} {et~al.}(2003){Paolillo}, {Fabbiano}, {Peres}, \&
  {Kim}}]{Paolillo-Fabbiano-Peres:03}
{Paolillo}, M., {Fabbiano}, G., {Peres}, G., \& {Kim}, D.-W. 2003, \apj, 586,
  850

\bibitem[{{Pellegrini} {et~al.}(2003){Pellegrini}, {Venturi}, {Comastri},
  {Fabbiano}, {Fiore}, {Vignali}, {Morganti}, \&
  {Trinchieri}}]{Pellegrini-Venturi-Comastri:03}
{Pellegrini}, S., {Venturi}, T., {Comastri}, A., {et~al.} 2003, \apj, 585, 677

\bibitem[{{Plagge} {et~al.}(2010){Plagge}, {Benson}, {Ade}, {Aird}, {Bleem},
  {Carlstrom}, {Chang}, {Cho}, {Crawford}, {Crites}, {de Haan}, {Dobbs},
  {George}, {Hall}, {Halverson}, {Holder}, {Holzapfel}, {Hrubes}, {Joy},
  {Keisler}, {Knox}, {Lee}, {Leitch}, {Lueker}, {Marrone}, {McMahon}, {Mehl},
  {Meyer}, {Mohr}, {Montroy}, {Padin}, {Pryke}, {Reichardt}, {Ruhl},
  {Schaffer}, {Shaw}, {Shirokoff}, {Spieler}, {Stalder}, {Staniszewski},
  {Stark}, {Vanderlinde}, {Vieira}, {Williamson}, \&
  {Zahn}}]{Plagge-Benson-Ade:10}
{Plagge}, T., {Benson}, B.~A., {Ade}, P.~A.~R., {et~al.} 2010, \apj, 716, 1118

\bibitem[{{Ponman} {et~al.}(1996){Ponman}, {Bourner}, {Ebeling}, \&
  {B{\"o}hringer}}]{Ponman-Bourner-Ebeling:96}
{Ponman}, T.~J., {Bourner}, P.~D.~J., {Ebeling}, H., \& {B{\"o}hringer}, H.
  1996, \mnras, 283, 690

\bibitem[{{Pope}(2009)}]{Pope:09}
{Pope}, E.~C.~D. 2009, \mnras, 494

\bibitem[{{Pratt} {et~al.}(2009){Pratt}, {Croston}, {Arnaud}, \&
  {B{\"o}hringer}}]{Pratt-Croston-Arnaud:09}
{Pratt}, G.~W., {Croston}, J.~H., {Arnaud}, M., \& {B{\"o}hringer}, H. 2009,
  \aap, 498, 361

\bibitem[{{Predehl} {et~al.}(2010){Predehl}, {Andritschke}, {B{\"o}hringer},
  {Bornemann}, {Br{\"a}uninger}, {Brunner}, {Brusa}, {Burkert}, {Burwitz},
  {Cappelluti}, {Churazov}, {Dennerl}, {Eder}, {Elbs}, {Freyberg}, {Friedrich},
  {F{\"u}rmetz}, {Gaida}, {H{\"a}lker}, {Hartner}, {Hasinger}, {Hermann},
  {Huber}, {Kendziorra}, {von Kienlin}, {Kink}, {Kreykenbohm}, {Lamer},
  {Lapchov}, {Lehmann}, {Meidinger}, {Mican}, {Mohr}, {M{\"u}hlegger},
  {M{\"u}ller}, {Nandra}, {Pavlinsky}, {Pfeffermann}, {Reiprich}, {Robrade},
  {Roh{\'e}}, {Santangelo}, {Sch{\"a}chner}, {Schanz}, {Schmid}, {Schmitt},
  {Schreib}, {Schrey}, {Schwope}, {Steinmetz}, {Str{\"u}der}, {Sunyaev},
  {Tenzer}, {Tiedemann}, {Vongehr}, \&
  {Wilms}}]{Predehl-Andritschke-Boehringer:10}
{Predehl}, P., {Andritschke}, R., {B{\"o}hringer}, H., {et~al.} 2010, in
  Presented at the Society of Photo-Optical Instrumentation Engineers (SPIE)
  Conference, Vol. 7732, Society of Photo-Optical Instrumentation Engineers
  (SPIE) Conference Series

\bibitem[{{Rasmussen} \& {Ponman}(2007)}]{Rasmussen-Ponman:07}
{Rasmussen}, J. \& {Ponman}, T.~J. 2007, \mnras, 380, 1554

\bibitem[{{Rasmussen} {et~al.}(2006){Rasmussen}, {Ponman}, {Mulchaey}, {Miles},
  \& {Raychaudhury}}]{Rasmussen-Ponman-Mulchaey:06}
{Rasmussen}, J., {Ponman}, T.~J., {Mulchaey}, J.~S., {Miles}, T.~A., \&
  {Raychaudhury}, S. 2006, \mnras, 373, 653

\bibitem[{{Reiprich}(2001)}]{Reiprich:01}
{Reiprich}, T.~H. 2001, PhD thesis, AA (Max-Planck-Institut f{\"u}r
  extraterrestrische Physik, P.O.~Box 1312, 85741 Garching, Germany),
  astro-ph/0308137

\bibitem[{{Reiprich}(2006)}]{Reiprich:06}
{Reiprich}, T.~H. 2006, \aap, 453, L39

\bibitem[{{Reiprich} \& {B{\"o}hringer}(2002)}]{Reiprich-Boehringer:02}
{Reiprich}, T.~H. \& {B{\"o}hringer}, H. 2002, \apj, 567, 716

\bibitem[{{Ricker} \& {Sarazin}(2001)}]{Ricker-Sarazin:01}
{Ricker}, P.~M. \& {Sarazin}, C.~L. 2001, \apj, 561, 621

\bibitem[{{Riemer-S{\o}rensen} {et~al.}(2009){Riemer-S{\o}rensen}, {Paraficz},
  {Ferreira}, {Pedersen}, {Limousin}, \&
  {Dahle}}]{Riemer-Sorensen-Paraficz-Ferreira:09}
{Riemer-S{\o}rensen}, S., {Paraficz}, D., {Ferreira}, D.~D.~M., {et~al.} 2009,
  \apj, 693, 1570

\bibitem[{{Ritchie} \& {Thomas}(2002)}]{Ritchie-Thomas:02}
{Ritchie}, B.~W. \& {Thomas}, P.~A. 2002, \mnras, 329, 675

\bibitem[{{Russell} {et~al.}(2007){Russell}, {Ponman}, \&
  {Sanderson}}]{Russell-Ponman-Sanderson:07}
{Russell}, P.~A., {Ponman}, T.~J., \& {Sanderson}, A.~J.~R. 2007, \mnras, 378,
  1217

\bibitem[{{Rykoff} {et~al.}(2008){Rykoff}, {Evrard}, {McKay}, {Becker},
  {Johnston}, {Koester}, {Nord}, {Rozo}, {Sheldon}, {Stanek}, \&
  {Wechsler}}]{Rykoff-Evrard-McKay:08}
{Rykoff}, E.~S., {Evrard}, A.~E., {McKay}, T.~A., {et~al.} 2008, \mnras, 387,
  L28

\bibitem[{{Sanderson} {et~al.}(2003){Sanderson}, {Ponman}, {Finoguenov},
  {Lloyd-Davies}, \& {Markevitch}}]{Sanderson-Ponman-Finoguenov:03}
{Sanderson}, A.~J.~R., {Ponman}, T.~J., {Finoguenov}, A., {Lloyd-Davies},
  E.~J., \& {Markevitch}, M. 2003, \mnras, 340, 989

\bibitem[{{Sanderson} {et~al.}(2006){Sanderson}, {Ponman}, \&
  {O'Sullivan}}]{Sanderson-Ponman-O'Sullivan:06}
{Sanderson}, A.~J.~R., {Ponman}, T.~J., \& {O'Sullivan}, E. 2006, \mnras, 372,
  1496

\bibitem[{{Sato} {et~al.}(2010){Sato}, {Kawaharada}, {Nakazawa}, {Matsushita},
  {Ishisaki}, {Yamasaki}, \& {Ohashi}}]{Sato-Kawaharada-Nakazawa:10}
{Sato}, K., {Kawaharada}, M., {Nakazawa}, K., {et~al.} 2010, \pasj, 62, 1445

\bibitem[{{Sato} {et~al.}(2009){Sato}, {Matsushita}, {Ishisaki}, {Yamasaki},
  {Ishida}, \& {Ohashi}}]{Sato-Matsushita-Ishisaki:09}
{Sato}, K., {Matsushita}, K., {Ishisaki}, Y., {et~al.} 2009, \pasj, 61, 353

\bibitem[{{Shang} \& {Scharf}(2009)}]{Shang-Scharf:09}
{Shang}, C. \& {Scharf}, C. 2009, \apj, 690, 879

\bibitem[{{Simionescu} {et~al.}(2011){Simionescu}, {Allen}, {Mantz}, {Werner},
  {Takei}, {Morris}, {Fabian}, {Sanders}, {Nulsen}, {George}, \&
  {Taylor}}]{Simionescu-Allen-Mantz:11}
{Simionescu}, A., {Allen}, S.~W., {Mantz}, A., {et~al.} 2011, Science, 331,
  1576

\bibitem[{{Snowden} {et~al.}(2008){Snowden}, {Mushotzky}, {Kuntz}, \&
  {Davis}}]{Snowden-Mushotzky-Kuntz:08}
{Snowden}, S.~L., {Mushotzky}, R.~F., {Kuntz}, K.~D., \& {Davis}, D.~S. 2008,
  \aap, 478, 615

\bibitem[{{Spavone} {et~al.}(2006){Spavone}, {Iodice}, {Longo}, {Paolillo}, \&
  {Sodani}}]{Spavone-Iodice-Longo:06}
{Spavone}, M., {Iodice}, E., {Longo}, G., {Paolillo}, M., \& {Sodani}, S. 2006,
  \aap, 457, 493

\bibitem[{{Stanek} {et~al.}(2006){Stanek}, {Evrard}, {B{\"o}hringer},
  {Schuecker}, \& {Nord}}]{Stanek-Evrard-Boehringer:06}
{Stanek}, R., {Evrard}, A.~E., {B{\"o}hringer}, H., {Schuecker}, P., \& {Nord},
  B. 2006, \apj, 648, 956

\bibitem[{{Subrahmanyan} {et~al.}(2003){Subrahmanyan}, {Beasley}, {Goss},
  {Golap}, \& {Hunstead}}]{Subrahmanyan-Beasley-Goss:03}
{Subrahmanyan}, R., {Beasley}, A.~J., {Goss}, W.~M., {Golap}, K., \&
  {Hunstead}, R.~W. 2003, \aj, 125, 1095

\bibitem[{{Sun} {et~al.}(2003){Sun}, {Forman}, {Vikhlinin}, {Hornstrup},
  {Jones}, \& {Murray}}]{Sun-Forman-Vikhlinin:03}
{Sun}, M., {Forman}, W., {Vikhlinin}, A., {et~al.} 2003, \apj, 598, 250

\bibitem[{{Sun} {et~al.}(2009){Sun}, {Voit}, {Donahue}, {Jones}, {Forman}, \&
  {Vikhlinin}}]{Sun-Voit-Donahue:09}
{Sun}, M., {Voit}, G.~M., {Donahue}, M., {et~al.} 2009, \apj, 693, 1142

\bibitem[{{Takizawa} {et~al.}(2010){Takizawa}, {Nagino}, \&
  {Matsushita}}]{Takizawa-Nagino-Matsushita:10}
{Takizawa}, M., {Nagino}, R., \& {Matsushita}, K. 2010, \pasj, 62, 951

\bibitem[{{Tokoi} {et~al.}(2008){Tokoi}, {Sato}, {Ishisaki}, {Ohashi},
  {Yamasaki}, {Nakazawa}, {Matsushita}, {Fukazawa}, {Hoshino}, {Tamura},
  {Egawa}, {Kawano}, {Ota}, {Isobe}, {Kawaharada}, {Awaki}, \&
  {Hughes}}]{Tokoi-Sato-Ishisaki:08}
{Tokoi}, K., {Sato}, K., {Ishisaki}, Y., {et~al.} 2008, \pasj, 60, 317

\bibitem[{{Tovmassian} \& {Plionis}(2009)}]{Tovmassian-Plionis:09}
{Tovmassian}, H.~M. \& {Plionis}, M. 2009, \apj, 696, 1441

\bibitem[{{Trinchieri} {et~al.}(2007){Trinchieri}, {Breitschwerdt}, {Pietsch},
  {Sulentic}, \& {Wolter}}]{Trinchieri-Breitschwerdt-Pietsch:07}
{Trinchieri}, G., {Breitschwerdt}, D., {Pietsch}, W., {Sulentic}, J., \&
  {Wolter}, A. 2007, \aap, 463, 153

\bibitem[{{Urban} {et~al.}(2011){Urban}, {Werner}, {Simionescu}, {Allen}, \&
  {B{\"o}hringer}}]{Urban-Werner-Simionescu:11}
{Urban}, O., {Werner}, N., {Simionescu}, A., {Allen}, S.~W., \&
  {B{\"o}hringer}, H. 2011, \mnras, 414, 2101

\bibitem[{{Ventimiglia} {et~al.}(2008){Ventimiglia}, {Voit}, {Donahue}, \&
  {Ameglio}}]{Ventimiglia-Voit-Donahue:08}
{Ventimiglia}, D.~A., {Voit}, G.~M., {Donahue}, M., \& {Ameglio}, S. 2008,
  \apj, 685, 118

\bibitem[{{Vikhlinin}(2006)}]{Vikhlinin:06}
{Vikhlinin}, A. 2006, \apj, 640, 710

\bibitem[{{Vikhlinin} {et~al.}(2009{\natexlab{a}}){Vikhlinin}, {Burenin},
  {Ebeling}, {Forman}, {Hornstrup}, {Jones}, {Kravtsov}, {Murray}, {Nagai},
  {Quintana}, \& {Voevodkin}}]{Vikhlinin-Burenin-Ebeling:09}
{Vikhlinin}, A., {Burenin}, R.~A., {Ebeling}, H., {et~al.} 2009{\natexlab{a}},
  \apj, 692, 1033

\bibitem[{{Vikhlinin} {et~al.}(2006){Vikhlinin}, {Kravtsov}, {Forman}, {Jones},
  {Markevitch}, {Murray}, \& {Van Speybroeck}}]{Vikhlinin-Kravtsov-Forman:06}
{Vikhlinin}, A., {Kravtsov}, A., {Forman}, W., {et~al.} 2006, \apj, 640, 691

\bibitem[{{Vikhlinin} {et~al.}(2009{\natexlab{b}}){Vikhlinin}, {Kravtsov},
  {Burenin}, {Ebeling}, {Forman}, {Hornstrup}, {Jones}, {Murray}, {Nagai},
  {Quintana}, \& {Voevodkin}}]{Vikhlinin-Kravtsov-Burenin:09}
{Vikhlinin}, A., {Kravtsov}, A.~V., {Burenin}, R.~A., {et~al.}
  2009{\natexlab{b}}, \apj, 692, 1060

\bibitem[{{Vikhlinin} {et~al.}(2005){Vikhlinin}, {Markevitch}, {Murray},
  {Jones}, {Forman}, \& {Van Speybroeck}}]{Vikhlinin-Markevitch-Murray:05}
{Vikhlinin}, A., {Markevitch}, M., {Murray}, S.~S., {et~al.} 2005, \apj, 628,
  655

\bibitem[{{Worrall} {et~al.}(2007){Worrall}, {Birkinshaw}, {Kraft}, \&
  {Hardcastle}}]{Worrall-Birkinshaw-Kraft:07}
{Worrall}, D.~M., {Birkinshaw}, M., {Kraft}, R.~P., \& {Hardcastle}, M.~J.
  2007, \apjl, 658, L79

\bibitem[{{Xue} \& {Wu}(2000{\natexlab{a}})}]{Xue-Wu:00b}
{Xue}, Y.-J. \& {Wu}, X.-P. 2000{\natexlab{a}}, \mnras, 318, 715

\bibitem[{{Xue} \& {Wu}(2000{\natexlab{b}})}]{Xue-Wu:00a}
{Xue}, Y.-J. \& {Wu}, X.-P. 2000{\natexlab{b}}, \apj, 538, 65

\bibitem[{{Yang} {et~al.}(2009){Yang}, {Ricker}, \&
  {Sutter}}]{Yang-Ricker-Sutter:09}
{Yang}, H., {Ricker}, P.~M., \& {Sutter}, P.~M. 2009, \apj, 699, 315

\bibitem[{{Zhang} {et~al.}(2011){Zhang}, {Andernach}, {Caretta}, {Reiprich},
  {B{\"o}hringer}, {Puchwein}, {Sijacki}, \&
  {Girardi}}]{Zhang-Andernach-Caretta:11}
{Zhang}, Y., {Andernach}, H., {Caretta}, C.~A., {et~al.} 2011, \aap, 526, A105+

\bibitem[{{Zhang} {et~al.}(2008){Zhang}, {Finoguenov}, {B{\"o}hringer},
  {Kneib}, {Smith}, {Kneissl}, {Okabe}, \&
  {Dahle}}]{Zhang-Finoguenov-Boehringer:08}
{Zhang}, Y., {Finoguenov}, A., {B{\"o}hringer}, H., {et~al.} 2008, \aap, 482,
  451

\bibitem[{{Zhang} {et~al.}(2010){Zhang}, {Okabe}, {Finoguenov}, {Smith},
  {Piffaretti}, {Valdarnini}, {Babul}, {Evrard}, {Mazzotta}, {Sanderson}, \&
  {Marrone}}]{Zhang-Okabe-Finoguenov:10}
{Zhang}, Y., {Okabe}, N., {Finoguenov}, A., {et~al.} 2010, \apj, 711, 1033

\bibitem[{{Zhang} {et~al.}(2009){Zhang}, {Reiprich}, {Finoguenov}, {Hudson}, \&
  {Sarazin}}]{Zhang-Reiprich-Finoguenov:09}
{Zhang}, Y., {Reiprich}, T.~H., {Finoguenov}, A., {Hudson}, D.~S., \&
  {Sarazin}, C.~L. 2009, \apj, 699, 1178

\end{thebibliography}
%
%
%
\appendix
%
\onecolumn
\section{Notes on Individual Groups}\label{sec:notes}
 \subsection*{A0160}
This group/poor cluster of galaxies was classified as ``intermediate type''
because there is a distinct sub-peak to the east of the center of extended
emission. However this sub-peak was excluded from our analysis of the ICM, as
it appears not to be interacting with the surrounding material. There is a
large degeneracy between temperature and normalization in the spectral fits,
so the total error on the surface brightness was overestimated. We corrected
for this unreasonably high uncertainty by choosing the smaller of the two
errors.

This system was investigated by Drake et al.\ (2000),
who found ``wake''-like structures in the X-ray emission, which they interpreted
as evidence for ram-pressure stripping of the cluster galaxies by the
surrounding ICM.

Jetha et al.\ (2005) also inspected radio observations for this object, and
found a spatial correlation between a wide angle tailed radio galaxy in the
BCG and structures in the X-ray emitting gas. They concluded
that the secondary peak (``source 1'') corresponds to a large elliptical at
slightly higher redshift than the cluster, which they found not to be
interacting with the cluster.
\subsection*{HCG62}
This group has been studied extensively, in radio, X-ray and optical bands. In
the innermost regions, the emission shows traces of dynamical activity,
although the overall shape is quite regular. We used only the newer two of the
four available \emph{Chandra} observations, because they are from a more
recent different background epoch and together have $\sim\unit[120]{ks}$ of
exposure, which is more than sufficient for our analysis.

Using \emph{ASCA} data, Fukazawa et al.\ (2001) and Nakazawa et
al.\ (2007) reported an excess of hard emission in the spectra of
this group. This hard excess could however not be confirmed by
a more recent \emph{Suzaku} observation (Tokoi et al.\ 2008).

Spavone et al.\ (2006) concentrated specifically on the brightest group
member, NGC4778, which shows distinct kinematical and morphological
peculiarities, which they argued is an indication for a possible recent minor
merger event. If this conclusion is correct, the group may still be dynamically
disturbed, even though the emission appears to be spherical at large radii.

Two X-ray cavities in this group were studied by Morita et al.\ (2006), but
found no clear explanation for their origin. These authors also found some
evidence for a departure from hydrostatic equilibrium in this system. They
used a triple $\beta$-model to describe the SBP.

Gu et al.\ (2007) identified an arc-shaped region spanning from south to northwest,
with an average metal abundance about a factor of two higher than the
surrounding regions. They concluded this may have been created by AGN activity,
or by a recent merger, in accordance with Spavone et al.

Gitti et al.\ (2010) combined X-ray data from \emph{Chandra} and
\emph{XMM-Newton} with \emph{GMRT} radio data and detect low-frequency radio
emission associated with the X-ray cavities.  
\subsection*{IC1262}
This poor cluster appears quite regular and relaxed in the overall emission
but shows some remarkable filamentary emission features close to the
center which are clearly visible in the \emph{Chandra} image.

Hudson et al.\ (2003) first reported a possible detection of a diffuse non-thermal
component, measured with \emph{Beppo-SAX}, which might also correspond to
excess emission in radio. Hudson \& Henriksen (2003) confirmed this
observation with \emph{Chandra} data, and concluded this cluster has recently
merged with a smaller subclump, triggering the non-thermal emission by
relativistic shocks.

On the other hand, Trinchieri et al.\ (2007) found the filamentary structures
to be actually cooler than the surrounding regions, and rejected the shock
front scenario proposed by Hudson \& Henriksen. They suggested these
structures may have been formed by ram pressure stripping of a nearby spiral
galaxy or a past radio source that has since faded and is now only a relic.
\subsection*{IC1633}
This system is the hottest ($T\sim\unit[3]{keV}$) and most massive
($M_{200}\sim16\cdot10^{13}~h^{-1}_{70}~ M_{\odot}$) object in
the sample presented here.
 \subsection*{IC4296}
This group is the faintest object in our sample, and in addition was observed
with only half of the area of the CCD chips (``1/2 subarray mode'') to avoid
pile-up, apparently because the observers were interested mostly in the giant
radio-active elliptical galaxy at the center (Pellegrini et
al.\ 2003). However, with this configuration large parts of the extended group
emission are lost, so both observations essentially are useless for
investigating the ICM. We nevertheless performed the analysis, but the object
was an obvious outlier in all of the investigated scaling relations, with too
high temperature and mass, clearly outside the normal scatter, so it was
ultimately excluded from the fits, although it still appears in some of the tables.
\subsection*{MKW4}
Using \emph{XMM-Newton} observations, O'Sullivan et al.\ (2003) studied this
poor \emph{HIFLUGCS} cluster and found a central temperature gradient lacking
strong cooling activity, as well as a sharp abundance peak at the cluster
core, located at the position of the central cD galaxy NGC4073. They argued
that this system is close to establishing hydrostatic equilibrium and has
developed a central temperature drop, but not yet a cooling flow.

Fukazawa et al.\ (2004) observed this group with \emph{Chandra} and also found
a central temperature drop and high central abundance. They attributed this
high abundance to the weakness of the AGN in the cD galaxy, which
allowed the ejected metals to remain near the cluster center.
\subsection*{MKW8}
This poor cluster is the brightest object in our sample, as well as the second
hottest. It is also included in the \emph{HIFLUGCS} sample. The X-ray map is
slightly asymmetric, and in the optical image there are two large galaxies
visible at the center.

Hwang et al. (1999) investigated the properties of a
small sample of groups and found MKW8 to be a dynamically complex system with
a relatively low velocity dispersion.
\subsection*{NGC326}
The radio-active central galaxy of this system has an inversion
symmetric (``Z''-shaped) morphology and has been studied in the radio regime
by Murgia et al.\ (2001).
\subsection*{NGC507}
This rich group/poor cluster shows some interesting dynamical features, with
quite irregular X-ray emission. The main X-ray peak is coincident with the
NGC507 galaxy, but the emission is extended in the north and east directions
towards another large galaxy (NGC508). In addition the cluster may be
contaminated by or even interacting with the nearby group NGC499, which is
located only about $\unit[15]{arcmin}$ to the north of NGC507, just outside
the region observed with \emph{Chandra}. This system is included in
\emph{HIFLUGCS}.

Paolillo et al.\ (2003) investigated this system and found two different
components in the X-ray emission, one belonging to the halo core and another
external, extended component. Accordingly, they used a bidimensional double
$\beta$-model to fit the surface brightness distribution. Kraft et al.\ (2004)
observed a sharp discontinuity in the surface brightness map and suggested this
is caused by a radio lobe within the galaxy NGC507, which is blowing higher
abundance gas out into the group halo.

Recently, Sato et al.\ (2009) re-observed this group with \emph{Suzaku} and
created radial temperature and metallicity profiles, as well as several metal
mass-to-light ratios and a 2-dimensional hardness map. They found it necessary
to use a two-temperature model to fit the ICM spectra, however they saw no
signs for a significant deviation from spherical symmetry.
 \subsection*{NGC1132}
Mulchaey \& Zabludoff (1999) have analyzed \emph{ASCA} observations of the
elliptical galaxy NGC1132 and found an extended X-ray halo. They concluded
this system is not an isolated galaxy but may be either the remnant of a
poor group which has in the past merged into a single giant elliptical (fossil
group) or has never managed to form a group due to a lack of other bright
nearby galaxies (``failed'' group).
 \subsection*{NGC1550}
The X-ray emission of this system is slightly elliptical in the central
regions. In the optical the group is dominated by the S0 galaxy NGC1550.

Kawaharada et al.\ (2003) used \emph{ASCA} data to estimate the mass of this
group and, comparing this to the optical luminosity, found an unusually high
mass-to-light ratio, which is closer to rich clusters than typical groups of
similar mass. They proposed this system may be classified as a ``dark group''
with an especially low baryon content. A study with \emph{XMM-Newton} data was
also published by Kawaharada et al.\ (2009), where they investigated the metal
distibution in the ICM and concluded that the group galaxies have in the past
been merged into the cD galaxy, so NGC1550 would qualify as a fossil group.

The temperature and entropy profiles of this group were investigated in detail
by Sun et al.\ (2003), using \emph{Chandra} observations. They found the
temperature profile to be quite similar to those of other groups, indicating a
universal profile. In the entropy profile they did not see evidence for the
``entropy floor'' predicted by simulations. Also, they studied the galactic
composition of this system and concluded NGC1550 is not a fossil group.

Sato et al.\ (2010) investigated the distribution of metals in this group
using \emph{Suzaku} data. They traced the abundances of several elements out
to a radial region of $\sim 0.5 r_{180}$, which would fit well the picture
of a fossil group whose central dominant galaxy has enriched the ICM out to large
radii.
 \subsection*{NGC4325}
This group has been studied by Russell et al.\ (2007), who used both \emph{ROSAT}
and \emph{Chandra} data. They found several indications for a disturbance in
the group core and the surrounding ICM, e.\ g.\ X-ray cavities. They concluded
that even though the group is radio-quiet, there is evidence for past AGN
activity and, judging by the steep entropy profile, another outburst may
be imminent.
 \subsection*{NGC5419 (Abell S753)}
Subrahmanyan et al.\ (2003) investigated this poor cluster in the radio regime
and presented several theories concerning the origin of the low surface
brightness radio source found in this system. It may be a relic of activity in
the dominant galaxy NGC5419 or of another cluster galaxy, or may have
been created as a lobe of a powerful radio galaxy, although there is no clear
optical counterpart.
 \subsection*{NGC6269 (AWM5)}
A small sub-peak is visible on the edge of the \emph{Chandra} field of view
which belongs to the group member NGC6265, which appears to be moving through
the ICM east of the cD galaxy NGC6269.

Baldi et al.\ (2009) have performed a detailed study of this group and
presented evidence for reheating of cool gas in the group core, probably due
to past AGN activity. They also investigated the ram pressure stripped tail of
NGC6265 and estimated the mass of this galaxy.
 \subsection*{NGC6338}
This system is very clearly in the process of merging, with a bright subclump just
to the north of the main group center, which is localized at the position of
the BCG. The subclump corresponds to a smaller galaxy apparently falling into
the gravitational potential of the group.
 \subsection*{NGC6482}
This group is both the coolest and least massive object in this sample. It was
studied in detail by Khosroshahi et al.\ (2004), who found evidence this is a
fossil group with short cooling times at the center but which nevertheless
shows now sign of a central temperature drop.
 \subsection*{RXCJ1022}
There is a secondary peak close to the group center, but no apparent change in
the ICM surrounding it. This may either be an infalling group that has not yet
effected on the ICM, or a chance alignment along the line of sight. There are
currently no redshift measurements available to check the latter
possibility. However in both cases it is advisable to remove this clump from
the group emission, as it is not (or not yet) part of the main group's
gravitational potential, so we exclude this region from the analysis.

This is the most distant object in the sample presented here ($z\sim0.05$).
 \subsection*{RXCJ2214 (3C442A)}
At the very center, this system shows signs of dynamical disturbance, where
two bright galaxies (NGC2736 and NGC2737) are in the process of merging.

Worrall et al.\ (2007) and Hardcastle et al.\ (2007) have studied the
interaction between the ICM and the radio lobes found in this system, which
probably have been created by past activity of the radio galaxy 3C442A. Both
concluded that the ICM disturbed by the merger has a strong impact on the
radio lobes, changing the dynamics and possibly even re-heating the radio plasma.
 \subsection*{S0463}
In this system there are two peaks of roughly the same size and luminosity, so
it is not immediately clear which should be the emission peak. The clump in
the south-western quadrant is more luminous, but the BCG lies at the position
of the other one, so we assume this to be the peak of emission as it should
correspond to the center of the group's gravitational potential.
 \subsection*{SS2S153 (NGC3411, USGC152)}
O'Sullivan et al.\ (2007) investigated this group in detail and reported a
remarkable feature in the temperature profile. Although the overall X-ray
emission is very regular and spherical, and the metallicity profile shows no
unusual features, there appears to be a shell of cool gas surrounding the
hotter group core. They concluded the most likely explanation is interplay
between the cool core and a past episode of heating from the currently
quiescent AGN.
%
\section{Tables}\label{app:tables}
\begin{table*}[!h]
\caption{Table of observations}
\begin{center}
\setlength\extrarowheight{2pt}
\begin{tabularx}{\textwidth}{XlXrrXXX}
\hline\hline
Target Name & \multicolumn{1}{c}{OBSID} & \multicolumn{1}{c}{Observation Date} & \multicolumn{1}{c}{Exposure} & \multicolumn{1}{c}{Good Time} & \multicolumn{1}{l}{PI} & \multicolumn{1}{l}{Pointing} & \multicolumn{1}{l}{Data Mode} \\ 
 & & \multicolumn{1}{c}{yyyy-mm-dd} & \multicolumn{1}{c}{ks} & \multicolumn{1}{c}{ks} & & & \\ \hline
A0160 & 3219 & 2002-10-18 & 59.27 & 59.10 & Ponman & ACIS-I & VFAINT \\ \hline
A1177 & 6940 & 2006-12-27 & 34.08 & 33.19 & Buote & ACIS-S & VFAINT \\ \hline
ESO552020 & 3206 & 2002-10-14 & 24.18 & 18.67 & Forman & ACIS-I & VFAINT \\ \hline
HCG62 & 10462 & 2009-03-02 & 68.04 & 66.38 & Rafferty & ACIS-S & VFAINT \\ 
 & 10874 & 2009-03-03 & 52.04 & 50.82 & Rafferty & ACIS-S & VFAINT \\ 
 & total &  & 120.08 & 117.19 &  &  &  \\ \hline
HCG97 & 4988 & 2005-01-14 & 58.13 & 31.89 & Vrtilek & ACIS-S & VFAINT \\\hline 
IC1262 & 2018 & 2001-08-23 & 31.13 & 26.97 & Trinchieri & ACIS-S & FAINT \\
 & 6949 & 2006-04-17 & 39.11 & 38.89 & Forman & ACIS-I & VFAINT \\ 
 & 7321 & 2006-04-19 & 38.05 & 37.84 & Forman & ACIS-I & VFAINT \\ 
 & 7322 & 2006-04-22 & 38.04 & 37.84 & Forman & ACIS-I & VFAINT \\ 
 & total &  & 146.33 & 141.54 &  &  &  \\ \hline
IC1633 & 4971 & 2003-11-16 & 25.12 & 24.37 & Forman & ACIS-I & VFAINT \\\hline 
IC4296 & 2021 & 2001-09-10 & 24.22 & 23.85 & Pellegrini & ACIS-S & FAINT \\ 
 & 3394 & 2001-12-15 & 25.40 & 17.61 & Pellegrini & ACIS-S & FAINT \\ 
 & total &  & 49.62 & 41.46 &  &  &  \\ \hline
MKW4 & 3234 & 2002-11-24 & 30.36 & 26.97 & Fukazawa & ACIS-S & VFAINT \\ \hline
MKW8 & 4942 & 2005-03-01 & 23.45 & 23.07 & Reiprich & ACIS-I & VFAINT \\ \hline
NGC326 & 6830 & 2006-09-02 & 95.67 & 95.40 & Worrall & ACIS-S & VFAINT \\ \hline
NGC507 & 317 & 2000-10-11 & 27.19 & flared & Murray & ACIS-S & FAINT \\ \hline
 & 2882 & 2002-01-08 & 44.21 & 43.55 & Murray & ACIS-I & VFAINT \\ 
 & total &  & 44.21 & 43.55 &  &  &  \\ \hline
NGC533 & 2880 & 2002-07-28 & 38.10 & 31.11 & Sarazin & ACIS-S & VFAINT \\ \hline
NGC777 & 5001 & 2004-12-23 & 10.17 & 9.75 & Murray & ACIS-I & VFAINT \\ \hline
NGC1132 & 801 & 1999-12-10 & 14.17 & 9.33 & Zabludoff & ACIS-S & FAINT \\ \hline
 & 3576 & 2003-11-16 & 40.17 & flared & Garmire & ACIS-S & FAINT \\ 
 & total &  & 14.17 & 9.33 &  &  &  \\ \hline
NGC1550 & 3186 & 2002-01-08 & 10.12 & 9.84 & Murray & ACIS-I & VFAINT \\ 
 & 3187 & 2002-01-08 & 9.77 & 9.59 & Murray & ACIS-I & VFAINT \\ 
 & 5800 & 2005-10-22 & 45.13 & 44.60 & Dupke & ACIS-S & VFAINT \\ 
 & 5801 & 2005-10-24 & 45.04 & 43.56 & Dupke & ACIS-S & VFAINT \\ 
 & total &  & 110.06 & 107.59 &  &  &  \\ \hline
NGC4325 & 3232 & 2003-02-04 & 30.48 & 29.03 & Ponman & ACIS-S & VFAINT \\ \hline
NGC4936 & 4997 & 2004-02-09 & 14.15 & 13.59 & Murray & ACIS-I & VFAINT \\ 
 & 4998 & 2004-02-19 & 15.15 & 14.77 & Murray & ACIS-I & VFAINT \\ 
 & total &  & 29.30 & 28.36 &  &  &  \\ \hline
NGC5129 & 6944 & 2006-04-13 & 21.18 & 20.74 & Buote & ACIS-S & VFAINT \\ 
 & 7325 & 2006-05-14 & 26.18 & 23.85 & Buote & ACIS-S & VFAINT \\ 
 & total &  & 47.36 & 44.59 &  &  &  \\ \hline
NGC5419 & 4999 & 2004-06-18 & 14.77 & 14.51 & Murray & ACIS-I & VFAINT \\ 
 & 5000 & 2004-06-19 & 15.01 & 10.37 & Murray & ACIS-I & VFAINT \\ 
 & total &  & 29.78 & 24.88 &  &  &  \\ \hline
NGC6269 & 4972 & 2003-12-29 & 40.16 & 39.67 & Forman & ACIS-I & VFAINT \\ \hline
NGC6338 & 4194 & 2003-09-17 & 47.94 & 46.90 & Ponman & ACIS-I & VFAINT \\ \hline
NGC6482 & 3218 & 2002-05-20 & 19.61 & 16.32 & Ponman & ACIS-S & VFAINT \\ \hline
RXCJ1022.0+3830 & 6942 & 2006-10-14 & 42.04 & 40.45 & Buote & ACIS-S & VFAINT \\ \hline
RXCJ2214.8+1350 & 5635 & 2005-07-27 & 28.57 & 27.22 & Kraft & ACIS-I & VFAINT \\ 
 & 6353 & 2005-07-28 & 14.17 & 14.00 & Kraft & ACIS-I & VFAINT \\ 
 & 6359 & 2005-10-07 & 20.15 & 19.95 & Kraft & ACIS-I & VFAINT \\ 
 & 6392 & 2006-01-12 & 33.13 & 32.77 & Kraft & ACIS-I & VFAINT \\ 
 & total &  & 96.02 & 93.94 &  &  &  \\ \hline
S0463 & 6956 & 2006-06-30 & 29.74 & 29.31 & Johnstone & ACIS-I & VFAINT \\ 
 & 7250 & 2006-06-30 & 29.47 & 29.03 & Johnstone & ACIS-I & VFAINT \\ 
 & total &  & 59.21 & 58.34 &  &  &  \\ \hline
SS2B153 & 3243 & 2002-11-05 & 29.91 & 18.41 & Vrtilek & ACIS-S & VFAINT \\ \hline
\end{tabularx}
\end{center}
\label{tab:obs}
\end{table*}
\clearpage
\renewcommand{\textfraction}{0.0001}
\renewcommand{\topfraction}{0.99}
\renewcommand{\bottomfraction}{0.99}
\renewcommand{\floatpagefraction}{0.98}
\section{Radial Prof\hspace{0mm}iles}\label{app:profiles}
\emph{Left to Right:} Temperature profile, metallicity profile, surface
brightness profile.
\begin{figure*}[h]
   \centering
   \includegraphics[width=0.26\textwidth]{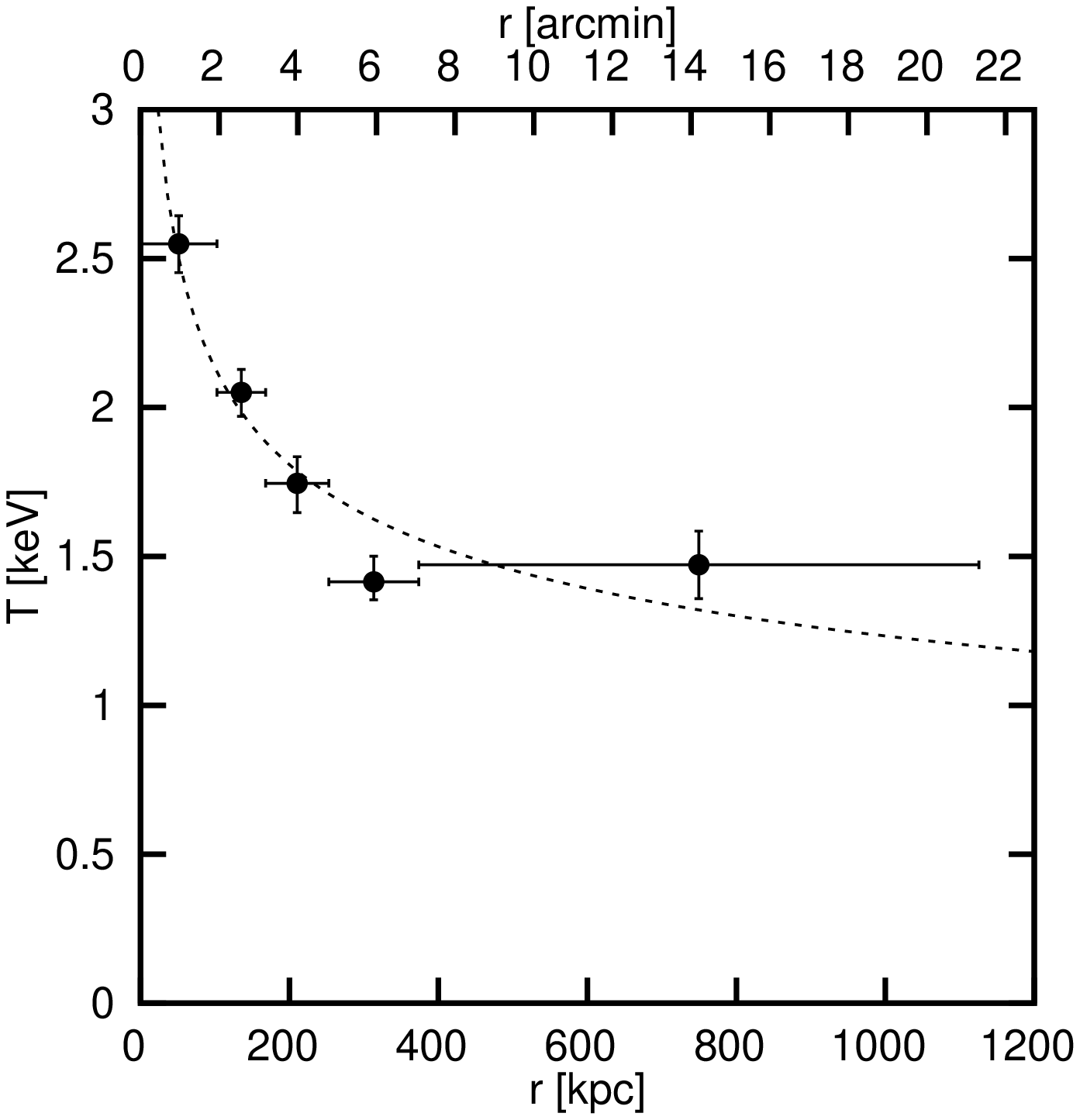}
   \includegraphics[width=0.26\textwidth]{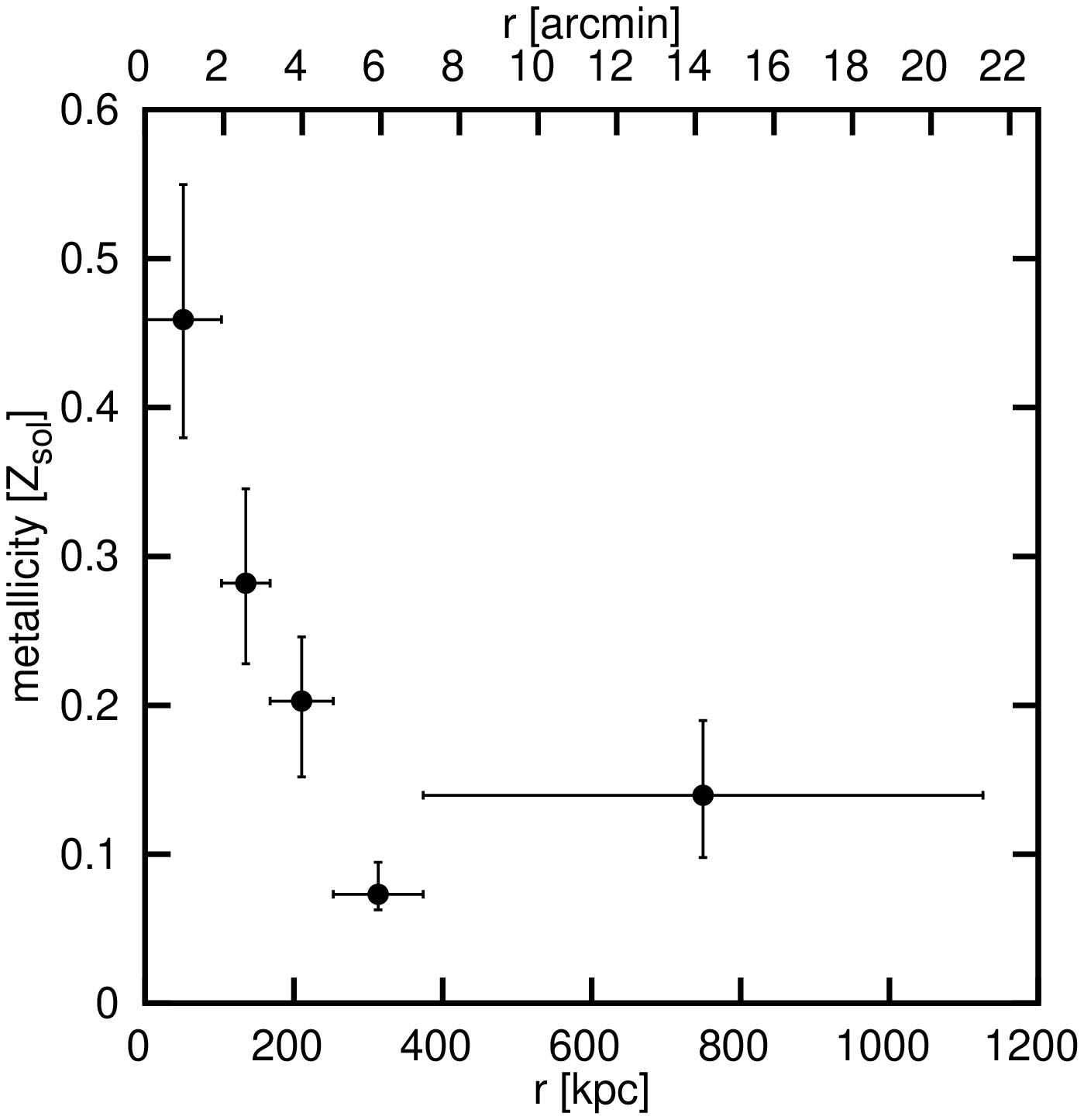}
   \includegraphics[width=0.26\textwidth]{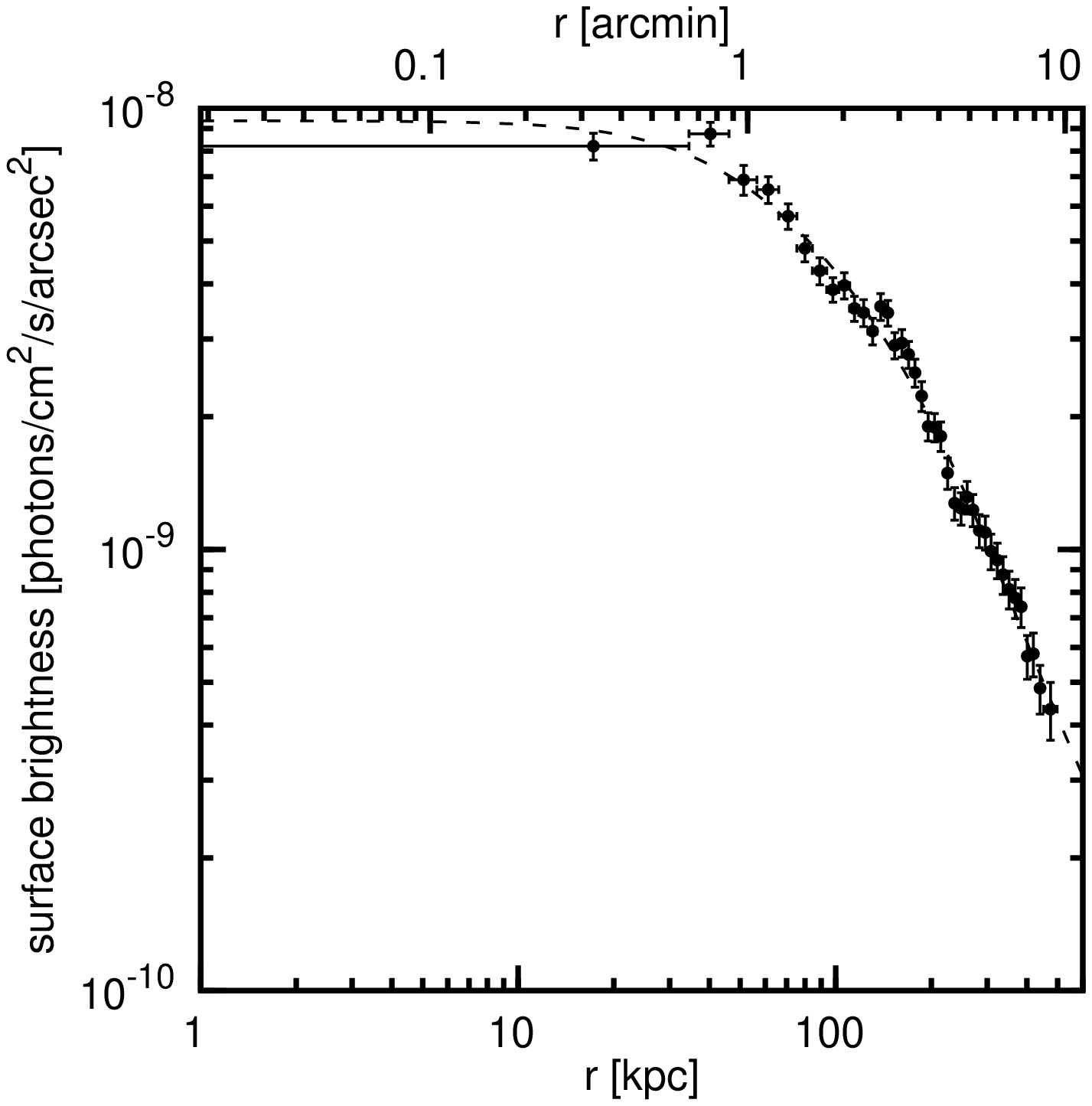}
   \caption{A0160}
   \label{fig:tprofa0160}%
\end{figure*}
\begin{figure*}[h]
   \centering
   \includegraphics[width=0.26\textwidth]{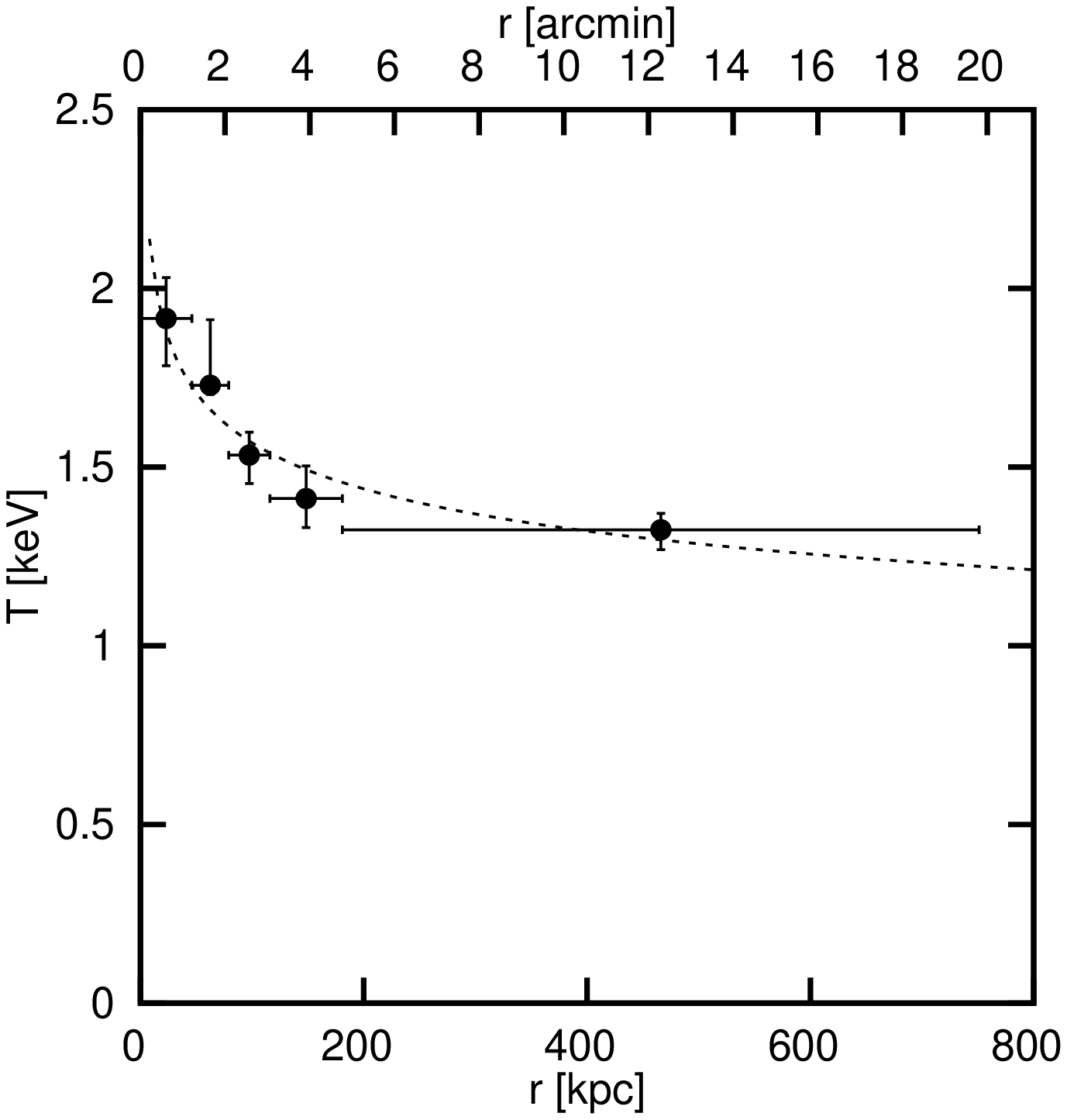}
   \includegraphics[width=0.26\textwidth]{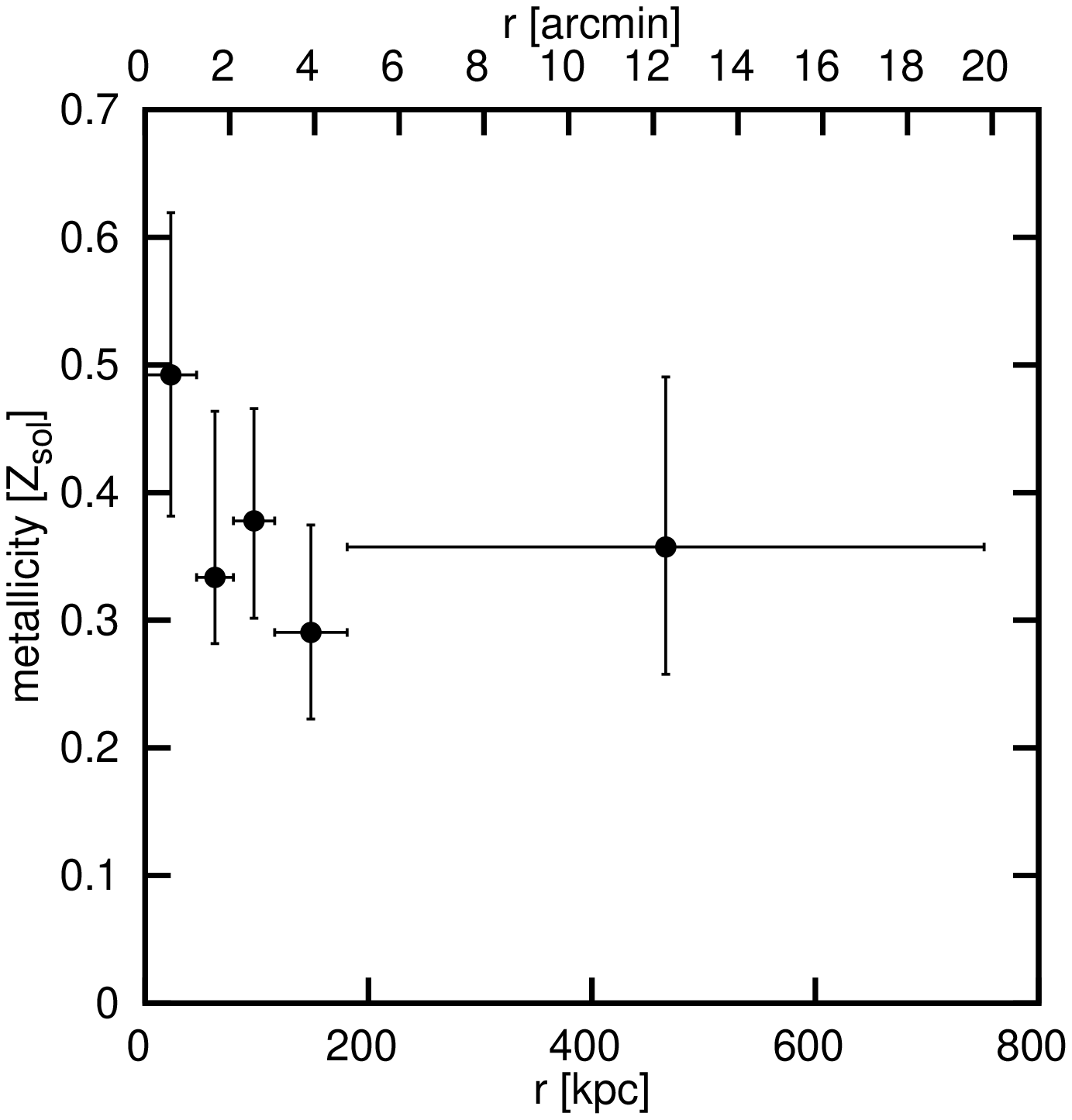}
   \includegraphics[width=0.26\textwidth]{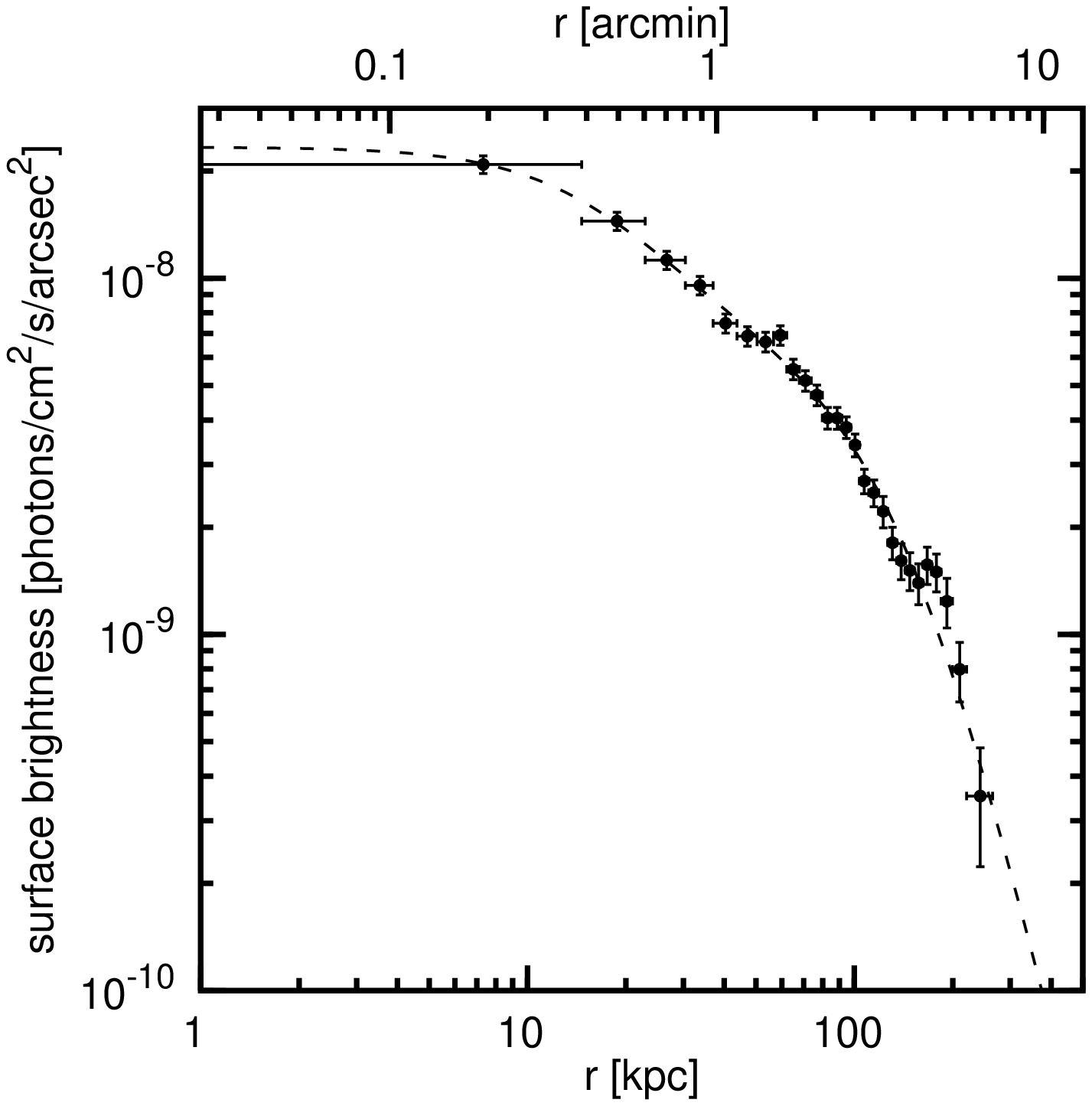}
   \caption{A1177}
   \label{fig:tprofa1177}%
\end{figure*}
\begin{figure*}[h]
   \centering
   \includegraphics[width=0.26\textwidth]{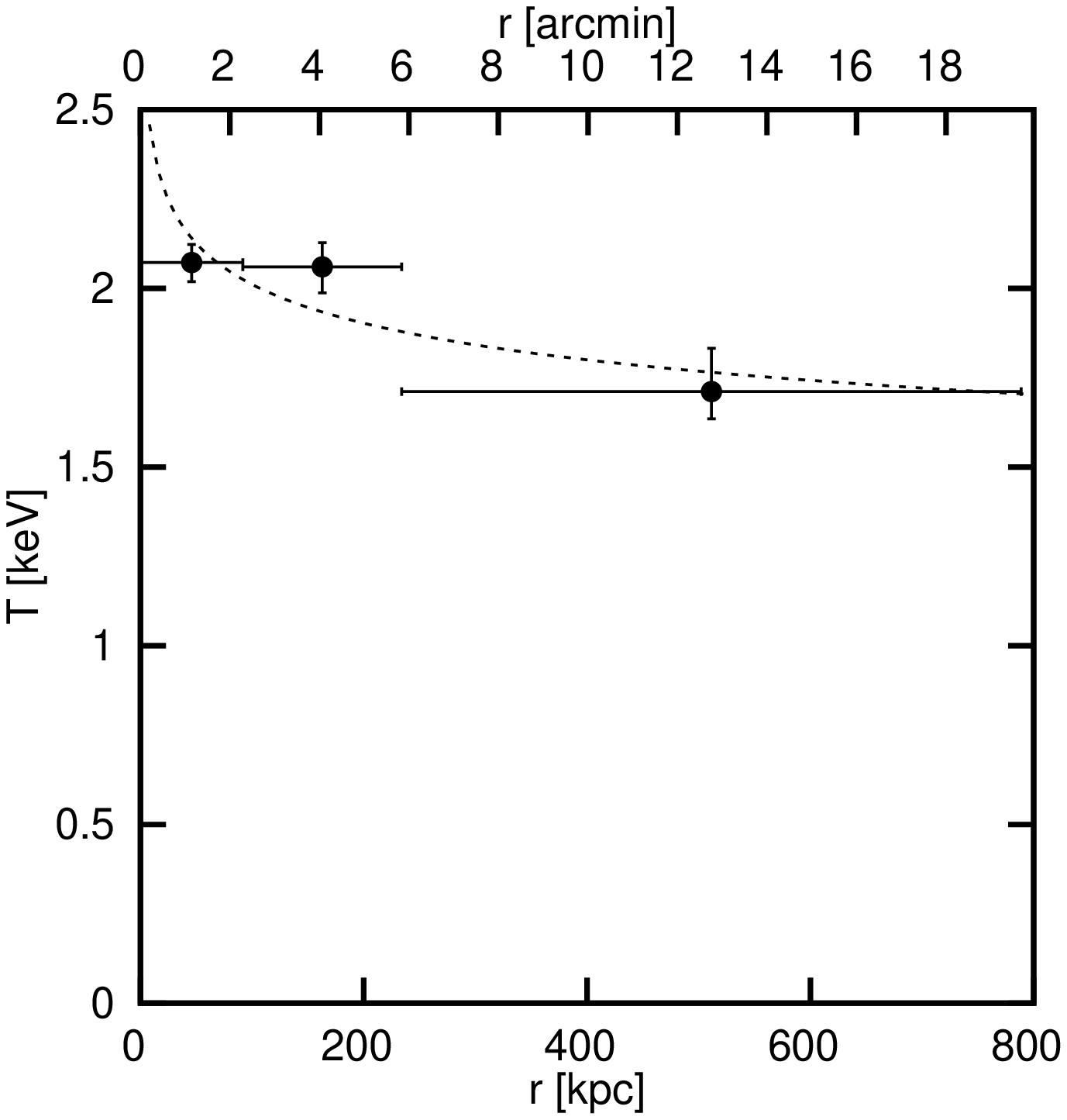}
   \includegraphics[width=0.26\textwidth]{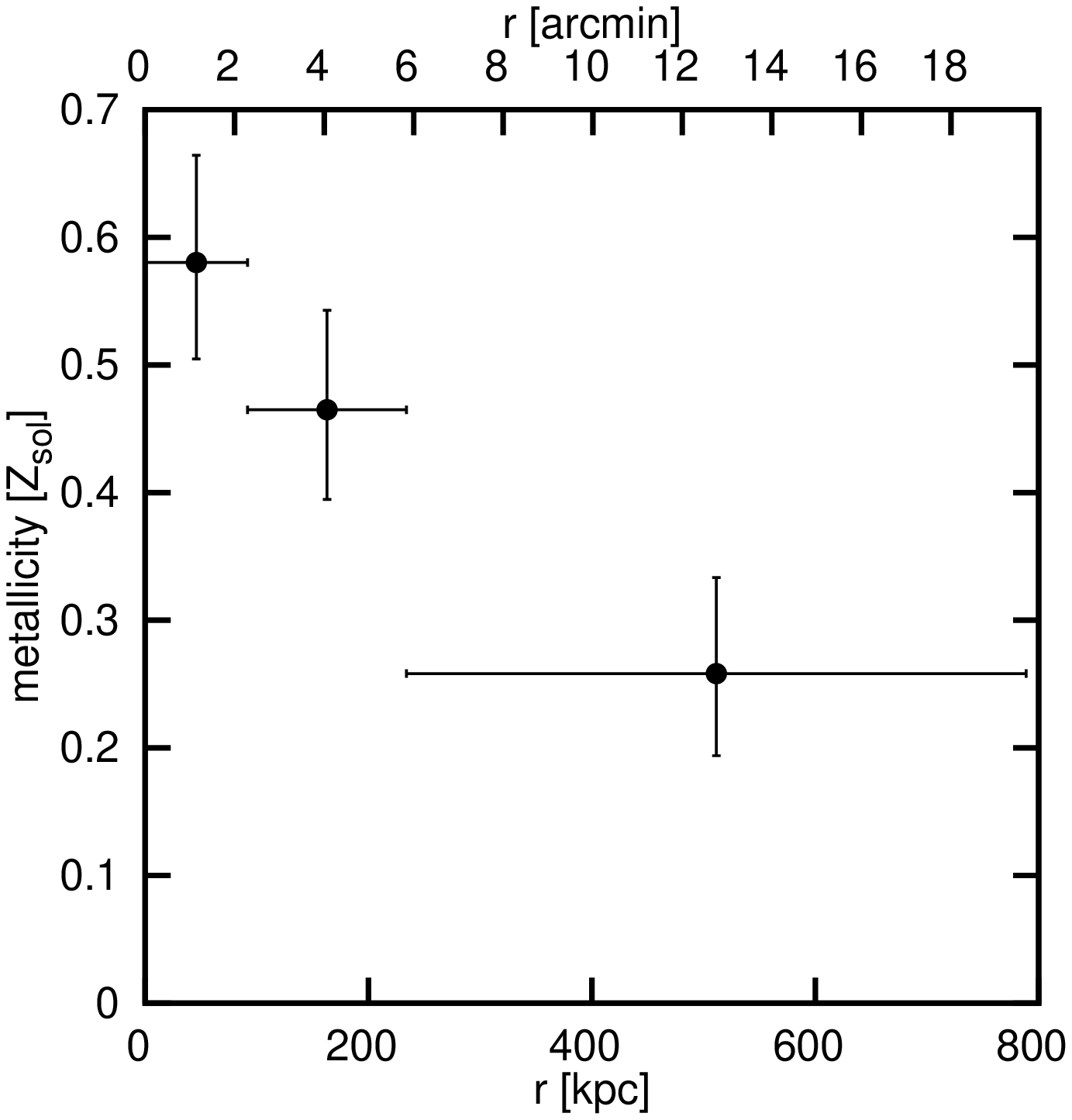}
   \includegraphics[width=0.26\textwidth]{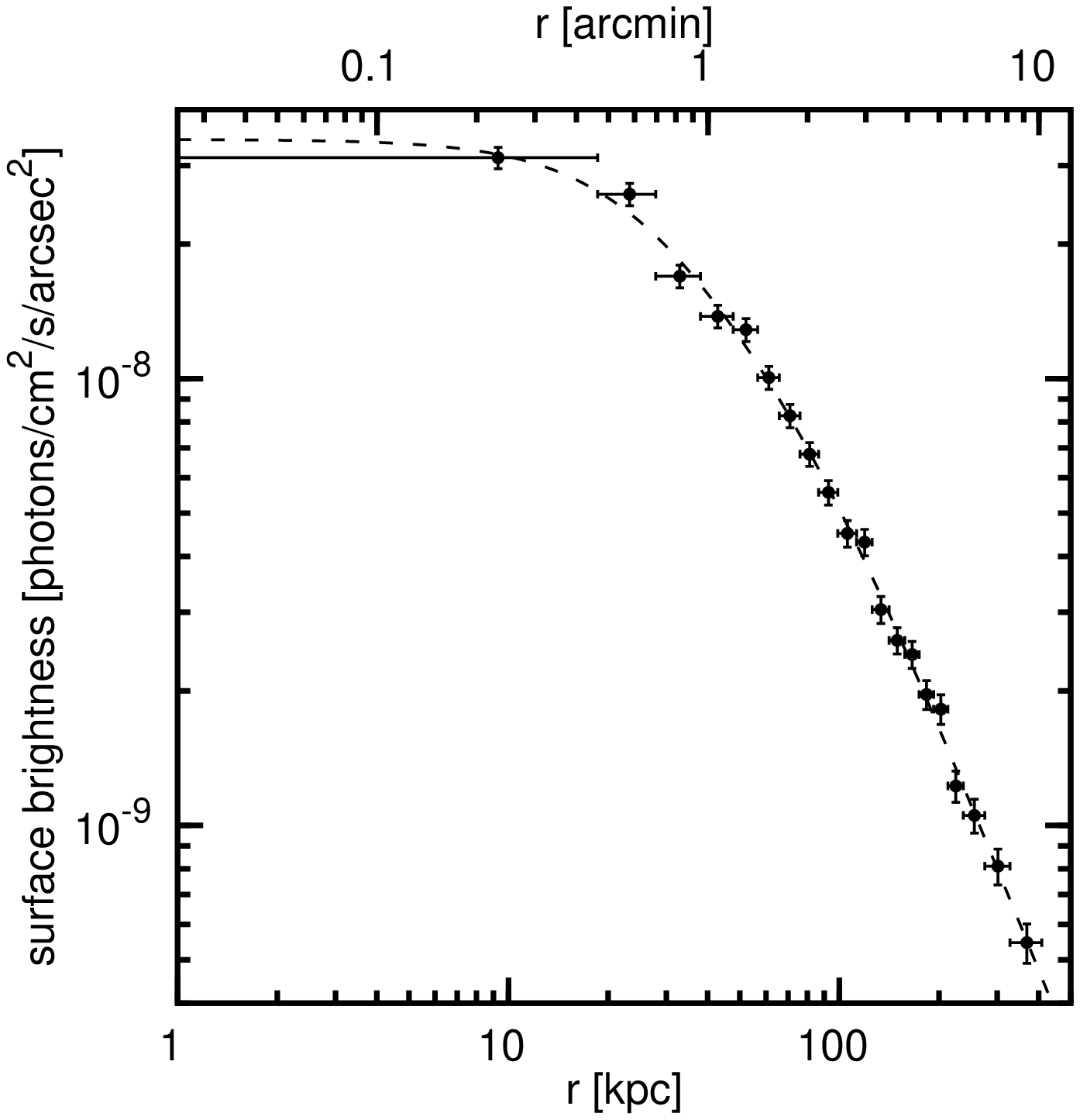}
   \caption{ESO552020}
   \label{fig:tprofeso55}%
\end{figure*}
\clearpage
\begin{figure*}[h]
   \centering
   \includegraphics[width=0.26\textwidth]{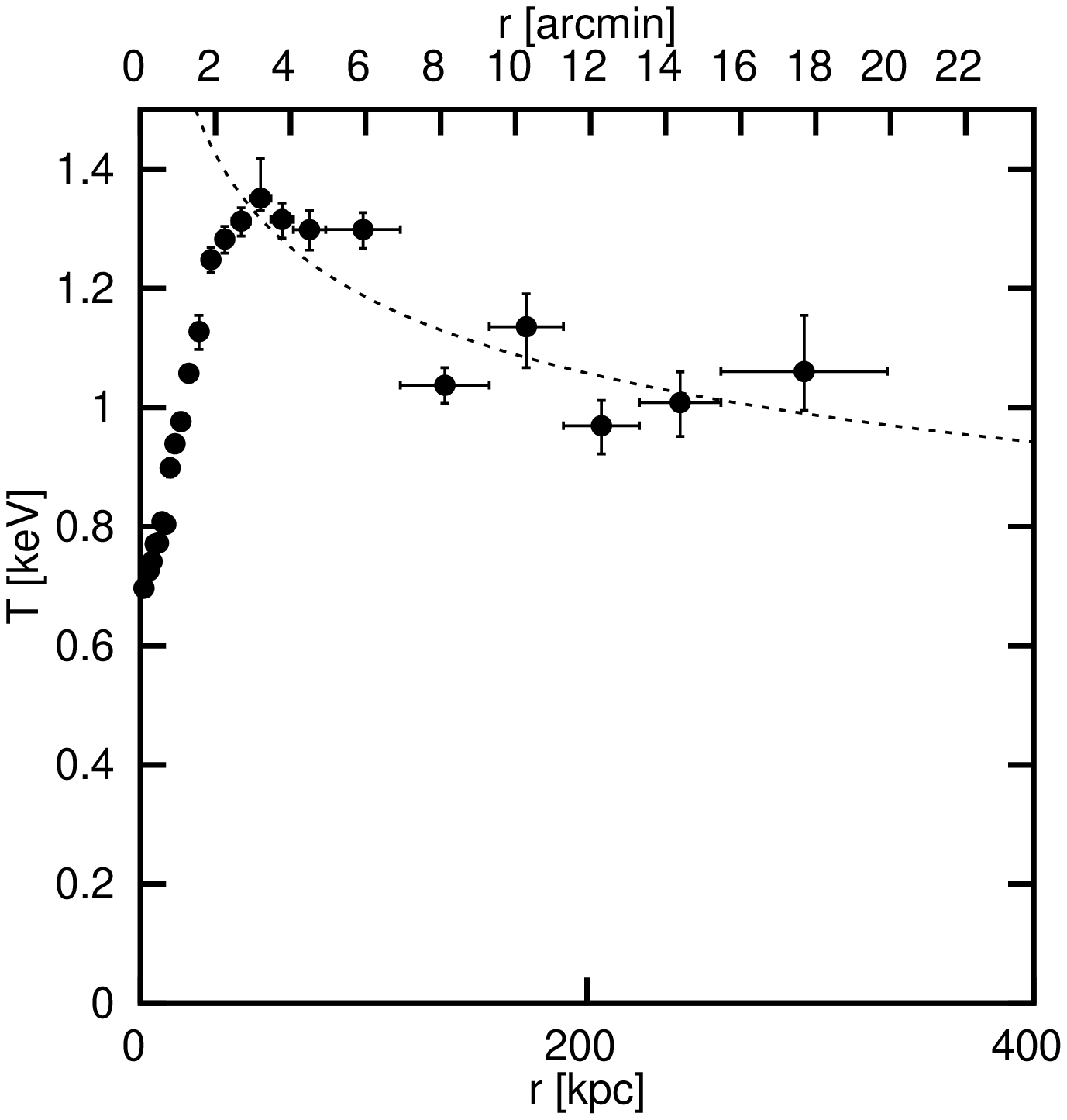}
   \includegraphics[width=0.26\textwidth]{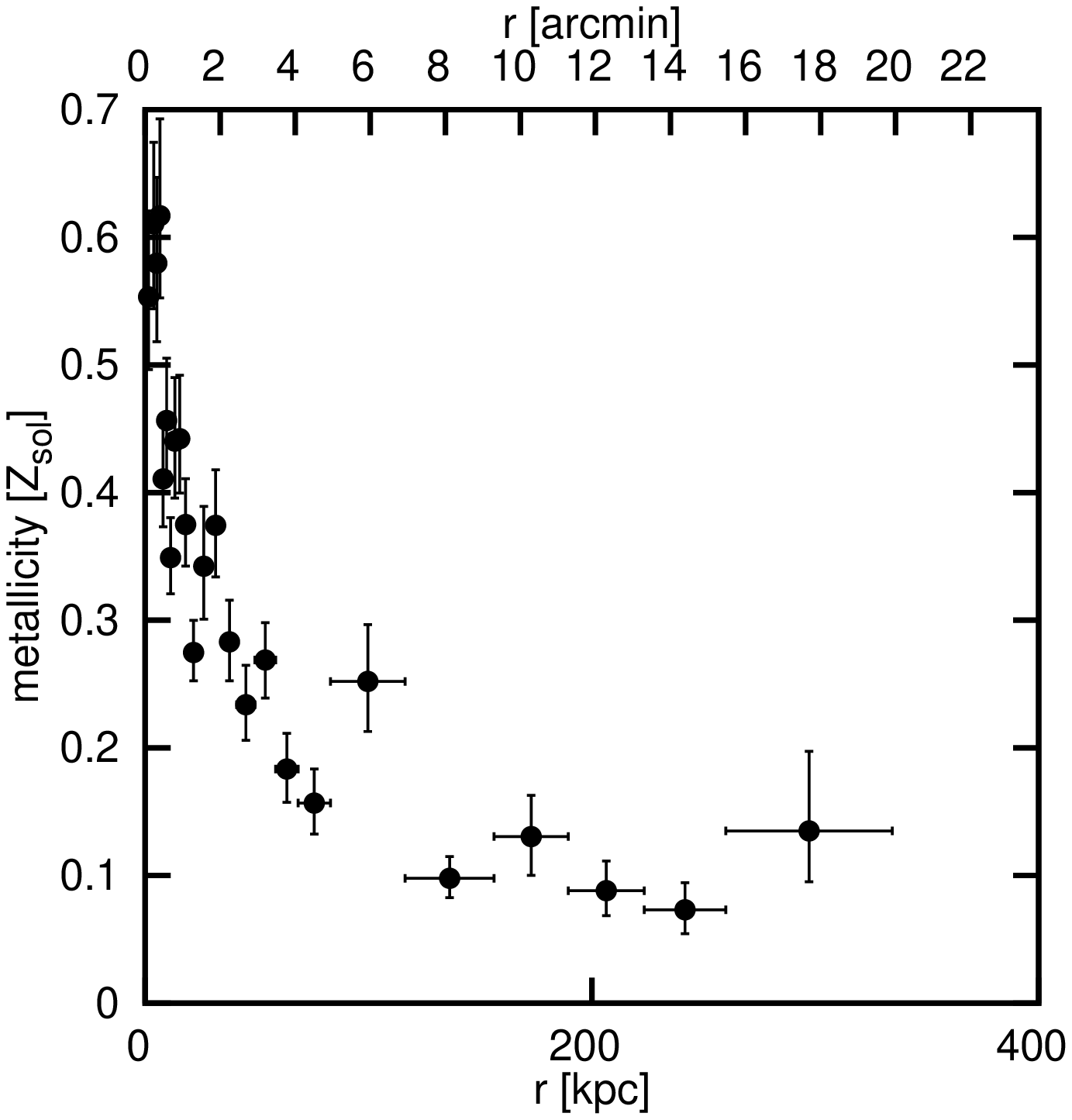}
   \includegraphics[width=0.26\textwidth]{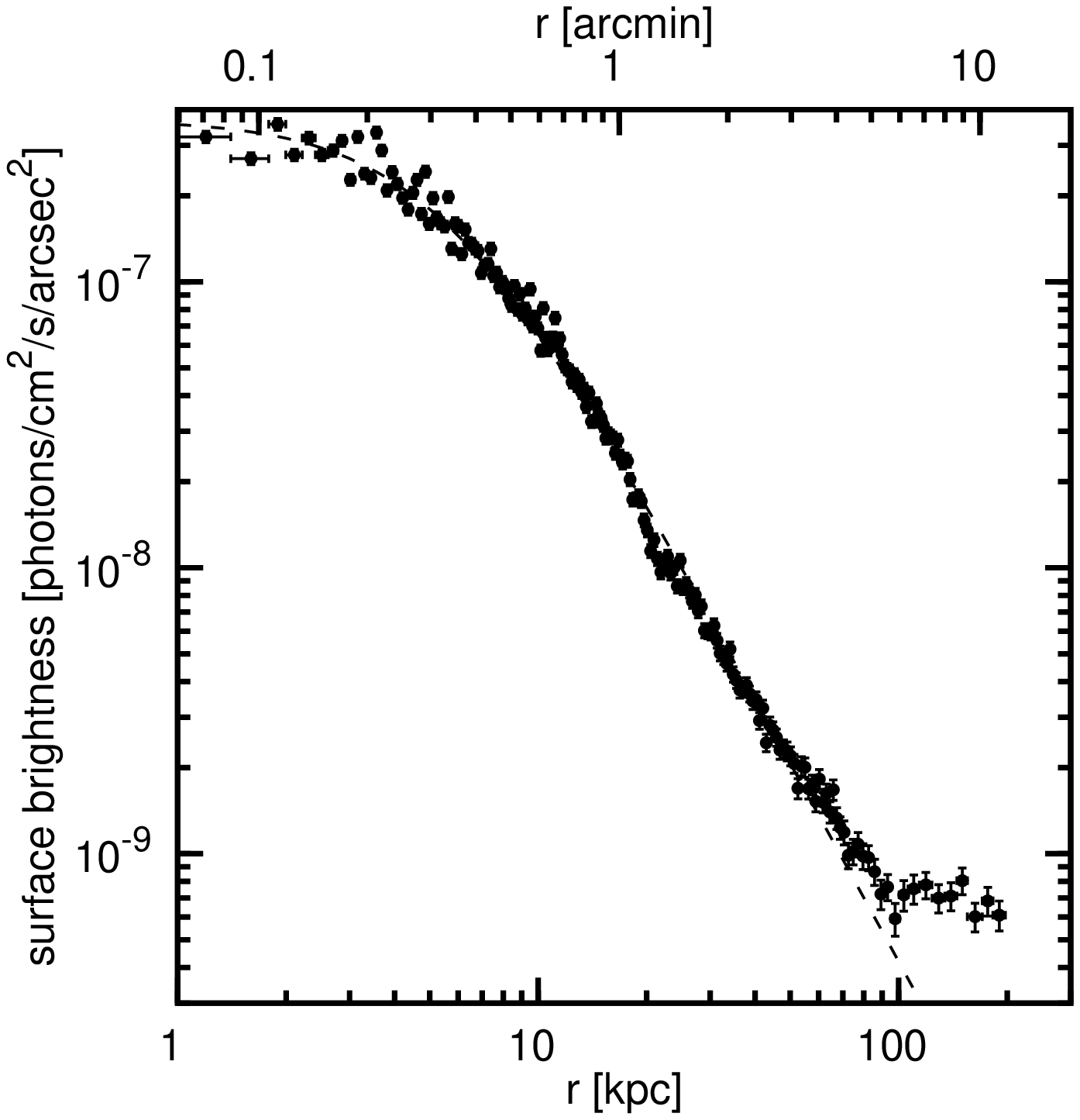}
   \caption{HCG62}
   \label{fig:tprofhcg62}%
\end{figure*}
\begin{figure*}[h]
   \centering
   \includegraphics[width=0.26\textwidth]{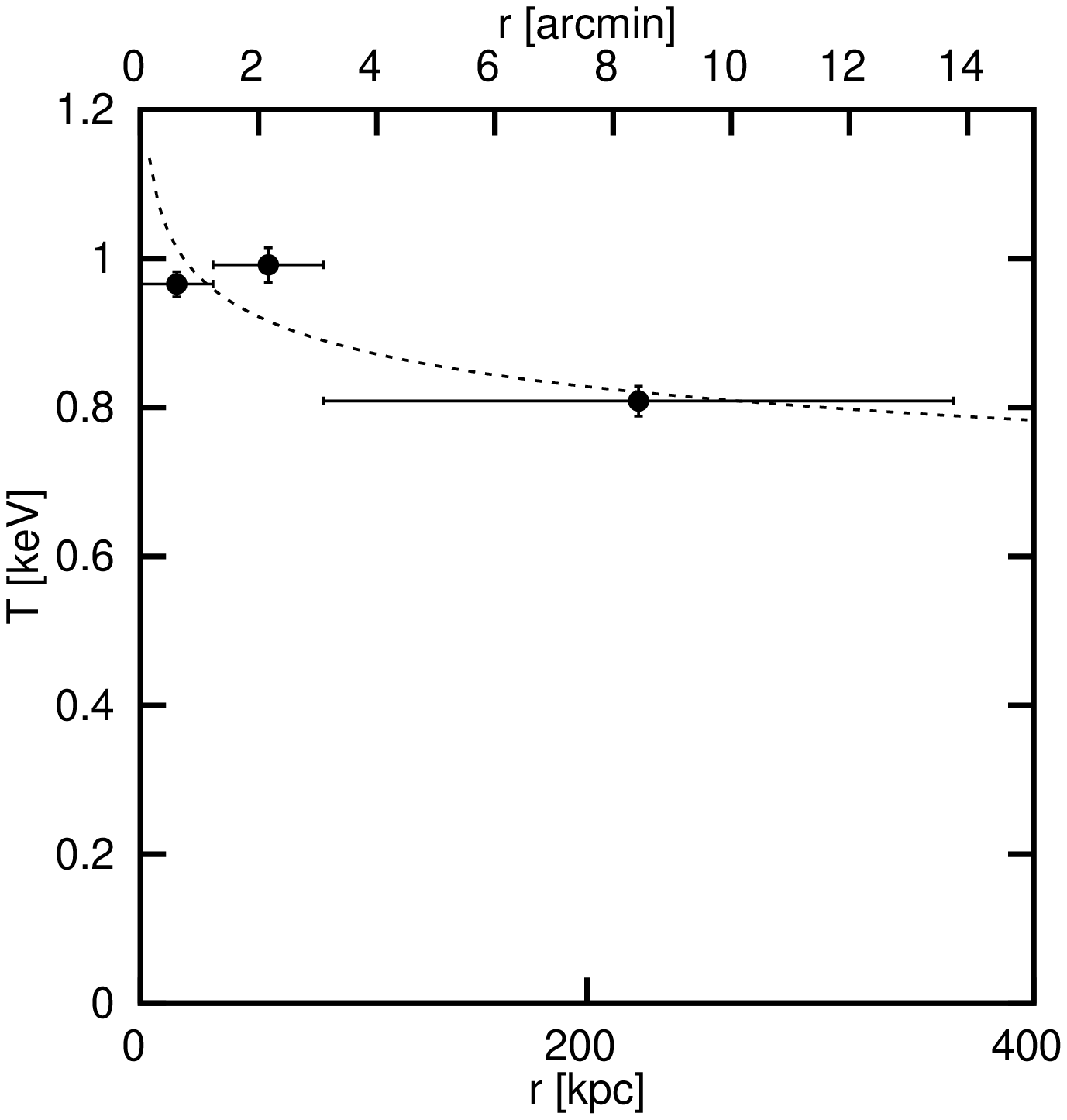}
   \includegraphics[width=0.26\textwidth]{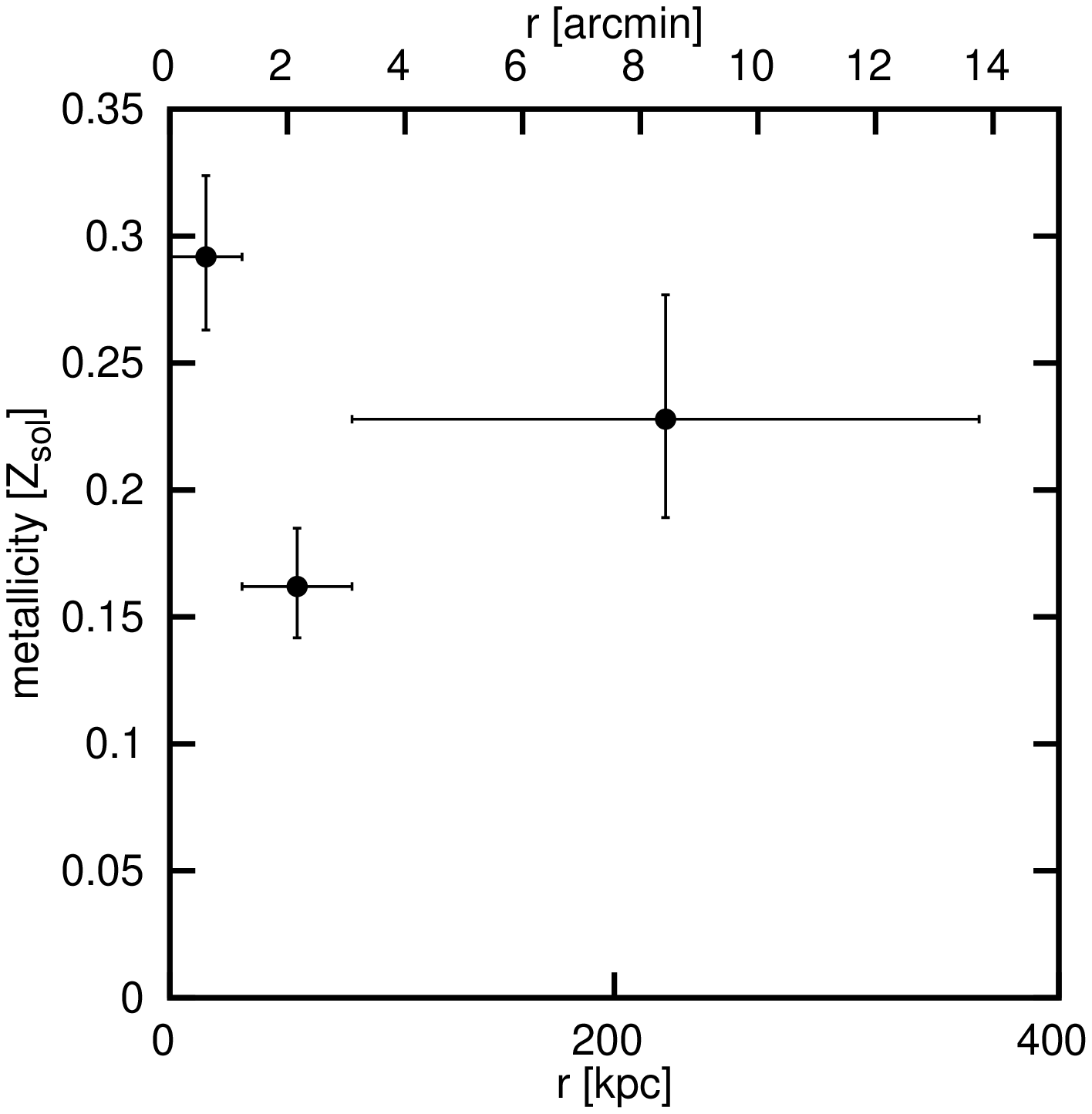}
   \includegraphics[width=0.26\textwidth]{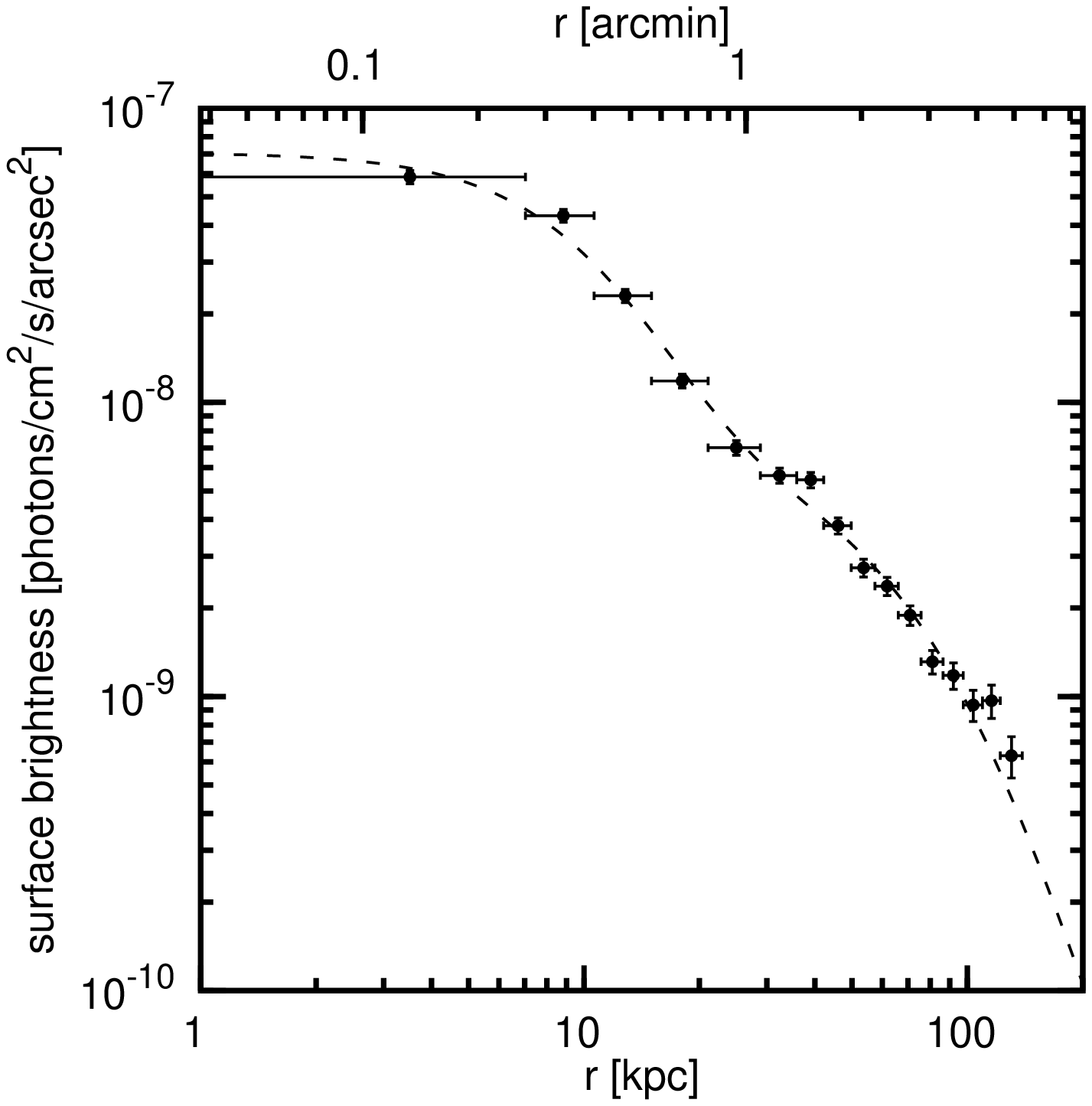}
   \caption{HCG97}
   \label{fig:tprofhcg97}%
\end{figure*}
\begin{figure*}[h]
   \centering
   \includegraphics[width=0.26\textwidth]{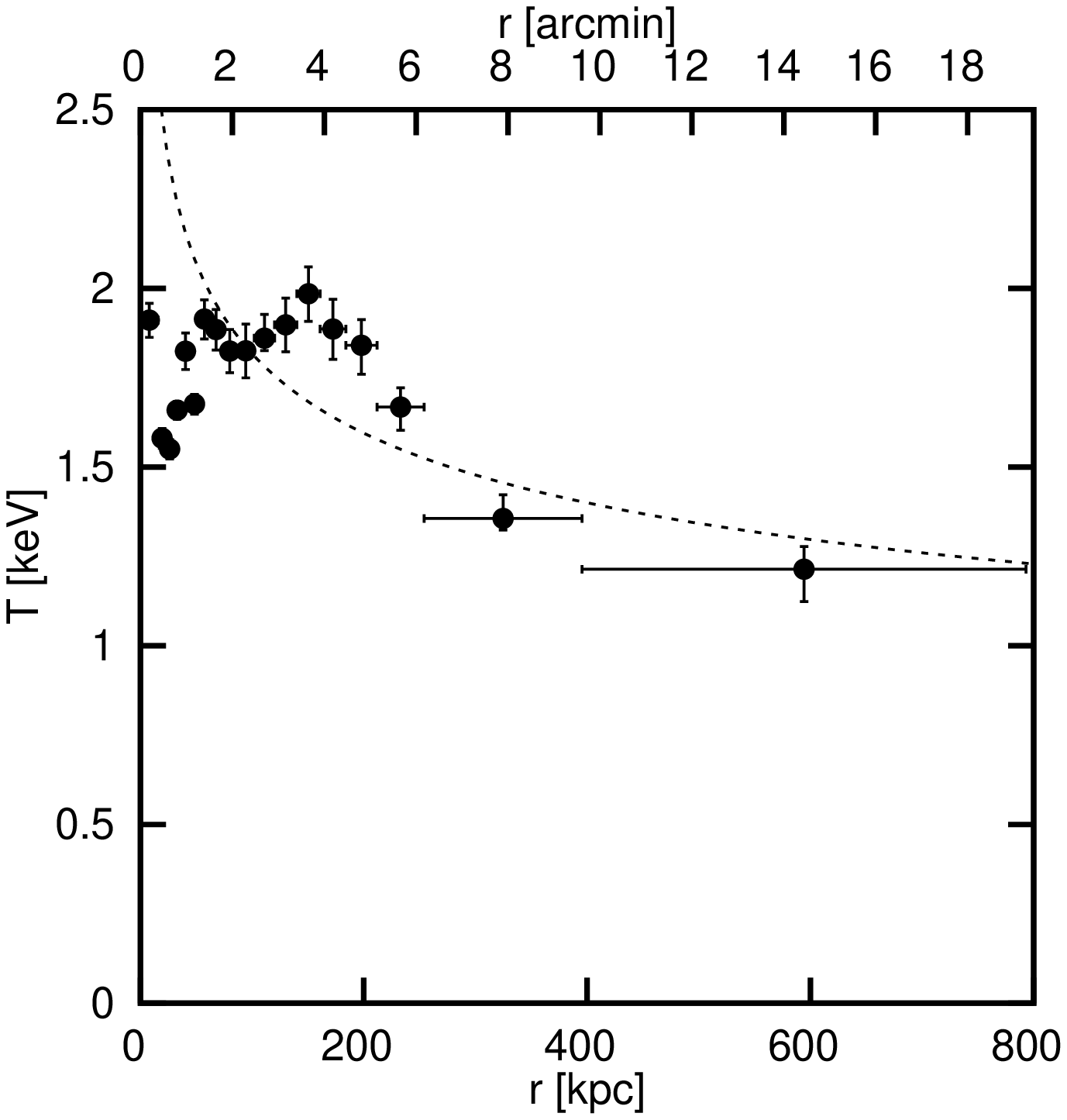}
   \includegraphics[width=0.26\textwidth]{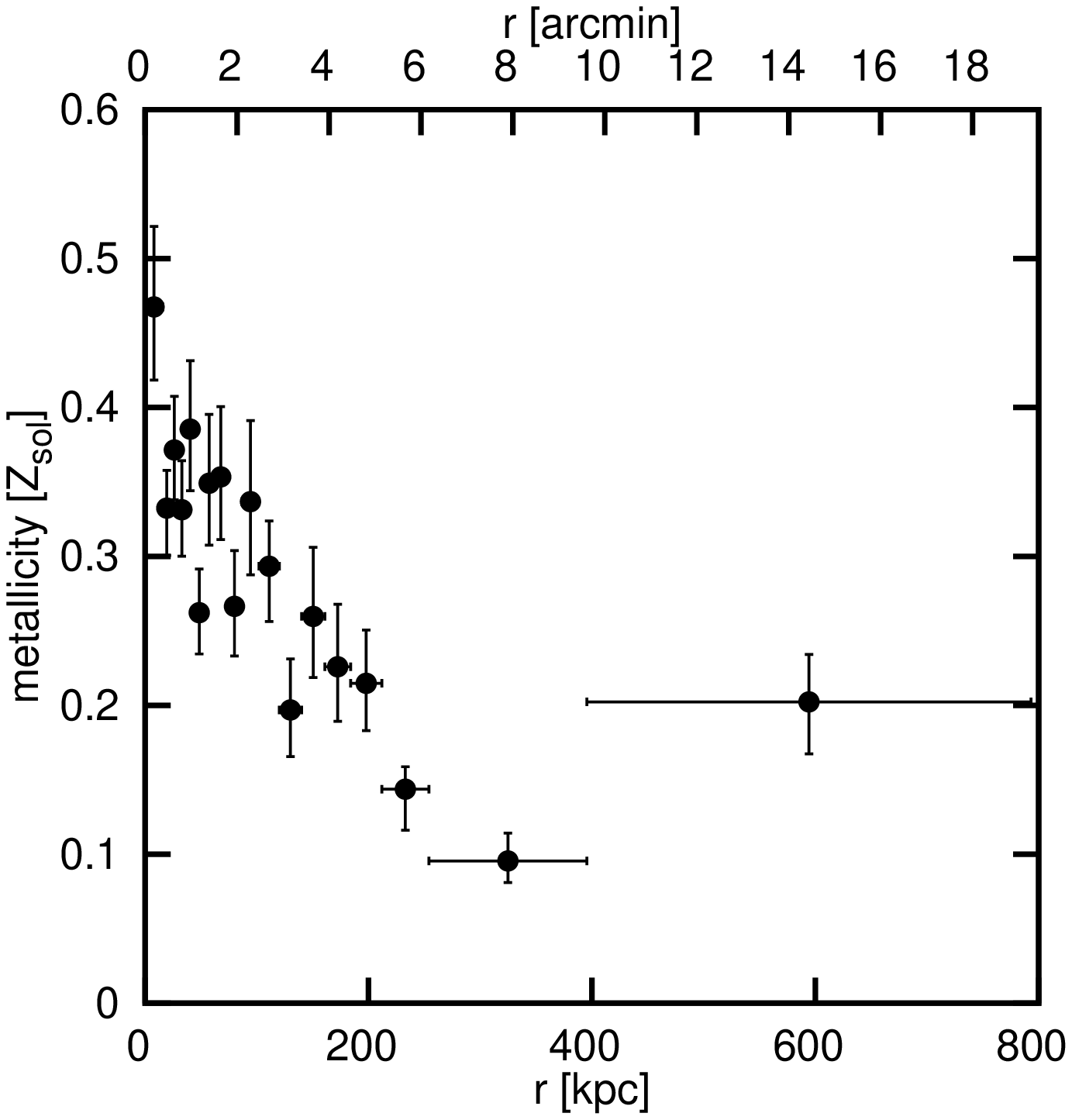}
   \includegraphics[width=0.26\textwidth]{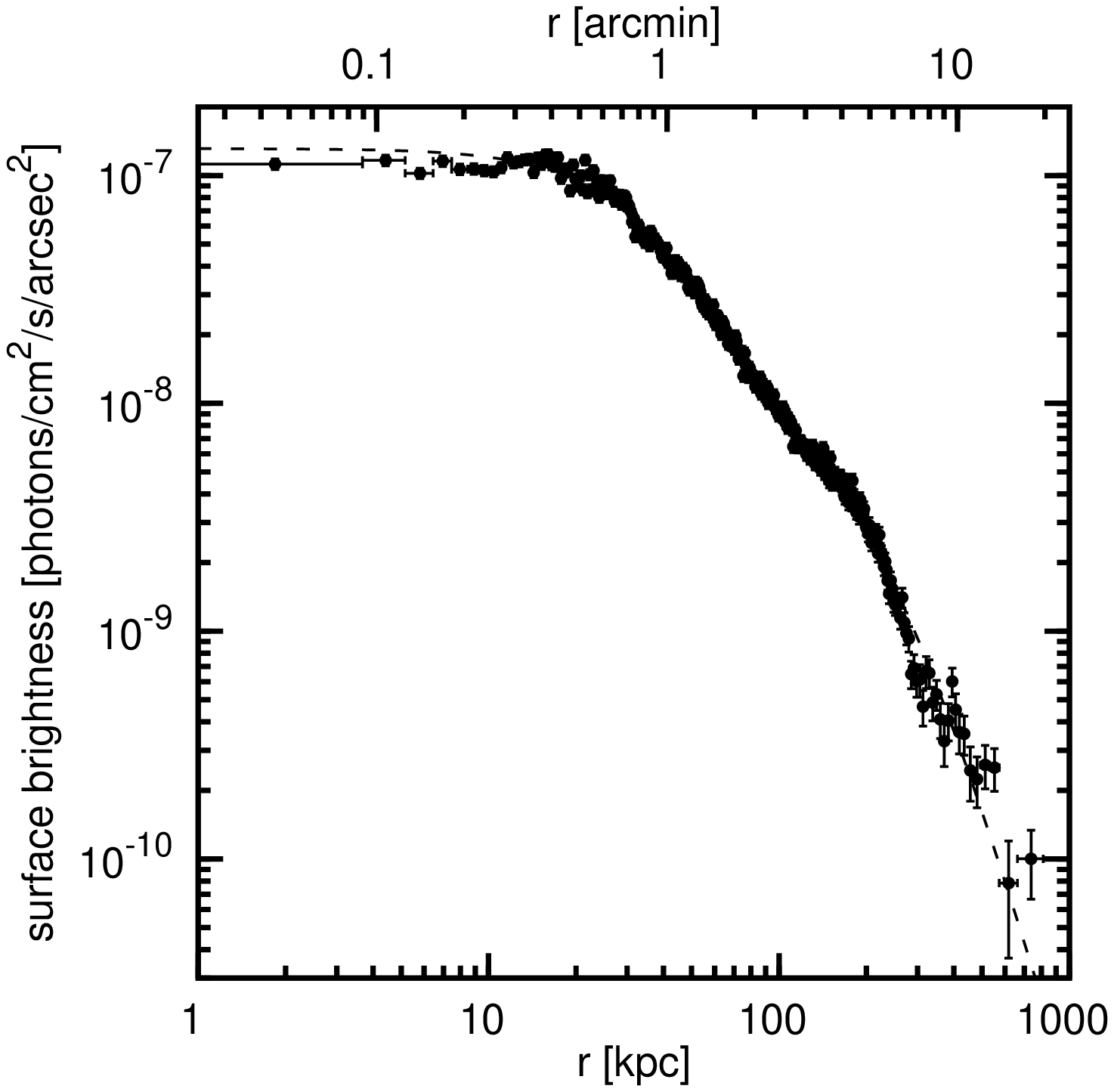}
   \caption{IC1262}
   \label{fig:tprofic1262}%
\end{figure*}
\begin{figure*}[h]
   \centering
   \includegraphics[width=0.26\textwidth]{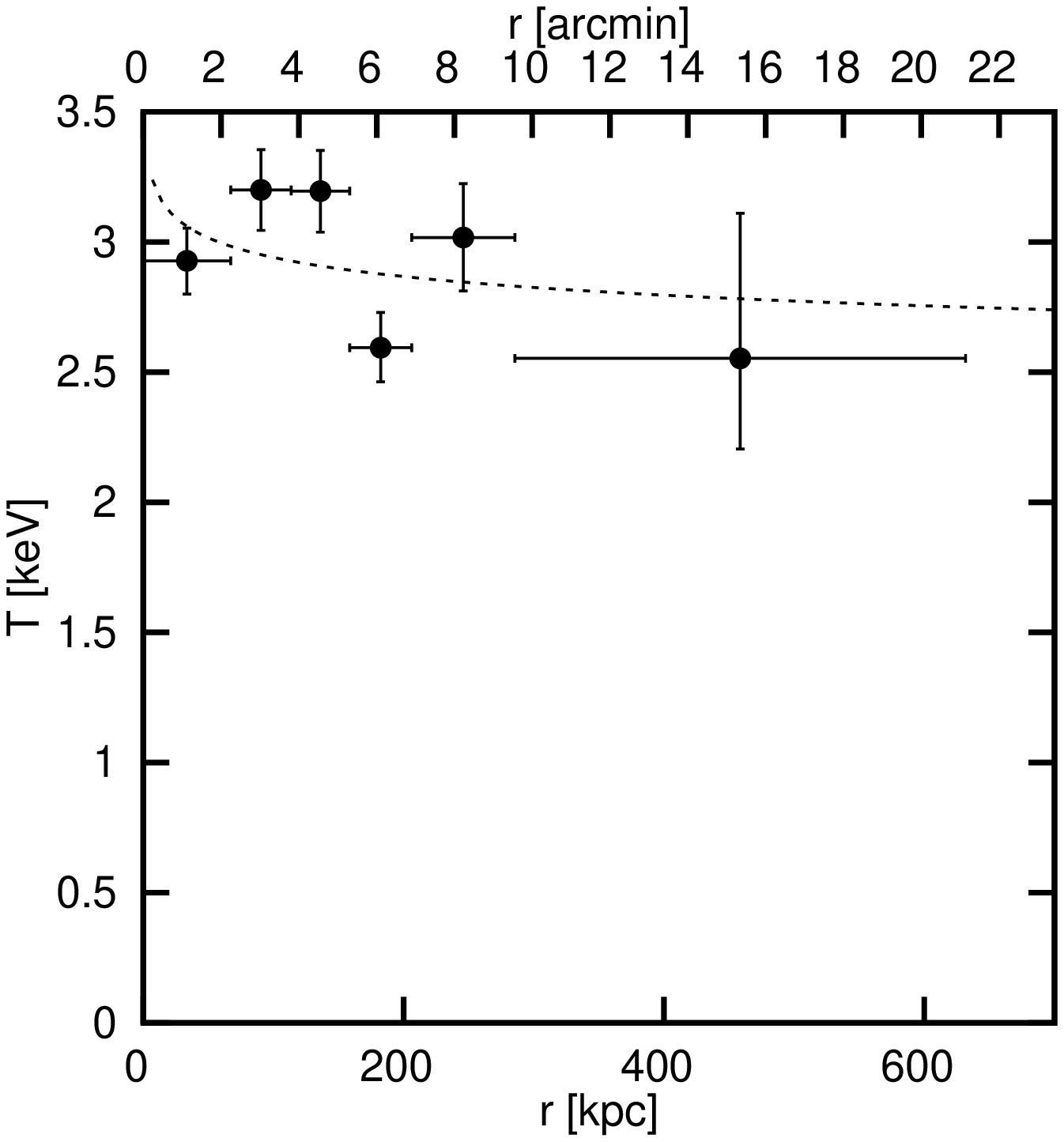}
   \includegraphics[width=0.26\textwidth]{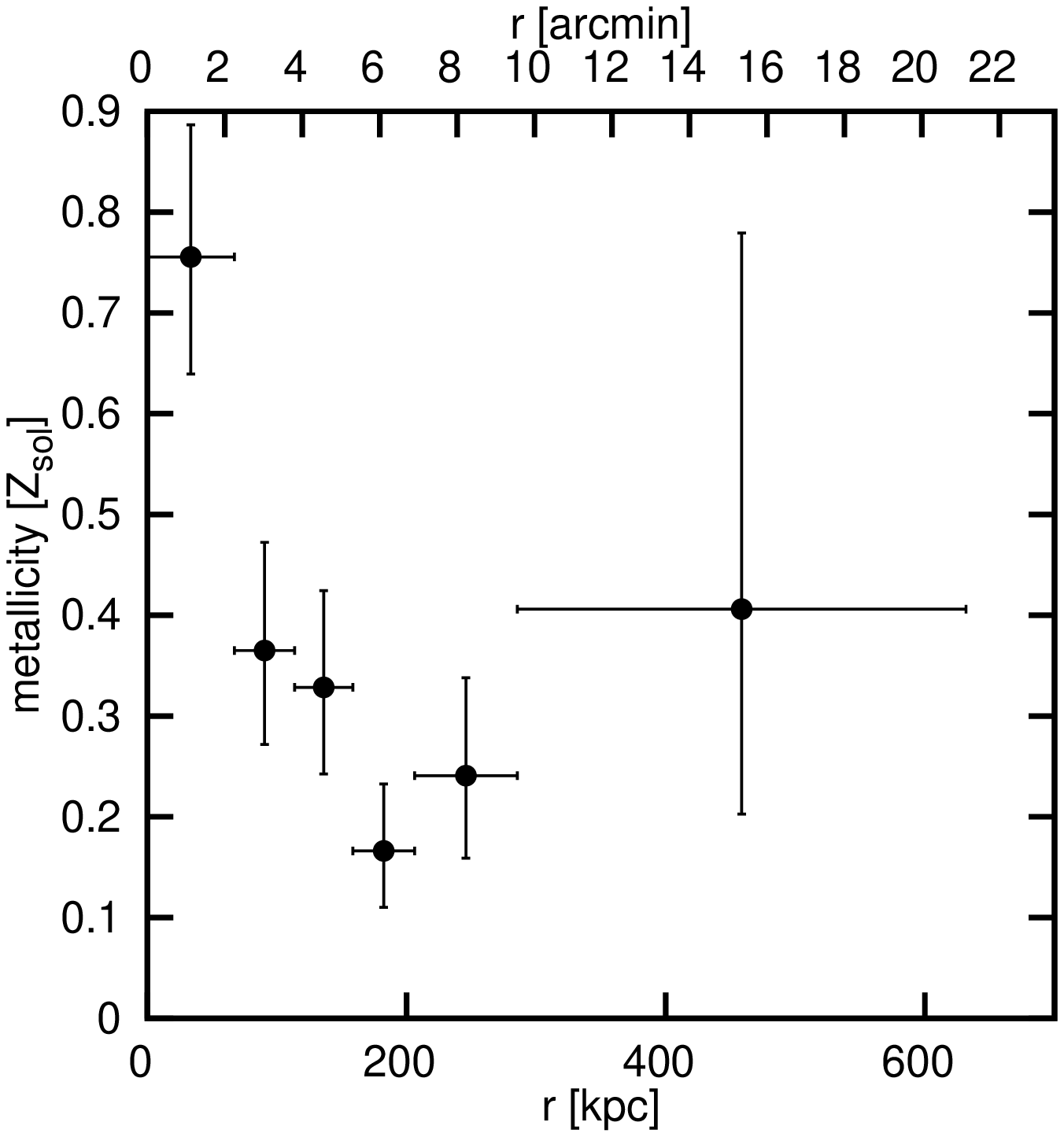}
   \includegraphics[width=0.26\textwidth]{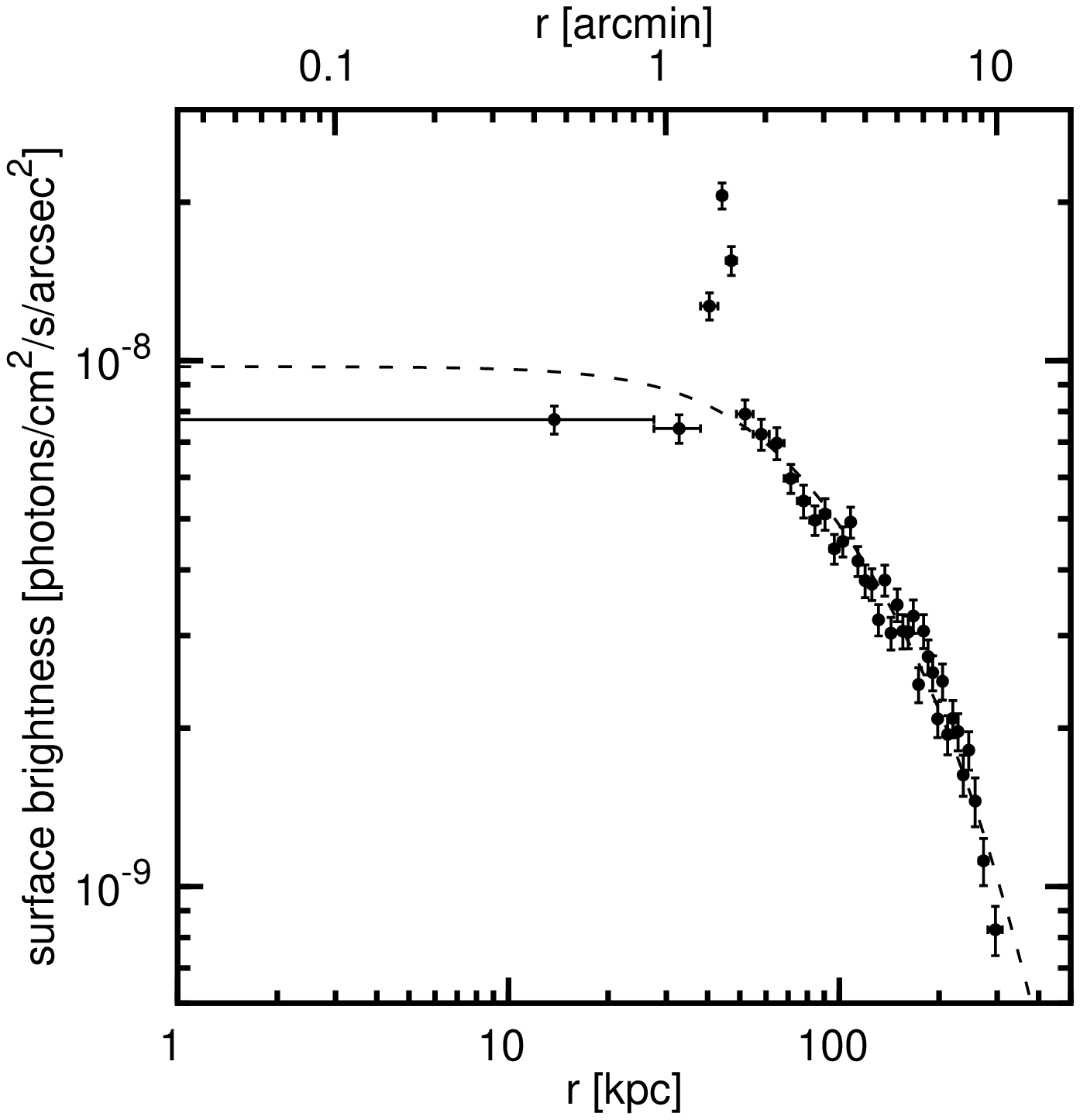}
   \caption{IC1633}
   \label{fig:tprofic1633}%
\end{figure*}
\clearpage
\begin{figure*}[h]
   \centering
   \includegraphics[width=0.26\textwidth]{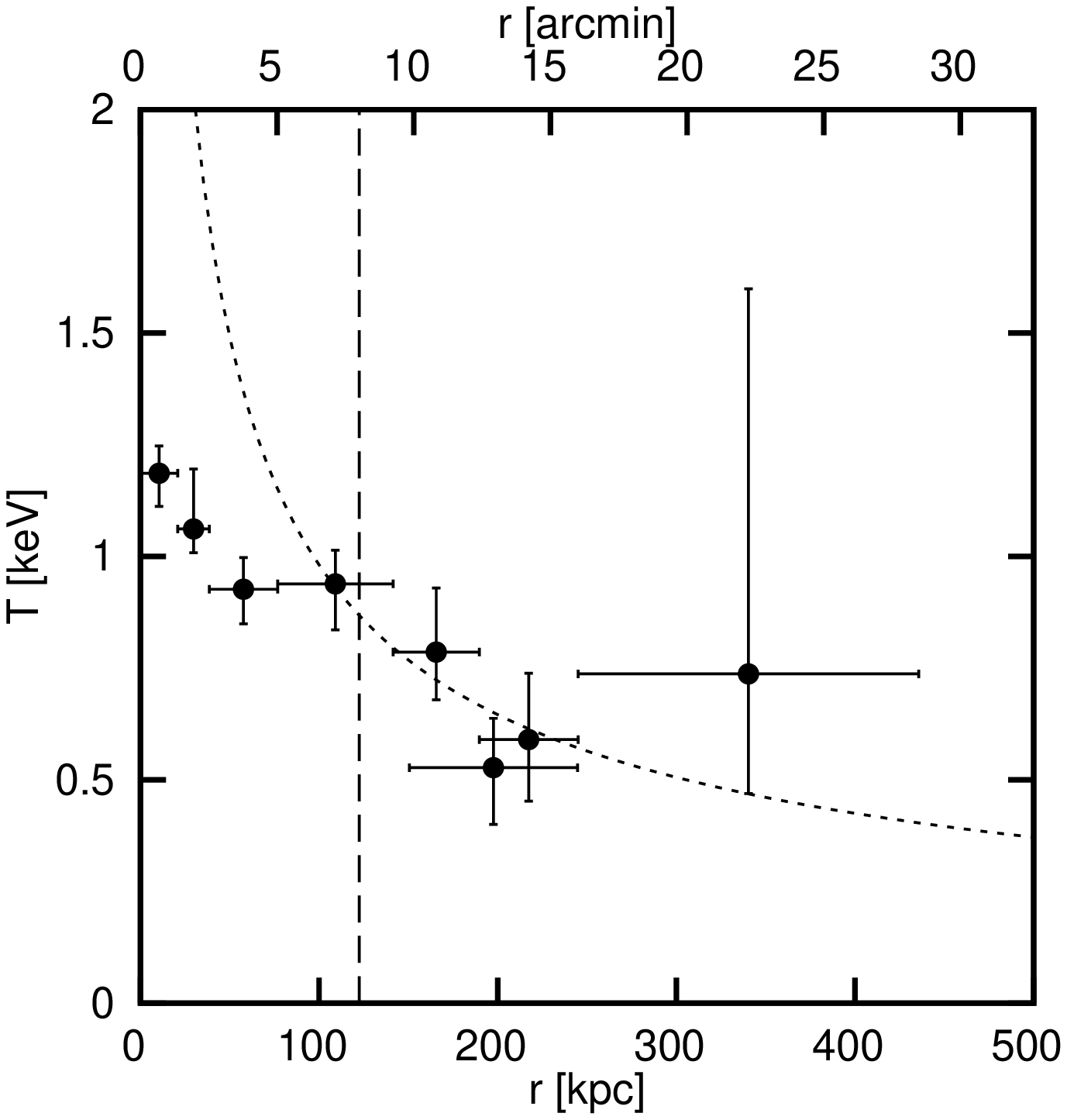}
   \includegraphics[width=0.26\textwidth]{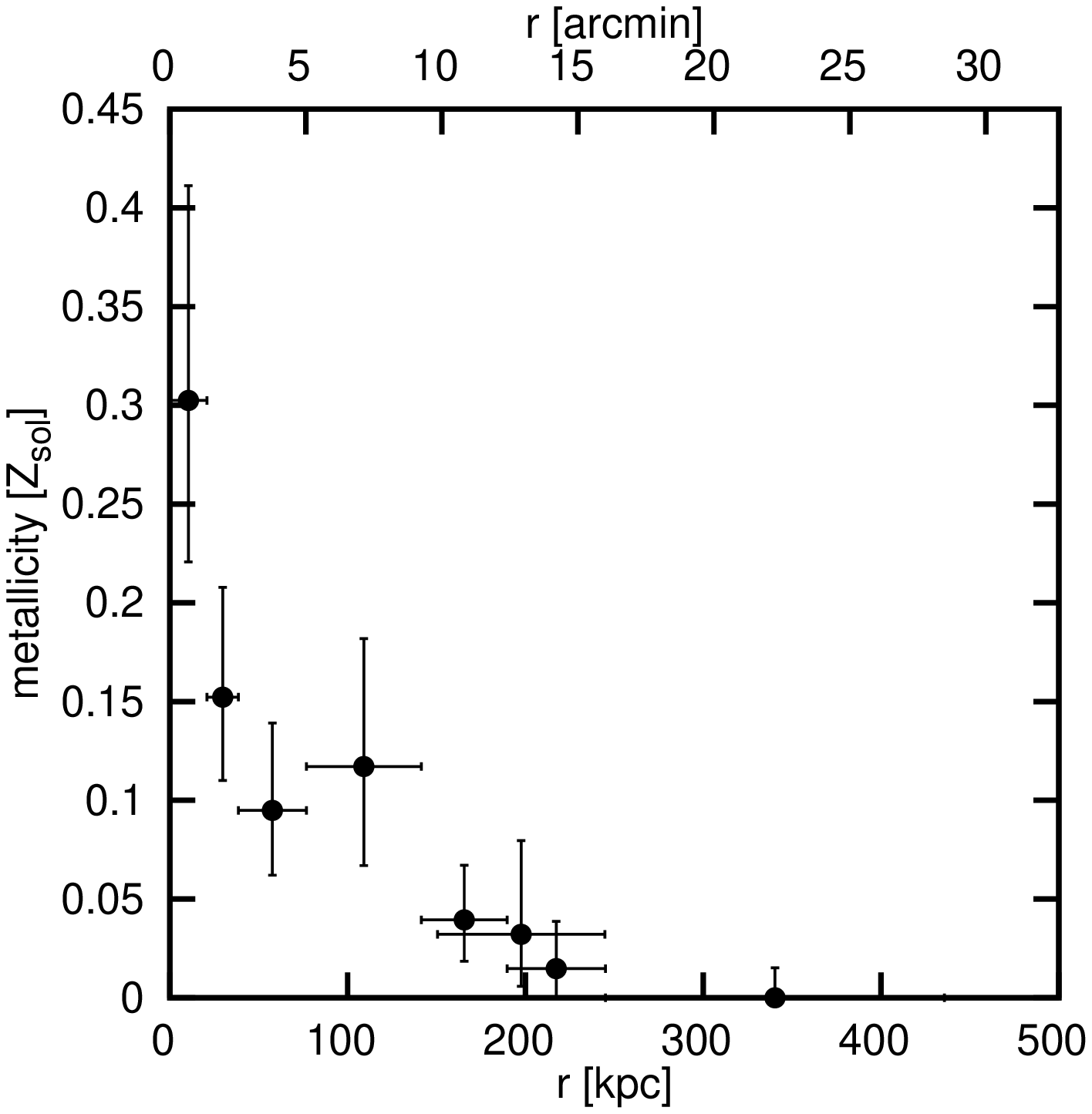}
   \includegraphics[width=0.26\textwidth]{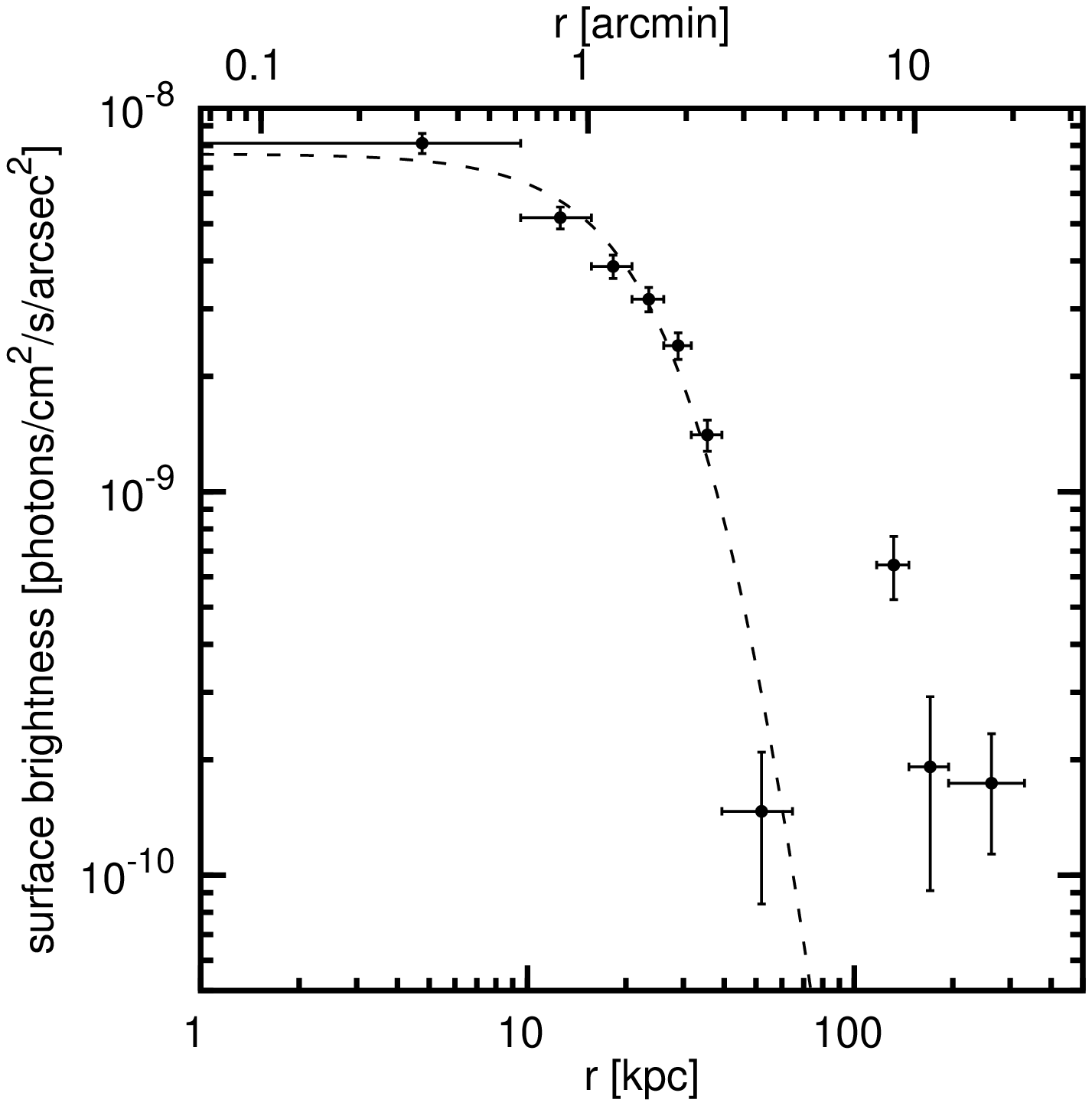}
   \caption{\emph{IC4296}}
   \label{fig:tprofic4296}%
\end{figure*}
\begin{figure*}[h]
   \centering
   \includegraphics[width=0.26\textwidth]{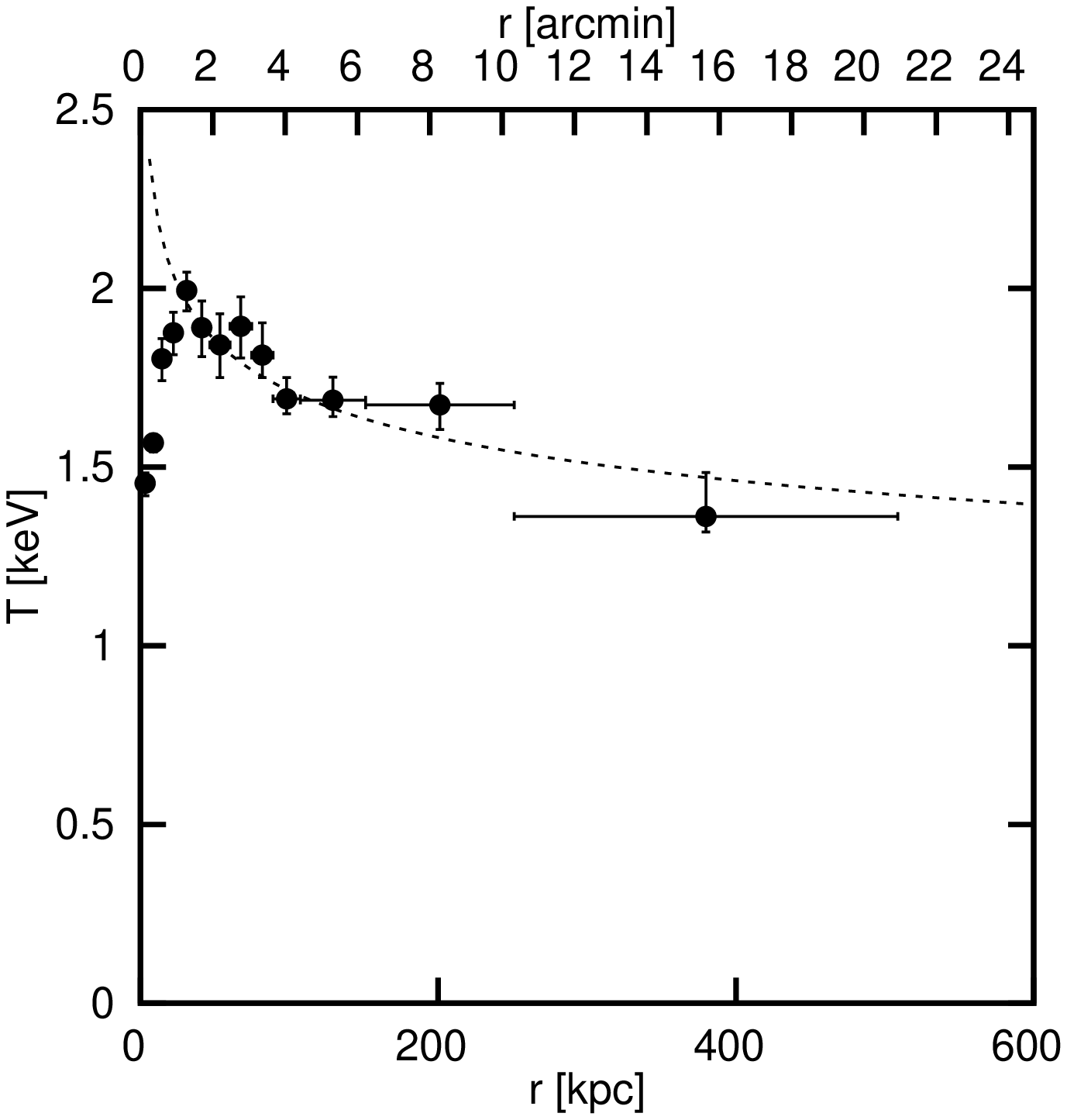}
   \includegraphics[width=0.26\textwidth]{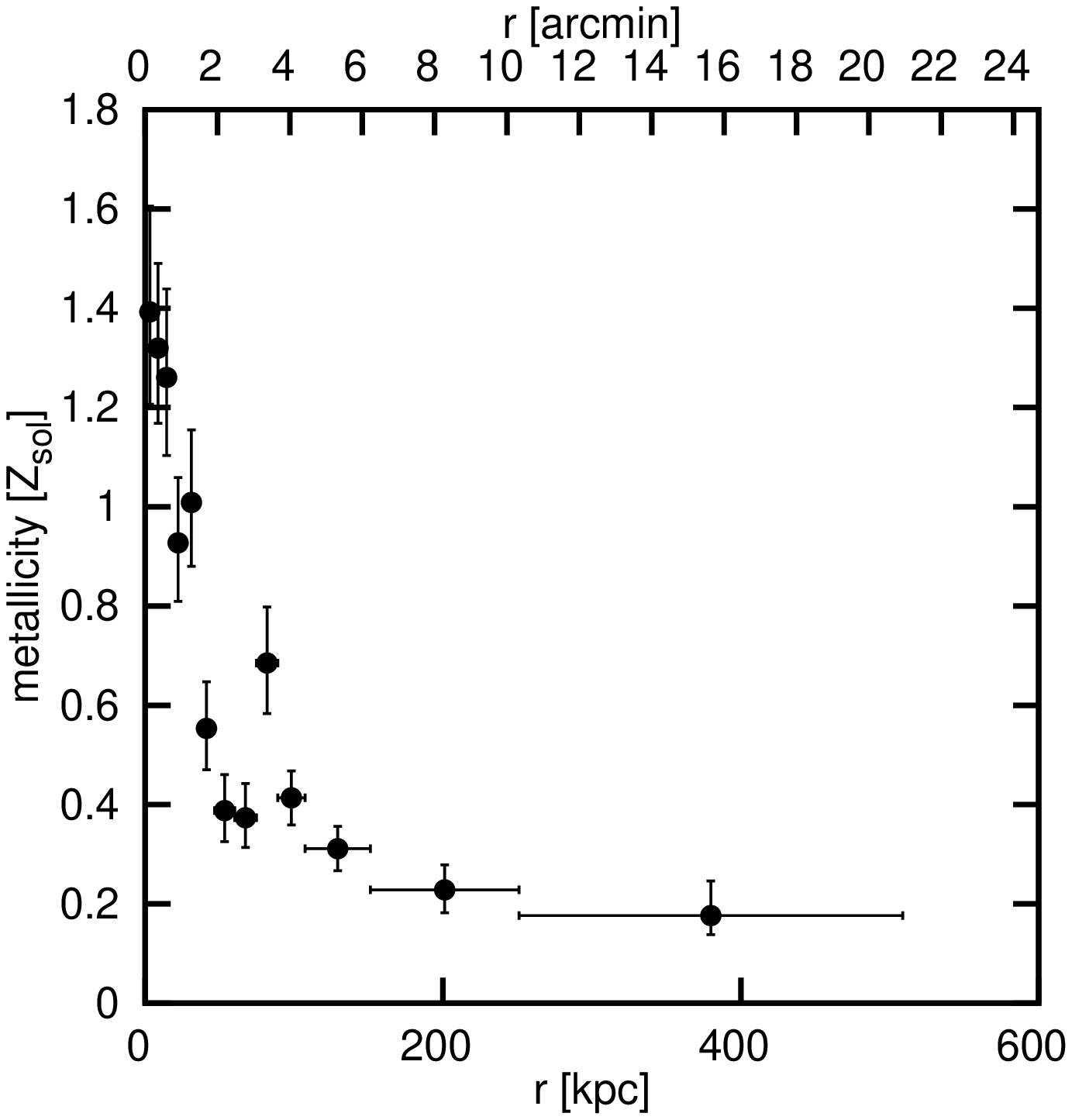}
   \includegraphics[width=0.26\textwidth]{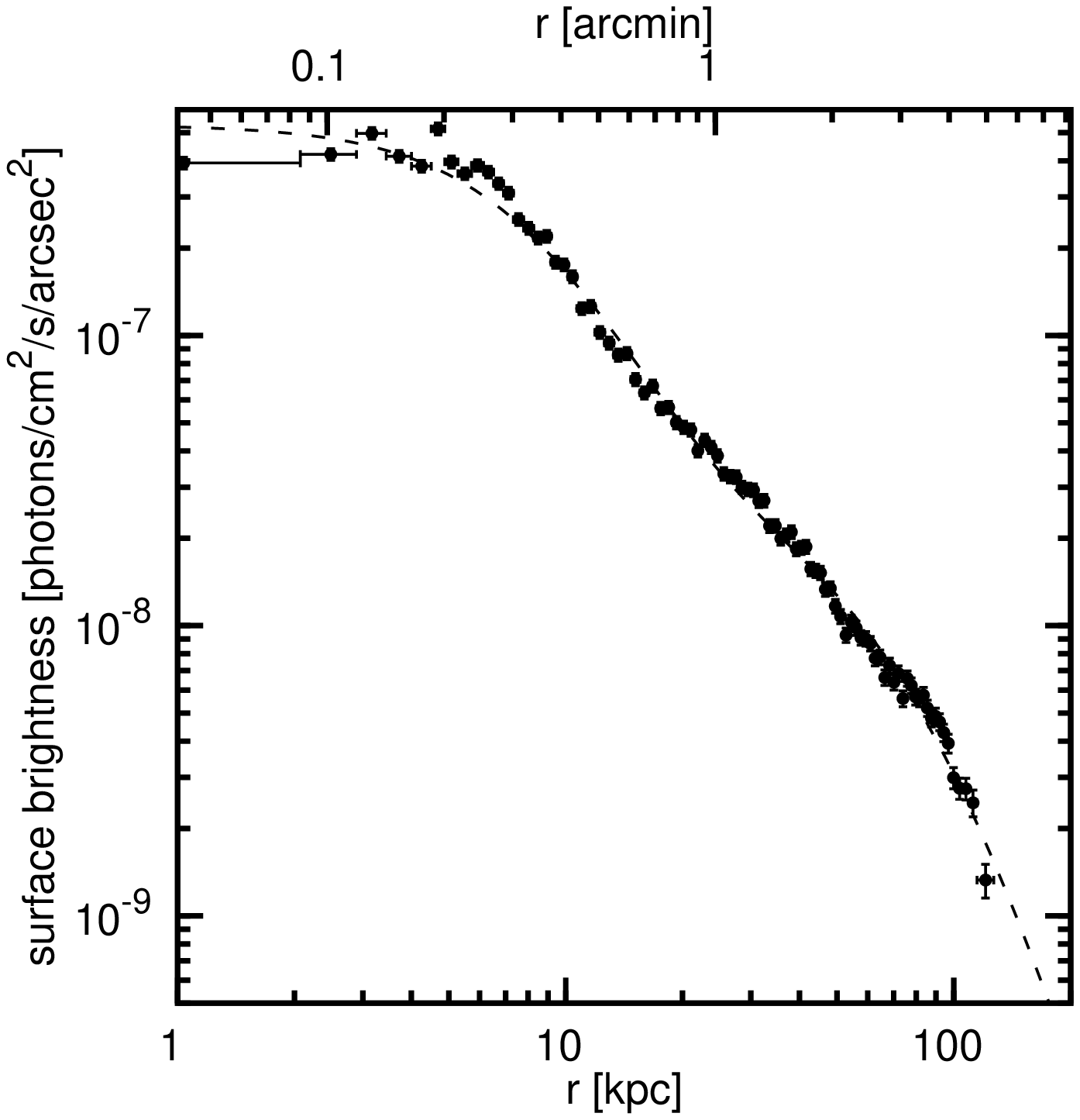}
   \caption{MKW4}
   \label{fig:tprofmkw4}%
\end{figure*}
\begin{figure*}[h]
   \centering
   \includegraphics[width=0.26\textwidth]{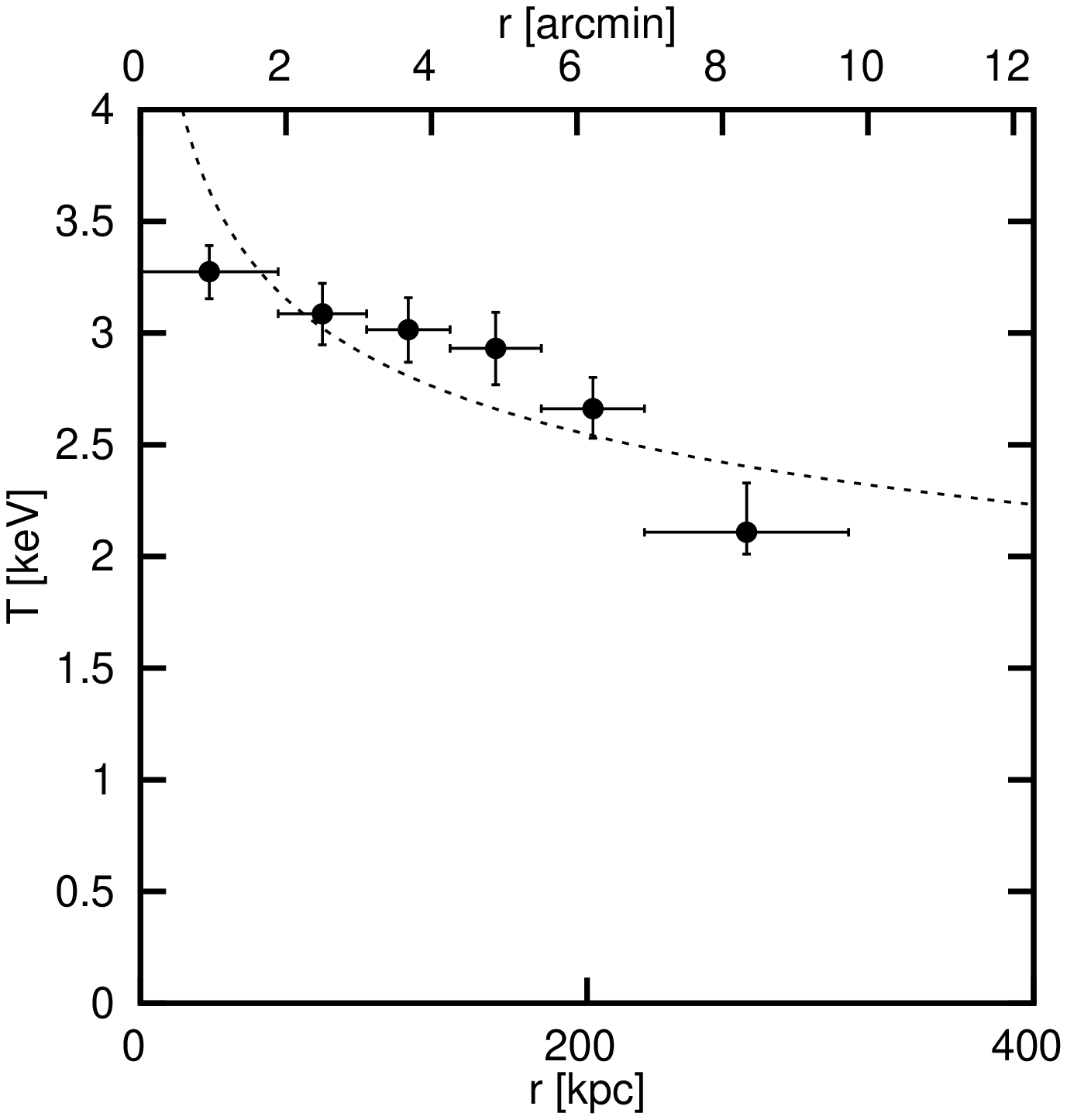}
   \includegraphics[width=0.26\textwidth]{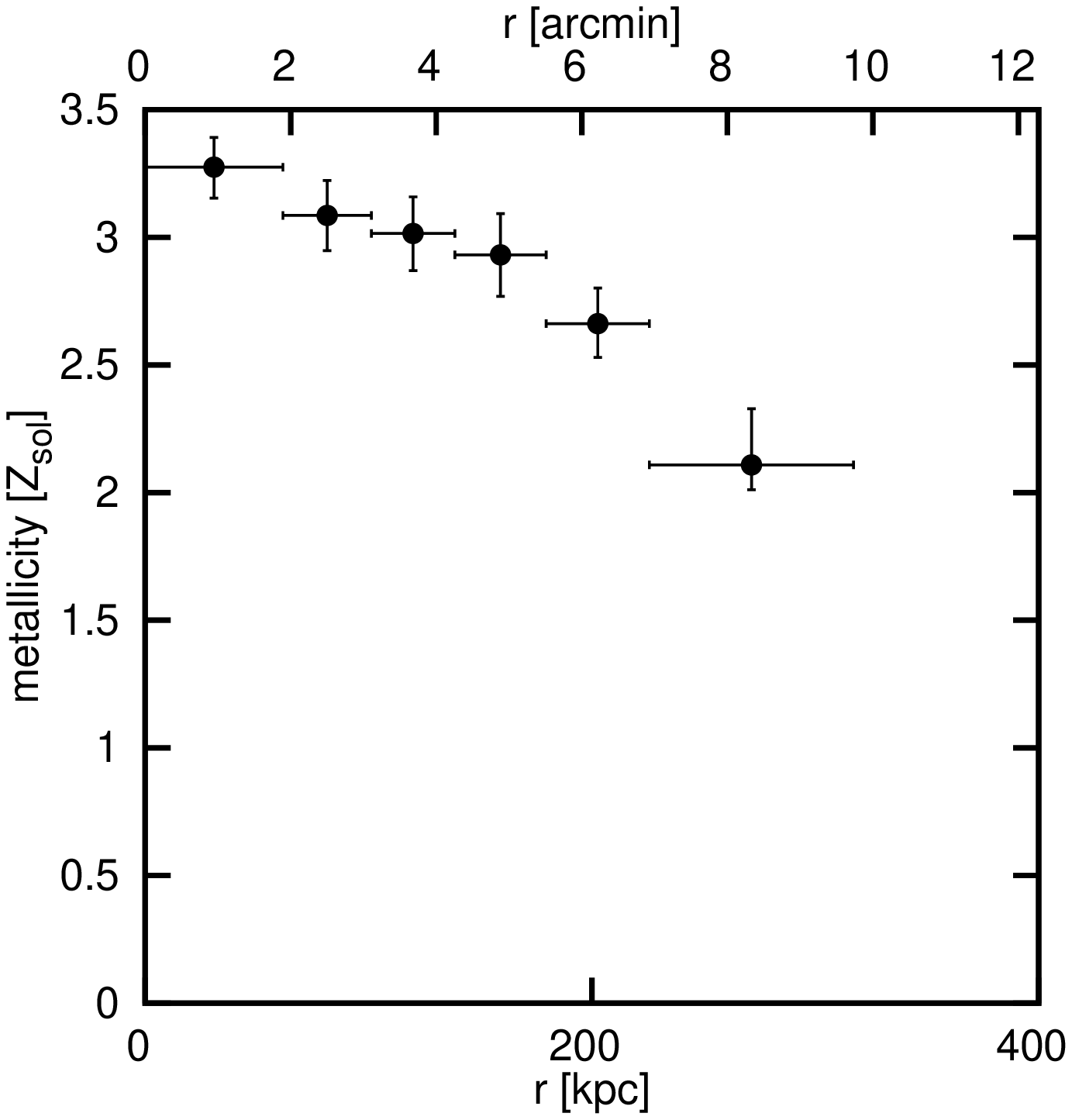}
   \includegraphics[width=0.26\textwidth]{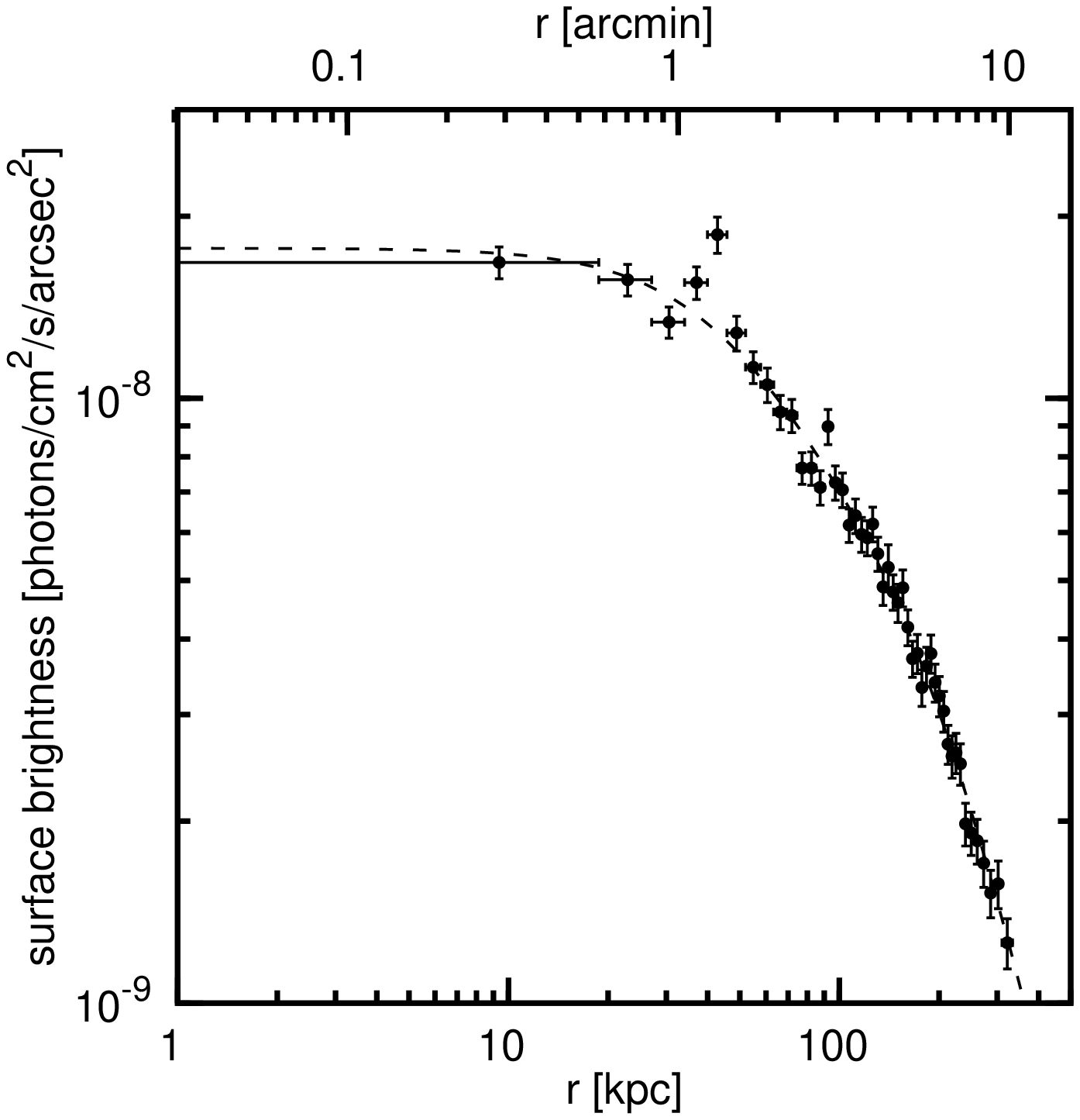}
   \caption{MKW8}
   \label{fig:tprofmkw8}%
\end{figure*}
\begin{figure*}[h]
   \centering
   \includegraphics[width=0.26\textwidth]{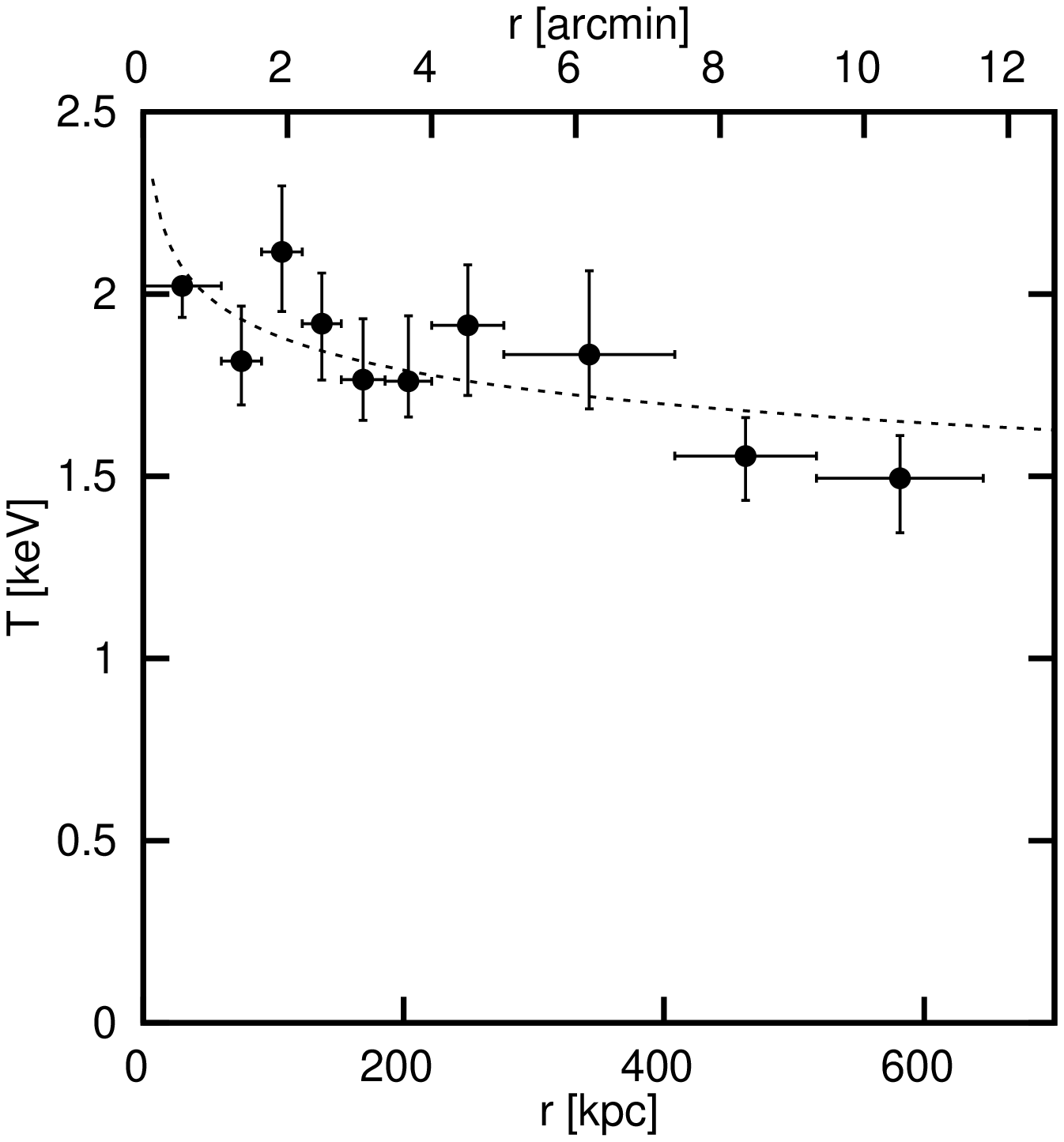}
   \includegraphics[width=0.26\textwidth]{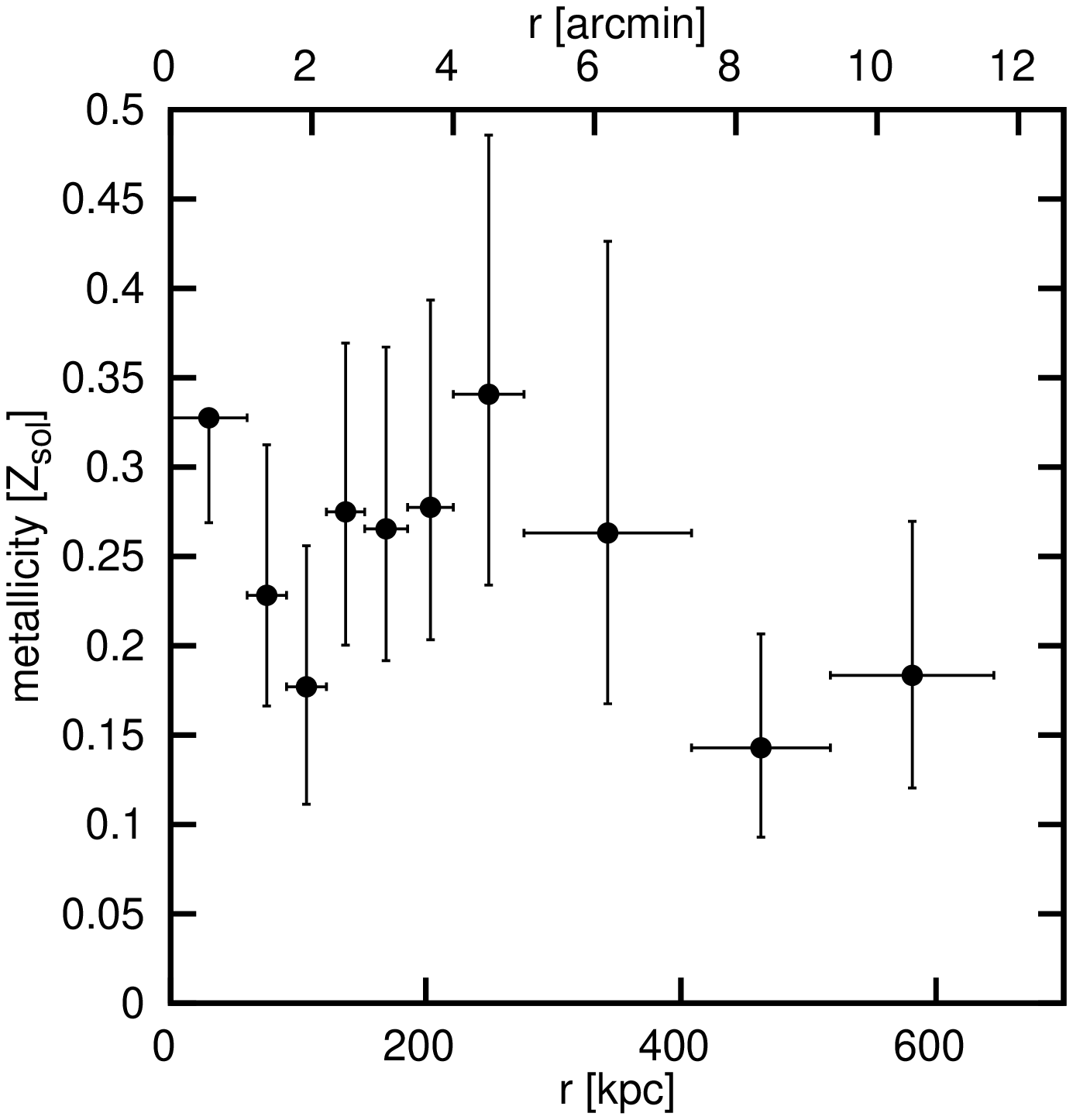}
   \includegraphics[width=0.26\textwidth]{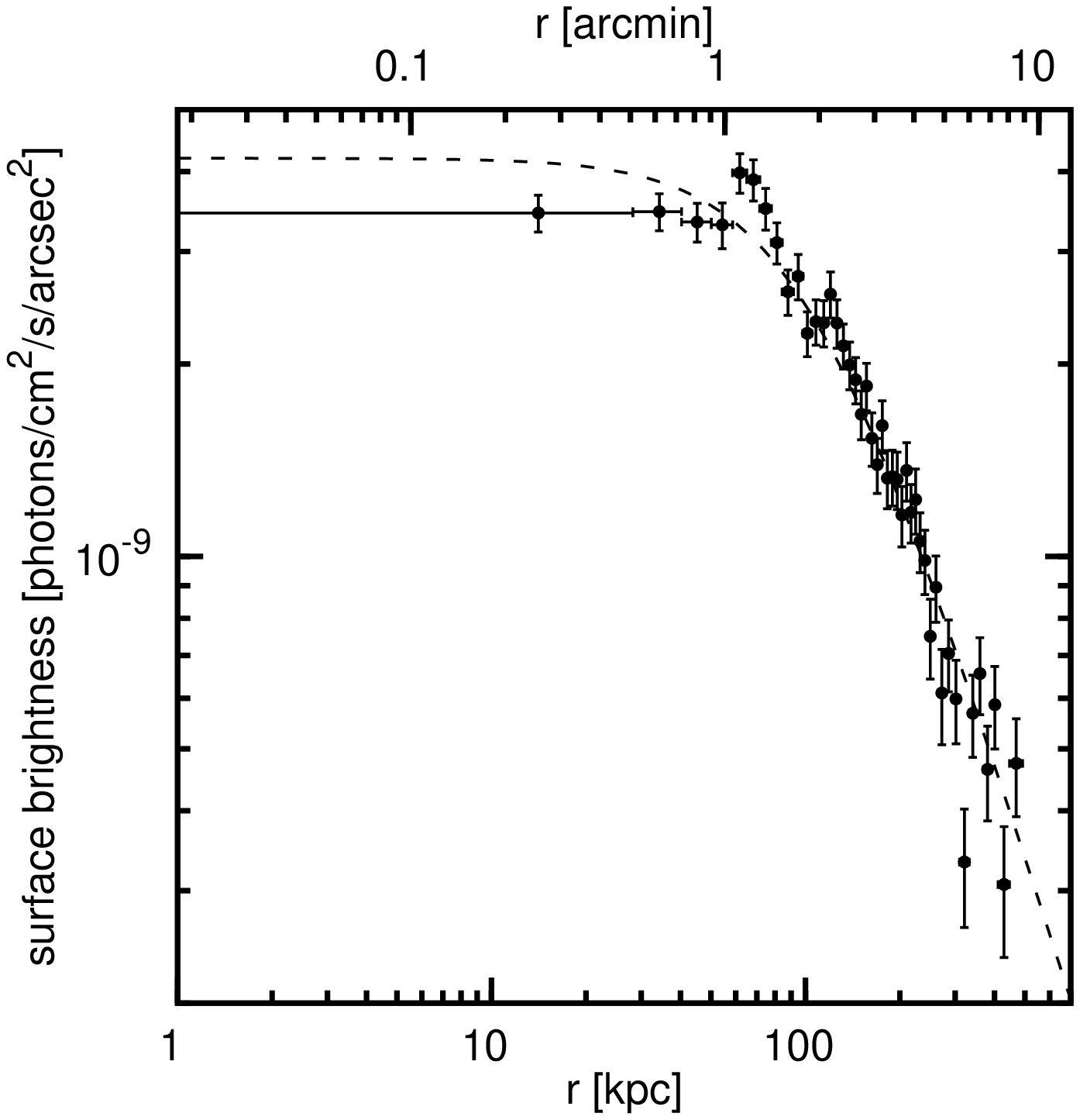}
   \caption{NGC326}
   \label{fig:tprofngc326}%
\end{figure*}
\clearpage
\begin{figure*}[h]
   \centering
   \includegraphics[width=0.26\textwidth]{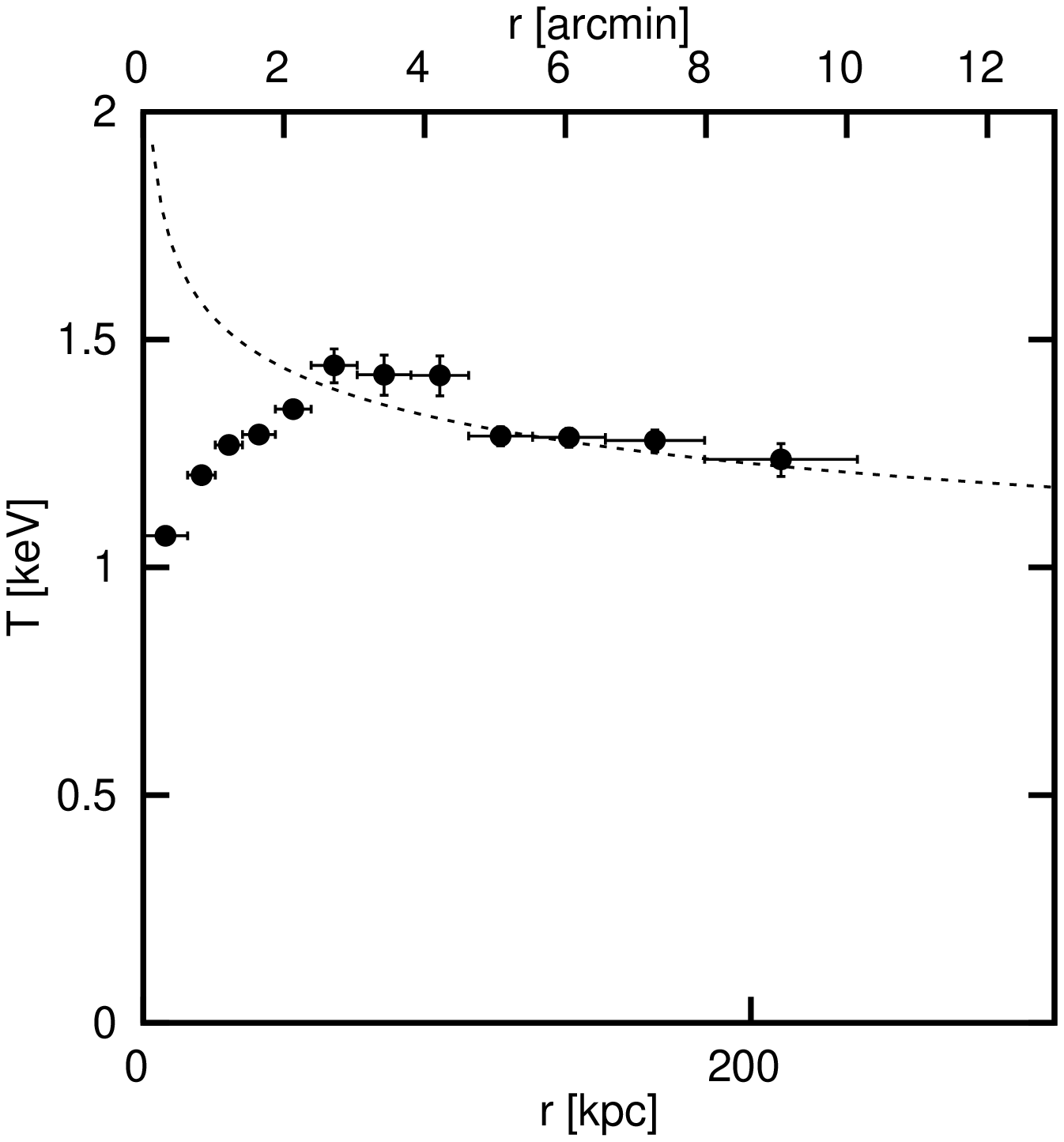}
   \includegraphics[width=0.26\textwidth]{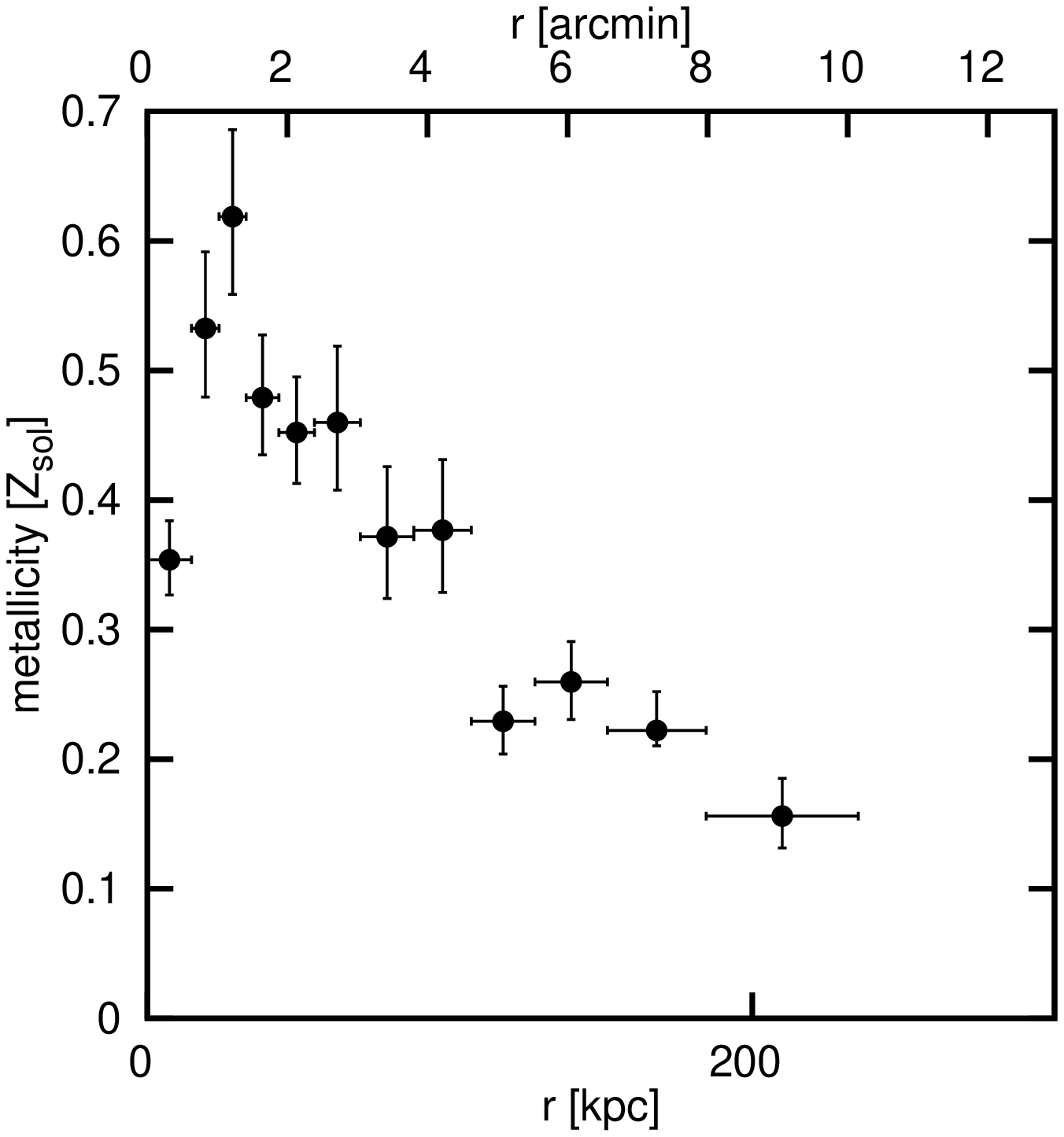}
   \includegraphics[width=0.26\textwidth]{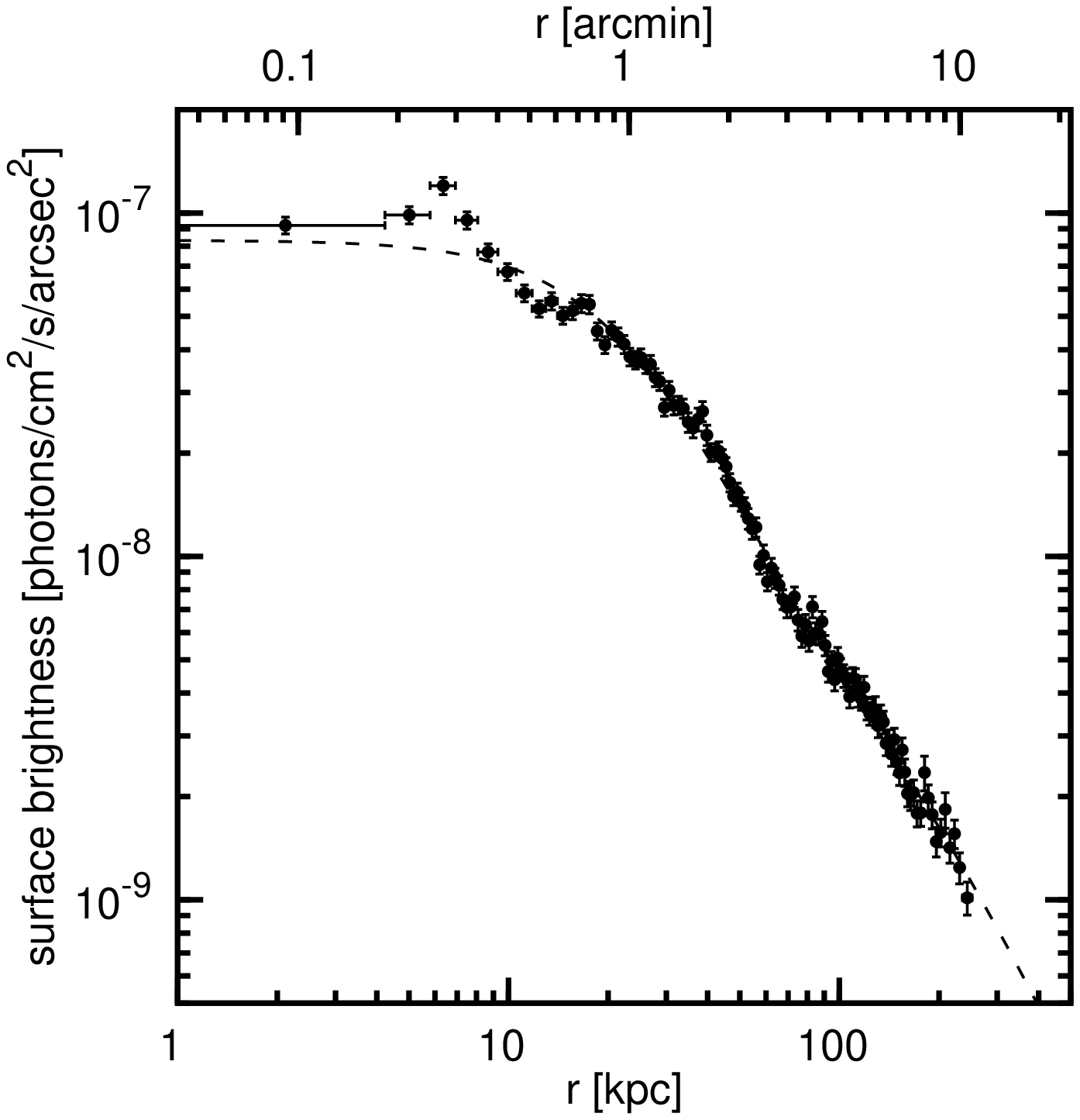}
   \caption{NGC507}
   \label{fig:tprofngc507}%
\end{figure*}
\begin{figure*}[h]
   \centering
   \includegraphics[width=0.26\textwidth]{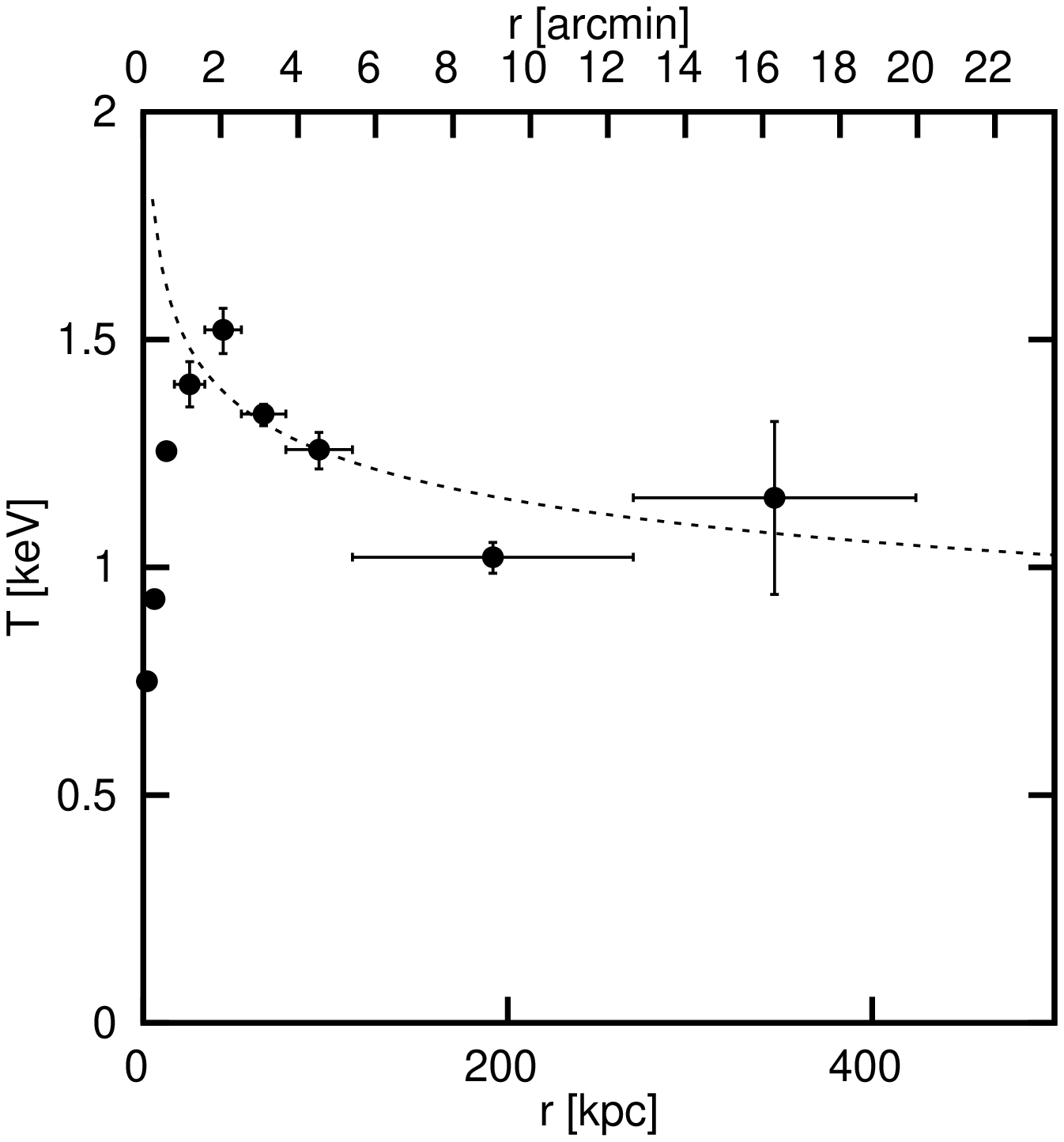}
   \includegraphics[width=0.26\textwidth]{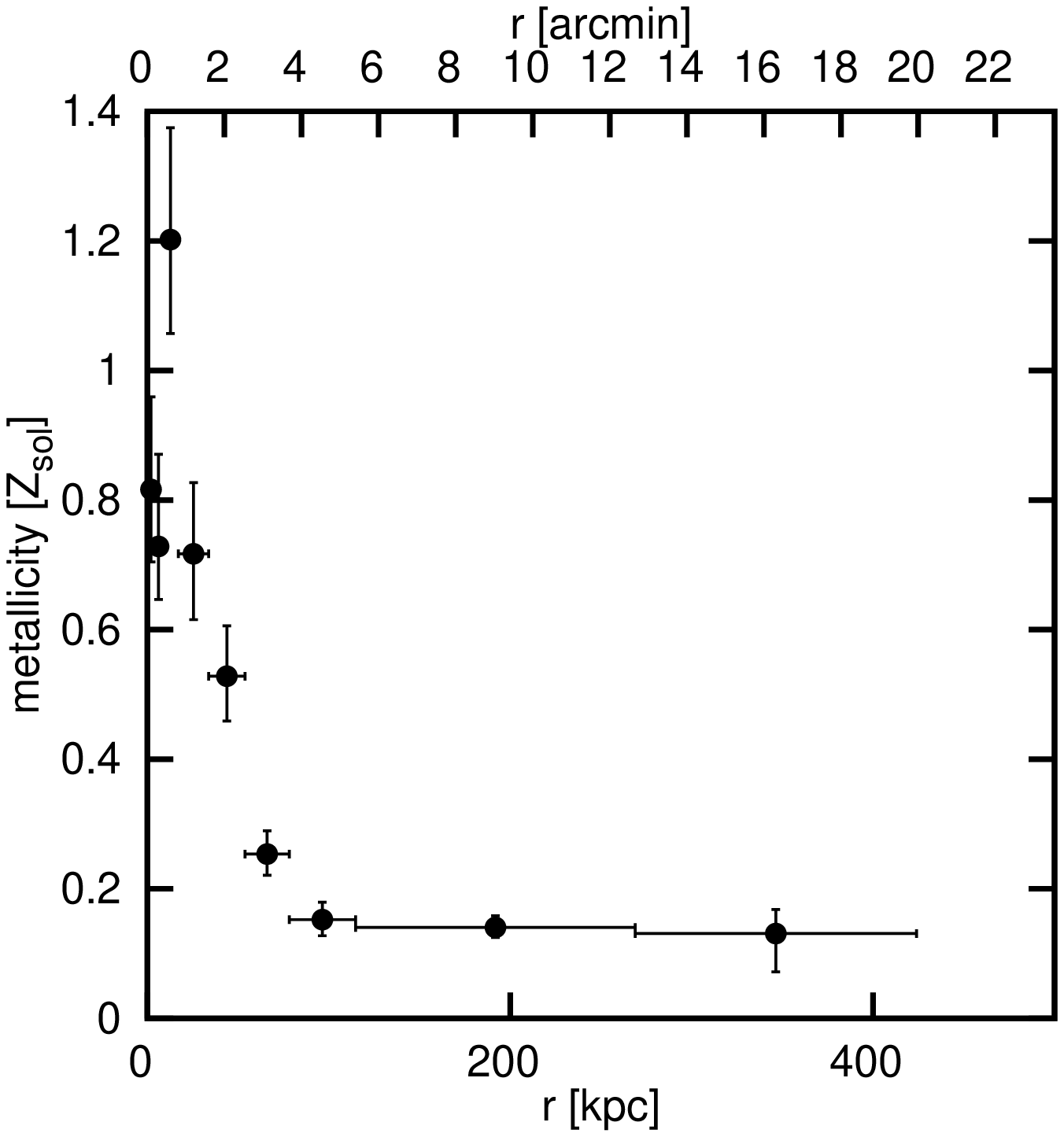}
   \includegraphics[width=0.26\textwidth]{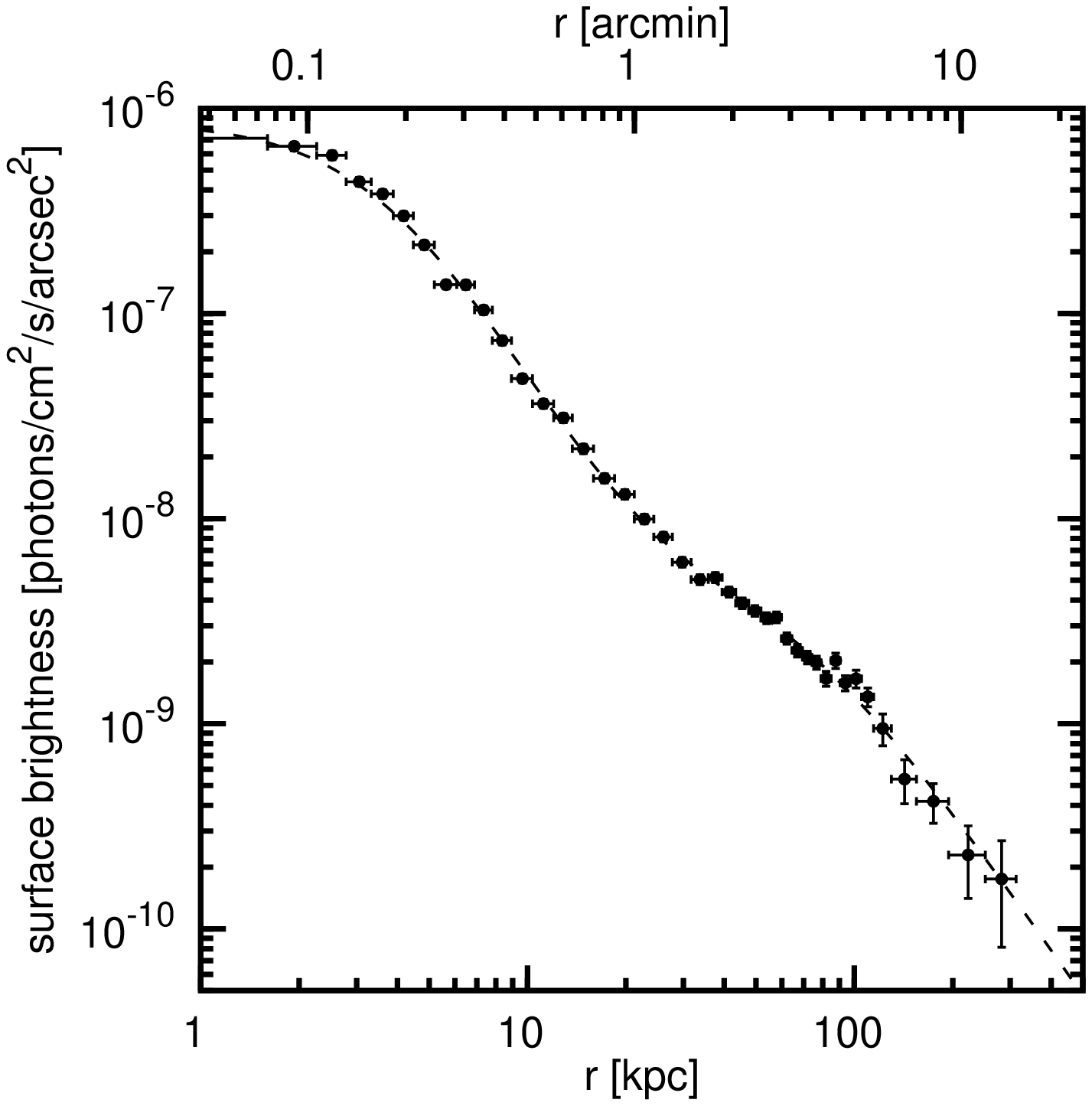}
   \caption{NGC533}
   \label{fig:tprofngc533}%
\end{figure*}
\begin{figure*}[h]
   \centering
   \includegraphics[width=0.26\textwidth]{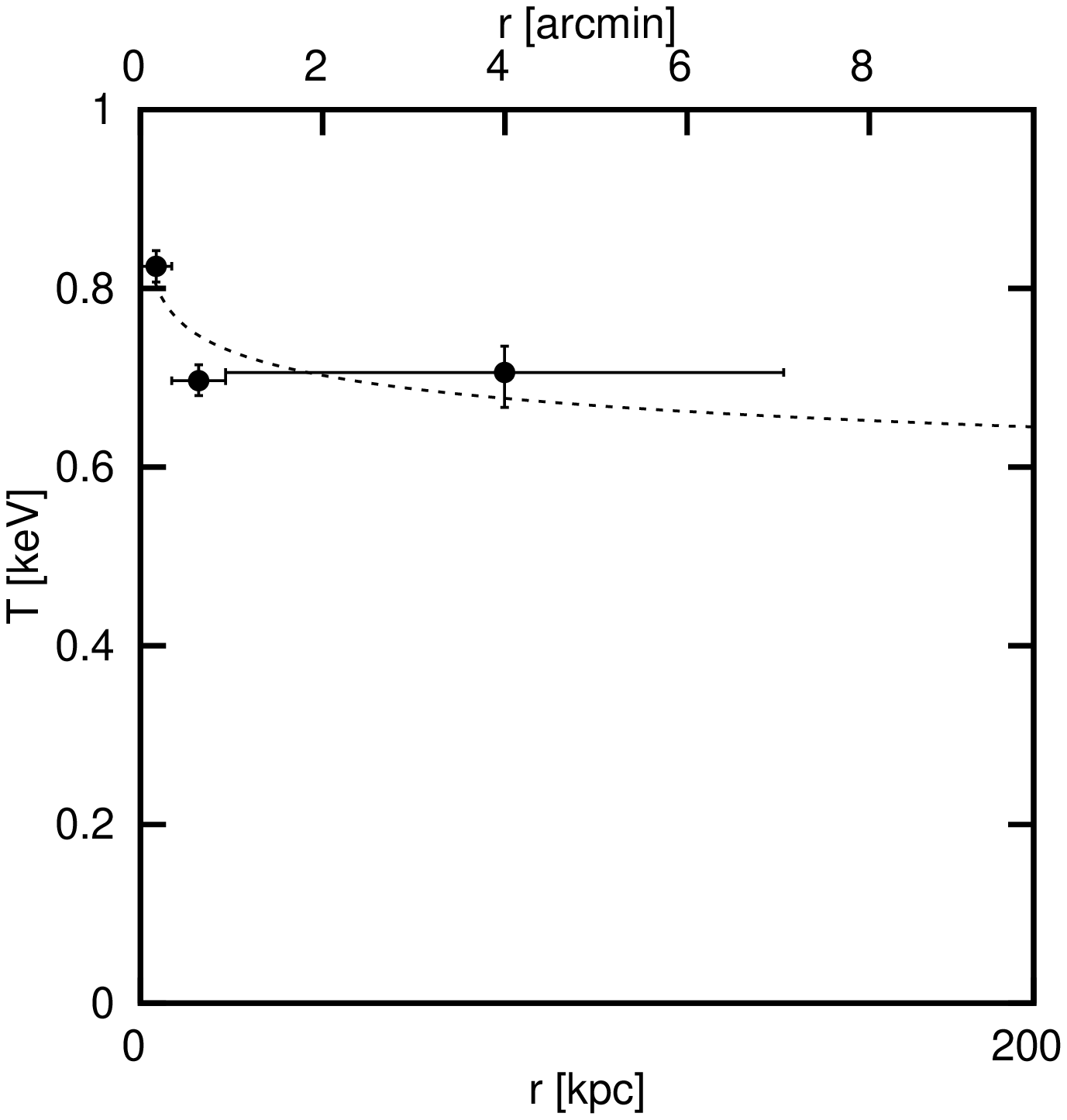}
   \includegraphics[width=0.26\textwidth]{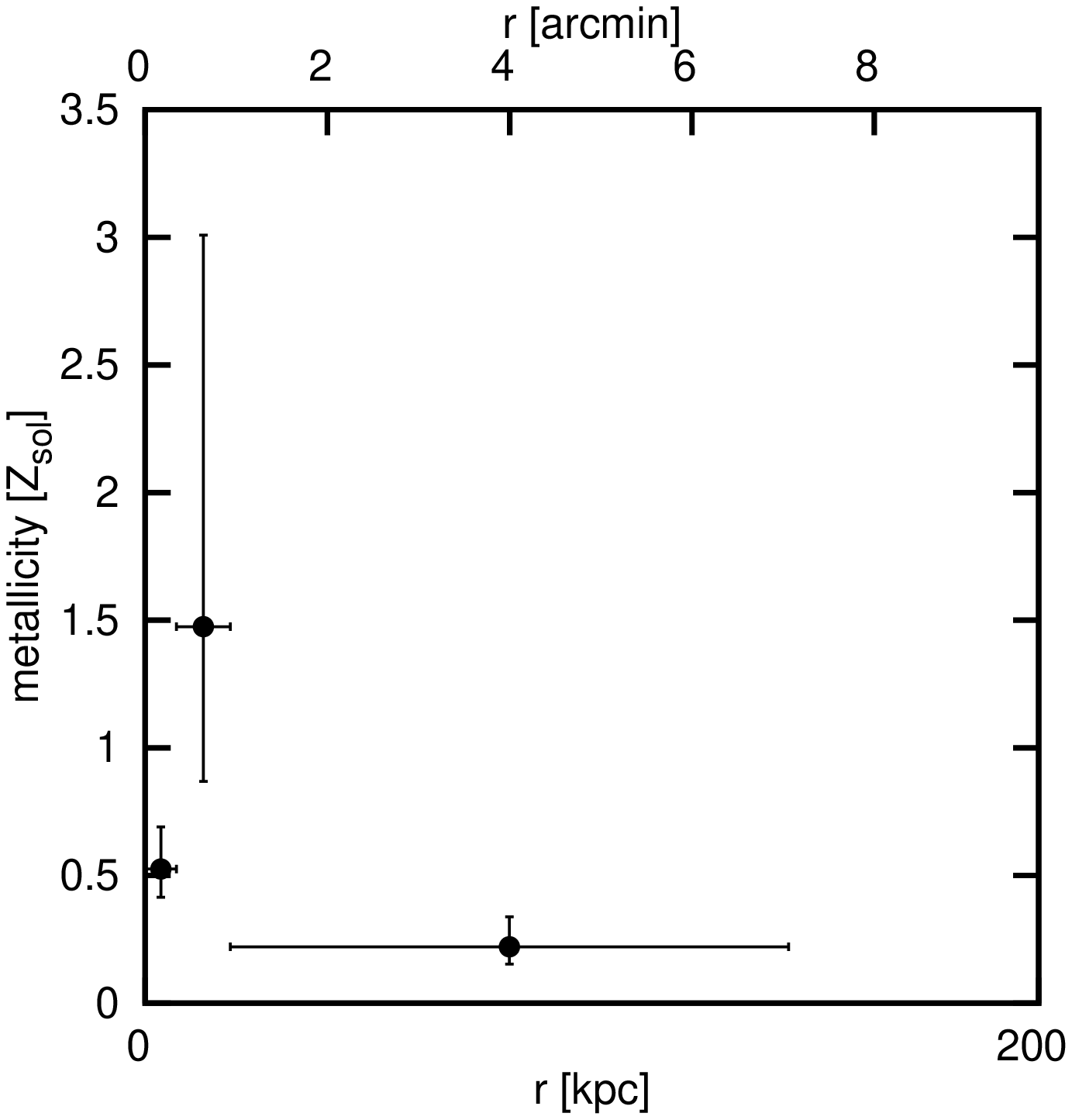}
   \includegraphics[width=0.26\textwidth]{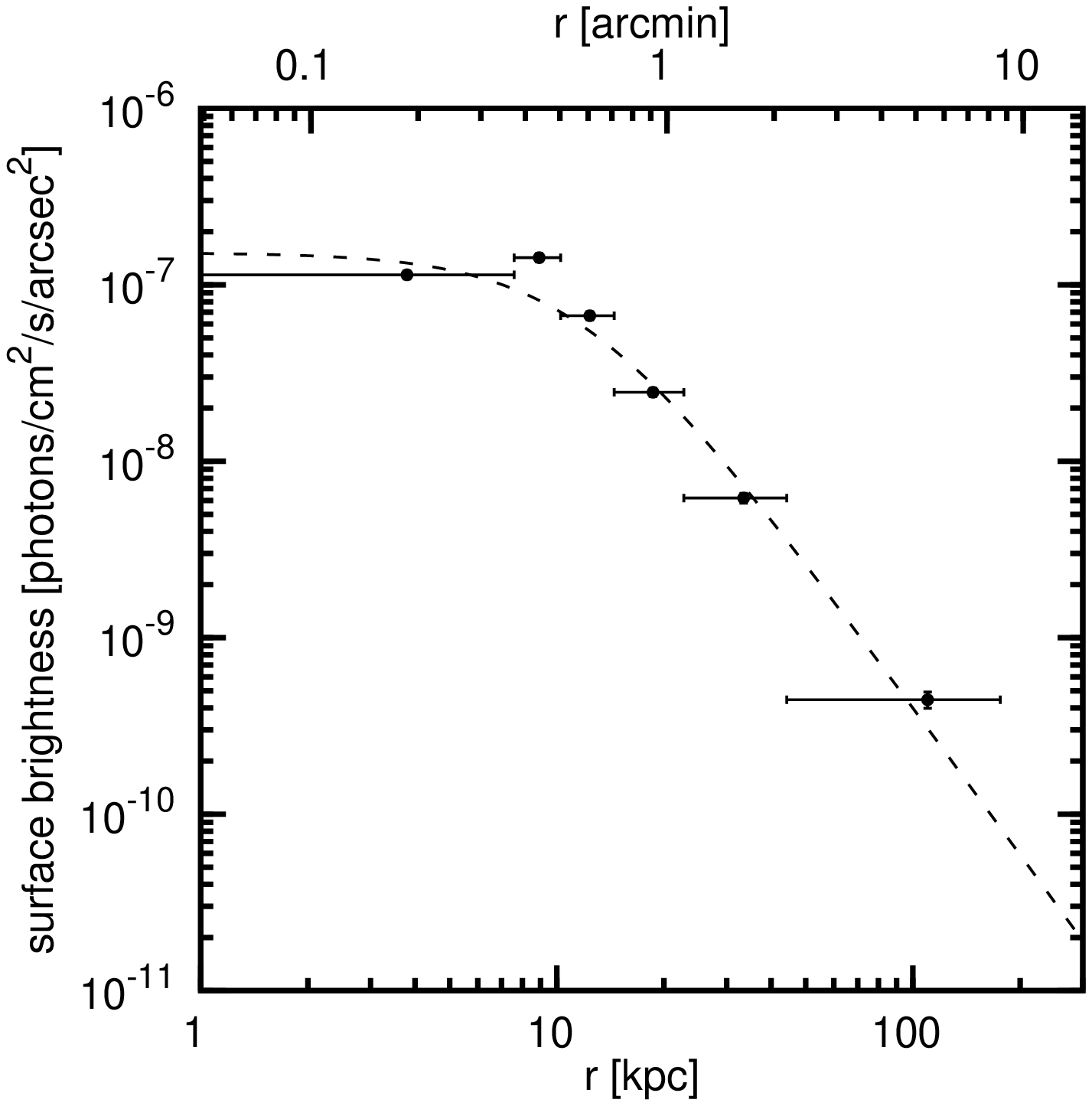}
   \caption{NGC777}
   \label{fig:tprofngc777}%
\end{figure*}
\begin{figure*}[h]
   \centering
   \includegraphics[width=0.26\textwidth]{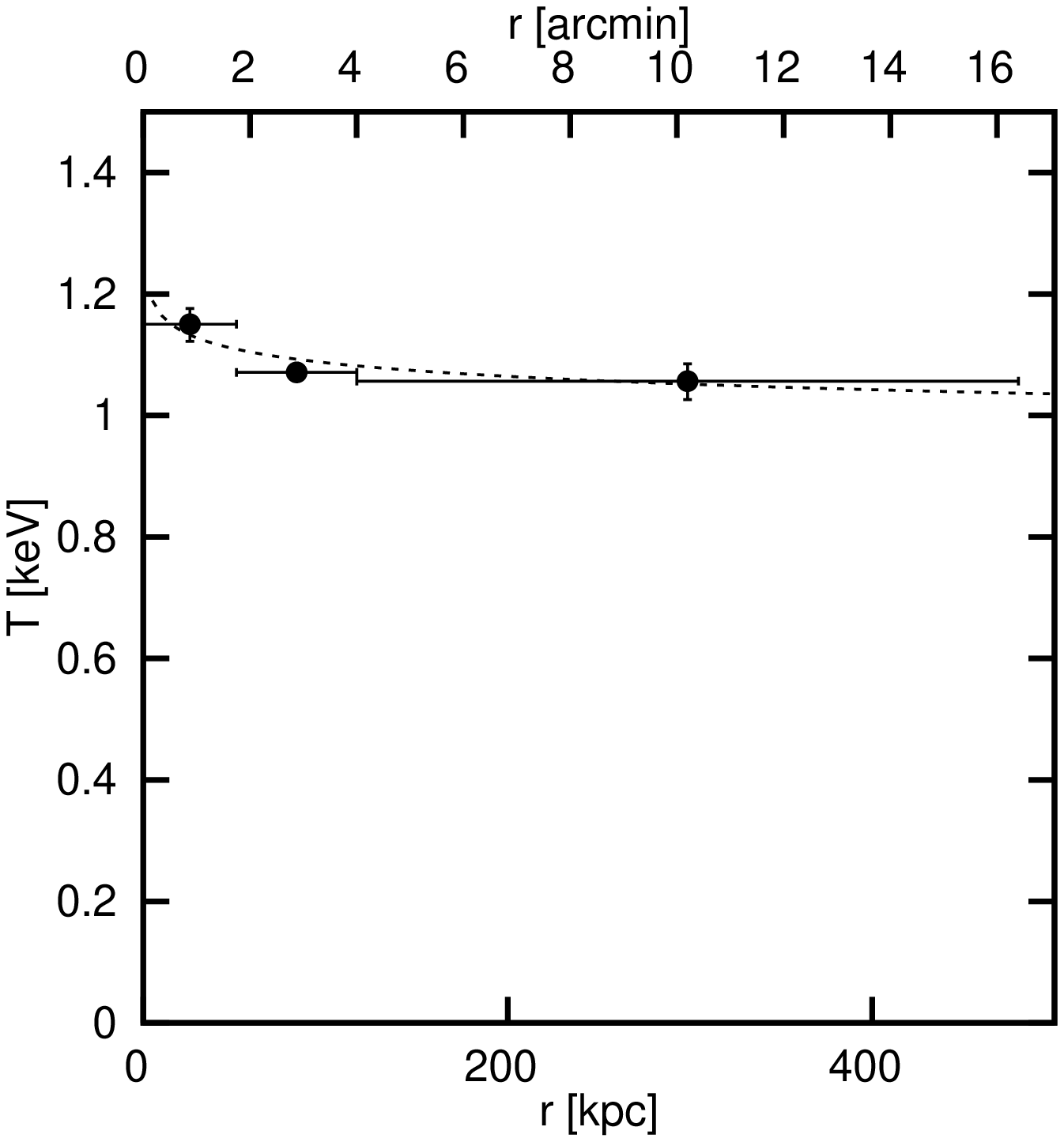}
   \includegraphics[width=0.26\textwidth]{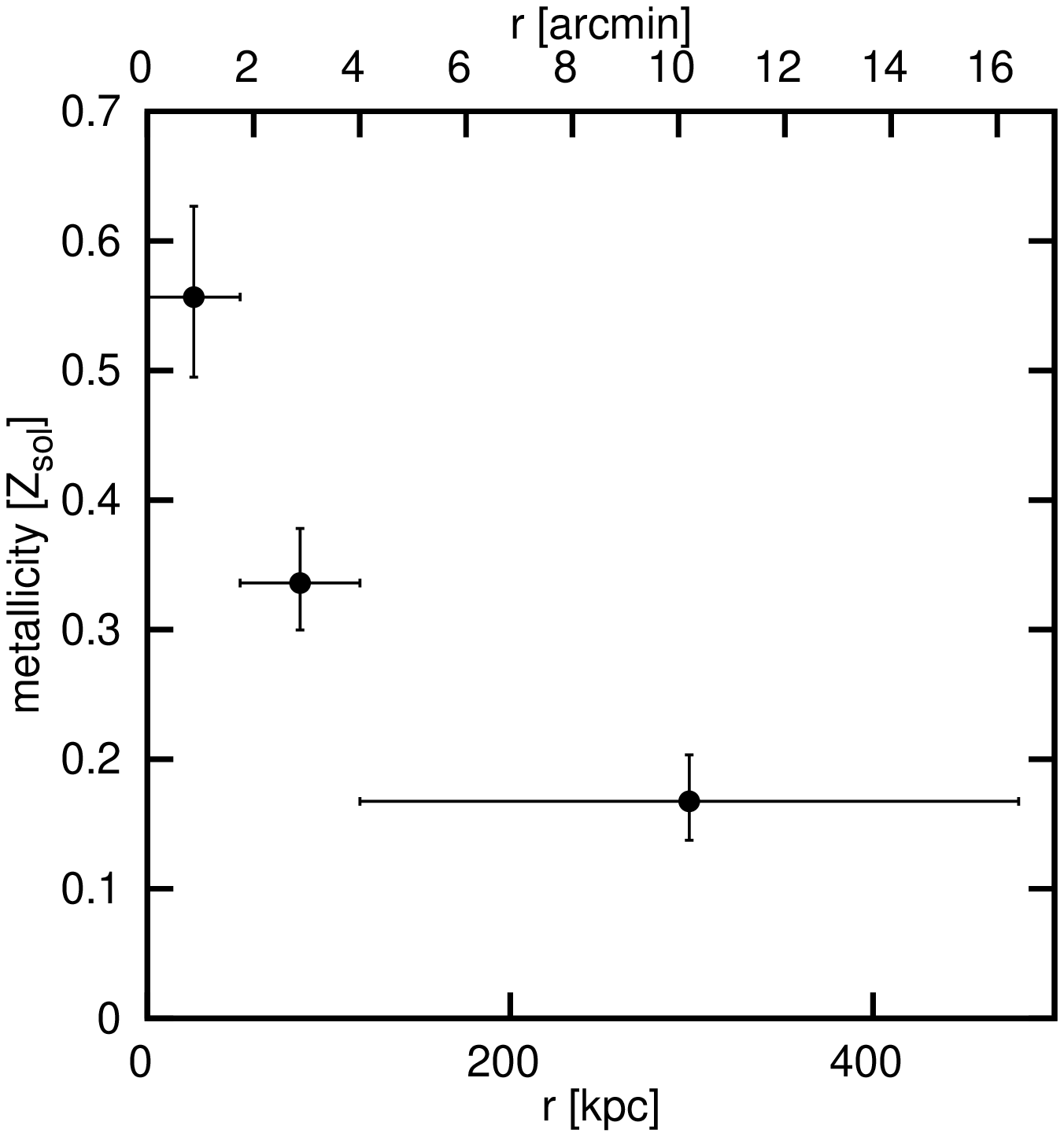}
   \includegraphics[width=0.26\textwidth]{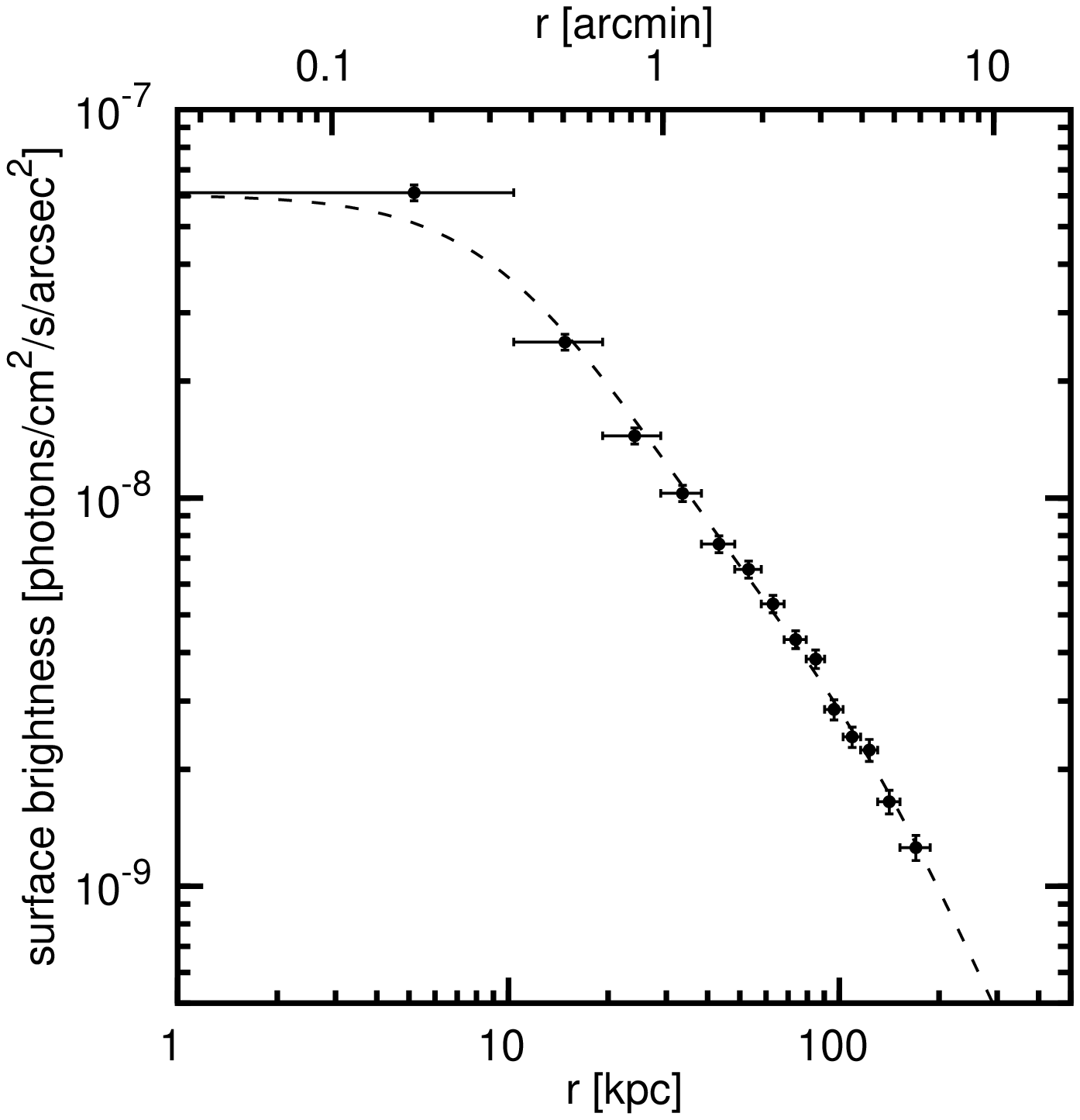}
  \caption{NGC1132}
   \label{fig:tprofngc1132}%
\end{figure*}
\clearpage
\begin{figure*}[h]
   \centering
   \includegraphics[width=0.26\textwidth]{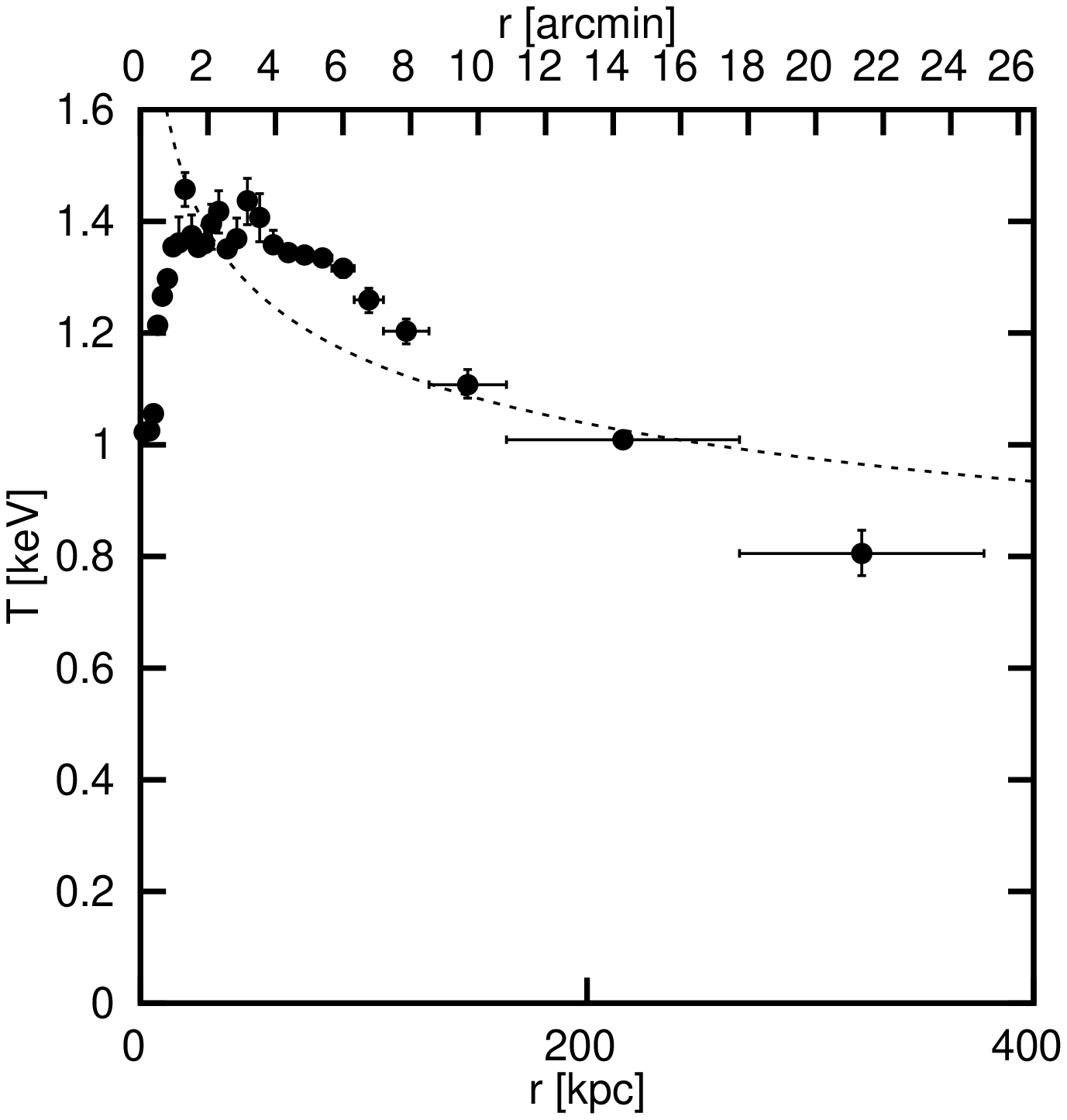}
   \includegraphics[width=0.26\textwidth]{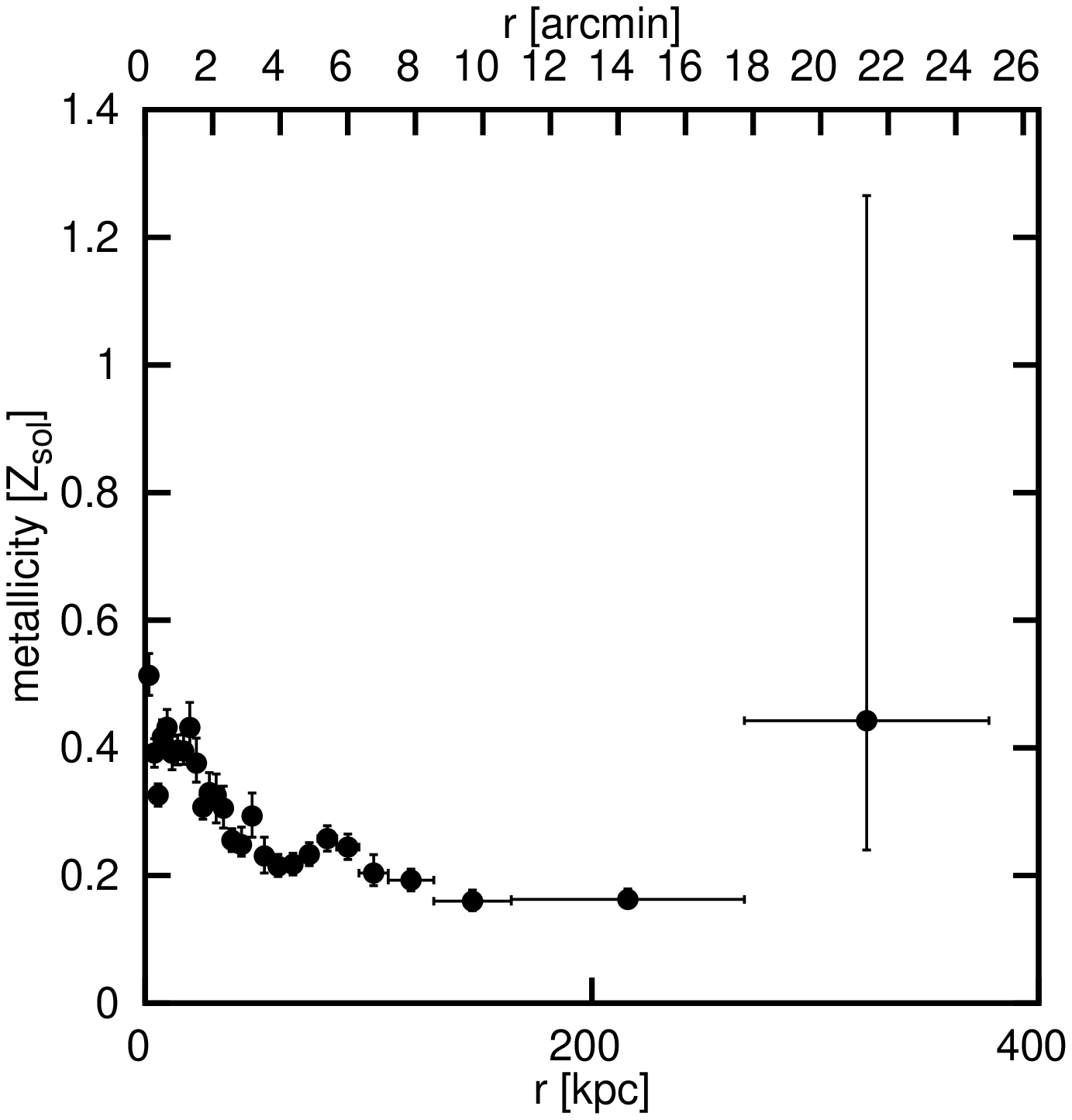}
   \includegraphics[width=0.26\textwidth]{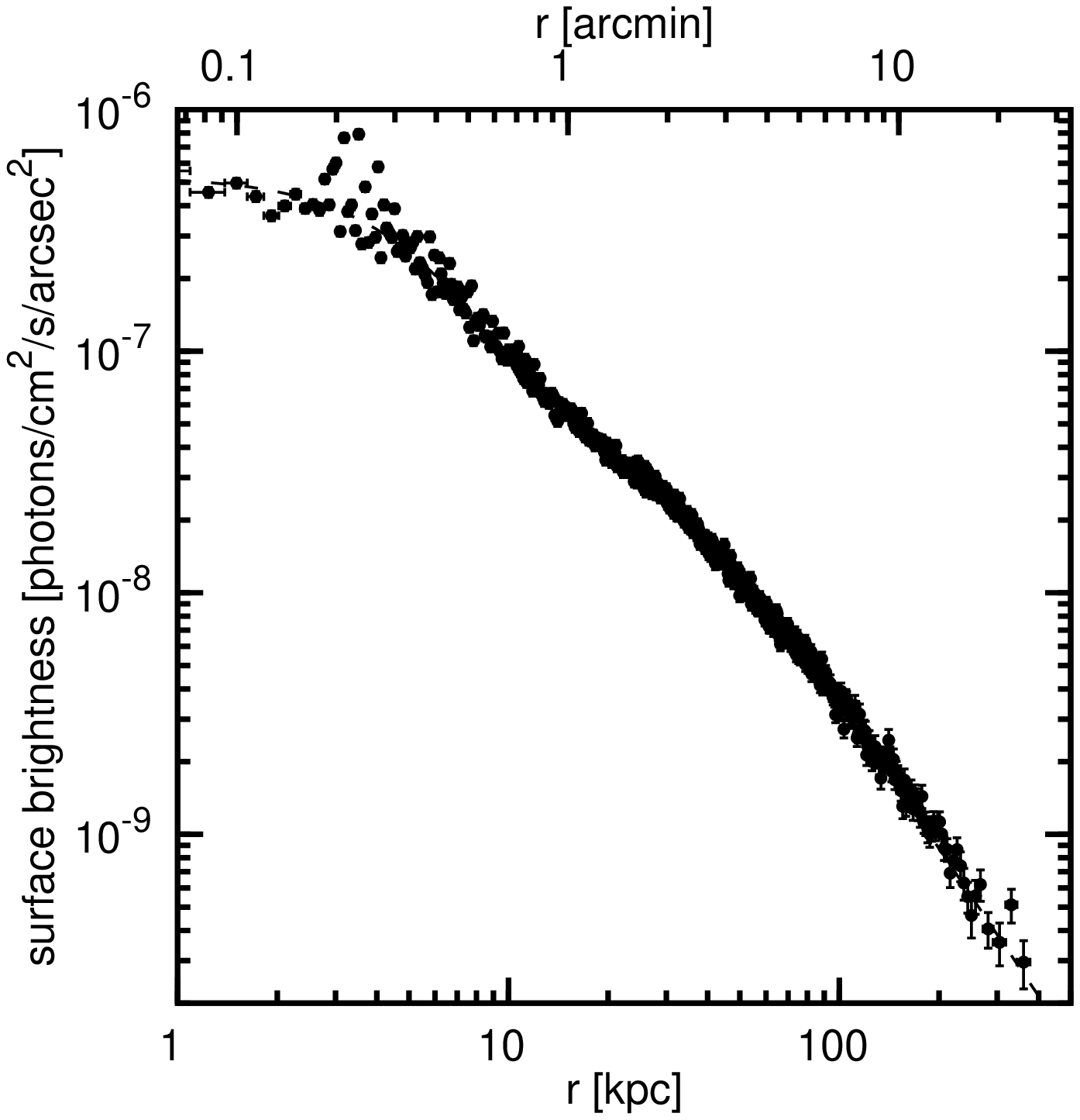}
   \caption{NGC1550}
   \label{fig:tprofngc1550}%
\end{figure*}
\begin{figure*}[h]
   \centering
   \includegraphics[width=0.26\textwidth]{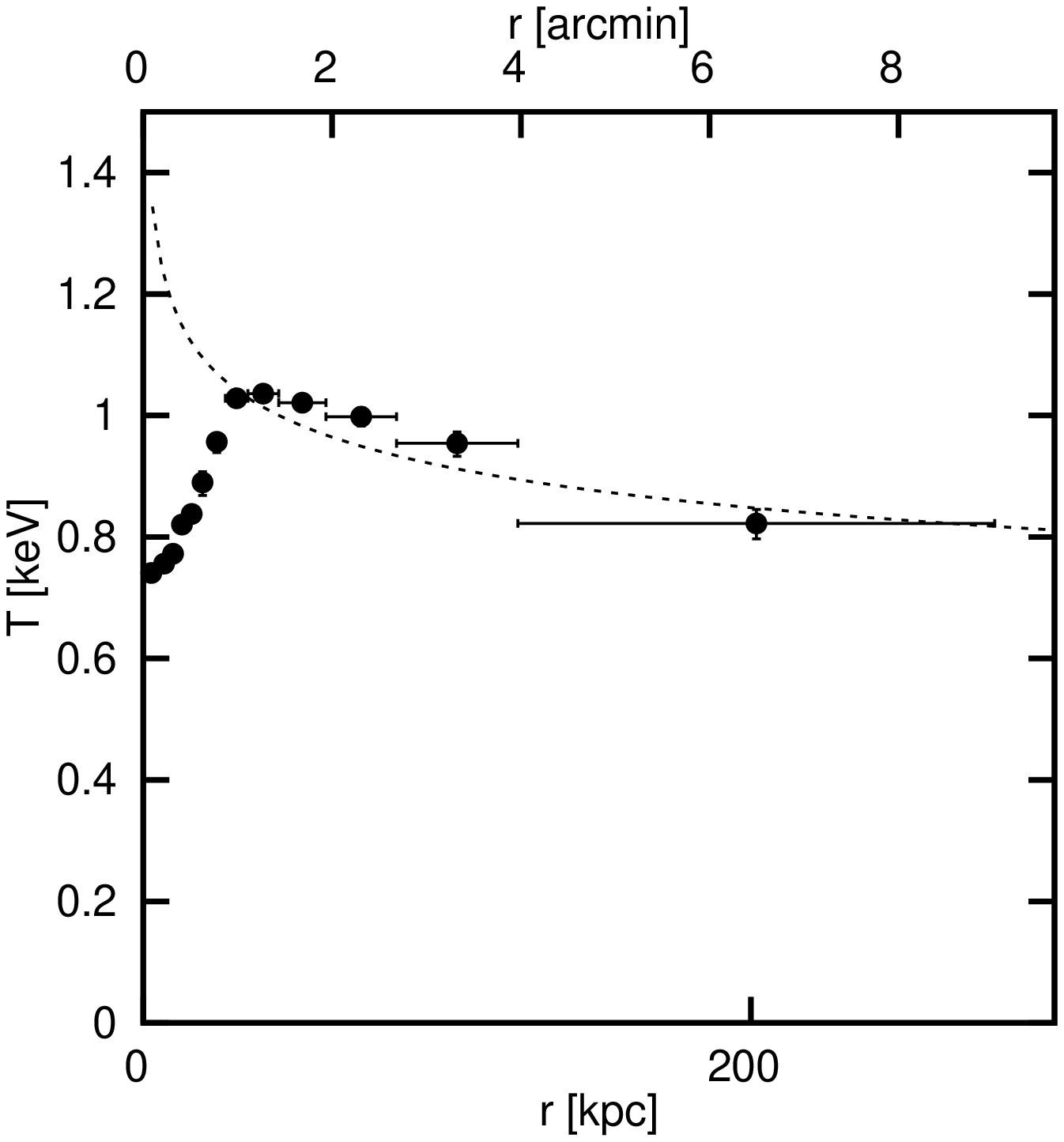}
   \includegraphics[width=0.26\textwidth]{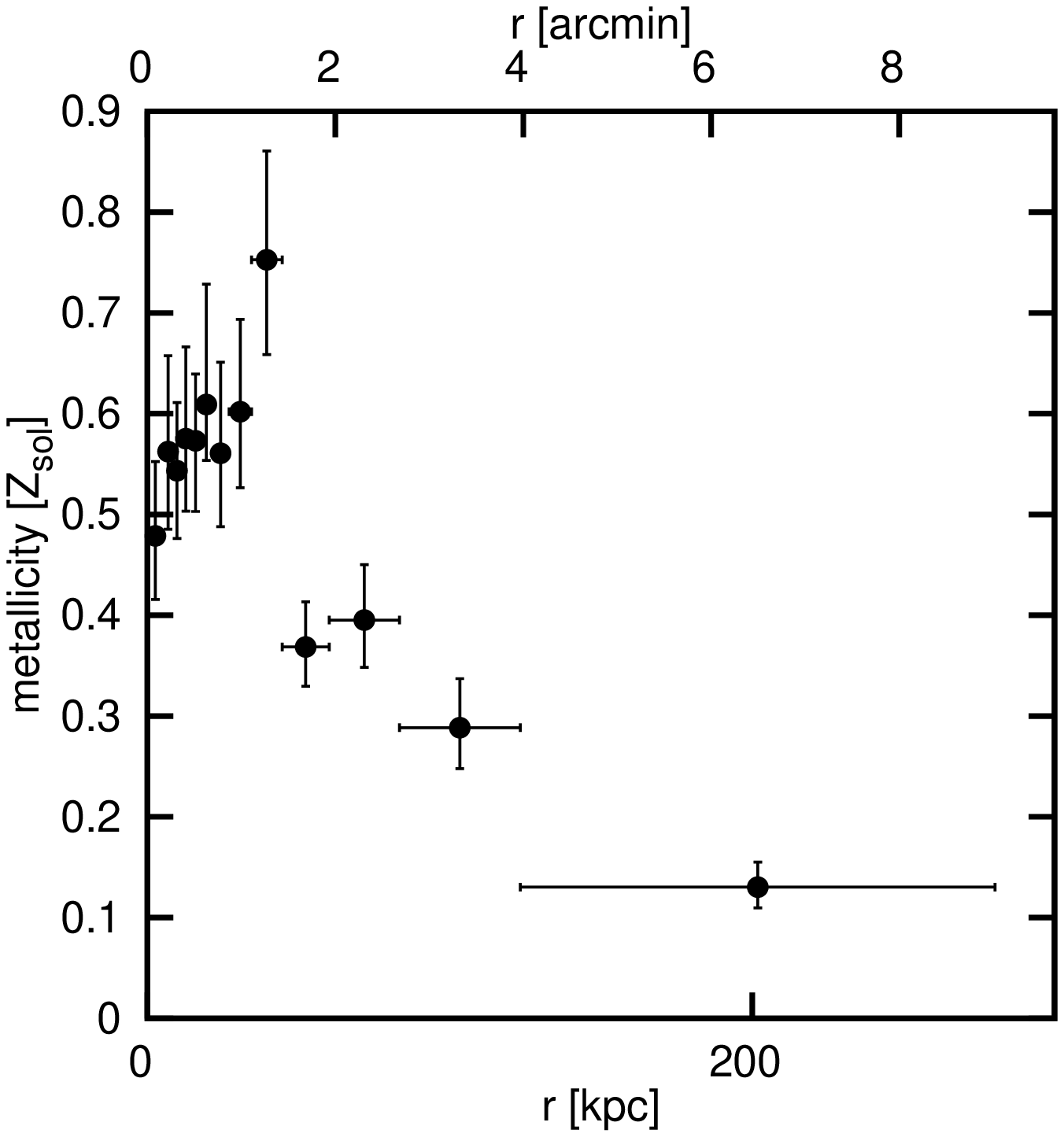}
   \includegraphics[width=0.26\textwidth]{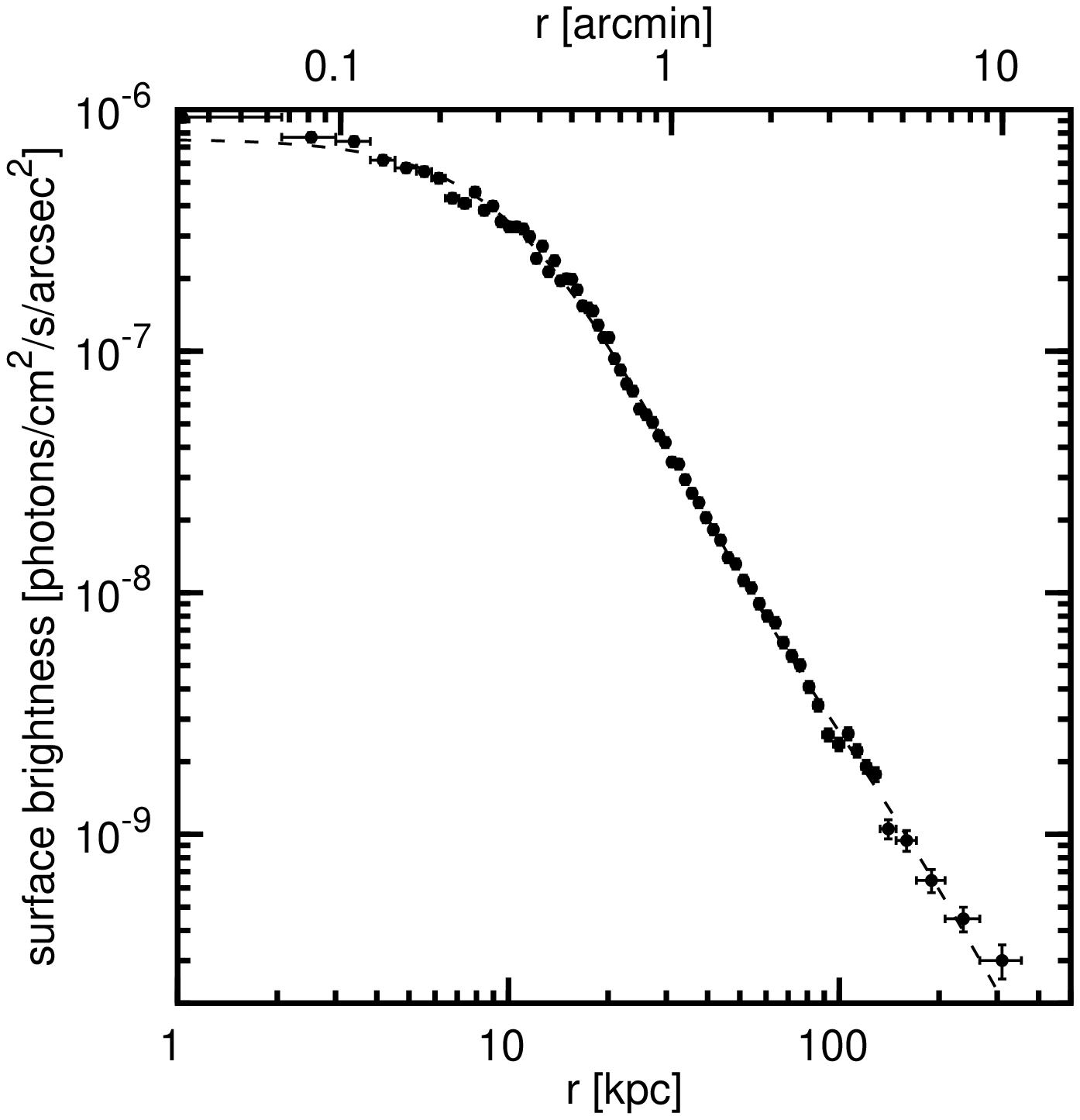}
   \caption{NGC4325}
   \label{fig:tprofngc4325}%
\end{figure*}
\begin{figure*}[h]
   \centering
   \includegraphics[width=0.26\textwidth]{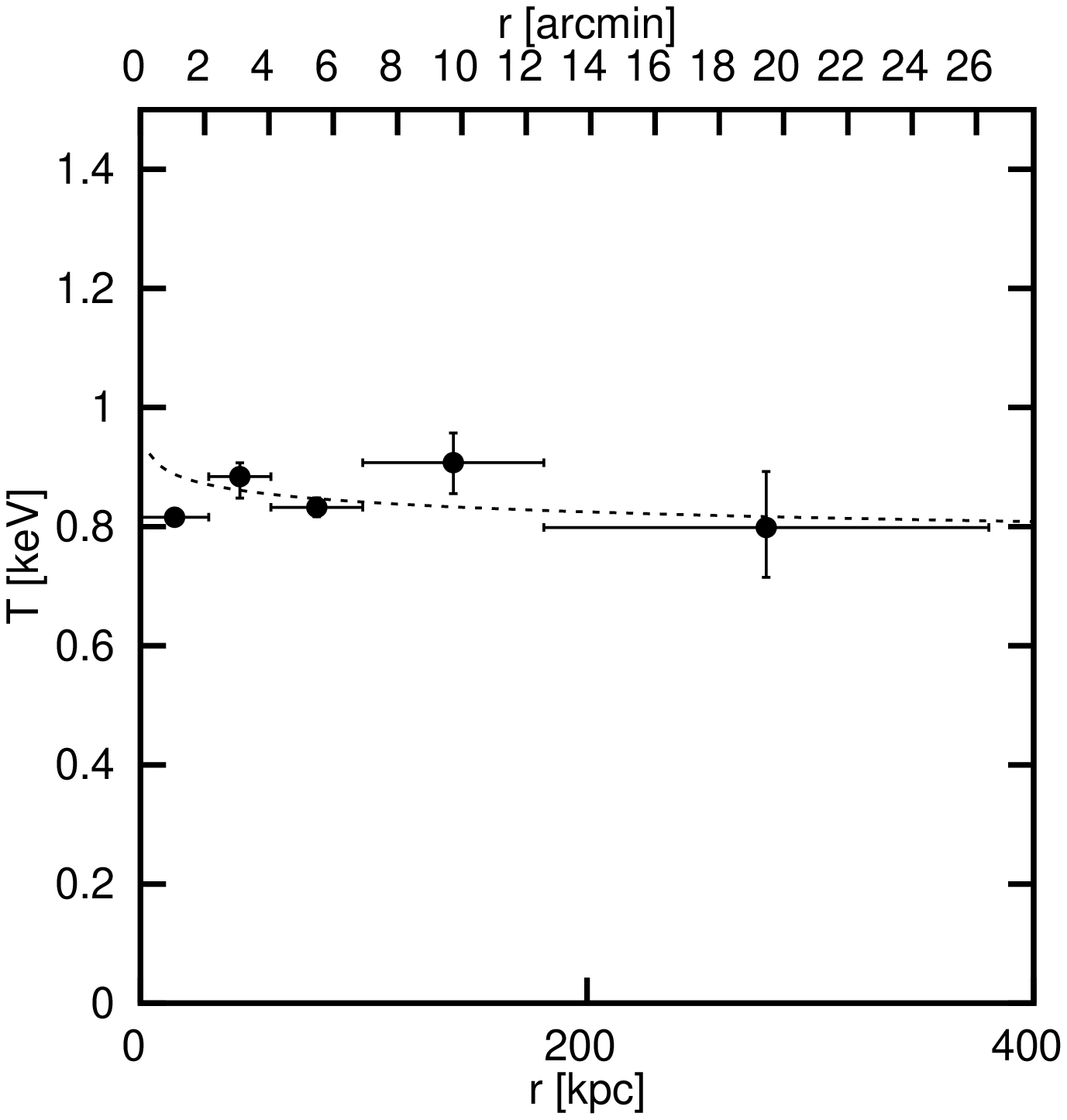}
   \includegraphics[width=0.26\textwidth]{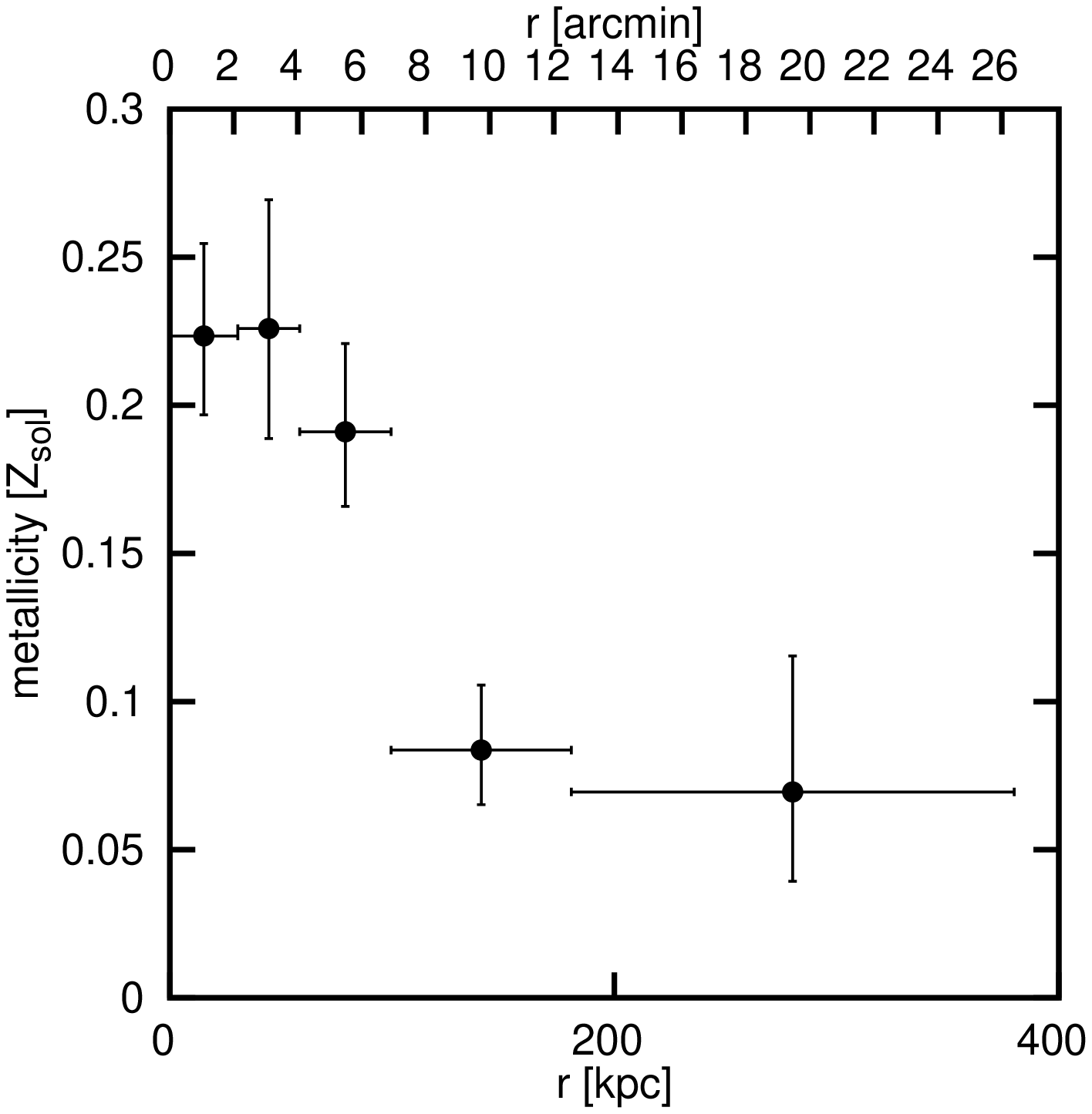}
   \includegraphics[width=0.26\textwidth]{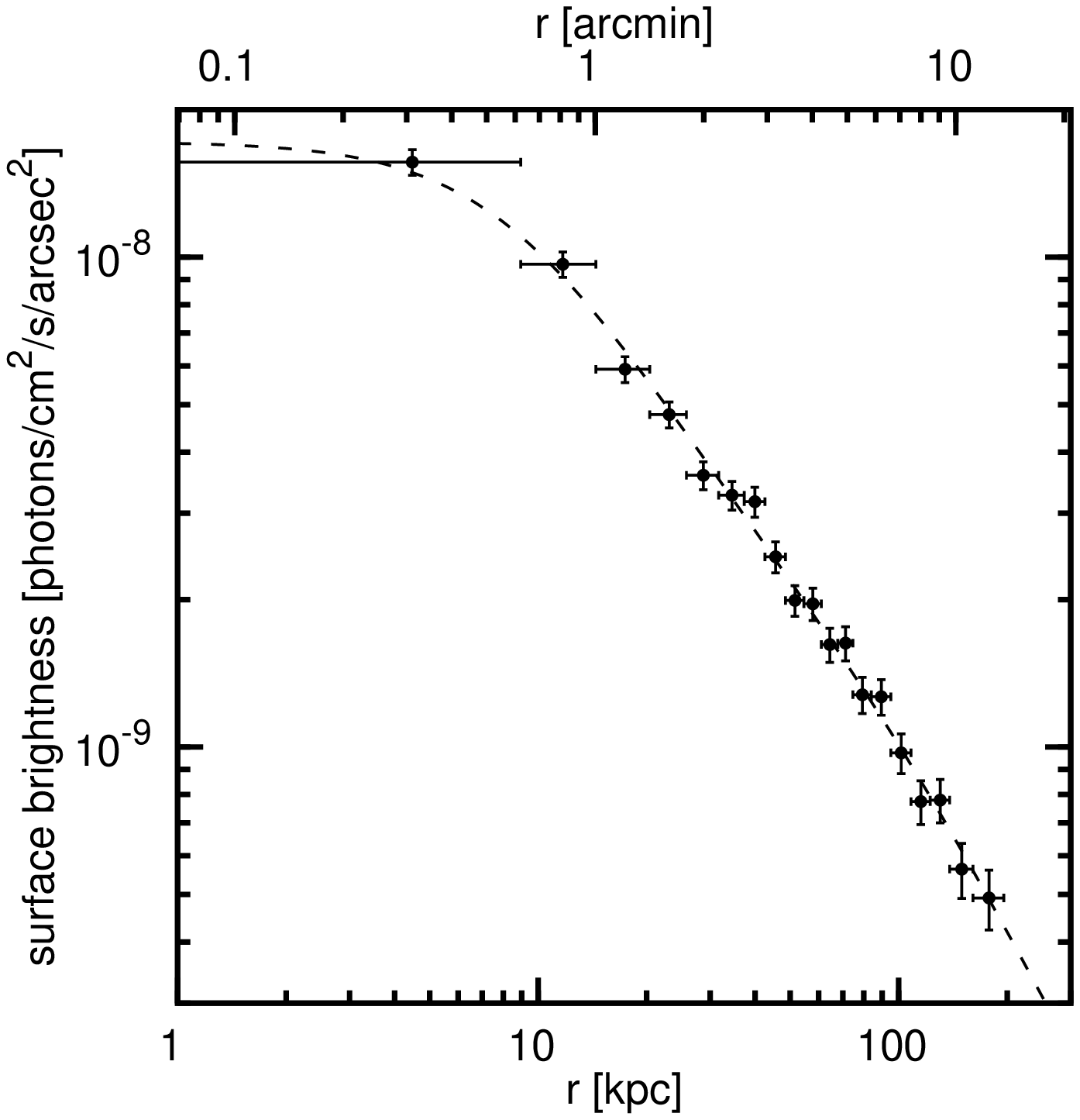}
   \caption{NGC4936}
   \label{fig:tprofngc4936}%
\end{figure*}
\begin{figure*}[!h]
   \centering
   \includegraphics[width=0.26\textwidth]{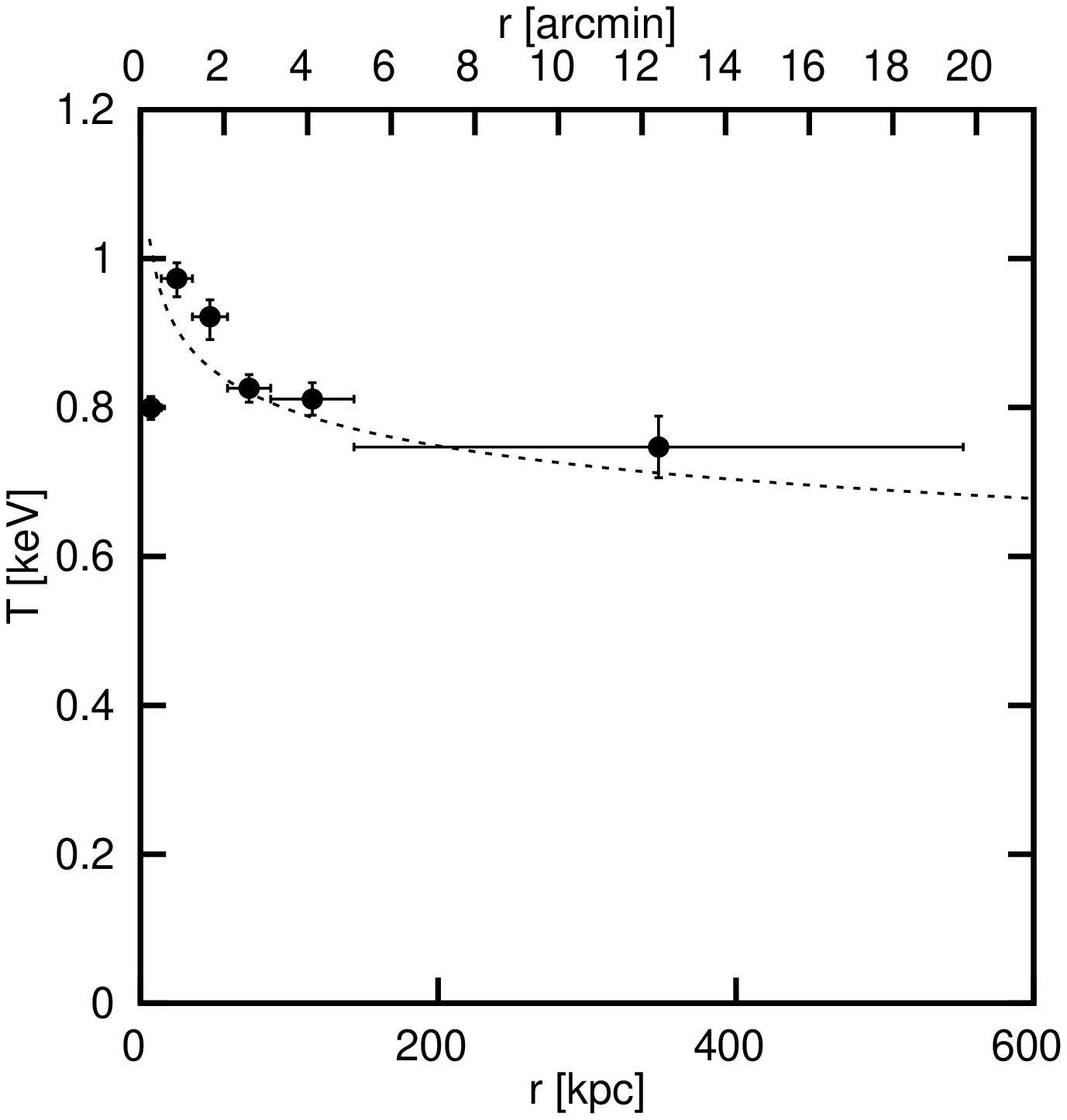}
   \includegraphics[width=0.26\textwidth]{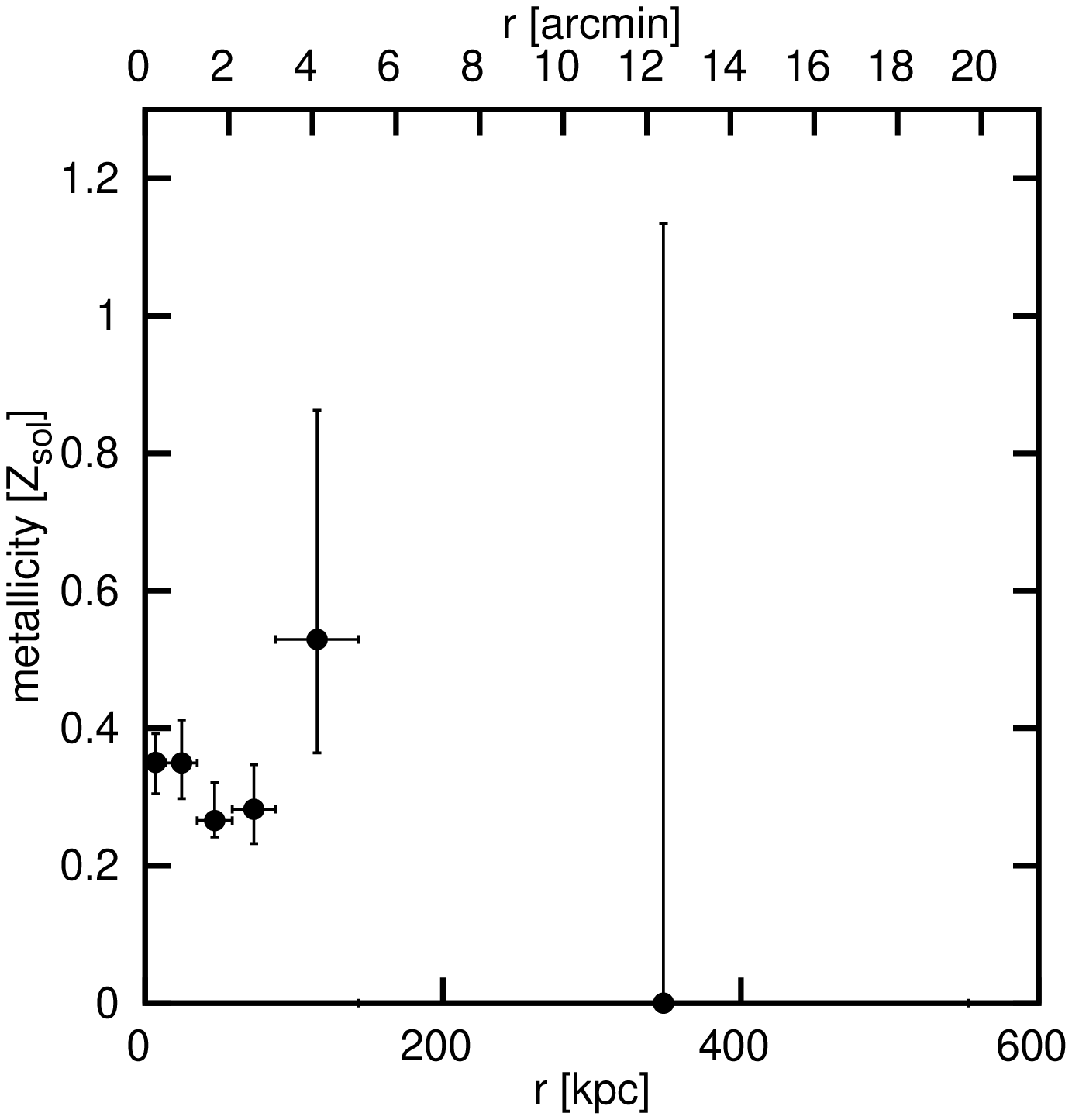}
   \includegraphics[width=0.26\textwidth]{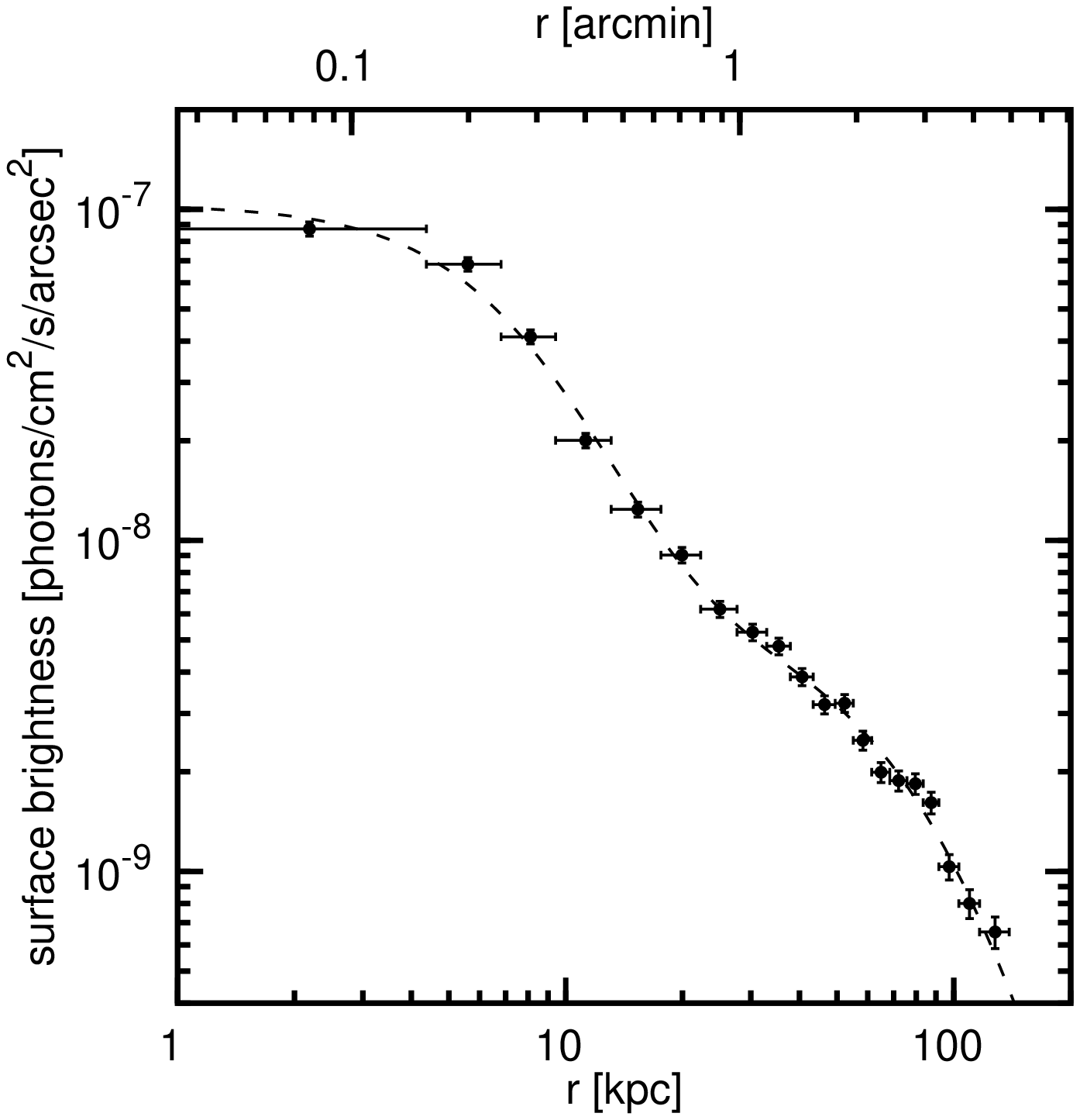}
   \caption{NGC5129}
   \label{fig:tprofngc5129}%
\end{figure*}
\clearpage
\begin{figure*}[h]
   \centering
   \includegraphics[width=0.26\textwidth]{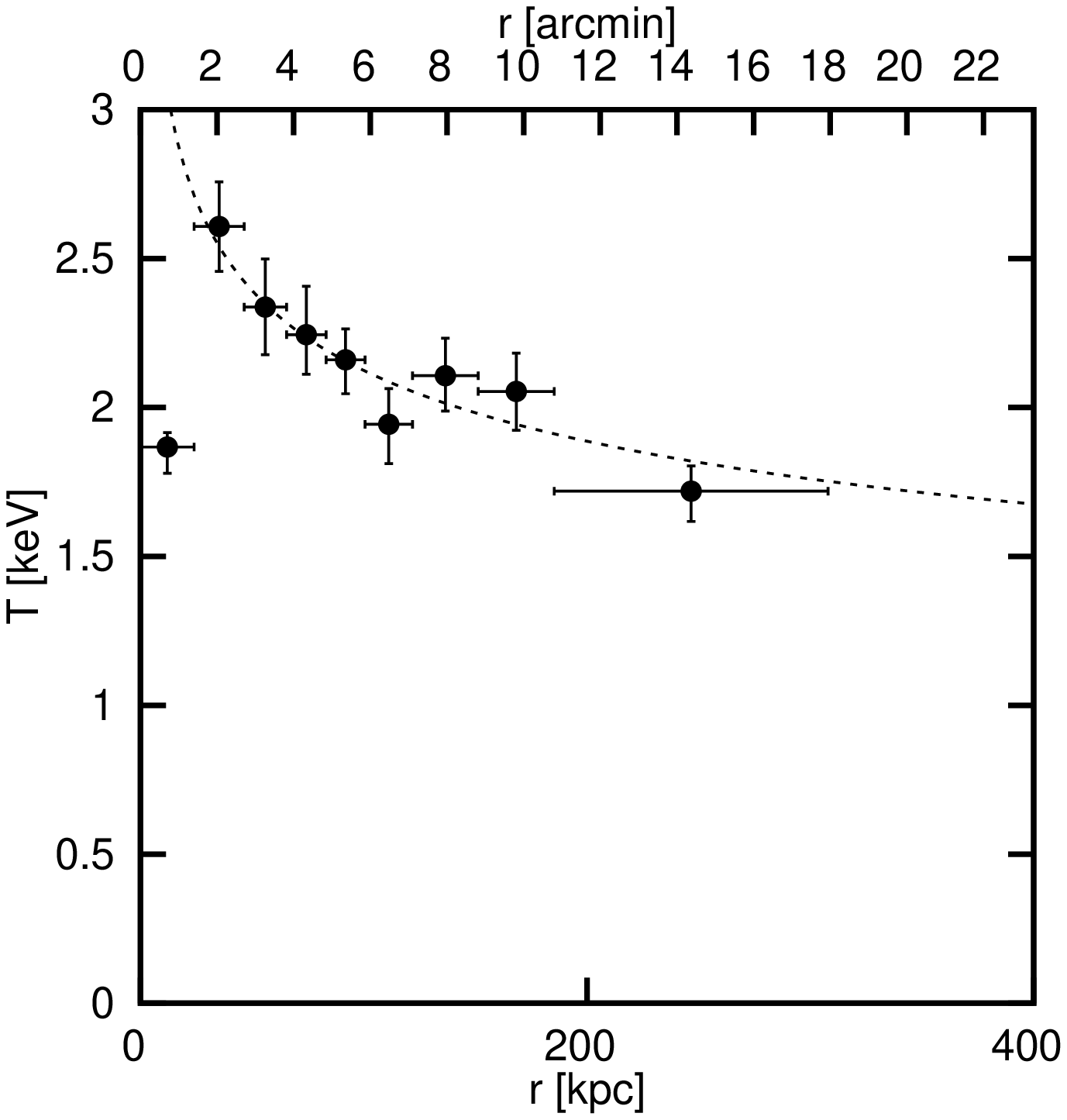}
   \includegraphics[width=0.26\textwidth]{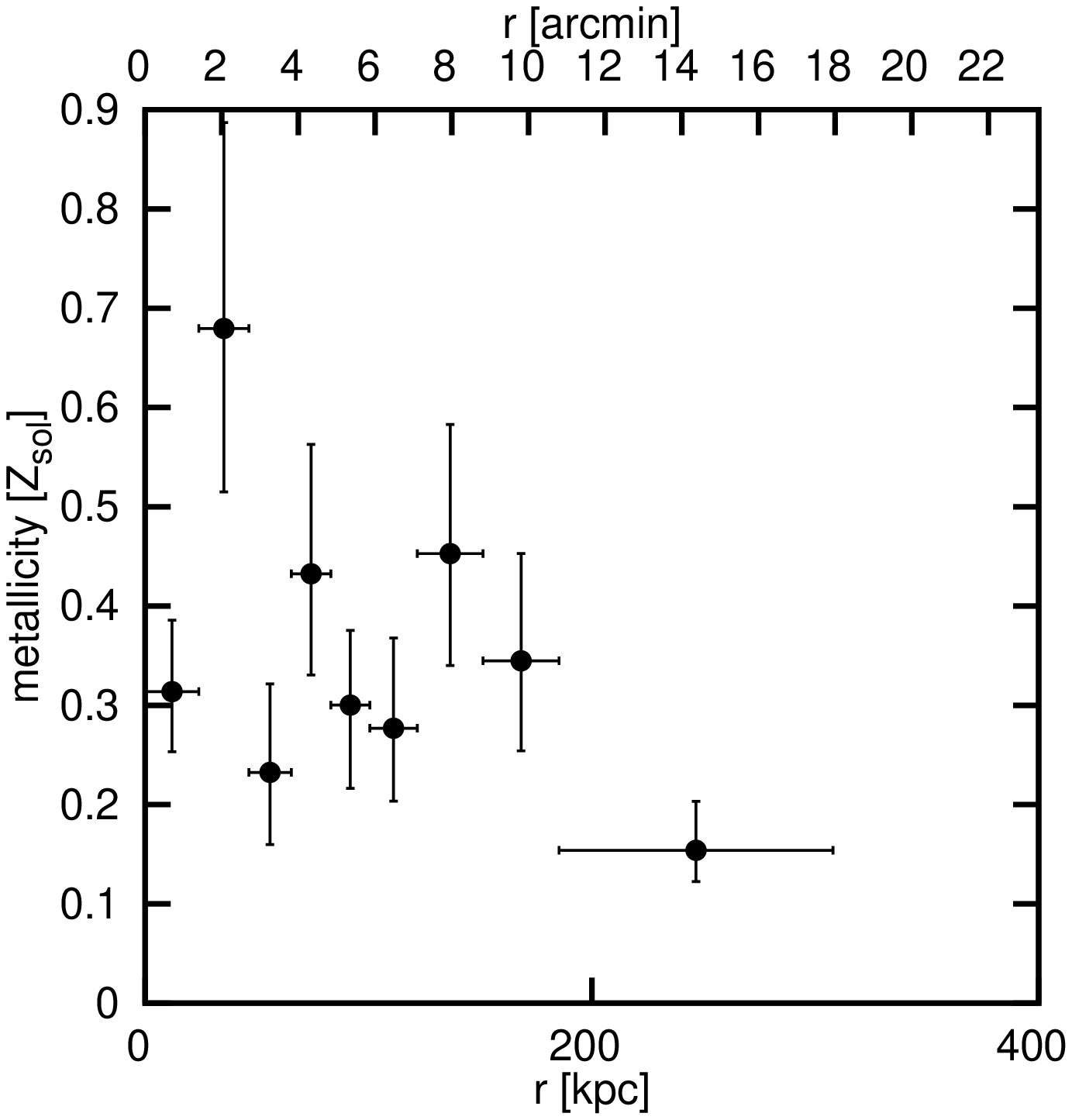}
   \includegraphics[width=0.26\textwidth]{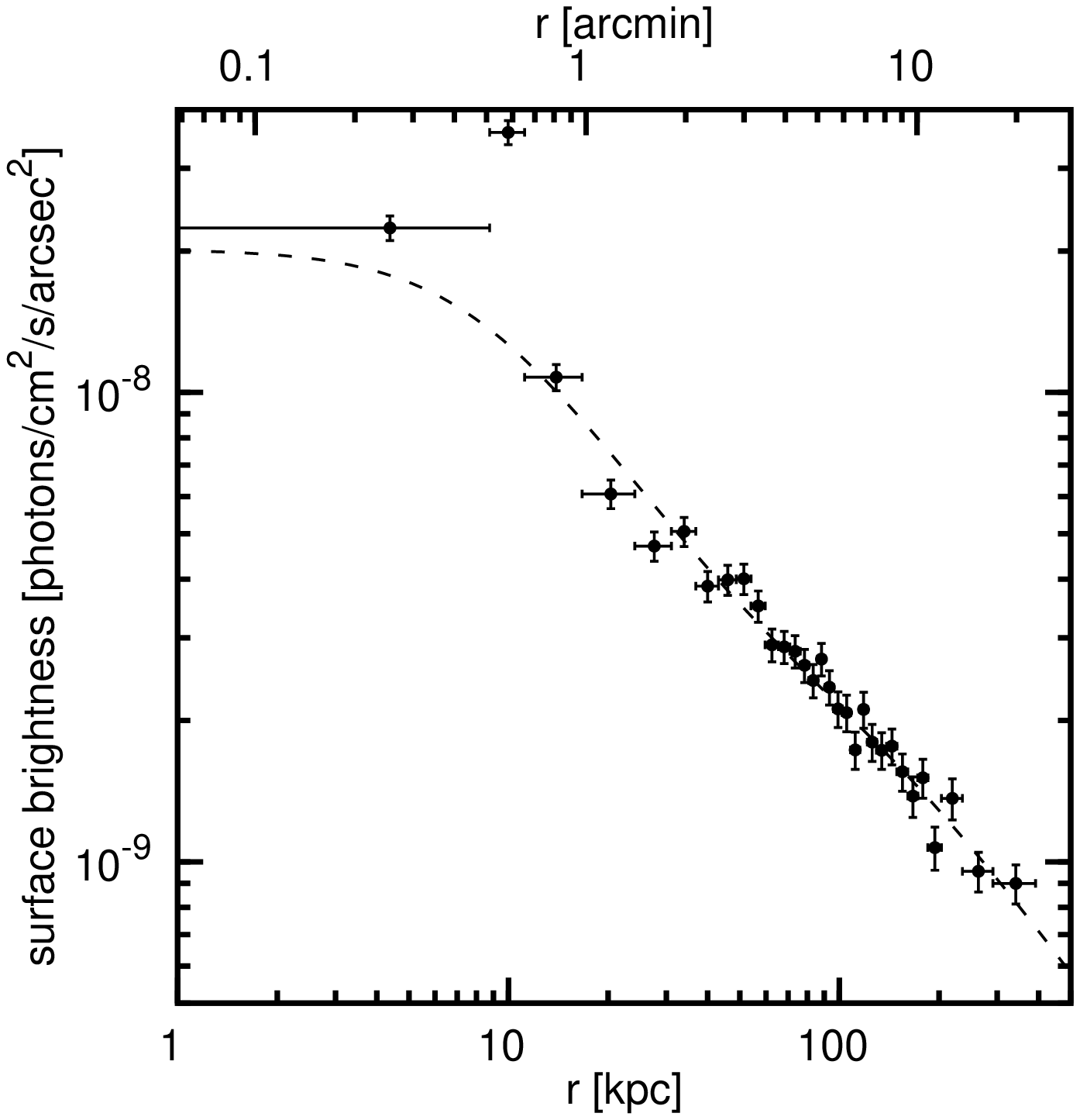}
   \caption{NGC5419}
   \label{fig:tprofngc5419}%
\end{figure*}
\begin{figure*}[h]
   \centering
   \includegraphics[width=0.26\textwidth]{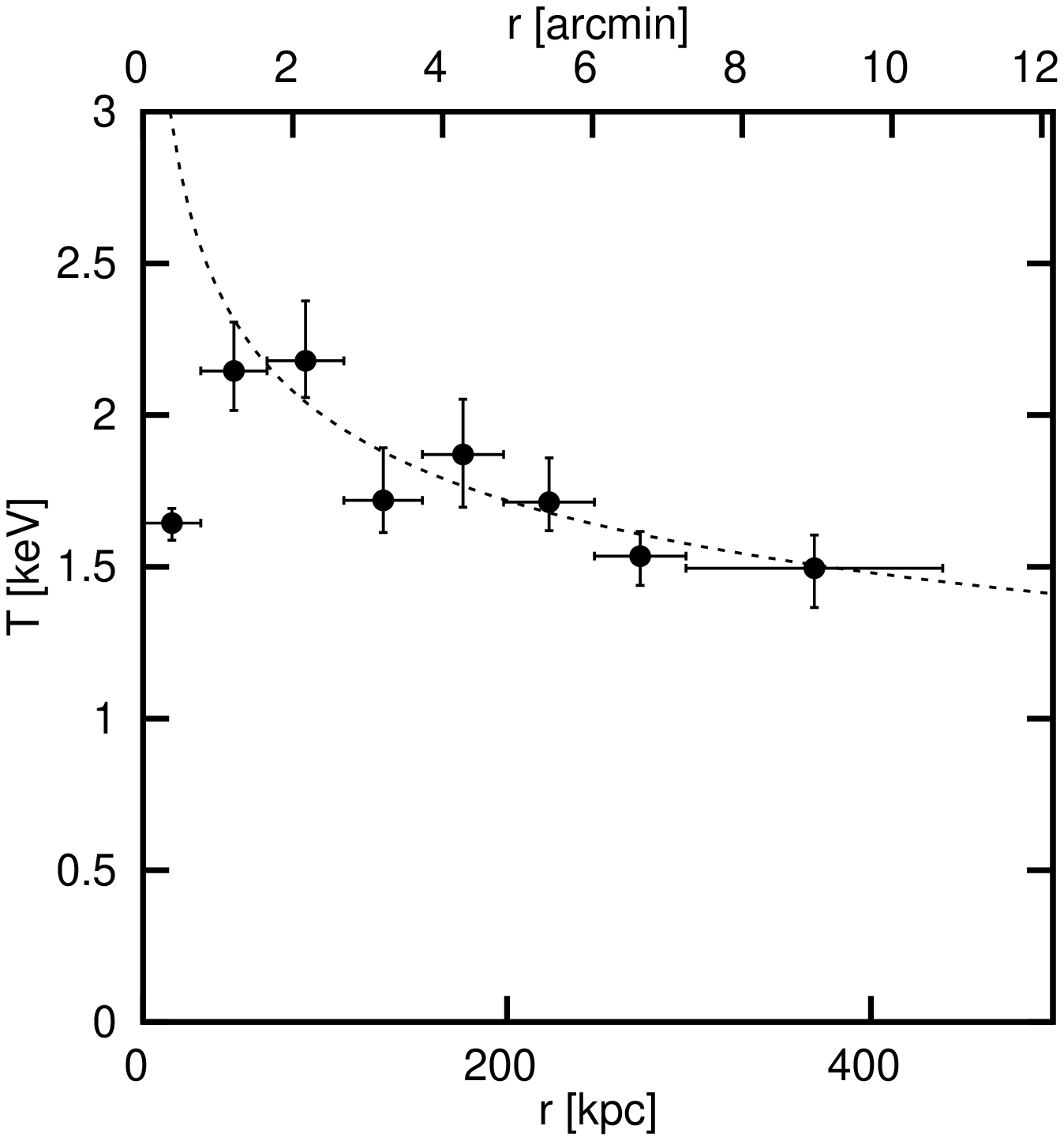}
   \includegraphics[width=0.26\textwidth]{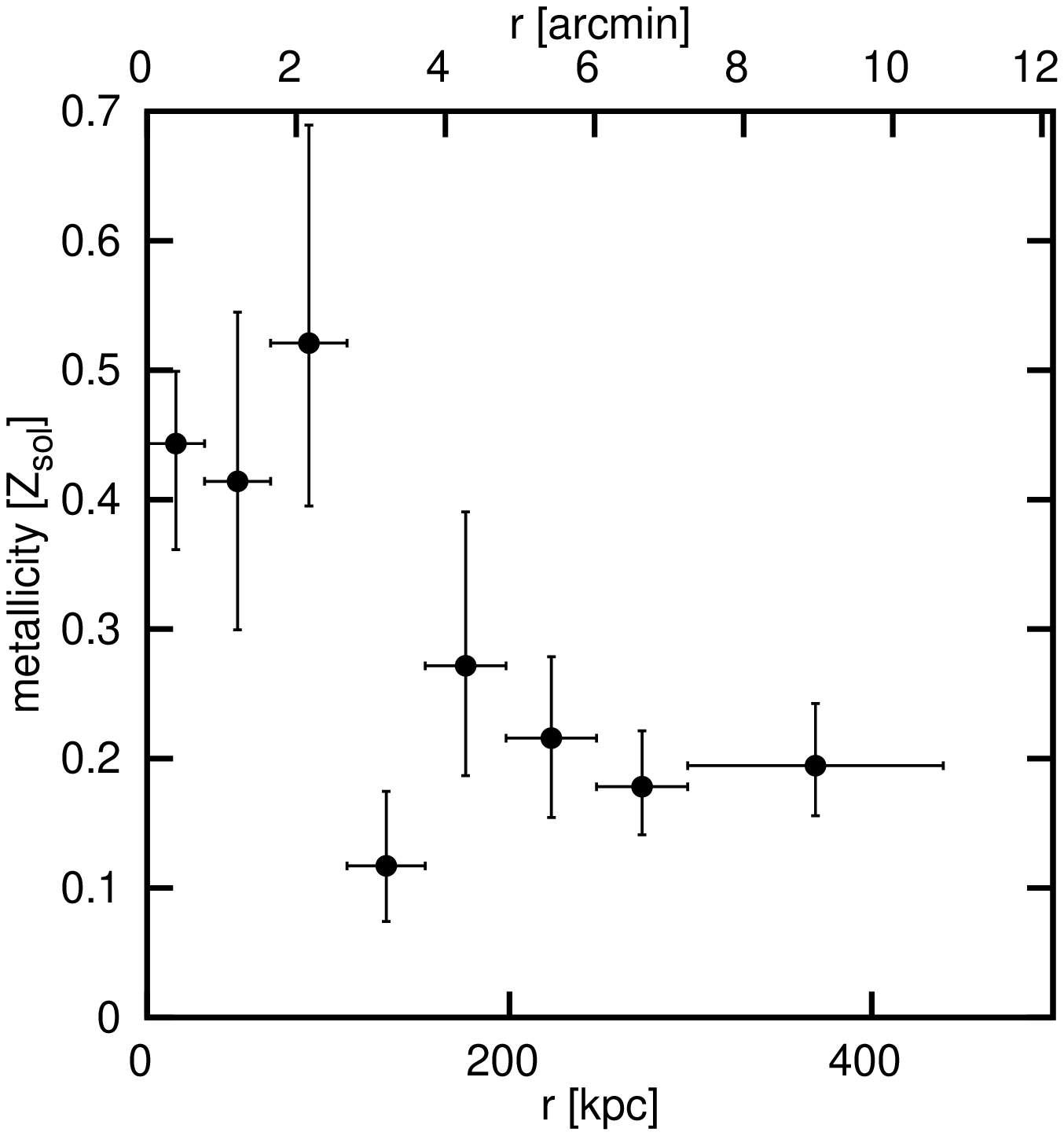}
   \includegraphics[width=0.26\textwidth]{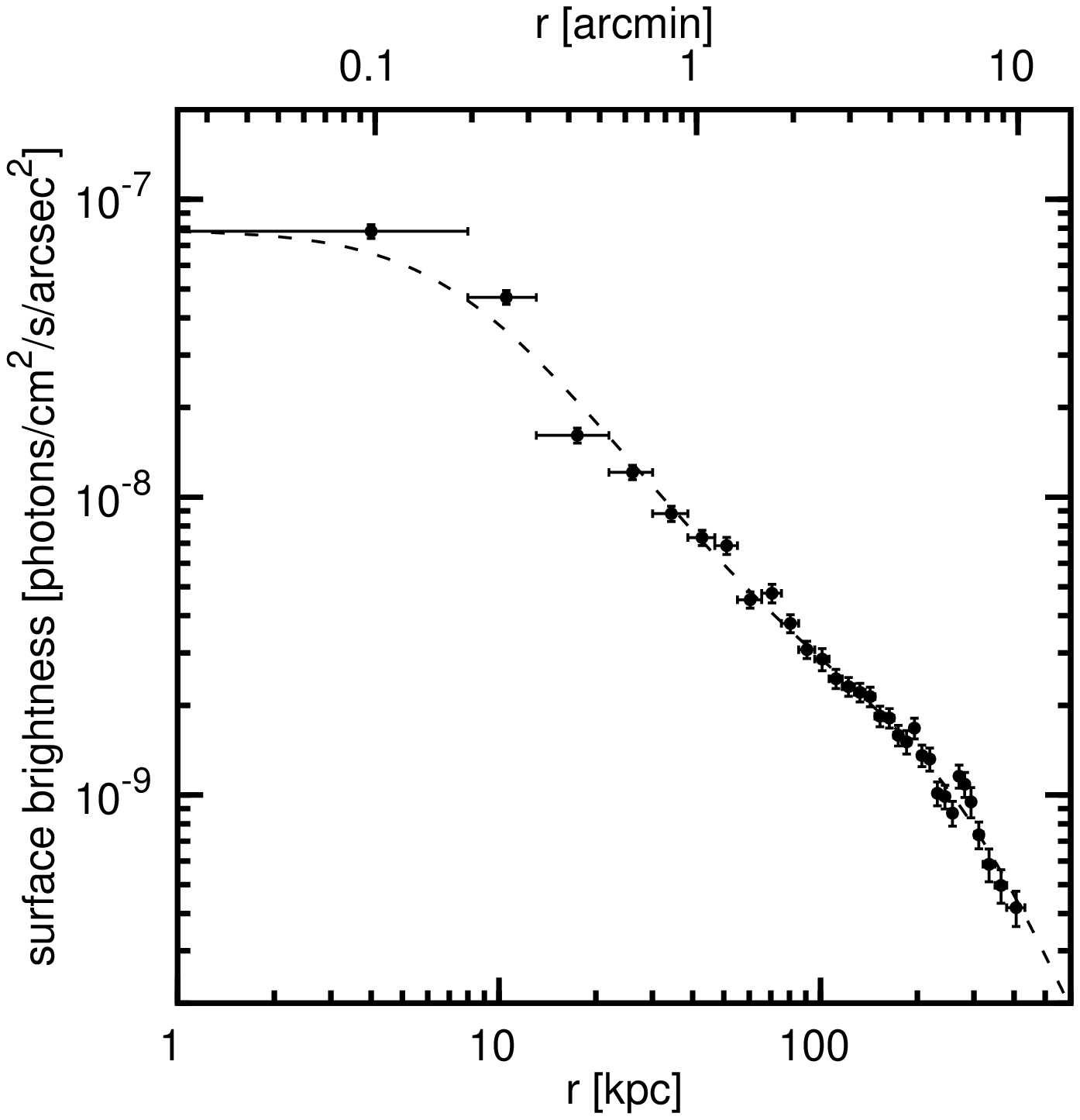}
   \caption{NGC6269}
   \label{fig:tprofngc6269}%
\end{figure*}
\begin{figure*}[h]
   \centering
   \includegraphics[width=0.26\textwidth]{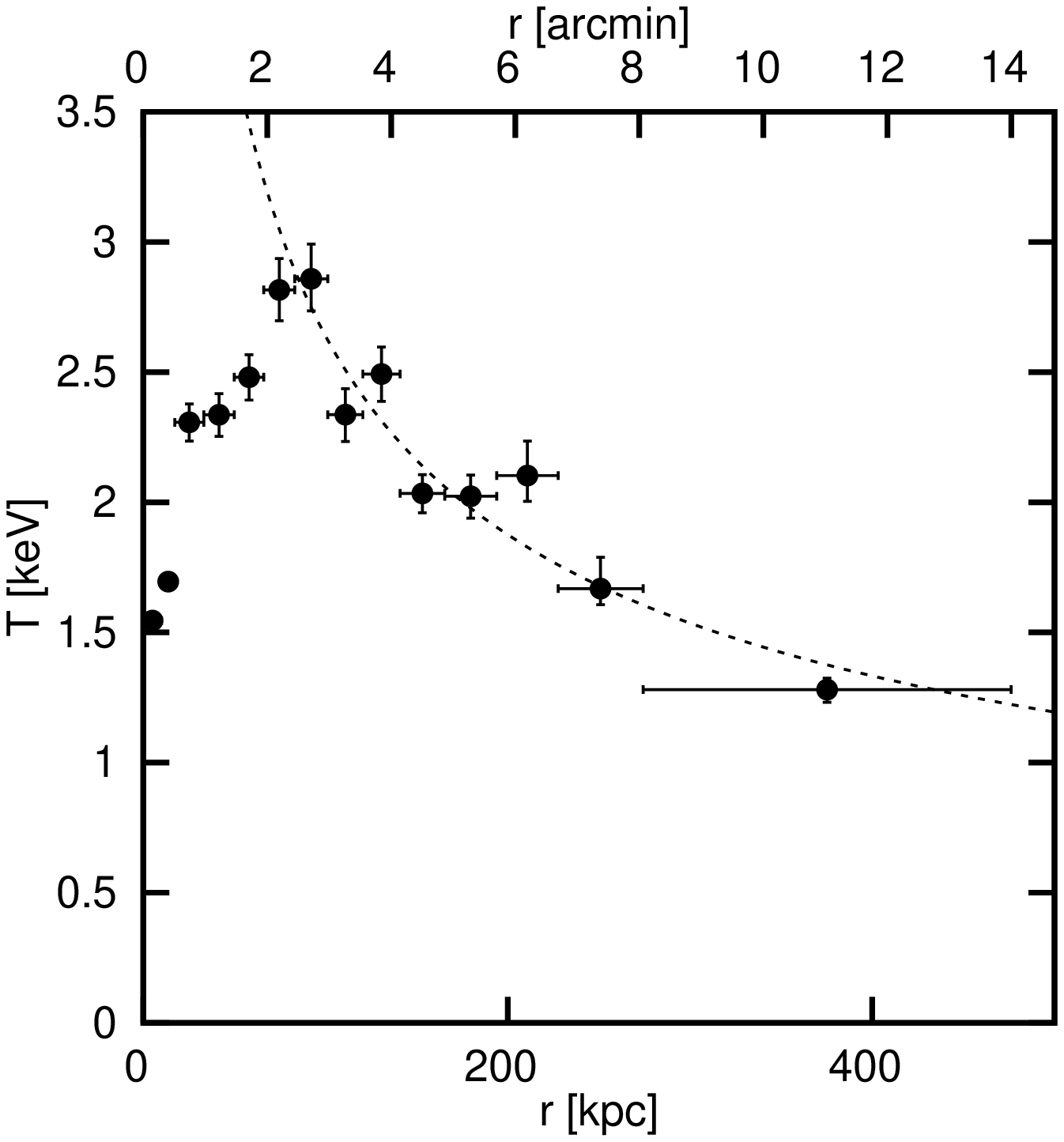}
   \includegraphics[width=0.26\textwidth]{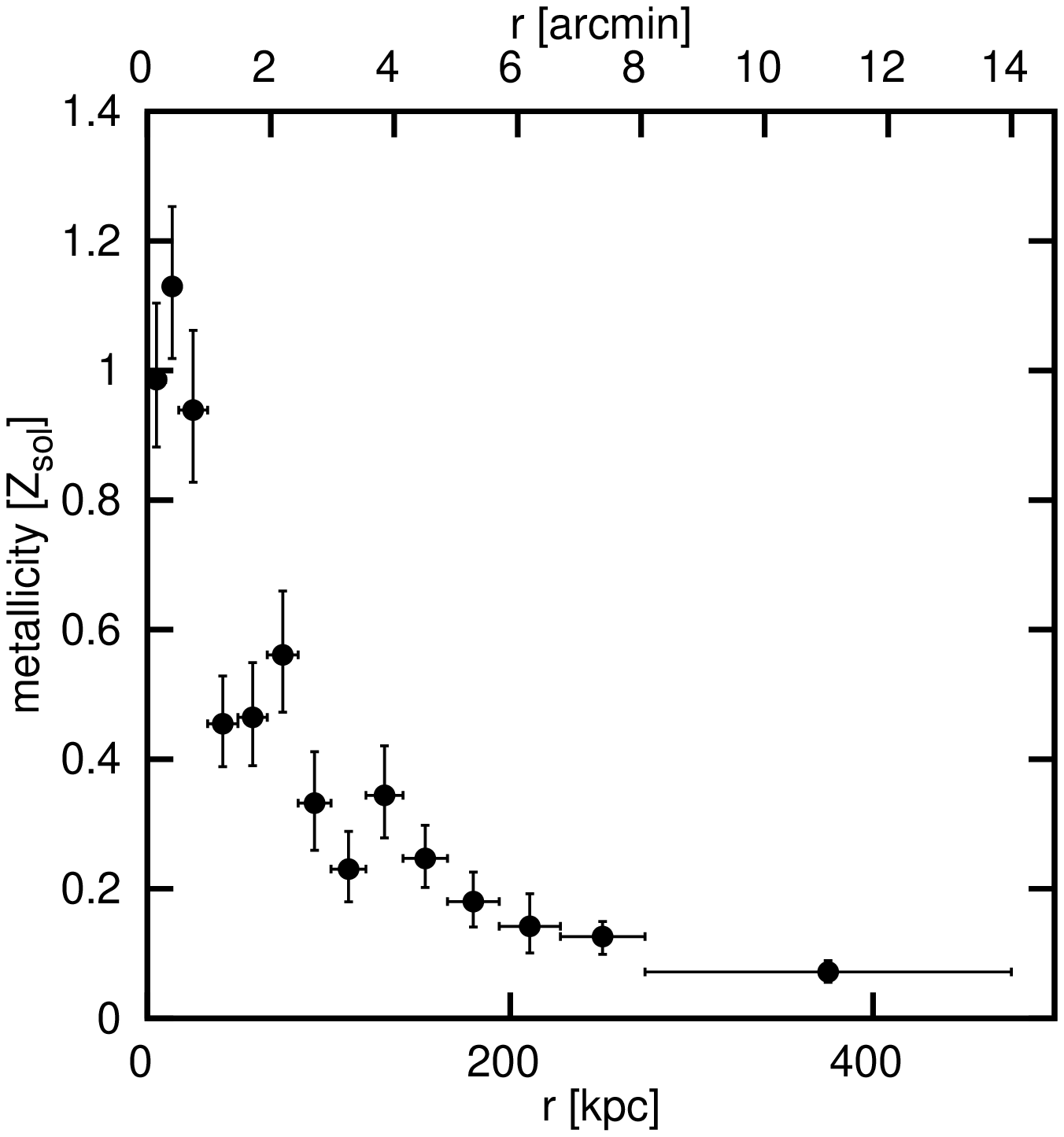}
   \includegraphics[width=0.26\textwidth]{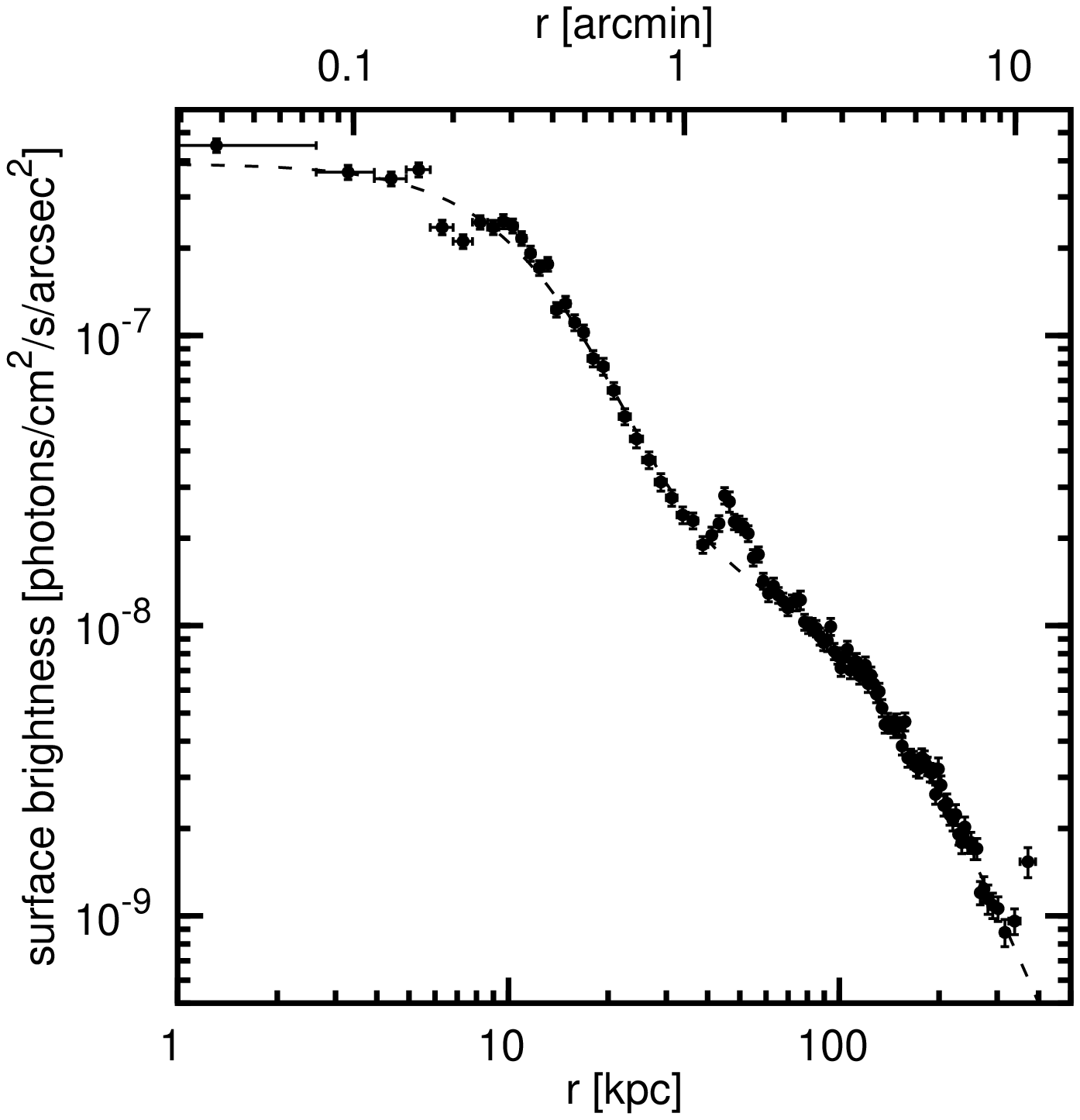}
   \caption{NGC6338}
   \label{fig:tprofngc6338}%
\end{figure*}
\begin{figure*}[h]
   \centering
   \includegraphics[width=0.26\textwidth]{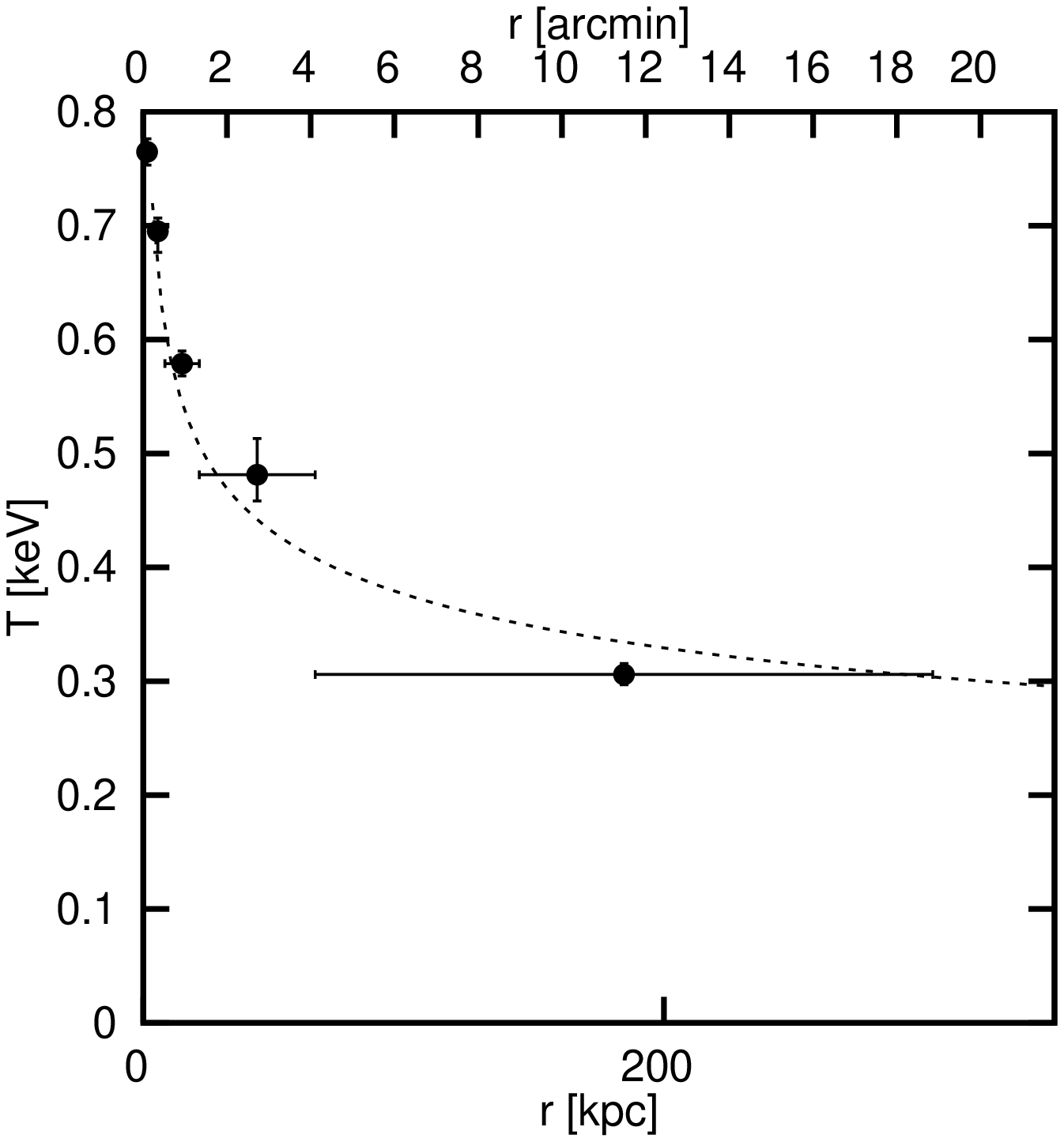}
   \includegraphics[width=0.26\textwidth]{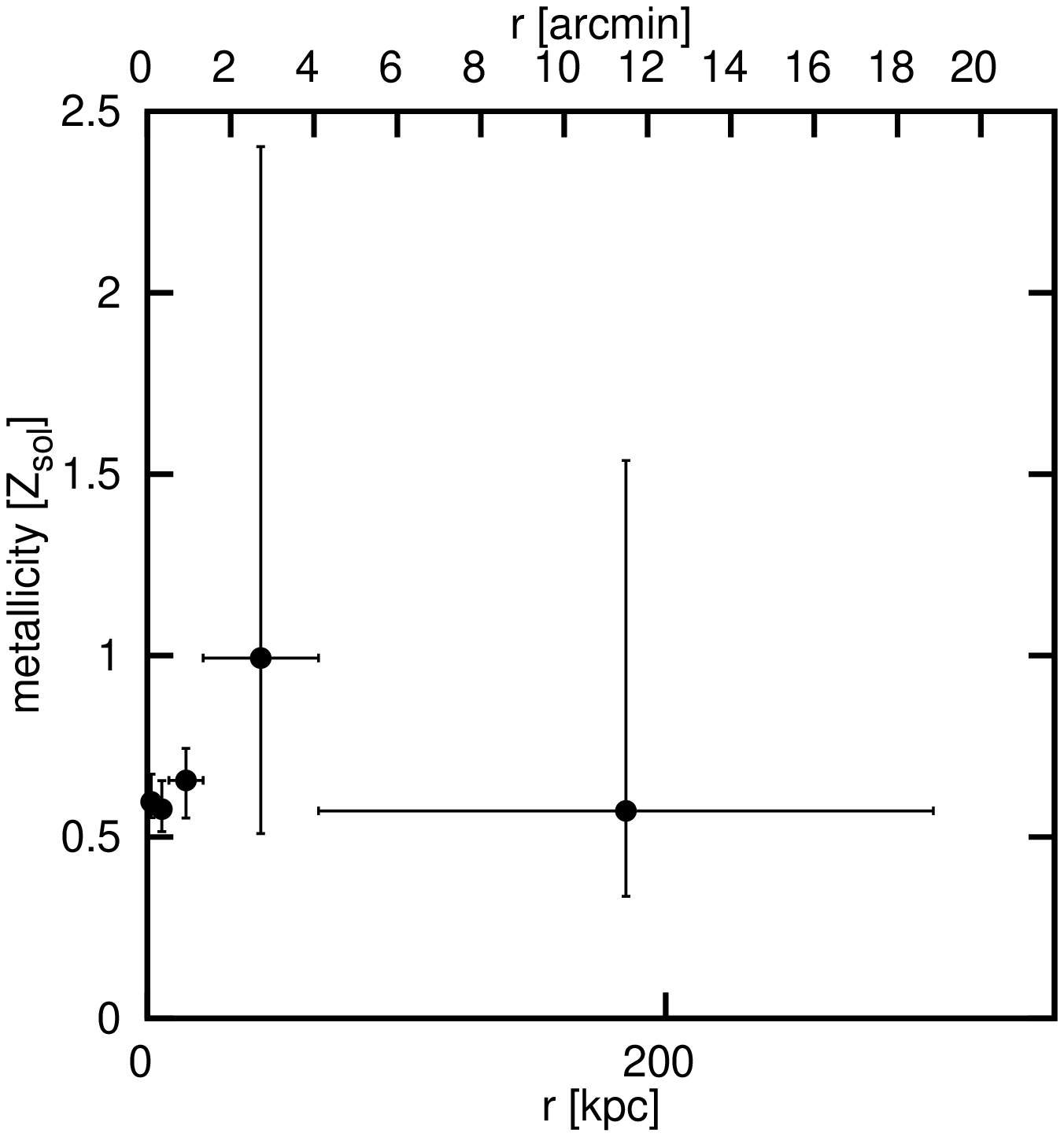}
   \includegraphics[width=0.26\textwidth]{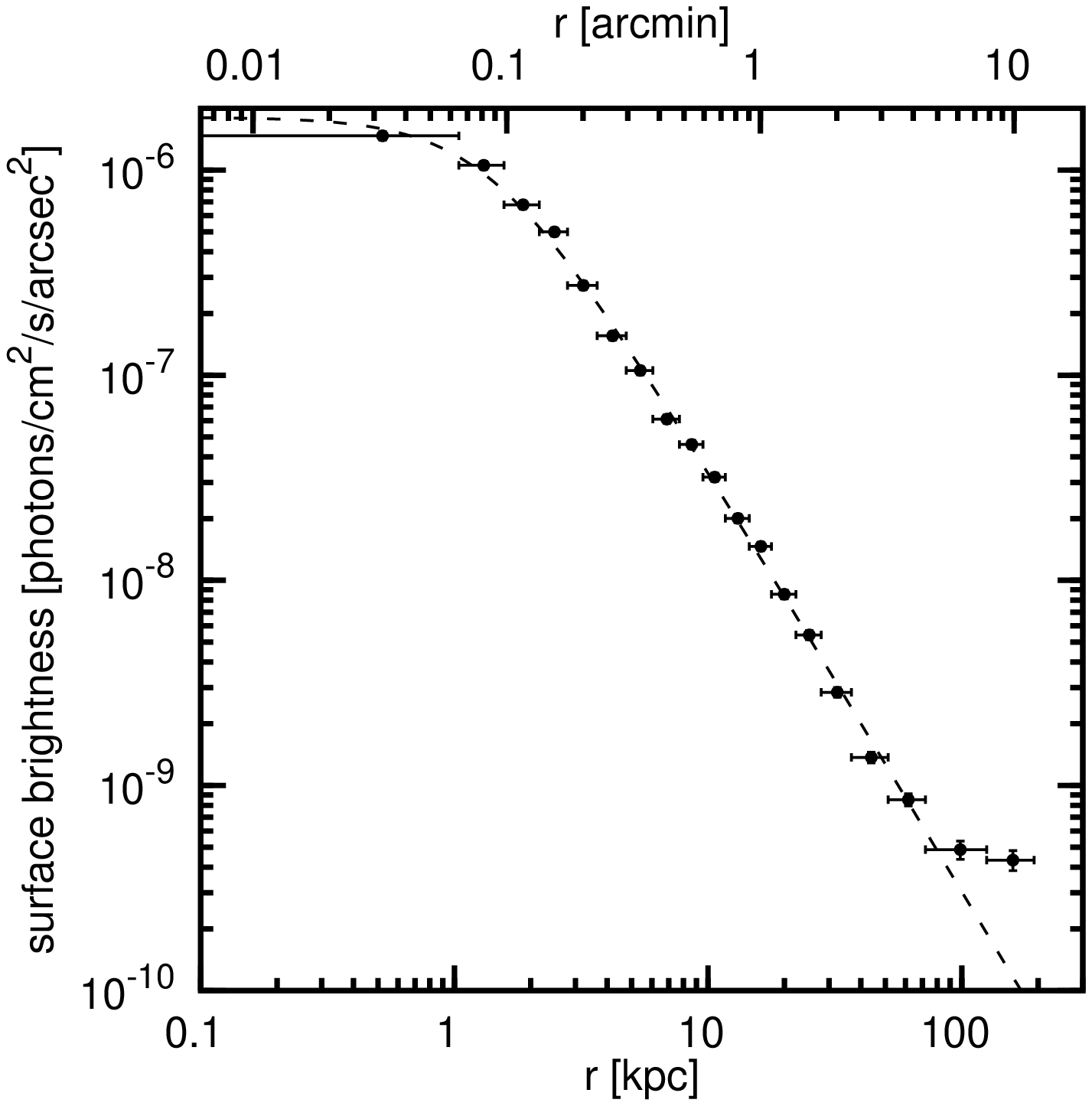}
   \caption{NGC6482}
   \label{fig:tprofngc6482}%
\end{figure*}
\clearpage
\begin{figure*}[h]
   \centering
   \includegraphics[width=0.26\textwidth]{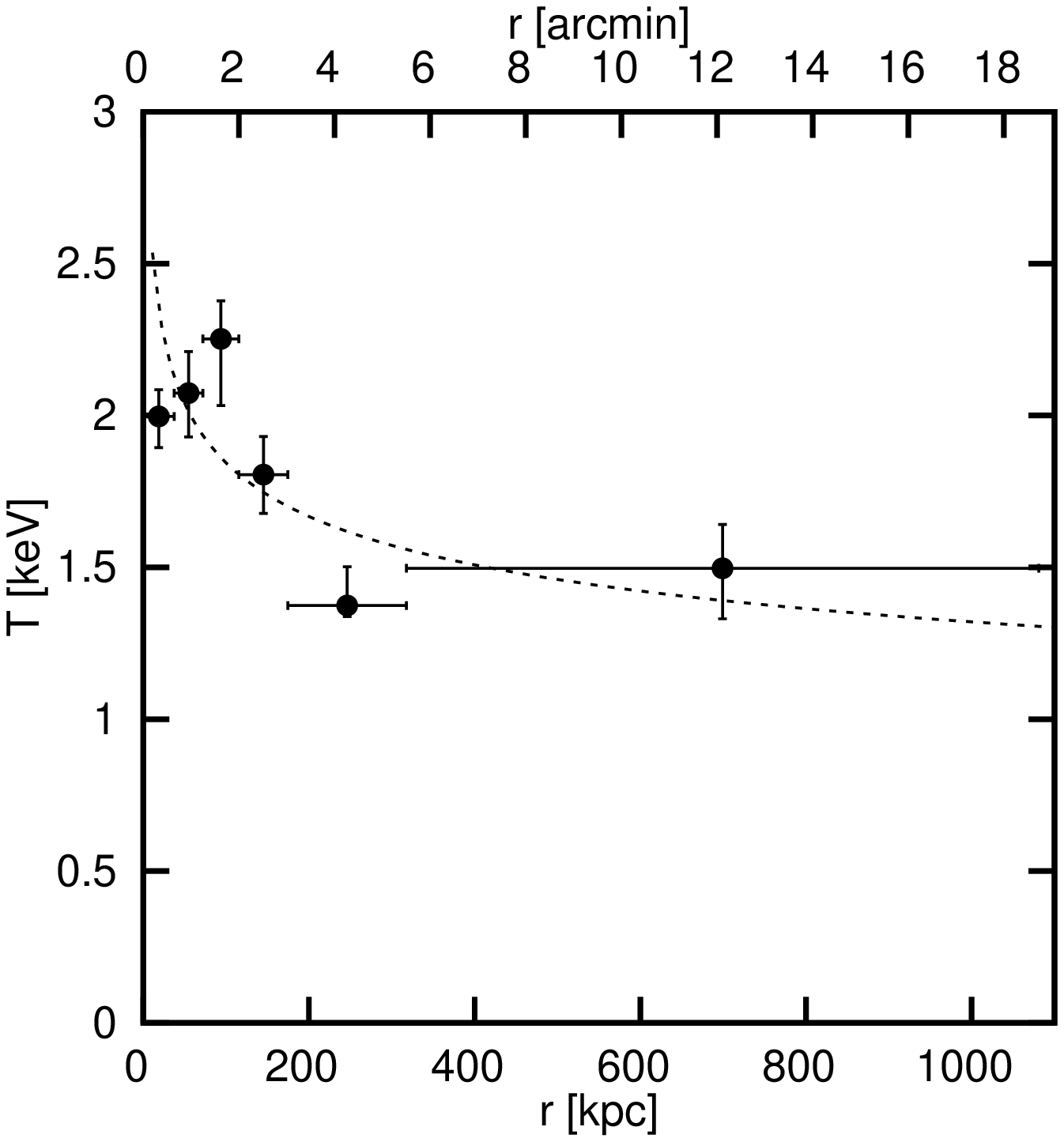}
   \includegraphics[width=0.26\textwidth]{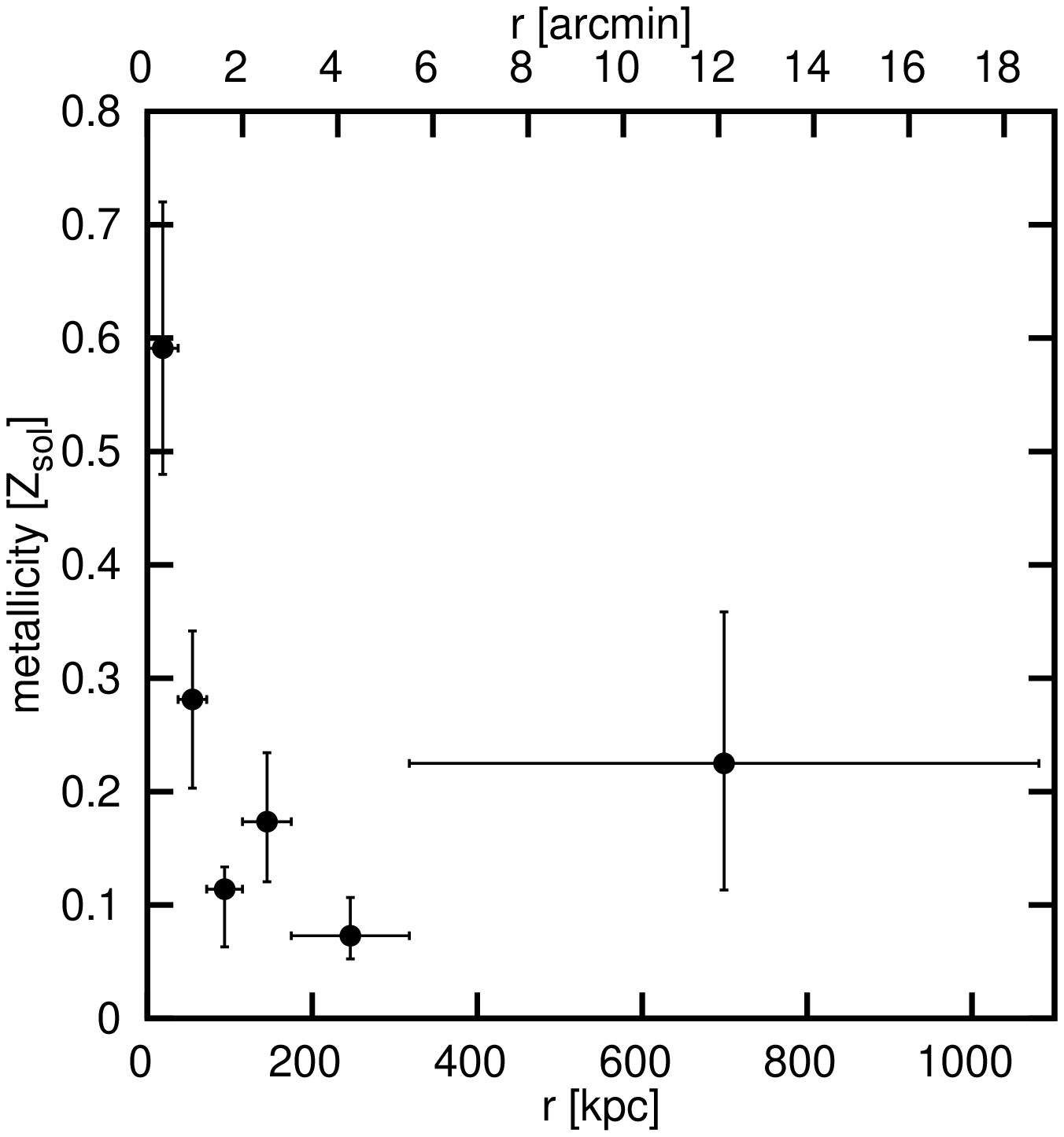}
   \includegraphics[width=0.26\textwidth]{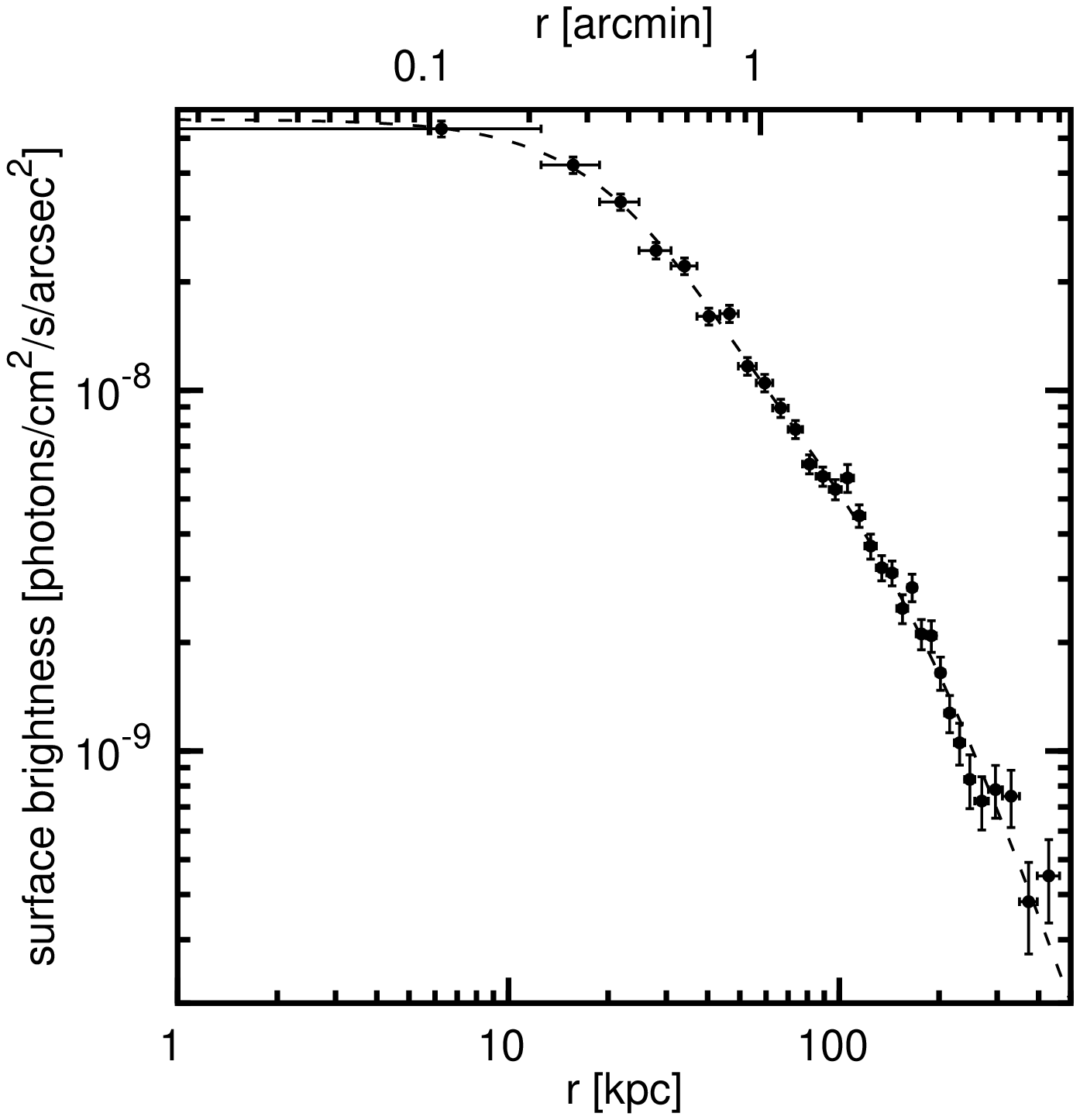}
   \caption{RXCJ1022}
   \label{fig:tprofrxcj1022}%
\end{figure*}
\begin{figure*}[h]
   \centering
   \includegraphics[width=0.26\textwidth]{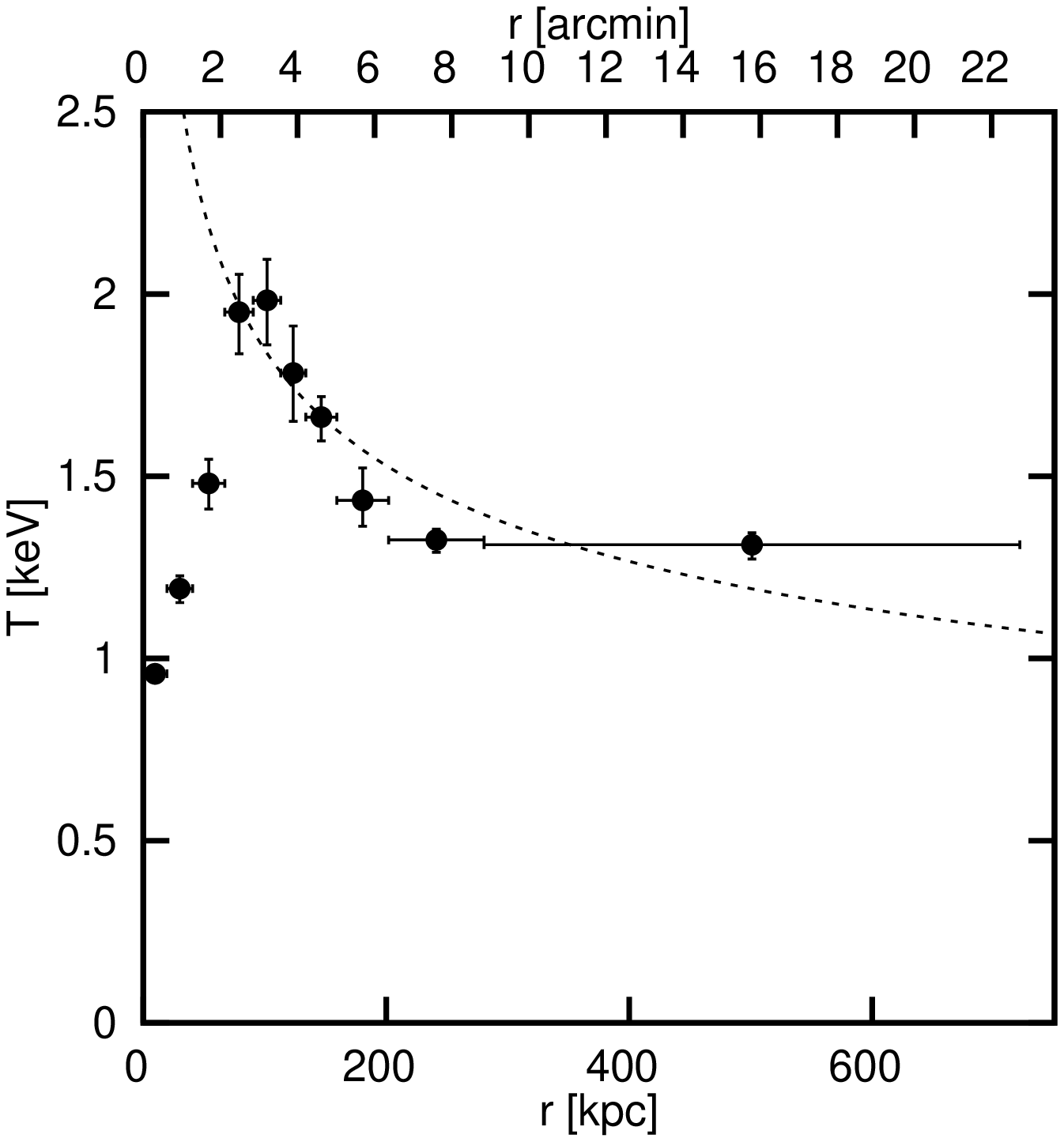}
   \includegraphics[width=0.26\textwidth]{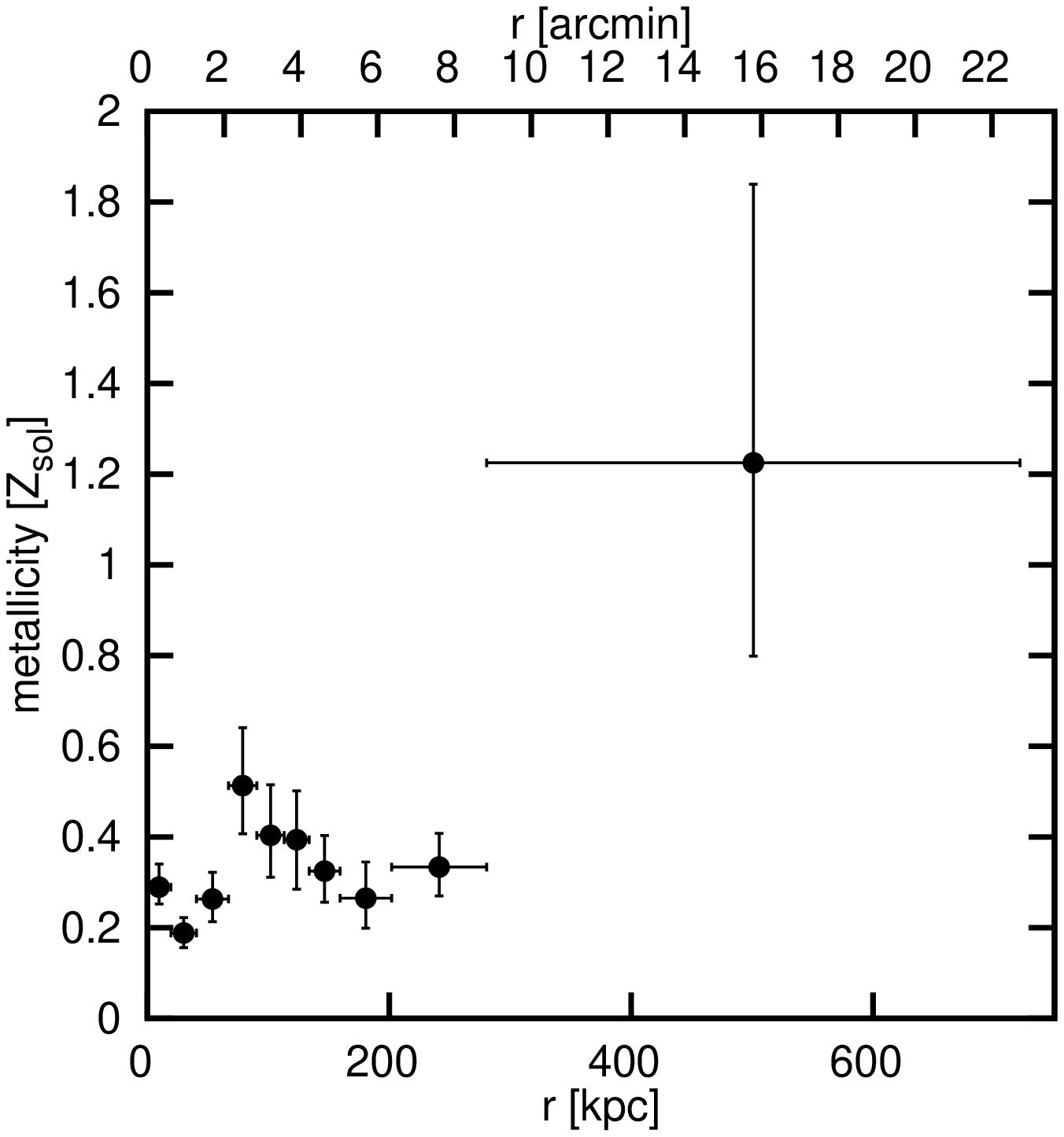}
   \includegraphics[width=0.26\textwidth]{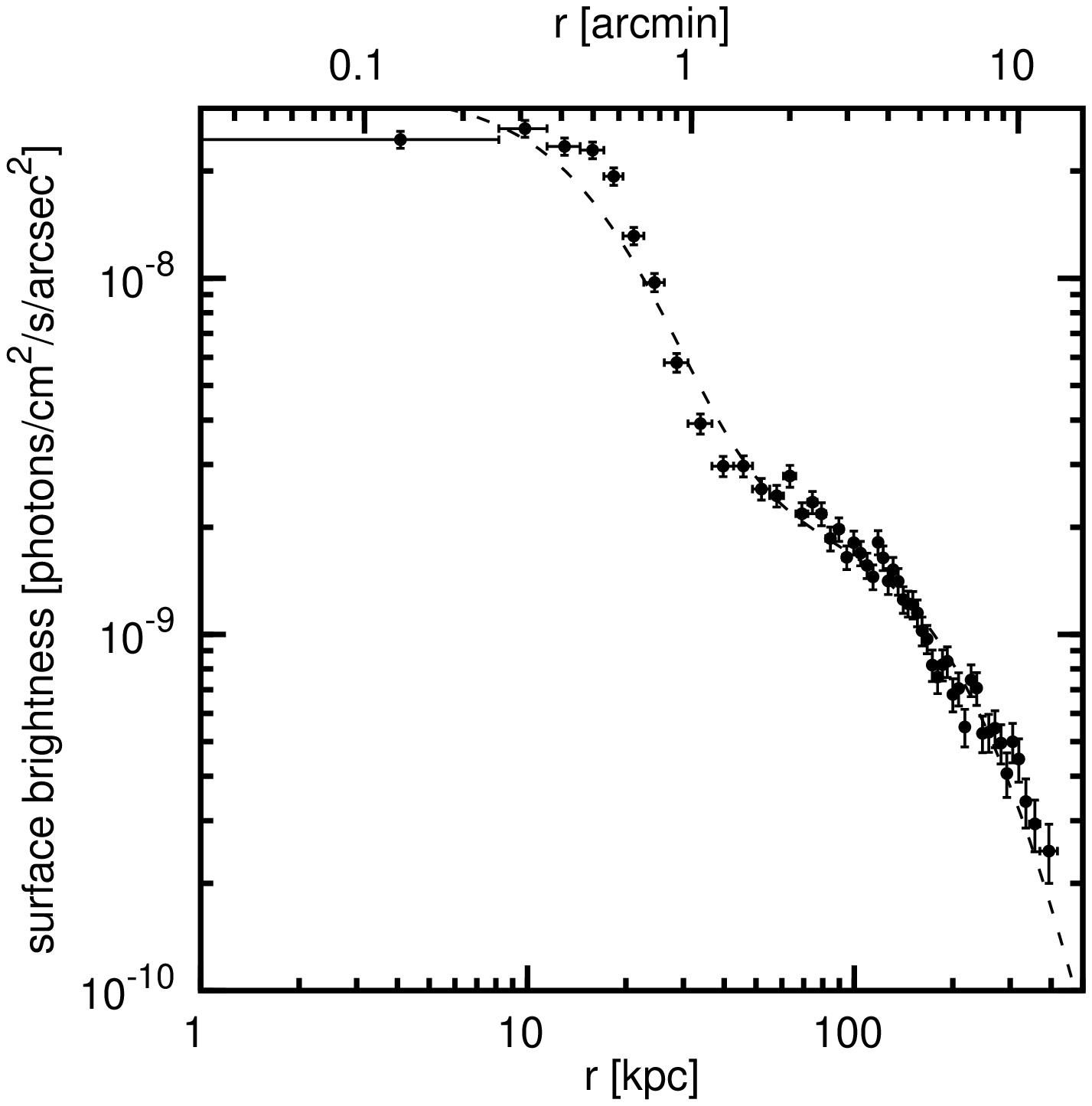}
   \caption{RXCJ2214}
   \label{fig:tprofrxcj2214}%
\end{figure*}
\begin{figure*}[h]
   \centering
   \includegraphics[width=0.26\textwidth]{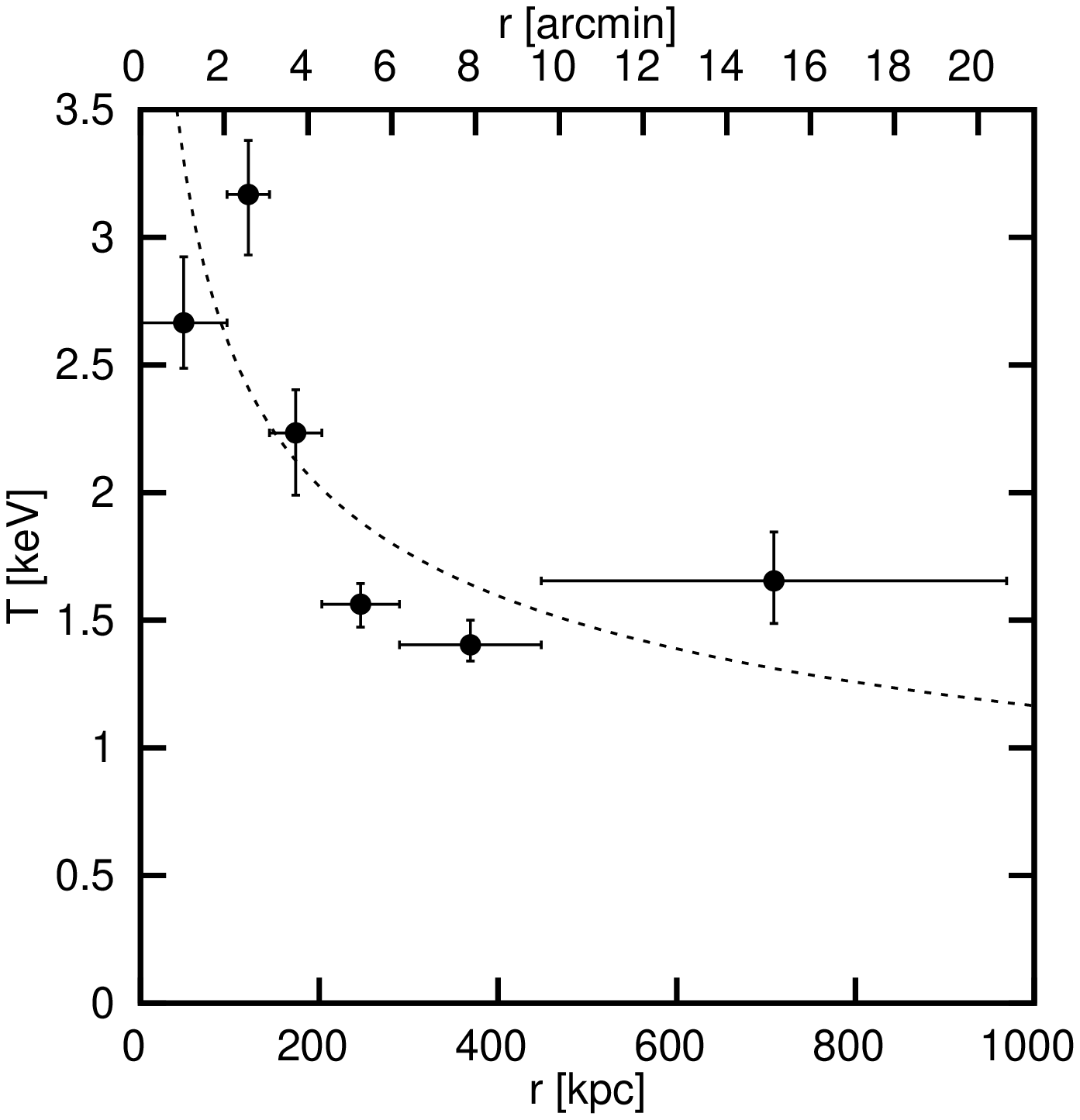}
   \includegraphics[width=0.26\textwidth]{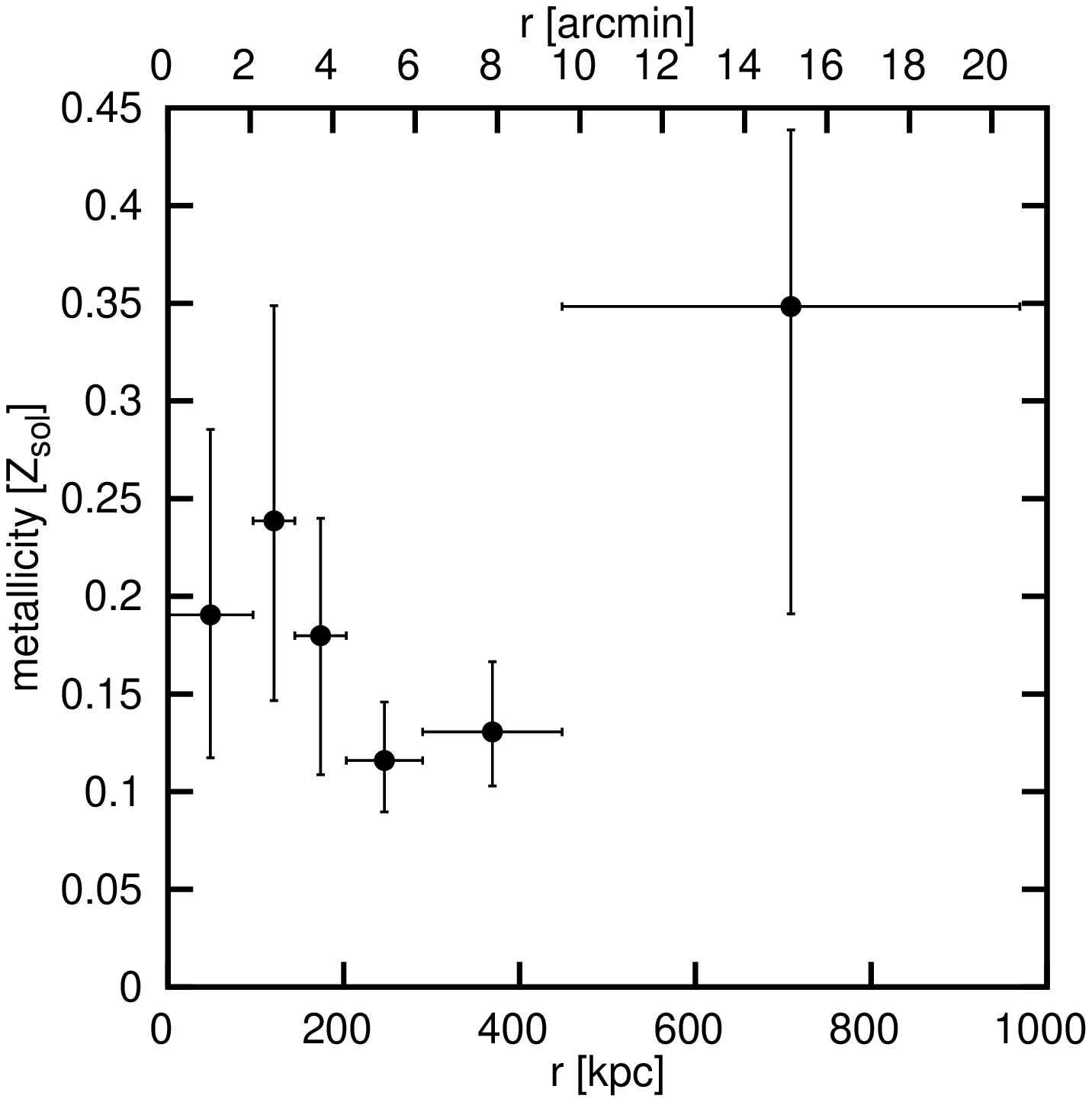}
   \includegraphics[width=0.26\textwidth]{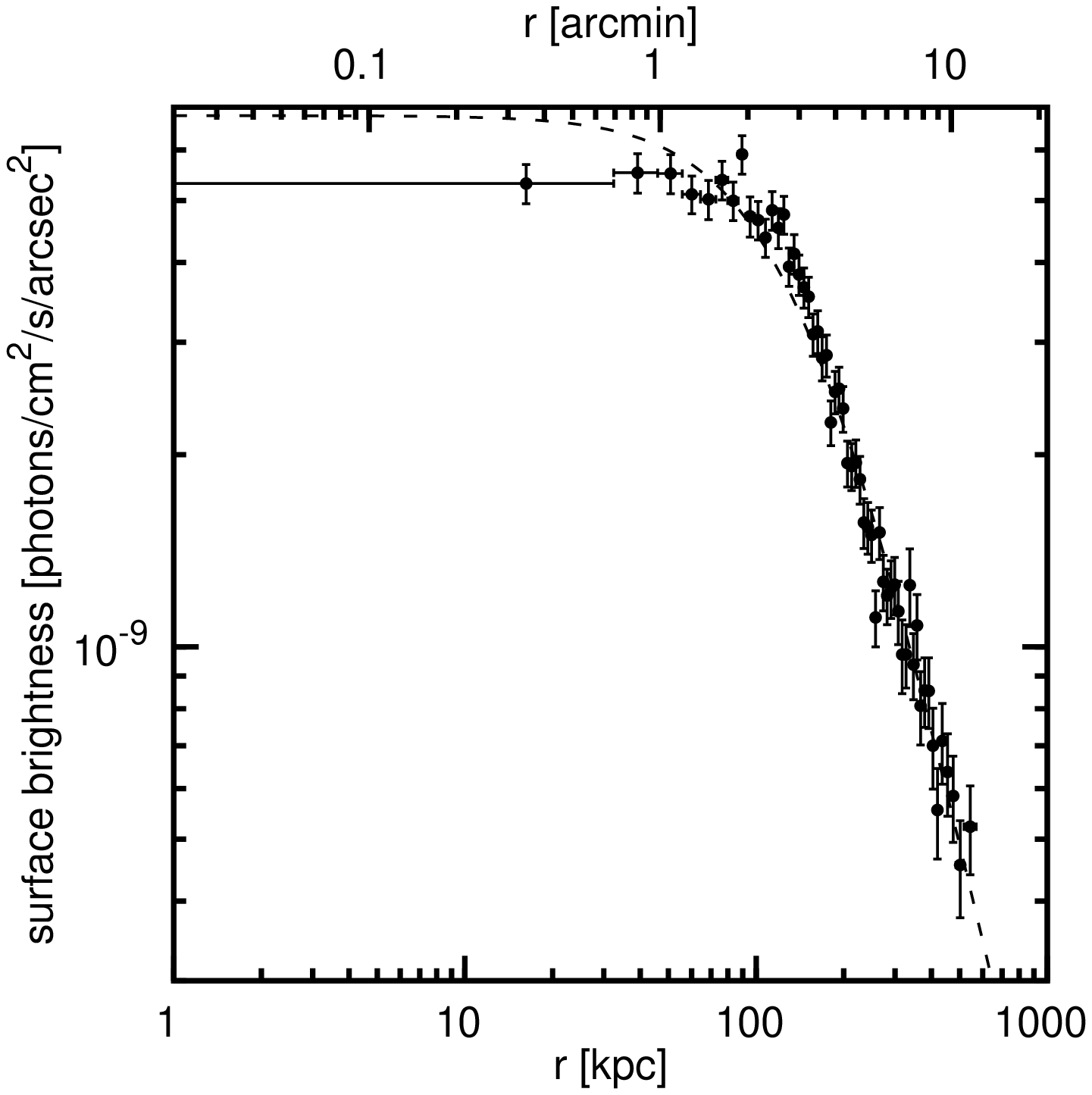}
   \caption{S0463}
   \label{fig:tprofs0463}%
\end{figure*}
\begin{figure*}[h]
   \centering
   \includegraphics[width=0.26\textwidth]{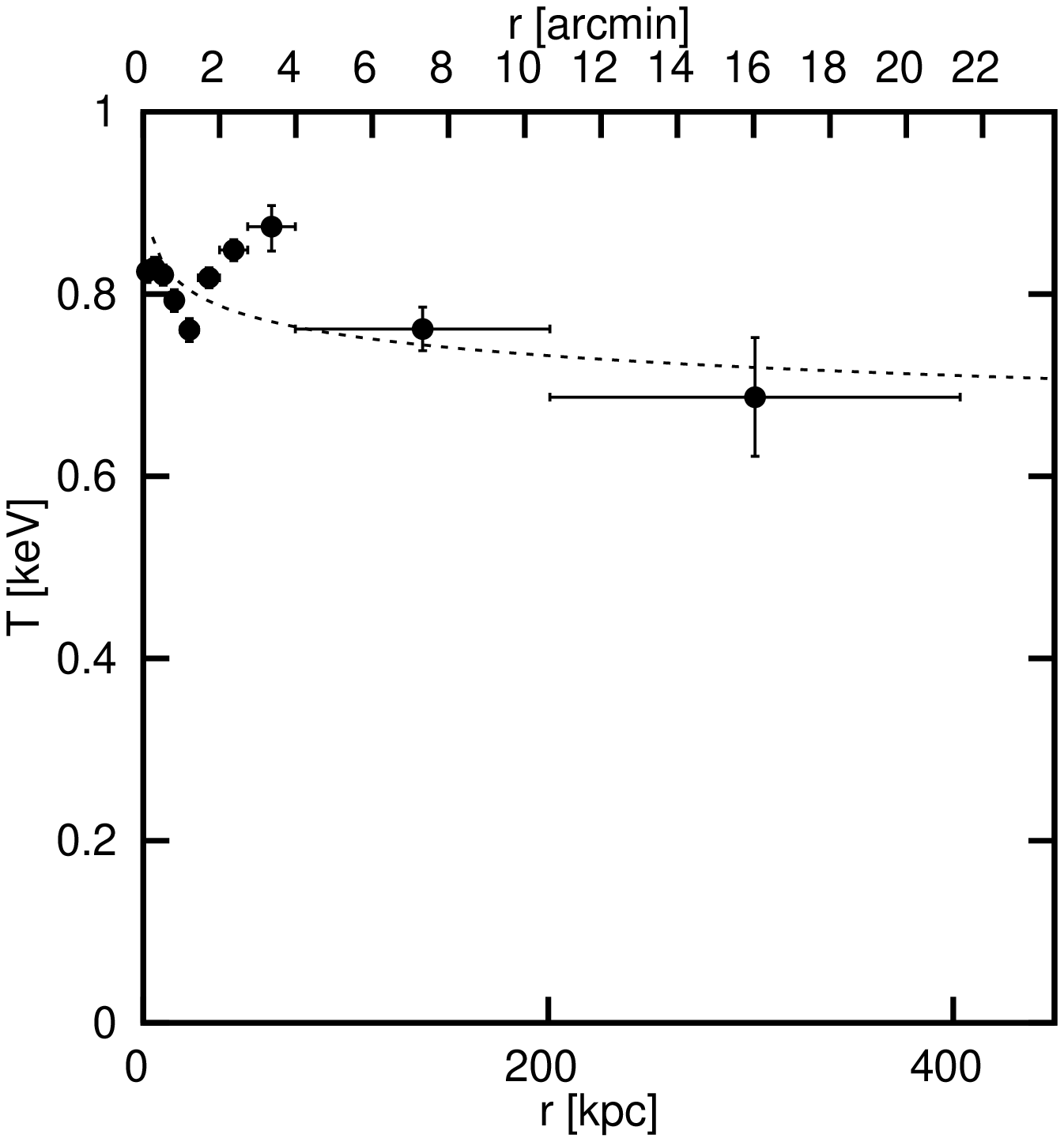}
   \includegraphics[width=0.26\textwidth]{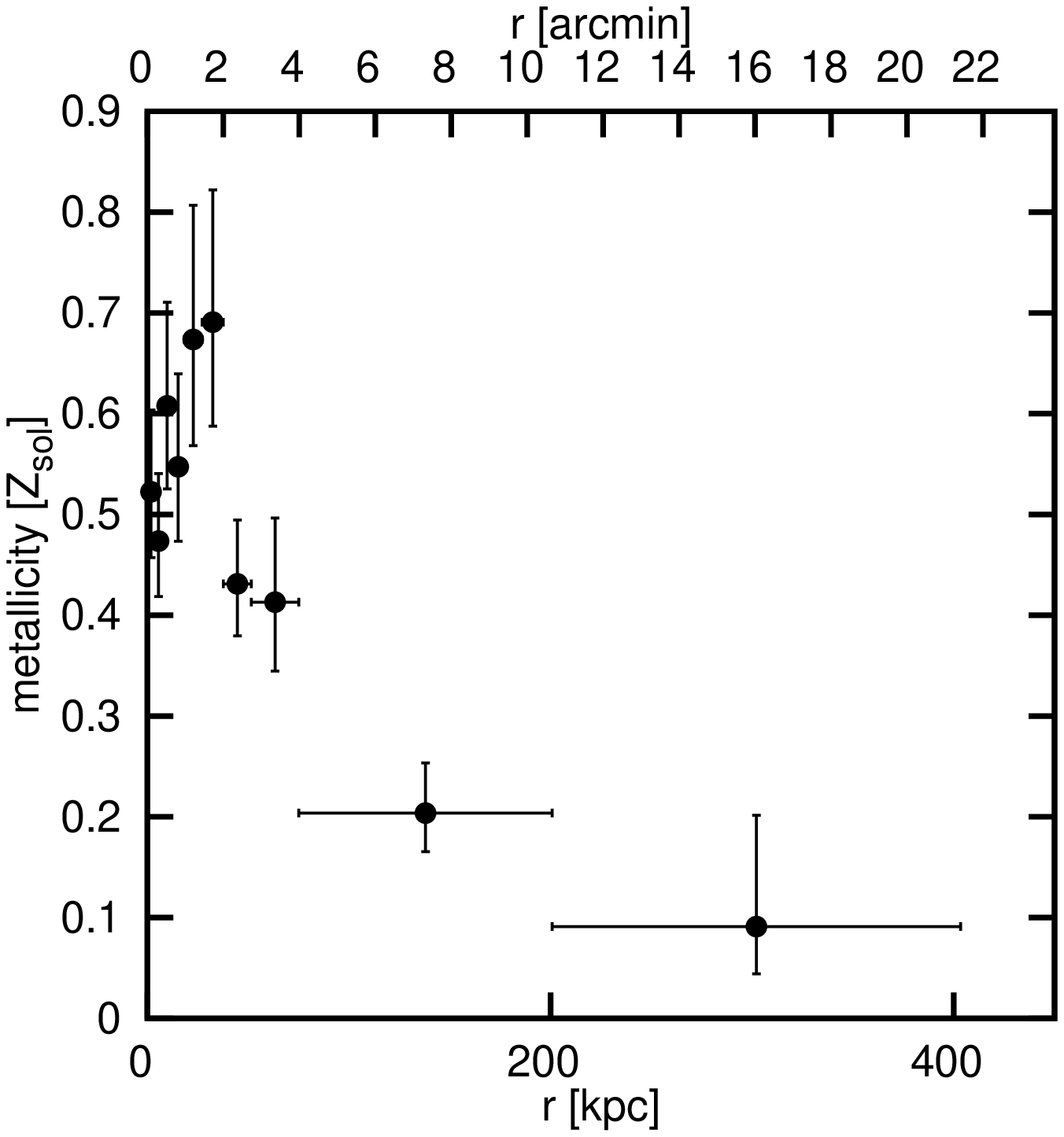}
   \includegraphics[width=0.26\textwidth]{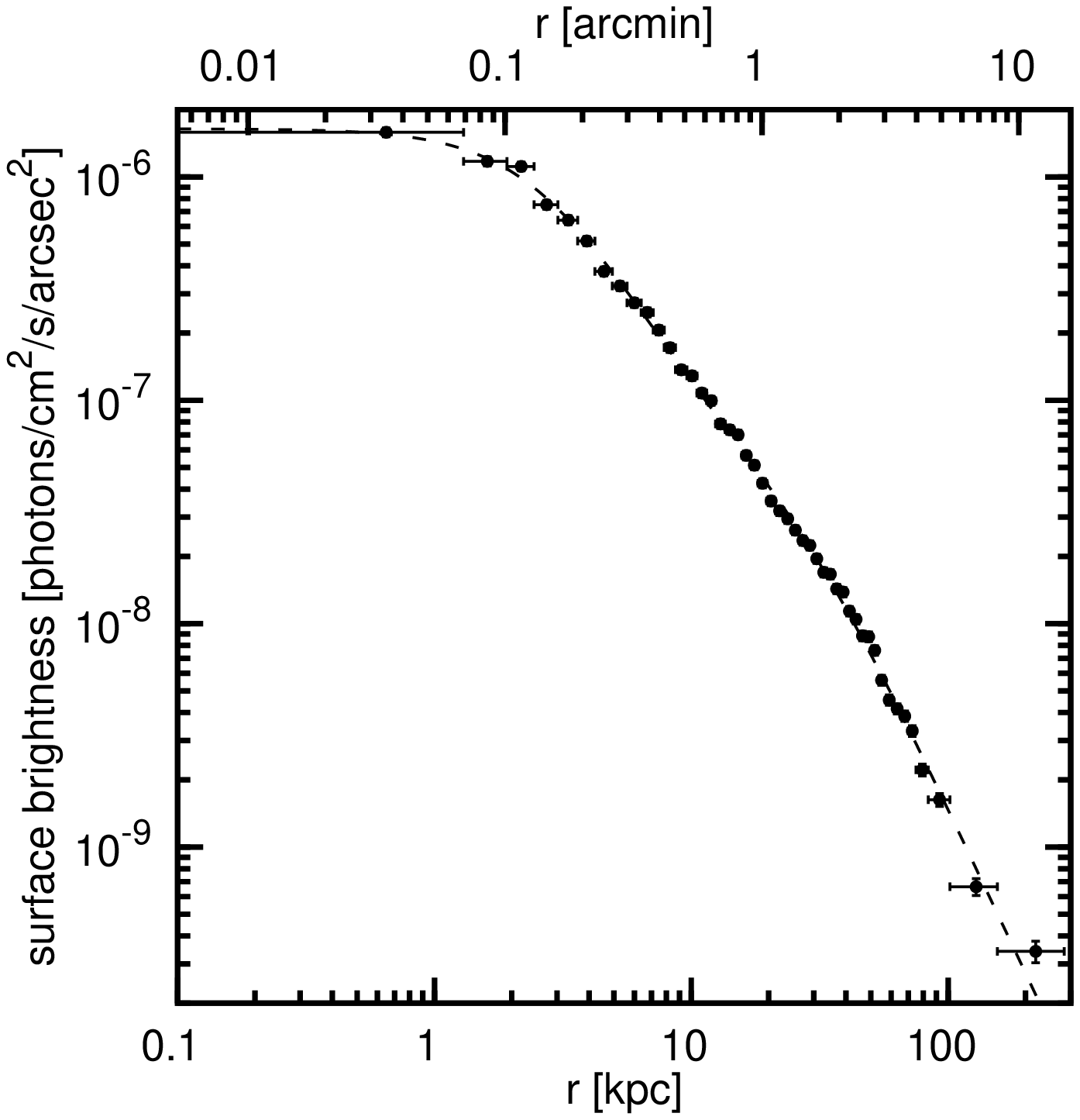}
   \caption{SS2B153}
   \label{fig:tprofss2b}%
\end{figure*}
\end{document}